# Beyond the Library Collections

*Proceedings of the 2022 Erasmus Staff Training Week at ULiège Library*

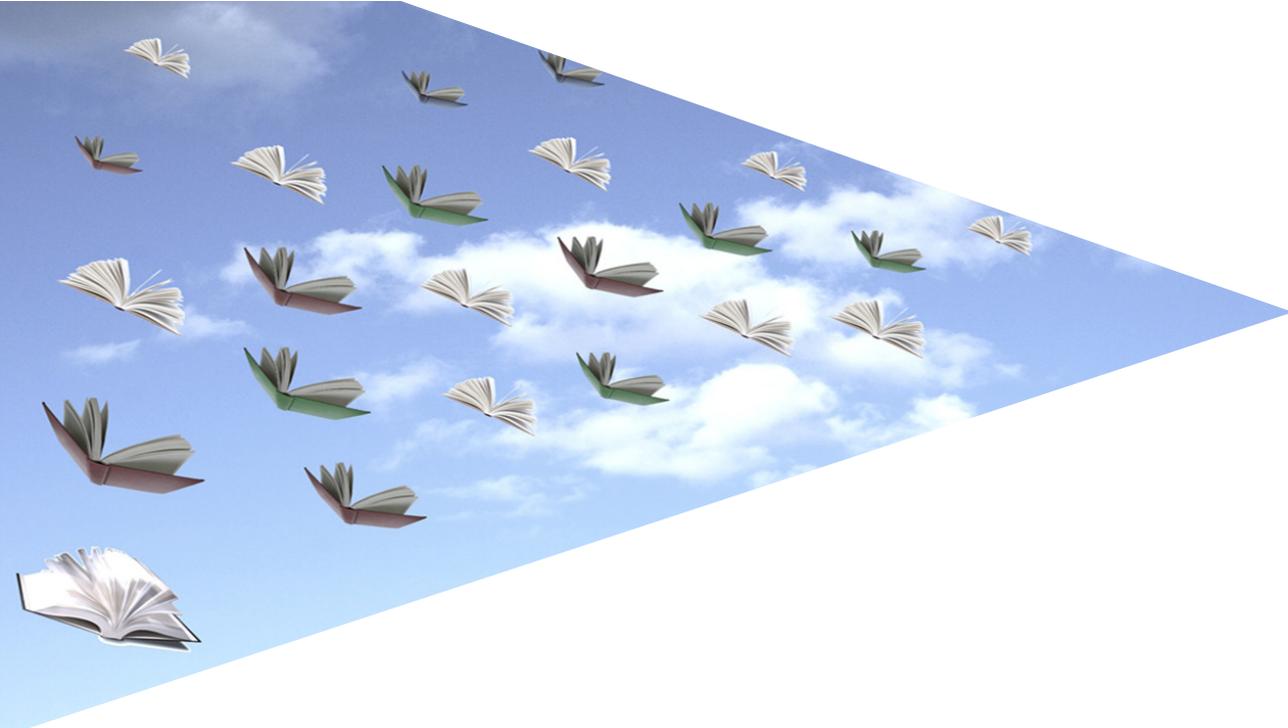 François Renaville & Fabienne Prosmans (Eds.)

**LIÈGE** université
**Library**

Beyond the Library Collections

# Beyond the Library Collections

*Proceedings of the 2022 Erasmus Staff Training Week at ULiège Library*

François Renaville and Fabienne Prosmans (Eds.)

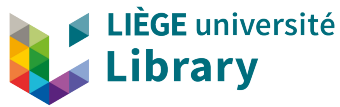

*To all access and delivery librarians –*
*your work is fundamental and valuable*

# Contents

# Preface

## Tina Baich

Resource sharing has long been a core library service, but throughout my relatively short seventeen years in the field, I've witnessed the rapid evolution of resource sharing to being truly essential. A single library collection cannot meet all the information needs of that library's constituents. In recognition of its centrality, resource sharing practitioners have simultaneously become even more user-centered and collaborative in developing and delivering services. Libraries are interdependent and must work together as a distributed network to support everyone's right to access and benefit from information.

In my mind, resource sharing has grown to encompass a number of library activities that make resources available across the international library network – from on demand acquisitions that meet immediate local needs, to shared print initiatives that ensure long-term access across cooperating libraries, to digital collections and repositories that make unique local content available to anyone with an internet connection. The papers in this collection represent this broad view of resource sharing. Several authors discuss the interlibrary loan management, workflows, and infrastructure that are foundational to resource sharing (Kassler & Fors; Forma; Lomba, Marzocchi, & Mazza; Guerra; Prosmans & Renaville), including purchase on demand programs (Van den Avijle & Maggiore; Favre & Velasco). Byström discusses shared print and digitization in Sweden, and several others highlight the importance of providing digital access, notably via ILL, to otherwise difficult to obtain materials through digital collections and institutional repositories (Preiß, Carette). While making digital surrogates available long-term is fundamentally resource sharing, this activity typically requires collaboration with other departments (Opisso).

Because of its user-centered focus, resource sharing is also necessarily adaptable and forward-looking. At the onset of the COVID-19 pandemic, many library services required adaptation, and two of the included papers discuss this necessity within resource sharing operations (Albo, Skalski). It is my belief that our commitment to our users and to equity led to changes that are, in some cases, for the long-term good of those we serve. For instance, Bouillet discusses the elimination of fees as both a practical business decision and a more equitable approach to library service, and Baeyens addresses two of the most pressing issues for resource sharing – e-book lending and controlled digital lending. Finally, Kirchmair describes an instance of adaptability and creativity in providing library resources in unforeseen circumstances.

As I consider this set of papers, my beliefs about resource sharing are reaffirmed. Sharing is universal. The techniques and services we use to share locally can be used globally. And, perhaps most importantly, through collaboration and knowledge sharing, we learn so much from each other. In a global information age, it is not enough to share within your city, state, country, or even region. Collaboration and sharing are absolutely necessary to the fulfillment of the library's mission to provide access to information. The authors of these papers represent a global community of resource sharing practitioners seeking to serve a global community of learners, researchers, and teachers. I'm privileged and proud to count myself as a member of their community.

## About the Author

Tina Baich, 2021-2023 Chair of IFLA Document Delivery & Resource Sharing Section Standing Committee
Indiana University–Purdue University Indianapolis
https://orcid.org/0000-0002-8046-2461

Tina Baich is the Senior Associate Dean for Scholarly Communication & Content Strategies at IUPUI University Library, Indianapolis, Indiana, USA. Before becoming a library administrator, she led the library's resource sharing efforts for 12 years. Tina is currently SPARC's Visiting Program Officer for the U.S. Repository Network and Chair of IFLA's Document Delivery & Resource Sharing Section Standing Committee.

# Introduction

**François Renaville and Fabienne Prosmans**

Erasmus+ is the EU programme for education, training, youth, and sport. It is a well-known fact that many of the initiatives are based on opportunities for students and recent graduates, but funding and experience opportunities also exist to help lecturers and non-academic staff develop their skills through job shadowing, teaching and training activities, thereby becoming better equipped with the knowledge, skills and competences needed in a dynamically changing society that is increasingly digital, mobile, and multicultural.

Taking part in an Erasmus week is a great way to pick up new ideas and practices from a different work environment and provides the opportunity to establish a relationship and to envisage collaborations with colleagues from other institutions in Europe.

The University of Liège Library has been organizing regular Erasmus staff training weeks since 2011. As of today, eight editions have taken place, most of them focusing on Open Access for academic libraries. So far, about 200 librarians and researchers in library sciences have had the opportunity to meet in Liège as part of an Erasmus week.

Starting with the 2019 edition, the Library wanted to encourage much more active participation from the attendees. Participants were therefore strongly invited to present a session related to the theme of the Week. The underlying idea was to encourage discussion and contact between participants by getting everyone to contribute concretely to the Week by sharing their own experiences, giving feedback, or communicating about ongoing or completed projects. The 2022 edition of the Erasmus Staff Training Week at ULiège Library continued this approach.

No library can buy or hold everything its users need. At a certain point, librarians need to pool their resources and collaborate to provide access to what they don't have. So the 2022 edition focused on services, projects, and policies that libraries can deploy and promote to increase and ease access to materials that do not belong to their print or electronic holdings. The theme was "Beyond the Library Collections. We don't have it? Here it is!". Some of the suggested topics for proposals were notably: centralized and shared collection storage with other libraries, long term loan projects between libraries, document delivery during/post COVID-19, challenges and impacts of new interlibrary loan policies and workflows, ILL and its impact on acquisitions policy, interfacing solutions for better document delivery, legal aspects (CDL, DRM, licences, etc.), and new partnerships and collaborations between libraries for better delivery and ILL services.

About 20 librarians, managers and researchers in information and library sciences took part to the 2022 edition. Their experiences and visions are reflected in this open access book. Libraries and the Erasmus programme have in common that collaboration, partnership, and solidarity are deeply embedded in their DNA, and the theme of the Week fully demonstrated this.

Part 1 of this book reports on various experiences related to collaboration between institutions and services in order to share collections and to guarantee preservation and access to heritage materials. Part 2 illustrates how DDA and EBA acquisition models can be used to fill gaps in collections and supply users with the expected resources. Parts 3 and 4 pay particular attention to interlibrary loan and resource sharing, sometimes focusing on experiences and workflows, sometimes highlighting successful examples of collaboration and networking. Finally, part 5 contains testimonies from libraries showing how document delivery and services have been affected by the COVID-19 pandemic.

It is our hope that you will find some inspiration in this book to start new collaboration or innovation projects in favour of access to information and research publications.

*April 2023*

## About the Authors

### François Renaville, Head of Library Systems
Université de Liège (University of Liège)
https://orcid.org/0000-0003-1453-1040

François Renaville (1976) studied linguistics, literature, and translation. After a two-year experience as a teacher in Finland, he joined the University of Liège Library as a subject librarian. He has been working on library systems since 2005. Since 2022, he has been member of the IGeLU (International Group of Ex Libris Users) Steering Committee and has been coordinating the "Systems & Data" unit at ULiège Library. He is interested in discoverability, integrations, delivery and user services. On a private level, he is a great coffee, chocolate and penguin lover.

### Fabienne Prosmans, Fulfilment and ILL Manager
Université de Liège (University of Liège)
https://orcid.org/0000-0002-1408-5207

Fabienne Prosmans (1972) studied mathematics at the University of Liège and obtained a PhD at the University of Paris Nord. Since 2005, she has been working at ULiège Library as a subject librarian in the fields of mathematics and applied sciences. She has been the fulfilment and ILL manager at ULiège Library since 2015 and is now coordinating the "User Services" unit.

PART I

# COLLABORATION FOR SHARED COLLECTIONS AND PRESERVATION

# Collection management collaborations in Sweden

**Karin Byström**

## Abstract

This paper will describe the background and progress on two collection management projects—the digitization of Swedish print materials and a collaboration on shared print. The author has many years' experience of working on both projects, and the paper gives a detailed picture of the ambitions, work and results. It concludes with some potential long-term effects of the projects that go beyond the initial scope, like the need for a shift from a local to a national and global perspective and for shared infrastructure.

## Keywords

Collection management; Library collaboration; Digitization; Shared print; Academic libraries; Sweden

## Article

## Introduction

This paper will present two ongoing collection management projects in Sweden—the digitization of Swedish print material and a national collaboration on shared print.

They might seem different at first, but they have a common goal—to provide easier access to material for users—today and in the future. Libraries work together to achieve the goal by increasing digital access and securing print copies in libraries.

## Digitization of Swedish print material

A digitization collaboration started between the National Library and the five largest university libraries (Lund, Uppsala, Stockholm, Göteborg and Umeå) in 2020. The goal was to make all Swedish print content easily accessible and freely available for research and to the public.

The libraries signed a "letter of intent" with the ambition to digitize and make all Swedish print available (Grönvall et al., 2020). It was a bold statement and the purpose was to draw attention to the fact that so little of this material is available in digital format.

The letter of intent stated that libraries would work on increasing the speed of digitization, and would develop better systems for access and tools to use the material. They would work for joint standards and strategies to work together in the same direction. They also committed to flagging the need for additional funding since it is not possible for libraries to achieve this within their current budgets.

The goal was a long-term national solution for funding and shared infrastructure for access and use, where libraries work together instead of from a local perspective. The collaboration based on the letter of intent was called DST—Digitization of Swedish Print (*Samverkansgrupper: Digitalisering, 2021*).

## *Start of the collaboration*

In February 2021 Karin Byström became project manager and set up five working groups. The working groups focused on:
- scanning and technical standards,
- access, metadata and copyright,
- collections and library collaboration,
- external collaboration and communications,
- financing.

In total, over 25 people from the six libraries were part of different working groups. The groups use a Confluence Wiki for collaboration. The head librarians of all university libraries and the deputy National librarian make up the steering committee. The project manager makes the yearly plan, reports on progress and raises strategic discussions with the steering committee.

During the first year, focus was to build a foundation for collaboration. One important task was to define "Swedish print" as material printed in Sweden (but also from areas that at some point in history used to be Sweden, i.e. Finland). The libraries also decided on a technical standard for scanning, based on international standards like FADGI.

All libraries also agreed on a joint selection priority. If there was room for some extra, internally funded digitization libraries should prioritize:
- journals 1850–1900,
- dissertations from their university,
- the oldest material from 1400–1500.

To increase the findability of digitized material in Libris (the Swedish national catalogue), the libraries worked together with the National library to set up a specific search for Swedish digitized print material. This makes the material easier to find on a metadata level.[1]

During the first year there was also a lot of external communication about the collaboration—a public webpage (*Informationssidor för digitaliseringen, 2021*), a newspaper article (Burman et al., 2022b), a radio interview and a webinar. There was also a seminar at the Gothenburg Book fair in September 2022, where representatives from the DST collaboration discussed researchers' needs for digitized material with Nina Tahmasebi, Associate Professor in Natural Language Processing at Gothenburg University (Burman et al., 2022a).

## No shared infrastructure

All of the libraries in the DST collaboration have their own digitization studios, established workflows and competent staff. Libraries work with digitizing collections as a way of making them more visible and usable. Digitization also reduces the need to handle and read the original material, so it is also a way to protect fragile material.

The libraries usually collaborate with researchers in funding specific digitization projects. Researchers can include funding for digitization in their research grants if they need digitized material. However, the libraries also fund digitization of selected collections, like dissertations from the university or special collections.

However, there is no shared infrastructure or user interface for digitized material, instead there are many different publishing platforms used by the libraries. Different platforms are used for different types of materials, so users need to search in many locations:

- Alvin—digitized cultural heritage, journals and monographs (Uppsala, Lund, Göteborg)[2]
- Diva—scholarly material (Uppsala, Lund, Umeå, Stockholm)[3]
- Gupea—scholarly material (Göteborg)[4]
- Digital collections platform (Nainuwa)—digitized cultural heritage (Umeå)[5]
- Open Journals (OJS)—digitized journals (Lund)[6]

---

1. https://libris.kb.se/deldatabas.jsp
2. https://www.alvin-portal.org
3. http://www.diva-portal.org
4. https://gupea.ub.gu.se
5. https://digital.ub.umu.se
6. https://journals.lub.lu.se

This means that it is not possible to search all digitized material in one platform.

## Not much digitized material

There have been some larger digitization projects, like the digitization of Swedish novels by Litteraturbanken,[7] Swedish newspapers by the National library[8] and dissertations done by many of the university libraries. However, only about 5–10% of the copyright-free, Swedish print is digitized, and a rough calculation showed that it might take up to 300 years to complete digitization if it continues at the current pace.

Many other countries have digitized much more or have even completed the process. One important explanation is that there is no government funding for digitization on a large scale. This lack of digitized Swedish material means that it is difficult to conduct research on Swedish conditions ([Burman et al., 2022b](#)). Digitization is an important part of making the print heritage visible and encouraging researchers and the public to use old sources. In this way, it functions as a support for a democratic society.

## Improving visibility and use of digitized material

One of the goals of the collaboration was to improve visibility and usability of digitized materials. It is important for libraries to be able to meet the new user needs coming from the research world.

The working group for Access, metadata and copyright conducted a small workshop to identify these user behaviors (Fig. 1) and user needs (Fig. 2). They visualized the broad spectrum of needs on the scale from doing a manual analysis to an automatic analysis, and working from a specific interest to doing an open exploration.

The result showed a very diverse and broad set of users and user behaviors—and it became clear that there is not *one* way of using digitized material. A researcher in digital methods needs large data sets that are easy to download, and a genealogist wants to search for a specific name or place in materials from many different sources, while a private person is looking for a famous quote or an old recipe.

7. [https://litteraturbanken.se](https://litteraturbanken.se)
8. [https://tidningar.kb.se](https://tidningar.kb.se)

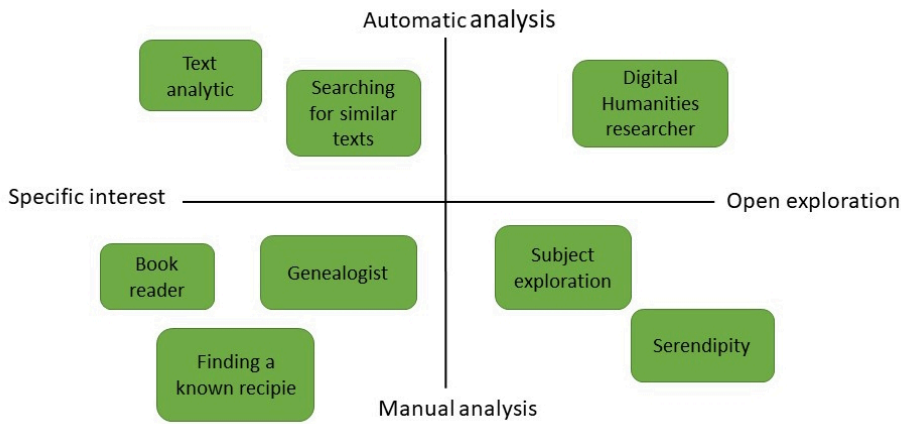

*Figure 1: User behaviors in using digitized material*

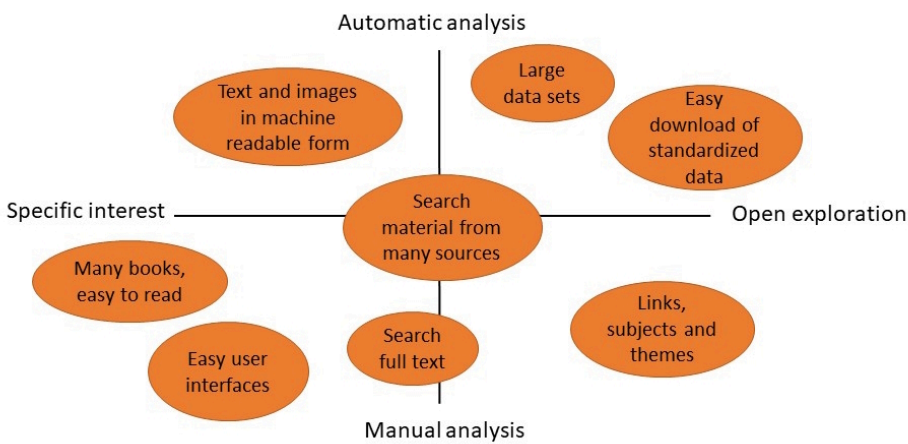

*Figure 2: User needs in using digitized materials*

While continuing to meet the traditional user needs, libraries must also support the new research methods that uses text as data, like digital analysis and text mining, which use machine-learning algorithms to analyze huge amounts of text. For this type of research, the original text itself almost loses its interest, and the focus is on language development and how trends and changes in society are reflected in texts. To meet these research needs, data must be open and connected with other digital collections and datasets. This requires standards for metadata and data files, use of open linked data and APIs for automatic transfer of data between systems and to users.

The Access working group listed the most high-priority user needs:
- to have a single user interface for digitized material from all libraries;
- to be able to do a full text search of digitized material;
- to have options for automatic download and analysis.

After the workshop and analysis, it became clear that the participating libraries do not meet these needs today, despite all their good work and well-developed platforms. The problem is not only the lack of specific functionality, like APIs and full text search, but also the fact that there are so many platforms, wich means that users can't get an overview of existing digitized material, let alone combine and use material from different platforms effectively.

The steering committee decided to investigate ways to gather all digital objects so that they are accessible regardless of the owing library or institution. The Access working group looked at different solutions, and found a possible solution in a new data platform developed by the National library, *data.kb.se*. During the spring of 2022, the group and project manager conducted an investigation regarding the possibility of all libraries using that platform. The focus was on the technical, economical and legal consequences for both the National library and the university libraries, as well as future collaboration and workflows.

The recommendation was to set up *data.kb.se* as a shared infrastructure based on the "minimal viable product" concept, where the first product is usable, but does not have all the desired functionality. This is a way to get the work started and to be able to develop and expand over time. The solution would build on the existing infrastructure for digital preservation and access used by the National Library, but be expanded to host digitized Swedish material from the other libraries too ([Byström, 2022](#)).

## Continued work

The steering committee reacted positively to the suggestions in the report, but there are many remaining challenges in setting up a new shared infrastructure. The questions of funding and collaboration for short- and long-term system development and maintenance must be further investigated, especially by the National library, who would be host of a new system. This will continue during 2023.

The working group for Scanning and technical standards are investigating options for outsourcing the scanning process and increasing the level of automation in metadata creation and file delivery. One example is to create automated bibliographic records for the digital version from the record of the print material. These changes in the workflow would mean that the libraries would spend less time on cataloguing and be able to scale up production.

In parallel to these activities, the working group for Collections and library collaboration is planning extended collaboration for sharing knowledge and experiences between the libraries. The DST collaboration will also establish contact with the research community for expert reference groups.

In a report from 2022, the Swedish Library Association highlights the need for digitization at scale and points to government funding, a clear commission to the National library and stronger library collaboration as important parts of the solution (Sjöström, 2022). The DST collaboration attempts to raise awareness of the digitization situation and to increase the dialog with university leadership, research funders and politicians. It is clear that libraries need additional funding for the work—both for the digitization itself and for system development. Even if the collaboration has not received any additional funding yet, the question of digitization is on the table and is discussed more than ever before in the library sector.

# Shared print

Another ongoing Swedish project is a national collaboration on shared print and joint weeding and preservation. It has not progressed as far to date, but has the potential to have a big impact on the long-term work with print collections.

## Background

Collection building at Swedish research libraries was investigated in a report from 2018 (Kungliga biblioteket, 2018). The report made it clear that many research libraries did not consider themselves as collection-building libraries—at least not for print collections. The report also showed that there was extensive weeding of print collections from research libraries. Statistics show that the print collections at university libraries decreased by 18% between 2015 and 2017. Much of the material that is weeded is the "half-old" material from the 1960s and 70s, that is not of high interest right now, and libraries make short-term weeding decisions based only on local needs and today's immediate requirements. As a collective, Swedish libraries are losing both breadth and depth in the collection.

The report also highlighted the unsecure situation for special libraries. They are often part of an institute or government agency, and when the parent organization has financial problems, the library risks budget cuts. There have been some closures of special libraries (the library of Statistics Sweden in 2019).

Sweden has a legal deposit law that regulates that two preservation copies should be kept at the National Library and Lund University. However, these copies are also for use, both in special reading rooms and for interlibrary loans, so they might be lost or torn. This means that there are no secured copies for the future, even for Swedish material. The new national Library strategy raises the need for a better strategy for handling and preserving physical collections (Regeringskansliet, 2022). At the moment there is no format for collaboration on collection building or guidelines for weeding.

With this background, the National Library started a project in 2019 to look at the possibility of creating a "National framework for weeding and preservation". Karin Byström was the leader of that project and there were project members from both the national library, university libraries, special libraries and the library deposit.

The group organized many workshops with libraries to understand their wishes and demands regarding a shared print solution. The group also looked at shared print solutions in other countries like the UK, USA and Norway to understand if there were any suitable options for Sweden.

A vision for a national collaboration was created: "research libraries collaborate in the long term and take joint responsibility for saving, preserving and making available material for the needs of today and the future for all their target groups". In order to achieve that vision there needs to be common principles for collaboration, preservation and access. A survey showed that almost 100% of higher education and special libraries would be interested in a collaboration based on the vision and that there was extensive interest from libraries in terms of working together more closely.

The final report was called "Everything for everyone, always"—a bit over the top, but it also symbolizes the ultimate vision—that libraries can give users what they want, now and in the future (Berglind et al., 2020).

### Start of collaboration

After the project, the National library organized round-table discussions with libraries to confirm the interest, and then set up the beginning of the collaboration in the spring of 2021 with a steering committee and a working group (*Samverkansgrupper: Nationellt, 2022*).

A letter of intent was developed to set a solid base for the collaboration. It states that print preservation and joint weeding are areas where libraries need to work together more closely (*Avsiktsförklaring, 2022*). The first libraries to sign were the nine libraries from the steering committee, and after a call to action, many more libraries signed. In August 2022, about thirty libraries had signed.

The purpose of the collaboration is clarified in the text: "With this letter of intent, libraries want to clarify their shared responsibility and the need for national coordination to ensure the long-term supply of information in the form of print material."

*National collection analysis*

The first concrete step for the working group will be to carry out a national collection analysis using metadata in the national catalogue Libris. The aim is to perform an analysis similar to one in the UK, called Strength in numbers ([Malpas & Lavoie, 2016](#)). That study found that only a few copies were held of a relatively large number of books, and that scarcity is common.

A national collection analysis will give a general overview of the national holdings of print monographs and journals and a better understanding of the possibilities and challenges. Interesting questions are how many records have few (or only one) holdings and records with many holdings. It will also be interesting to see how the collection is distributed between library types and regions in Sweden. From the collection analysis, the working group hopes to identify areas of collaboration on preservation and weeding. The first preliminary results and metadata problems are described in a recent article in Library Management ([Byström et al., 2022](#)).

## Conclusion

Both projects show new ways for libraries to work together with collections in Sweden. They are both large and demand united resources—no one library can do it alone. Libraries do this by resource sharing and by sharing the burden. By sharing collections, competencies and staff as well as systems and standards, libraries can create something beyond the reach of a single library.

In the future, there will be great opportunities for synergy effects between the projects. Digitization could be a solution to increase access to last copies. Additionally, duplicate copies of journals could be used for quick and cheap digitization.

There are also some positive long-term effects of the collaboration that go beyond the scope of shared print or digitization. Both projects show the need for a shift from a local focus to national collaboration and the need for shared infrastructure for collection management and resource sharing.

## Bibliography

*Avsiktsförklaring kring bevarande och gallring.* (2022). Retrieved October 31, 2022, from [https://www.kb.se/samverkan-och-utveckling/biblioteksutveckling/samverkansgrupper/avsiktsforklaring-kring-gallring-och-bevarande.html](https://www.kb.se/samverkan-och-utveckling/biblioteksutveckling/samverkansgrupper/avsiktsforklaring-kring-gallring-och-bevarande.html)

Berglind, J., Byström, K., Ericson, C., Hallnäs, H., Sjöström, M., & Ängkvist, M. (2020).

*Allt åt alla för alltid. Förslag till ett nationellt ramverk för gallring och bevarande*. Kungliga Biblioteket.

Burman, L., Byström, K., Ilshammar, L., & Tahmasebi, N. (2022a, September 22). *Nu digitaliserar vi hela det svenska trycket: Men hur, när och varför?* [Video]. Bokmässan. https://bokmassan.se/programs/nu-digitaliserar-vi-hela-det-svenska-trycket-men-hur-nar-och-varfor/

Burman, L., Carlsson, H., Ilshammar, L., Palmqvist, M., Sjögren, M., & Widmark, W. (2022b, March 28). *Satsa på storskalig digitalisering av det svenska trycket*. https://www.kb.se/samverkan-och-utveckling/nytt-fran-kb/nyheter-samverkan-och-utveckling/2022-03-28-satsa-pa-storskalig-digitalisering-av-det-svenska-trycket.html

Byström, K. (2022). *Förstudie gällande användande av data.kb.se som gemensam data-plattform för DST-samarbetet.*

Byström, K., Isaksson, A., Thordstein, A., & Undorf, W. (2022). First steps towards shared print collaboration in Sweden. *Library Management.* https://doi.org/10.1108/LM-11-2022-0106

Grönvall, K., Ilshammar, L., Palmqvist, M., Nylander, E., Widmark, W., Sjögren, M., & Burman, L. (2020). *Avsiktsförklaring om det digitaliserade svenska trycket*. https://www.kb.se/download/18.7bef1f3916eca87a61b1094/1623150278073/Avsiktsfork-laring-svenskt-tryck.pdf

*Informationssidor för digitaliseringen av det svenska trycket*. (2021). Retrieved October 31, 2022, from https://wiki.ub.uu.se/pages/viewpage.action?pageId=114891791

Kungliga biblioteket. (2018). *Samlingsbyggandet på svenska forskningsbibliotek: Nuläget och förslag på nationella samarbeten kring fysiska samlingar*. https://urn.kb.se/resolve?urn=urn:nbn:se:kb:publ-340

Malpas, C., & Lavoie, B. F. (2016). *Strength in numbers: The Research Libraries UK (RLUK) collective collection*. OCLC Research. https://www.oclc.org/content/dam/research/publications/2016/oclcresearch-strength-in-numbers-rluk-collective-collec-tion-2016-a4.pdf

Regeringskansliet. (2022). *Strategi för ett starkt biblioteksväsende 2022–2025*. https://www.kb.se/download/18.2dd50321183eebadf40832/1666677483654/strategi-for-ett-starkt-biblioteksvasende-20222025.pdf

*Samverkansgrupper: Digitalisering av det svenska trycket*. (2021). Retrieved October 31, 2022, from https://www.kb.se/samverkan-och-utveckling/biblioteksutveck-ling/samverkansgrupper.html#item-4623d586417be5cfa5ac4bd

*Samverkansgrupper: Nationellt ramverk för bevarande och gallring*. (2022). Retrieved October 31, 2022, from https://www.kb.se/samverkan-och-utveckling/bib-lioteksutveckling/samverkansgrupper.html#item-479c86e0d1833619d76c15ab0

Sjöström, E. (2022). *Kunskap för en ny ti: Om digitalisering av det tryckta kulturarvet*

*på universitetsbiblioteken i Sverige.* Svensk Biblioteksförening. [https://wwwbiblioteks-for.cdn.triggerfish.cloud/uploads/2022/12/kunskap-for-en-ny-tid.pdf](https://wwwbiblioteks-for.cdn.triggerfish.cloud/uploads/2022/12/kunskap-for-en-ny-tid.pdf)

## About the Author

Karin Byström, Librarian
Uppsala universitetsbibliotek (Uppsala University Library)
[https://orcid.org/0000-0002-1912-106X](https://orcid.org/0000-0002-1912-106X)

Karin Byström (1973) is a librarian at Uppsala University Library in Sweden where she works as a coordinator at the Media division. She has extensive experience of working with many parts of collection development – cataloguing and acquisition, e-resources and open access, digitization and shared print. Karin has an interest in library collaboration and she has been involved in many Swedish library collaboration projects. She is currently the project manager of the collaboration for Digitization of Swedish print and part of the working group for the Swedish shared print initiative. Since 2019 Karin has been a member of the standing committee of the IFLA Acquisition and Collection Development section, where she currently holds the position as secretary.

# Access to academic heritage: ENC theses available online

**Camille Carette**

## Abstract

Led by the library of the French School École nationale des chartes (ENC), ThENC@ is a multi-layered project whose aim is to digitize and upload onto a single online platform academic works, the particularity of which is that these theses written by the School archivists-paleographers are not actually held by the School library. This project addresses the issues of digitization and restoration of documents, constitution of a corpus when the physical documents are not stored in a single place, copyright and open access, metadata work, and creation of a digital library. It shows the important role of libraries in accessing both old and recent academic works online.

## Keywords

Theses; Academic works; Digitization; Digital library; Archives

## Article

## Introduction

In November 2021, the library of the École nationale des chartes (ENC) launched a website called ThENC@.[1] Acting as a digital library, this site provides access to some of the theses produced by students at the School. However, the project that led to the creation of this website began years earlier and encountered multiple obstacles—the main one being that none of these theses are held at the School itself or in its library.

---

1. https://bibnum.chartes.psl.eu/s/thenca/

# Why aren't the theses in the School library?

## A school specializing in history and heritage

Founded in 1821 and located at Sorbonne University for a long time, before moving near the Louvre and the National Library of France a few years ago, the École nationale des chartes is a two-hundred-year-old school that mainly teaches historical and philological sciences and trains future heritage professionals.[2] It has about 170 students: archivists-paleographers, master's students specializing in digital technologies and digital humanities, and doctoral students.

The archivist-paleographer degree is the School's historic curriculum. After a demanding entrance exam, these students become trainee civil servants and follow a four-year program, at the end of which they must write a thesis.[3] After their studies, they mainly become researchers, archive curators or library curators.

## A complex situation: theses not so easy to read

The specificity of the theses written by archivists-paleographers is that they are not PhD theses, but "school theses". More importantly, they are considered as private documents belonging to their authors, and therefore have the status of private archives and cannot be read without their permission. Consequently, they do not fall under the same obligations intended to facilitate their consultation by the research community, nor are they held in a university library as it is the case with doctoral theses.

Theses produced before 1961 can be found in many places: in the institution where the author may have worked, such as municipal libraries or departmental archives, in the family archives of their descendants, in a learned society, or even in the possession of someone to whom they have donated a physical copy. The only physical copy of a particular thesis was recently found at Harvard University, for example!

In 1961, an agreement between the School and the French National Archives made the deposit of a copy in the National Archives mandatory. Having all the copies located in the same place was a good thing, but despite this, it did not make the consultation of these documents much easier, as the authorization of the author remained necessary. Therefore, even today, users must contact the National Archives, which in return contact the author or their legal claimants to ask permission.

2. https://www.chartes.psl.eu/en/rubrique-ecole/institution-au-service-histoire-du-patrimoine-1821
3. https://www.chartes.psl.eu/en/cursus/the-diploma-of-archiviste-paleographe

As of today, we do not know the location of 38% of the 3,000 theses that have been written since 1849, the year of the first thesis defence at the École nationale des chartes, but one thing is certain: with the curious exception of one of them, none of them are in the School library. This leads to a confusing situation for library users, who often think they can consult those documents there.

## What could be done? Asking authors, managing legal difficulties

One way to solve these consultation problems was naturally to launch a digitization project, led by the School library. But before we could start to digitize anything, we had to get permission from the authors—and finding them was no mean feat. The task became even more complicated when the authors had died, and their heirs had to be found. So whilst the first authorizations were collected in 2011 and 2012, it was only years later that the first documents could be digitized. As of today, it is possible that authors who might well agree to participate in the project are still not aware of its existence.

Another key point of the project was its legal aspect: it was clear from the start that not all theses could be put online in open access. Indeed, several scenarios complicated the situation. In a first case, an author could authorize the online publication of their thesis, but limit the access to library readers only. In another case, they could agree to their thesis being posted online in open access, but it turned out that a printed version of the thesis had been published years earlier, which made the publisher's permission mandatory as well. In a last case, it was the content of the thesis that could pose a problem: if all the illustrations used were not free of rights, then not all the pages of the thesis could be put online in open access.

These difficulties had to be considered very early on. It was up to the library to ensure that the illustrations used did not cause any problems, or that a published version of the thesis did not exist. But above all, even without knowing how those theses were going to be published and valued online, it involved thinking about a future solution that would make it possible to control access to certain documents or parts of them. Therefore, using national or international open access platforms was not a suitable option for us.

# From finding the documents to putting them online

## Locating the documents: an inter-institutional work

As mentioned, physical documents prior to 1960 are scattered across France—and sometimes beyond its borders. The help of other professionals was essential in this quest, and it was very often librarians, archivists or researchers who made the presence of a thesis in an institution known. The ThENC@ project was therefore an opportunity to forge partnerships with other institutions: libraries, of course, but also archive centres or learned societies.[4]

One of those major partnerships is, of course, with the French National Archives, where more than 57% of theses are held. A very beneficial partnership with the IRHT also developed.[5] When the theses available in other institutions were not directly digitized by these institutions, the IRHT staff members digitized them during their visits there. Collaboration between institutions was therefore, and still is, one of the pillars of this very particular project, in which bridges are made between archives and libraries in order to reconstitute a corpus.

## Finally, digitizing!

The preservation and physical preparation of documents is important if we want to be able to obtain any digital data. In order to prepare the documents for digitization, on-site checks were necessary—when possible, that is when the documents were located at the National Archives. The oldest and most damaged ones have been restored in the library of the École nationale des chartes. Sometimes, it took weeks to restore a single thesis (Vielliard & Gaudemer, 2021).

Thanks to funding from PSL University, all thesis summaries—called "positions de thèse" in French—were digitized in 2018. In 2019, more than 300 theses were then digitized, and ocrized when possible, which marked the real start of the project (Université Paris Sciences et Lettres, n.d.). In 2020 and 2021, the project also received help from the DIM-MAP,[6] under the project name "Ouvrir le patrimoine académique : les thèses ENC accessibles en ligne" (Postec & Mathis, n.d.). This made it possible to continue digitizing documents, as well as to create a website to publish these academic works and to make them available online.

## Finding the technical solution

A first corpus of digitized documents was created, and the next step was to make it accessible to the public. Creating a platform from scratch was quickly discarded, due to the obvious difficulties in maintaining in-house tools over the long term. The library encourages the authors of the most recent theses to upload them in priority on platforms such as the French open access platform HAL,[7] on which about thirty theses are available to date, and was even at the origin of the creation of a new type of document on this platform: "institution thesis".[8] However, national platforms of this kind for submitting thesis or research work could not be considered for the whole corpus, as the majority of the documents were not native PDFs—and that would not have solved the question of the documents which had to be published in restricted access!

---

4. https://bibnum.chartes.psl.eu/s/thenca/page/partenaires

5. *Institut de recherche et d'histoire des textes*, a French research unit (CNRS) that mainly works on manuscripts.

6. *Domaine d'Intérêt Majeur "Matériaux anciens et patrimoniaux"* (DIM MAP))—a research network dedicated to the study of ancient and patrimonial materials.

Creating a digital library dedicated to these theses appeared to be the best and most satisfying solution. Thus, ThENC@ is currently powered by Omeka S,[9] an open-source content management system making it possible to manage digital collections stored in an SQL database, and to display them on a website. The choice of this tool was guided by several factors: a long-standing preference of the School and its library for open-source software, the presence of a large community of users in France, and Omeka S having interesting functionalities for the project. Among them, the possibility of managing restricted access for non-logged-in users, or that of using the Apache SolR search engine.

## Improve user experience by improving metadata

### *Creating records for all theses*

The scattering of theses across the country meant one thing: no one had a complete list of all the theses written since 1849. Some theses were catalogued, others not. At the beginning of the project, the School librarians catalogued the theses held in the French National Archives since 1961—but it was not all the theses. Even then, different metadata standards were used in two French online catalogues for university libraries: Unimarc records were created in the union catalogue Sudoc,[10] and XML-EAD was used in Calames,[11] the public catalogue for collections of manuscripts and archives at higher education and research institutions. The lack of a full list made it more difficult for researchers looking for a specific thesis.

7. https://hal.science/
8. https://hal-enc.archives-ouvertes.fr/search/index/?q=%2A&docType_s=ETABTHESE
9. https://omeka.org/about/project/
10. http://www.sudoc.abes.fr
11. http://www.calames.abes.fr/

One of the roles of ThENC@ is therefore to provide the complete list of theses, including those which are not digitized. This list was compiled thanks to the thesis summaries mentioned earlier, which let us know that a thesis existed even though we may never be able to find it.

Although based on the detailed records mentioned above, ThENC@ records are simple and use Dublin Core properties—ideal for data interoperability, and for exports using the OAI-PMH protocol. This choice was influenced by the choice of software, as Omeka S and Dublin Core work together very well. By doing this, the library, in addition to providing access to documents that are not in its physical collections, also acts as a collection creator, aiming to federate the data into a single place.

## Enriching metadata

This project was, in many ways, an opportunity to enrich the existing metadata sets. One of the major operations was the creation of authority records for all archivists-paleographers who did not already have one in the IdRef platform,[12] in collaboration with the bibliographic agency for higher education (Abes).

In order to help end users, other metadata enrichments have taken place directly in the ThENC@ digital library. One of them was the addition of the location of the printed copy, when known, to help people to find it when the thesis is not available online or digitized. In order to improve the user experience, there is also an ongoing project to classify theses by thematic collections. The more thoughts there are on how to help researchers in their quest for these hard-to-find documents, the more ideas for refining the metadata are generated.

## Creating bibliographies related to the theses

The truth must be accepted: some theses will never be online, either due to lack of authorization from the author, or for lack of knowledge as to the whereabouts of the thesis. So, what can be done to help users access the information those documents may contain? One of the solutions is to report all the work of an author relating to their thesis: articles, conferences... sometimes even another version of the thesis itself, published as a printed book.

Zotero bibliographies have been created and linked to the records. It is a long and still ongoing project—as well as a collective one, which involved students from the School. By studying in detail the background and the career of a class of archivists-paleographers from 1900, they were able to find three theses, all located in a different place ([Ceccarelli, 2021](#)). On the other hand, they found no trace of some authors from this year: the older those former students, the harder it is to find information about them and their work, particularly if they did not pursue a career in heritage, teaching or public institutions. Some theses are probably lost for good.

## Conclusion: where are we now?

Today, ThENC@ has been online for a year and contains more than 500 theses, most of which are searchable PDFs. The project is still ongoing: not only can metadata and bibliographies always be improved, but new digitizations are arriving each year!

ThENC@ is a multi-layered project, which involves the entire library staff and a wide variety of skills. From searching for documents and authors across France to launching a digital library, through restoring old documents, digitizing, creating records, or dealing with open access issues, this project shows, more than ever, the role of libraries in accessing old and recent academic works online, including when these documents are outside their collections.

## Bibliography

Ceccarelli, G. (2021, April 8). *La promotion 1900 : un chantier du projet ThENC@.* https://chartes.hypotheses.org/7415

Postec, A., & Mathis, R. (n.d.). *ThENC@. Ouvrir le patrimoine académique : les thèses ENC accessibles en ligne.* Retrieved February 1, 2023, from https://www.dim-map.fr/projets-soutenus/thenca/

Université Paris Sciences et Lettres. (n.d.). *Thenc@. Thèses ENC accessibles en ligne.* Retrieved February 1, 2023, from https://explore.psl.eu/fr/ressources-et-savoirs-psl/projets-psl-explore/thenc-theses-enc-accessibles-en-ligne

Vielliard, F., & Gaudemer, L. (2021, November 26). *Du manuscrit à l'imprimé. Le thèse de Léopold Delisle: aspects matériels et intellectuels.* [Conference session]. ThENC@. Ouvrir le patrimoine académique : les thèses d'École des chartes accessibles en ligne, Paris, France. https://hal-enc.archives-ouvertes.fr/hal-03482814v1

# About the Author

Camille Carette, Library Systems Manager
École nationale des chartes (The École nationale des chartes)
https://orcid.org/0000-0001-7123-1584

After studying history at the Panthéon-Sorbonne University, Camille Carette (1994) trained in the use of digital technologies in heritage and research. Since 2019, she has worked at the library of the École nationale des chartes, first exclusively on the digitization of the theses of archivists-paleographers, then as library systems manager from 2020. Among other things, she is in charge of the ILS, the discovery tool and the digital library. In 2023, she joined the organizers of AUFO, the association of French-speaking users of Omeka.

# When David and Goliath work together: The impact of a cooperation agreement on library strategy

**Susanne Kirchmair**

## Abstract

When MCI was founded in 1994, there was no way to know what it would become more than 25 years later, namely a university of applied sciences. As this development was unplanned and unforeseen, MCI had to devise an unconventional solution for providing library resources. As a result, a cooperation agreement was formulated between the University Library and MCI. I will describe the evolution of this collaboration and its impact on library strategy in general. I will also discuss the cooperation's advantages, disadvantages, and challenges. Finally, I will provide insights into recent developments and the future outlook.

## Keywords

Library; Cooperation; Collaboration; Strategy; Impact

## Article

## Introduction

MCI | The Entrepreneurial School® is located in Innsbruck, the capital of the County of Tyrol in Western Austria. Innsbruck has 160,000 inhabitants and approximately 35,000 students. Due to its relatively small size – one can easily walk from one end of the city to the other within two hours – and bustling city and cultural life, it is popular amongst young people. Another attraction is the location. The mountains surrounding MCI allow for various outdoor activities all year round. Therefore, MCI is a favourite for winter sports enthusiasts who can enjoy lectures and skiing or snowboarding on the same day, as the slopes are very close to the city centre and can be reached within half

an hour. Hiking, biking, and other activities can also be enjoyed during the warmer seasons.

## Institutions involved

Five higher education institutions are located in Innsbruck with many schools for all levels. Two of those institutions are actively involved in a cooperation, which will be discussed later. The two partners are MCI | The Entrepreneurial School® (MCI) and the Leopold-Franzens-Universität Innsbruck (University of Innsbruck). Both parties are discussed below to provide background information.

### MCI and MCI library services

MCI is a subsidiary of the University of Innsbruck that started as a continuous education offer for the university's alumni. As the so-called "speedboat" of the University of Innsbruck, it was created to ensure its alumni would stay in the education loop even though they had left their alma mater.

In 1993, a new law (Republik Österreich, 1993) allowed the foundation of *Fachhochschulen* (i.e. universities of applied sciences) in Austria. Until then, it was only possible to receive an academic degree at public or private universities or to attend other higher education institutions to be trained as a teacher, nurse or social worker. Therefore, the emergence of a new type of academic institution fundamentally disrupted the Austrian higher education landscape and challenged such new universities to build the infrastructure for higher education from scratch.

Following the aforementioned new law, the Management Center Innsbruck was founded in the academic year 1995–1996. As its name suggests, it mainly focused on providing continued education in business. After offering courses and seminars solely to a managerial and professional target group, MCI started taking in regular students for the part-time business and management programme, concluding with a diploma.

MCI is a public institution with a broad consortium of ownership (Fig. 1). It offers a variety of programmes and services across a wide span of topics within the range of management, society, technology, and life sciences.

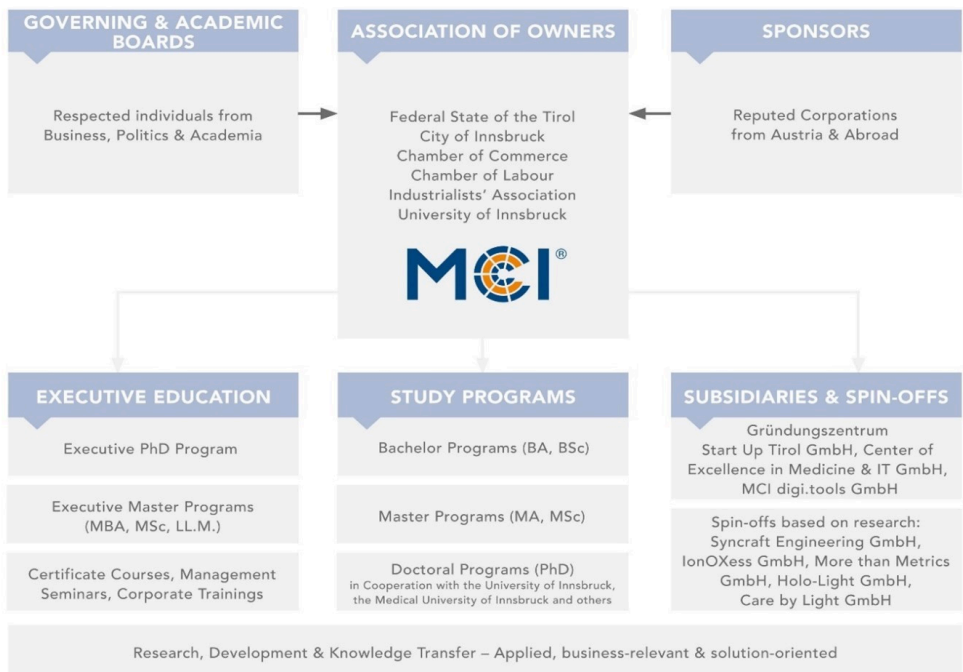

*Figure 1: MCI Consortium (MCI, 2021)*

MCI receives federal funding for allocated study places and charges regular tuition fees of EUR 363 per semester. It has also received several accreditations, such as AACSB and FIBAA.

MCI offers 29 study programmes with a total of 3,634 students, has 304 partner universities across the globe, and, so far, is proud to have educated 14,572 graduates (MCI, n.d.-d) who are well prepared for and welcomed in the labour market. Furthermore, MCI has about 400 employees from 37 nations, including 300 internal faculty members and 887 external faculty members, due to its practice-oriented approach (MCI, 2022, p. 3).

The MCI campus consists of five central locations within walking distance of each other in Innsbruck. It offers all the necessary "infrastructure with attractive and well-equipped lecture halls, computer labs, and group working rooms, with state-of-the-art workstations and research facilities, student hostels in the vicinity, a wide choice of good places for eating out, convenient parking, and excellent public transport" (MCI, n.d.-b).

Recently, MCI Management Center Innsbruck was rebranded as MCI | The Entrepreneurial School® to strengthen the well-known brand and adapt to the international market. MCI's mission is to provide a "meaningful connection between universities, grandes écoles, business schools, universities of applied sciences, and the world of business. It represents a strong international brand that successfully combines academic teaching and training, solution-oriented research and development, impactful knowledge transfer, [and] innovative start-ups" (MCI, n.d.-a).

It is widely known that the "availability of research facilities and a place for quiet reflection in combination with efficient access to scientific literature is fundamental to a modern university infrastructure" (MCI, n.d.-c), and MCI meets these criteria. The primary goal of MCI's library services is to support all study, research, and teaching activities.

Therefore, MCI provides libraries and study areas with a wide selection of books, journals, magazines, and recent newspapers. In the three main MCI campuses, libraries are equipped with a printer and copier, a book scanner, reading desks, and personal computer workstations. These libraries are purely reference collections.

MCI library services offer access to an extensive digital collection of e-books, e-journals, and other research from many different subject areas. In total, we have licensed 20 databases. This collection is constantly being expanded to meet the needs of the staff and students. All the online resources are easily accessible on and off the MCI campus. To access the licensed online content remotely, we use the EZProxy server from OCLC.

The MCI has a cooperation agreement with the University of Innsbruck and the State Library of Tyrol (ULB) to complement its library services and provide the opportunity to borrow books and use interlibrary loans and reading rooms.

In October 2011, the first academic librarian began with the new formation and expansion of MCI's library services. Since 2020, all library services at MCI have been handled by a team of two full-time academic librarians. They are responsible for administration, communication, teaching, training, and development.

## University of Innsbruck and ULB

On 15 October 1669, Emperor Leopold I approved the levying of the "Haller Salzauschlag", the special tax to finance a Tyrolean state university. This measure was decisive for the establishment of today's University of Innsbruck (University of Innsbruck, n.d.-b). In 2019, the University celebrated 350 years since its foundation and is, therefore, one of Austria's oldest universities (University of Innsbruck, n.d.-a). The University has 28,106 students in total; of these, 13,439 are international students, and every semester, about 4,000 lectures are held (University of Innsbruck, n.d.-c).

The University has its own library, which was founded in 1745. As the largest library in western Austria, it supports research, teaching, and study. Moreover, as the Tyrolean Regional Library, it also reflects the regional literary landscape. It offers space for encounters and exchange and builds bridges between the past and the present (Universitäts- und Landesbibliothek Tirol, n.d.-a). Its full name is Universitäts- und Landesbibliothek Tirol, which translates to the University and State Library of the County of Tyrol. Its name signifies that the library is not only meant for university affiliates but is also open to the public.

The library has seven branches across the city of Innsbruck, organised by topic related to the specific location. Its holdings comprise about 3.5 million printed media and 80 electronic databases. The library has approximately 200 employees across several faculty libraries and service departments ([Universitäts- und Landesbibliothek Tirol, n.d.-c](#)).

## Cooperation agreement

After MCI was founded, it soon became evident that academic education needed to be accompanied by the respective library resources.

In the beginning, these resources comprised printed books and journals that had to be provided on-site and some space to use them. These materials also had to be administrated in some way. However, the number of students and study programmes was small, and no significant growth was planned initially. Additionally, MCI was located in a city with an extensive university library within walking distance which was also open to the public. Therefore, there was no need to plan for its own library unlike other Austrian universities of applied sciences, which were founded to be stand-alone higher education institutions, often in locations without any university infrastructure.

### *Development of the cooperation*

The cooperation that emerged from the historical circumstances described above is unique in Austria. The following milestones explain the development of the cooperation:

- 1995: Start of the first diploma study programme in business and management at MCI
- 1998: Opening of the new ULB branch "SoWi-Library" (faculty library of social and economic sciences) at campus Universitätsstraße
- 1999: MCI moves to campus Universitätsstraße (now the headquarter)
- 2000: First contract between MCI and "Verein zur Förderung der SoWi-Bibliothek" (association for the advancement of the SoWi-Library)
- 2003: First amendment to the original contract
- 2008: Second amendment to the original contract
- 2015: Change of cooperation partner from the association to ULB

*Details of the cooperation agreement*

To use the ULB as the MCI library, a payment from MCI was agreed upon in exchange for student and staff access to the infrastructure and services of the whole University library, especially the SoWi-Library (Chen et al., 2000). Ideas about the future development of the cooperation and further collaboration were also noted, some of which could not be executed due to legal reasons (e.g. the joint use of electronic resources). The contract accounted for other facts as well, such as the foundation of an MCI subsidiary for executive education and the extension of the library's services to this subsidiary (Niedermair & Altmann, 2003). Payment calculated at a fixed yearly rate was determined for expanding the library's opening hours, and a handling fee was levied for every book bought by MCI (Niedermair & Altmann, 2008).

After a change in the leadership of the ULB, the new administration lifted the cooperation to a more official level within the University library. This change also ensured that the agreed-upon payments went to a general account for third-party funds for the University library. The changes were retroactively enacted on 1 January 2014 (Frasnelli & Altmann, 2015).

The main features of the modified cooperation are as follows:
- Free use of the MCI student and staff IDs as ULB library cards with the same rights as University affiliates
- Use of all library facilities of the ULB
- Expansion of opening hours (evenings, Saturdays, and Sundays)
- Acquisition, cataloguing, and other services for printed books
- Additional purchase of highly demanded (text) books

Today, the MCI actively advertises the cooperation as an addition to its own library services, which provides MCI's researchers, lecturers, and students with full access to the university library facilities and services (MCI, n.d.-c).

## Impact on library strategy

The emergence of the cooperation has severely impacted MCI's strategic decisions regarding the library.

The collections, services, and facilities provided by the ULB are fully available for MCI affiliates, and vice versa. As MCI always buys books according to its needs, sharing printed resources and facilities is easy and beneficial for both institutions.

The most severe impact is that if MCI does not need a library due to the ULB fulfilling that purpose, there will also be no need for a librarian and library facilities. Moreover, there will not only be any need for library software and resources but also no clear development strategy for its collection and services.

However, MCI provides reference collections on its campus for quick use by staff and students. Several study areas are open to the public too. Additionally, there is also a need for the provision of online resources due to legal and practical reasons. Licensing contracts only extend to one academic institution. As MCI is not a full subsidiary of the University of Innsbruck, it has to purchase its own databases. Furthermore, as remote access is mandatory, this access can also only be guaranteed for affiliates of the licensing institution.

## Advantages

The advantages of the arrangement described above are clear, as it allows MCI to use the enormous resources, facilities, and services of the ULB.

These benefits also include borrowing materials from all ULB branches, inter-library loans, and document delivery services. Therefore, there is no need for MCI to provide extensive physical spaces for books and reading rooms, physical and digital processing of printed books, circulation, library software, and a discovery system. Consequently, there is no need for specialised human resources or a library department, and one individual at the interface of MCI and ULB can adequately manage the cooperation, exchange of books, and reference collections at MCI locations.

## Disadvantages

Despite the advantages of the cooperation outlined in the previous section, many previously hidden disadvantages have become more evident and pressing in recent years.

These disadvantages include, first and foremost, that there was no information professional at MCI for the first 15 years of its existence. Therefore, MCI heavily relied on the ULB, which led to the belief amongst many students and staff that it did not have its own library. This belief was fuelled by creative wording such as "Studienlandschaft" (which roughly translates to "study landscape") instead of "Library" for the rooms containing the reference collections. Consequently, the library was invisible within MCI.

There were also no specialised services for students, faculty, and researchers, as there was little to no MCI-specific knowledge at ULB. MCI also prides itself on its service orientation, which was practically non-existent at ULB.

All communication at MCI is always targeted towards a German- and English-speaking public. Unfortunately, many ULB websites are only available in German ([Universitäts-und Landesbibliothek Tirol, n.d.-c), which is a significant problem, especially for international students.

At MCI, there was a prevalent notion that ULB handles all matters regarding library resources, and, therefore, there was improper handling of institutional licenses and subscriptions. Some journal subscriptions still exist in the names of individuals who no longer work at MCI. This problem persists since some of the former subscriptions have not been identified.

The strategic development of library stock, facilities, and services at MCI has never taken place. MCI's library grows according to actual needs, such as the emergence of new study programmes. However, this strategy is challenging as the subjects that need to be covered range from social sciences and business to technology and life sciences. Furthermore, as staff and student numbers grow, the physical space available for them and the resources needed on campus seem to shrink. As a "real" library never existed at MCI because the ULB was our library, there was never a real space for it.

Although Innsbruck is relatively small, there are no large-scale libraries at the different MCI locations. As a result, one would always have to go to one of the ULB branches throughout town according to their specialisations. Due to licensing restrictions, remote access is not available for ULB's online resources. However, MCI affiliates can use the allowed walk-in user access on ULB locations.

A lack of communication on the part of the ULB has made it difficult to effectively serve MCI affiliates. This issue became evident, for example, when news regarding services or opening hours was seldom communicated in advance, even when such information severely impacted MCI.

Finally, several components of the cooperation agreements were never fully executed, possibly due to a lack of close collaboration, so much so that MCI was not even listed as a cooperation partner of the ULB on its official website (Universitäts- und Landesbibliothek Tirol, n.d.-b). However, MCI strongly promotes the agreement in its official communication (MCI, n.d.-c).

## Challenges

Even though the unique cooperation agreement is a significant asset, we are facing major challenges due to the development of MCI and higher education and libraries in general.

As this cooperation can only go so far in terms of sharing resources, MCI faces massive expenditures for online resources. There is an increasing need for e-books and e-journals with a remote access prerequisite. At the same time, there is a demand for 24/7 access to resources and facilities for on-campus studying. However, as we are constantly growing, reading areas with round-the-clock access are not our priority, and, irrespective of this, there is a constant battle for space. Therefore, although we strive to become an almost complete e-library, this task is impossible with budget restrictions.

There is also a strong need for adequately equipped library facilities in additional MCI locations and the adaptation of existing libraries and study areas. Those spaces are far too small for the resources and unattractive for students to spend time in.

We also witness the constant evolution and expansion of study programmes, which make it difficult to provide sufficient resources on every subject. For example, MCI is now offering a programme in medical technologies that requires expensive resources in medicine. Many medical journals are already covered in our existing database subscriptions, but we struggle to get affordable access to technical standards for only a small group of people.

There is an ever-growing need for services and training, such as literature research, bibliographic management, and academic writing, which we are happy to provide to all study programmes. However, our capacities are extremely limited as we are only two full-time librarians who have to cover everything from administration to teaching.

We also see topics such as copyright issues, online programmes, publication services, and open sciences as services for other departments and study programmes. There is a growing need for support, and the library services are the right place for these questions. However, as we are only a small team, we are limited by the domains we can cover adequately.

Furthermore, there are other issues, such as the growth of online programmes, which have their own unique challenges regarding the supply and use of literature. The recent focus on research outcomes in universities of applied sciences also poses new questions and demands more support for junior researchers. Moreover, topics such as the global pandemic, energy shortage, and the political situation in Europe also significantly impact future strategies.

## Recent developments and outlook

Library strategy at MCI as well as the agreement will likely change soon. The main reasons for these changes may be attributed to the following two recent developments.

First, MCI is in the process of constructing a new building. For the first time, all five MCI locations, with three libraries containing reserve collections and small study areas, will move into a single building next to our current headquarters. However, as is already known, this new building will not be able to contain all departments. Therefore, the building has been planned without a library, which is usually a central point in any university building. As the plans currently state, there will be a new Library & Learning Centre in our current headquarters. This centre will consist of study areas, rooms for group work, and printed materials for all study programmes. The centre will be a reserve col-lection as we expect to continue the cooperation and, therefore, still outsource tasks such as circulation, user management, and book handling, including cataloguing.

Second, there is a new department head of our closest partner, the SoWi-Library. In March 2023, she took over from the former director who retired in November 2022 and is now in the process of getting acquainted with her new library and team. For MCI, significant discussion is necessary as the processes and finer details of the cooperation are tacit knowledge. Additionally, we expect the contract to change and will look into the details during the upcoming months.

## Conclusion

After more than 20 years, it has been proven that the cooperation agreement between David (MCI) and Goliath (ULB) continues to be beneficial for both parties involved.

ULB benefits through the funding of the extended opening hours of one of its branches, which is beneficial for University and MCI students. It also profits from payment for tasks performed within ordinary working routines. This advantage gives ULB some extra budget for the purchase of resources.

On the other hand, MCI benefits tremendously by outsourcing tasks and processes that would otherwise be extremely expensive and time-consuming. In addition, the use of the ULB infrastructure generates considerably more space for students than MCI could ever provide on its own.

Therefore, we are eagerly awaiting future developments regarding the cooperation and are convinced that even two organisations that are so different in size, such as MCI and the University of Innsbruck, can mutually benefit from this kind of collaboration. We hope the cooperation might also be an exemplary model for other institutions and encourage them to collaborate with their neighbours and even competitors.

## Bibliography

Chen, J., Niedermair, K., & Altmann, A. (2000, February 26). *Memorandum zur Kooperation zwischen dem Verein zur Förderung der SoWi-Fakultätsbibliothek der Universität Innsbruck und der MCI-Management Center Innsbruck GmbH*. Unpublished internal document. Innsbruck. Verein zur Förderung der SoWi-Bibliothek; MCI Management Center Innsbruck.

Frasnelli, E., & Altmann, A. (2015, June 15). *Abänderung der zwischen der SoWi-Fakultätsbibliothek sowie dem Verein zur Förderung der SoWi-Fakultätsbibliothek und der MCI Management Center Innsbruck – Internationale Hochschule GmbH) abgeschlossenen*

*Vereinbarung vom 26. Februar 2000 bzw. 30. September 2003 sowie der am 31. Dezember 2008 vorgenommen Abänderung.* Unpublished internal document. Innsbruck. Universitäts- und Landesbibliothek Tirol; MCI Management Center Innsbruck.

MCI Management Center Innsbruck. (n.d.-a). *About us.* Retrieved March 28, 2023, from https://www.mci.edu/en/university/the-mci/about-us

MCI Management Center Innsbruck. (n.d.-b). *Campus & locations.* Retrieved March 28, 2023, from https://www.mci.edu/en/university/the-mci/campus

MCI Management Center Innsbruck. (n.d.-c). *Library.* Retrieved March 28, 2023, from https://www.mci4me.at/en/services/library

MCI Management Center Innsbruck. (n.d.-d). *Study at the Management Center Innsbruck.* Retrieved March 28, 2023, from https://www.mci.edu/en/

MCI Management Center Innsbruck. (2021, June 18). *Consortium.* https://www.mci.edu/en/download/category/11-allgemein?download=164:patron-structure

MCI Management Center Innsbruck. (2022, May 5). *Facts & figures 2021.* https://www.mci.edu/en/download/category/19-allgemeine-broschueren?download=101:facts-figures

Niedermair, K., & Altmann, A. (2003, September 30). *Vereinbarung zwischen der SoWi-Fakultätsbibliothek sowie dem Verein zur Förderung der SoWi-Fakultätsbibliothek und der MCI-Management Center Innsbruck GmbH.* Unpublished internal document. Innsbruck. SoWi-Fakultätsbibliothek; Verein zur Förderung der SoWi-Bibliothek; MCI Management Center Innsbruck.

Niedermair, K., & Altmann, A. (2008, December 31). *Abänderung der zwischen der SoWi-Fakultätsbibliothek sowie dem Verein zur Förderung der SoWi-Fakultätsbibliothek und der MCI Management Center Innsbruck – Internationale Hochschule GmbH (vormals: MCI Management Center Innsbruck GmbH) am 30.09.2003 abgeschlossenen Vereinbarung.* Unpublished internal document. Innsbruck. SoWi-Fakultätsbibliothek; MCI Management Center Innsbruck.

Republik Österreich. (1993, May 28). *Bundesgesetz über Fachhochschul-Studiengänge: FHStG.* https://www.ris.bka.gv.at/Dokumente/BgblPdf/1993_340_0/1993_340_0.pdf

Universitäts- und Landesbibliothek Tirol. (n.d.-a). *Über Uns.* Universität Innsbruck. Retrieved March 28, 2023, from https://web.archive.org/web/20221007224718/https://www.uibk.ac.at/ulb/ueber-uns/

Universitäts- und Landesbibliothek Tirol. (n.d.-b). *Kooperationen.* Universität Innsbruck. Retrieved March 28, 2023, from https://web.archive.org/web/20230314173652/https://www.uibk.ac.at/ulb/ueber-uns/kooperationen.html

Universitäts- und Landesbibliothek Tirol. (n.d.-c). *Universitäts- und Landesbibliothek Tirol.* Universität Innsbruck. Retrieved March 28, 2023, from https://www.uibk.ac.at/ulb/

University of Innsbruck. (n.d.-a). *350 Years of the University of Innsbruck*. Universität Innsbruck. Retrieved March 28, 2023, from https://www.uibk.ac.at/350-jahre/

University of Innsbruck. (n.d.-b). *University history*. Universität Innsbruck. Retrieved March 28, 2023, from https://www.uibk.ac.at/350-jahre/jubilaeum/geschichte.html

University of Innsbruck. (n.d.-c). *Welcome at the alpine-urban campus*. Universität Innsbruck. Retrieved March 28, 2023, from https://www.uibk.ac.at/en/

## About the Author

Susanne Kirchmair, Head of Library Services

MCI | Die Unternehmerische Hochschule (MCI | The Entrepreneurial School)

https://orcid.org/0000-0002-7339-5192

Susanne Kirchmair (1975) has an academic background in International Business & Economics from the University of Innsbruck (A), Aarhus Business School (DK), and University of Groningen (NL), as well as Library & Information Studies from the University of Vienna (A). She started her career as a librarian at the University and State Library of Tyrol. She worked in circulation, acquisition, and training in the Social and Economic Sciences faculty library. Since 2011, she has been responsible for Library Services at MCI, implementing resources, tools, and services to enhance the user experience and provide high-quality content for studying and research. She also teaches in various study programs and continuing staff education to support the academic success of students and faculty.

# How the interlibrary loan service can collaborate with other libraries and university services

**Isàvena Opisso**

## Abstract

The Interlibrary Loan Service of the Universitat Autònoma de Barcelona has always been in constant renewal. In this paper, three new projects in collaboration with other university departments are analysed. These are (i) the analysis of interlibrary loan requests with the collection-management department; (ii) the uploading of public-domain documents in the repository with the institutional-repository department; and (iii) the digitization of PhD theses with the thesis department.

## Keywords

## Article

## Introduction

The Universitat Autònoma de Barcelona (UAB) is a public university with more than 40,000 students mostly located on-campus. After 50 years of existence, the UAB has consolidated its position amongst the 200 best universities in the world within the main university rankings.

The Library Service at the Universitat Autònoma de Barcelona has certified its quality system through ISO 9001 since 2000.

The methodology used to reinforce quality is through ongoing improvement actions that call for constant renewal. The working process of the UAB Library Service facilitates and enhances the implementation of improvement procedures and work instructions. The working method is by means of a wiki site in which all the library members can make proposals.

Most of the actions proposed by the Interlibrary Loan (ILL) Services over recent years have been focused on improving the service itself. During the COVID lockdown, the ILL team carried out a global rethink of both the workflow and requirements for supporting students, researchers, and lecturers.

New services were created in collaboration with other departments in the library and in university departments.

Three new projects in collaboration with other university departments are analysed in this paper. These are (i) the analysis of interlibrary loan request with the collection management department; (ii) the uploading of public domain documents in the repository with the institutional repository department; and (iii) the digitization of PhD theses with the thesis department.

## Analysis of interlibrary loan requests with the collection management department

The mission of the UAB Library Service is to guarantee the information resources necessary for the University to achieve its objectives; resources are selected following the Collection Management Plan.[1] The library collection increases regularly with new physical and digital items. Most of them are acquired, but the collection also increases through donations and the exchange of publications published by the UAB with documents from other libraries. The summary of this annual increase in the collection is reflected in the annual report of the Library Service.[2]

Additionally, the University Library has offered the possibility to make suggestions for purchases with an acquisition form through the library web page[3] where members of the UAB can purchase acquisitions of monographs or journals subscriptions. Purchase applications are evaluated by libraries and before the evaluation, requesters are informed of the process that their request has followed.

Users request documents that are not available in the UAB library collection by means of an ILL request. Each interlibrary loan request responds to a specific informational need, but the analysis of many requests provides an opportunity to obtain useful data regarding user needs that can be relevant to make decisions pertaining to the collection management.

Data that can be obtained with the systematic analysis of a significant number of requests is that of the titles requested by various UAB users. That is, distinct users make a request at distinct moments for a document that is not in the UAB library collection.

---

1. https://www.uab.cat/web/our-collections/collection-development-policy-1345738248491.html
2. https://ddd.uab.cat/record/29
3. https://www.uab.cat/web/our-collections/suggestions-for-purchase-1345738248359.html

The methodology used to analyse ILL requests consists of analysing, every six months, those requests received over the previous calendar year.

The request information is exported to a spreadsheet programme, the list of items is divided between items that have ISBN or ISSN number, as they are considered unique identifiers, and items that don't have unique identifiers.

The duplicate detection of items that have ISBN or ISSN is done with the option to detect duplication in the spreadsheet programme.

The duplicate detection of items that do not have a unique identifier cannot be done with the searching duplicates tool in the spreadsheet programme. The titles of the requests come mainly from information collected by users in bibliographic citations and in different library catalogues. This means that small variations in the title are common, mainly in the titles of rare books and references that come from bibliographic citations.

The procedure used is to sort alphabetically and duplicates are searched by reviewing the alphabetical list title by title.

Once the information has been analysed, a list of duplicates is drawn up and given to the collection managers in the library, who can then decide to purchase and include the title within the library collection.

As expected, the duplicate average is remarkably low (Table 1).

|  | 2$^{nd}$ semester 2020 | 1$^{st}$ semester 2021 | 2$^{nd}$ semester 2021 | 1$^{st}$ semester 2022 |
|---|---|---|---|---|
| Number of duplicates | 2 | 3 | 0 | 0 |

*Table 1: Number of duplicates*

With four analyses carried out, between zero and three duplicates have been detected in each analysis. This fact confirms the good quality of the Collection Management Plan.

Additionally, it must be taken into consideration that this is a procedure that requires approximately two hours of work every six months and can provide valuable and comparable data across collection management data and, furthermore, offers the possibility of detecting titles that should be incorporated into the library collection.

The analysis of data that can be extracted from the requests will lead to future lines of research that can contribute to improving the service. A possible line of analysis could be to find out which research discipline an ILL requester belongs to. With this information, a global rethink could be made working in collaboration with possible providers or universities that can provide loans or digital copies to meet the needs of interlibrary loan.

# Uploading public domain documents in the repository with the institutional repository department

The process used in the Interlibrary Loan Service was the same for rare books and special collections in the public domain as for documents with copyright.

When another institution sent a request, the request was processed, digitized, and uploaded to a secure webpage where the requester could download the digital document. Once it had been downloaded, the digital document was then deleted.

The need to access digitized documents during the COVID lockdown, when access to the libraries was not possible, prompted us to think about uploading to the institutional repository those documents that had been digitized for a given requester, and that are in the public domain .

Spanish laws regarding copyrights surpass the life of the author and are maintained until 70 years after the death of the author. As a general rule, once this period has passed, the work goes on to form part of the public domain.[4]

The Spanish National Library facilitates information on Spanish authors whose works can be found at the Spanish National Library and are in the public domain, according to the information in the Library's catalogues, so they can be published, reproduced or disseminated publicly.[5] This information facilitates the identification of works that are potentials for inclusion in the repository.

Documents published in the Digital Repository that are in public domain can also be consulted through the library discovery system, increasing their visibility. Those items are also harvested in different collectors such as Europeana[6] and Hispana,[7] which is the access portal to Spanish digital heritage and the national content aggregator to Europeana.

UAB repository gathers the list of collectors, which includes different types of materials. The complete list can be consulted in the collectors' section of the library service website.[8]

The aim of this specific best practice is to make available, in open access, those documents that are already digitized and that can be held in the institutional repository because they are in the public domain.

A procedure was carried out to identify the material requested that is in the public

4. https://www.boe.es/buscar/doc.php?id=BOE-A-2006-12308
5. https://bnelab.bne.es/dato/autores-espanoles-en-dominio-publico/
6. http://www.europeanaconnect.eu/
7. https://hispana.mcu.es/es/inicio/inicio.do
8. https://www.uab.cat/web/our-collections/collectors-1345836940616.html

domain and to define the tasks for uploading the relevant digitized items to the repository.

The main points on which this procedure is based are the following:

- Only complete documents like books, theses and issues will be uploaded to the repository. Digitization of articles, chapters or partial books are disregarded.
- The interlibrary loan staff identifies as likely to be uploaded to the repository all those requests whose author died more than 70 years ago. Before proceeding with the digitization, the university's expert staff in copyright verifies that the document is in public domain.
- Once the document is digitized, it is made available to the user and, in turn, the library staff arranges for it to be uploaded to the repository.

## Digitization of PhD theses with the thesis department

PhD theses presented at the UAB are published in the Theses and Dissertations Online Repository (TDX), which acts as the repository for the UAB. The UAB library keeps a paper copy of the PhD theses that were not integrated into TDX at the time of the defence. Spanish doctoral theses prior to 2011 that are not digitized cannot be digitized without the author's express permission. Several restrictions on consultation and photocopying also apply.

As the Interlibrary Loan Service does not have permission to digitize or loan these PhD theses, the request for theses must be denied.

There is usually only one paper copy of the PhD thesis lodged at the university where the author presented their doctorate. Users have very limited possibilities to access a PhD thesis if the university that owns the only copy denies their request. However, PhD holders can upload their PhD theses to the institutional repository.[9]

A working instruction has been created to try to locate the thesis authors. If the author is located, the Interlibrary Loan Service informs them that a user is interested in their thesis and offers support for carrying out the administrative procedure needed to upload this doctoral thesis to the institutional repository.

Locating the author is not always possible, as this mainly depends on whether the author is still a university member or not. Library staff checks if the author is still studying or working in the University. If they are not a University member, the contact details provided when they presented their doctorate at the University are used.

The author is contacted by email explaining the possibility of digitizing their thesis and

9. https://www.uab.cat/web/studies/phds/published-theses-1345688357239.html

uploading it to the repository in open access. It is specified that the costs of digitization are at the expense of the user making the request.

The author is provided with an authorization form to publish the thesis in open access in the repository. The author must also specify under which open access licence they want the work to be available. The author is also provided with the document of the Open Access Commission[10] of the UAB that indicates which Creative Commons licences are recommended for the different types of publications. In the case of a thesis, the Creative Commons licence recommended is CC-BY-SA,[11] but authors can always decide to publish their thesis in the repository under another open access licence.

Once the author signs the self-certification, the user is informed of the digitization budget, the document is digitized, sent to the user, uploaded to the repository, and the author is informed that the document is now available in open access.

It cannot be known a priori whether PhD theses uploaded to the repository will receive substantial numbers of requests; the only data known is that at least one researcher has shown an interested in it.

The digital version of the theses can also be retrieved through the library discovery system, which increases visibility.

The statistics of views and downloads of doctoral theses uploaded to the institutional repository, after the procedure of a request in the Interlibrary Loan Service, shows an average of 3.2 visits per month and two downloads per month (Table 2).

The statistical figures highlight the importance of carrying out this digitization procedure of doctoral theses when a researcher makes a request, since the interest in its content goes far beyond of the interest of the researcher that has made the ILL request.

Since the process of uploading doctoral theses to the repository began, theses from different knowledge disciplines and in different languages (Catalan and Spanish) have been incorporated.

Statistics show that all disciplines and languages have views and downloads beyond the interest of the requesting researcher. The document with lowest impact has an average of 1.2 monthly views and 0.3 monthly downloads and a total of 29 visits and seven downloads (Table 2), a figure that more than justifies the work involved in managing the digitization of these doctoral theses.

The statistics also show that a significant number of visits and downloads of these documents are performed outside Spain. As a curiosity, analysing the digitization of the doctoral thesis "Burgos Rincón, Javier. Printing and book culture in eighteenth-century Barcelona (1680-1808)", it can be observed that 91% of the visits received in this case come from outside Spain.

10. https://www.uab.cat/web/research/open-access-uab/open-access-board-1345692601011.html
11. https://ddd.uab.cat/record/129205

After analysing this procedure, and seeing the results, the possibility of systematically digitizing the doctoral theses of which we only keep a print version is worthy of consideration.

| Author | Added into the repository on | Total Views | Total Downloads | Avg Views per month | Avg Downloads per month |
|---|---|---|---|---|---|
| Burgos Rincón, Javier[12] | 22/09/2022 | 11 | 7 | 2.4 | 1.6 |
| Coll, Araceli[13] | 16/04/2021 | 69 | 27 | 3.2 | 1.2 |
| Borràs, Josep[14] | 15/03/2021 | 80 | 39 | 3.5 | 1.7 |
| Bacaria, Jordi[15] | 24/02/2021 | 29 | 7 | 1.2 | 0.3 |
| Moliner Rodríguez, Jordi[16] | 09/02/2021 | 90 | 61 | 3.8 | 2.6 |
| Martínez Ribas, Ricardo[17] | 11/11/2020 | 144 | 130 | 5.4 | 4.8 |
| **Total Average** | | **70.5** | **45.2** | **3.2** | **2.0** |

*Table 2: Statistics of views and downloads of the theses published in the repository (as of 13 February 2023)*

## Conclusion

Collaboration with other departments both in the library and in the University offers diverse opportunities to create new value-added services for the interlibrary-loan department and for the ability to work together to detect users' needs, and—consequently—to be able to provide a better service.

---

12. https://ddd.uab.cat/record/265579/usage?ln=en
13. https://ddd.uab.cat/record/239021/usage?ln=en
14. https://ddd.uab.cat/record/237646/usage?ln=en
15. https://ddd.uab.cat/record/237051/usage?ln=en
16. https://ddd.uab.cat/record/236534/usage?ln=en
17. https://ddd.uab.cat/record/233784/usage?ln=en

# About the Author

Isàvena Opisso, Librarian

Universitat Autònoma de Barcelona | Universidad Autónoma de Barcelona (Autonomous University of Barcelona)

https://orcid.org/0000-0003-3337-014X

Isàvena Opisso (1973) became university librarian at the Universitat Autònoma de Barcelona in 2019. Prior to this position, she worked as a university librarian at the Universitat de Barcelona (2009-2018) and public library librarian in the public libraries of Olot and La Garriga (2008-2009). Her current work focuses on interlibrary loan and article processing charges.

# PART II

# DDA AND EBA AS OPPORTUNITIES

# Purchase on demand via interlibrary loan: Analysis of the first four years at University of Antwerp Library

**Jennifer Van den Avijle and Luca Maggiore**

## Abstract

A library cannot and should not purchase every book ever published. It should however nurture a collection development policy guided by the needs of the patrons and environment in which it operates. Interlibrary loan offers patrons an affordable, easy and quick way to meet the most pressing needs where hiatuses in the collection are concerned. However, international loan requests for books can be costly and have a high delivery runtime with only a short lending period. Purchasing these titles rather than requesting them can turn out to be more cost- and time effective. In addition, adding requested titles to the collection is a sure way of developing the collection based on actual needs. The University of Antwerp Library implemented a Purchase on Demand pilot in 2018. This paper presents our findings after four years of POD-ILL as an additional acquisition model and partial alternative to international ILL. We provide a short introduction on the concept of POD-ILL and take a closer look at the collection purchased through POD-ILL. As expected, most purchases were made for the faculties of Law and Social Sciences and the departments of History and Literature. As an alternative to international ILL, the main benefit is the extended loan due to an owned versus lent copy. Our yearly budget was more than sufficient to cover the cost, which has provided some leeway in the purchase strategy and parameters. We have recently opted for inclusion of study books that are in high demand.

## Keywords

Purchase on Demand; POD; Interlibrary loan; ILL; Demand-driven acquisition; DDA; University of Antwerp

# Article

## Purchase on Demand

Every library had its own collection development budget, strategies and goals. At the University of Antwerp, each faculty provides a collection development budget that is managed by the library department Acquisition and Metadata. Collection development decisions are made by the Faculty Library Advisor, on the advice of our E-info and Acquisitions library departments. Some faculties have a clear strategy and sufficient budget and develop their collection adequately. Some faculties have the budget, but no clear strategy and other faculties invest almost exclusively in research and therefore simply have no library budget, leaving students and researchers without recent and relevant literature, be it print or electronic.

Regardless of budget and strategies, the library collection can never satisfy all patron needs. Document delivery services and interlibrary loan offer affordable, easy and quick ways to fulfil the most pressing needs if certain items are not held in the collection. For articles and book chapters specifically, electronic document delivery is a very affordable and efficient alternative to purchase.

ILL of print items is an amazing service that makes collections all over the globe available. International ILL of print items however might not always be the best way to satisfy patrons' urgent needs. It regularly takes a long processing and delivery time and the loan period is relatively short, even more so for research purposes (Chan, 2004, p. 27). The patron is often forced to copy or scan large parts of the book before returning it. International requests can be quite expensive. In many cases, the patron reimburses only a part of the delivery cost and the library has to further supplement the delivery fee.

This is where Purchase on Demand through Interlibrary Loan comes in. Instead of lending a requested book through ILL, the book is purchased from an online vendor with speedy delivery such as Amazon (Ward, 2002, p. 96). The book is catalogued and added to the collection before or after lending it to the requesting patron. In effect, it's an additional acquisition model that doesn't replace, but rather supplements the existing collection development strategies and offers an affordable alternative for international, expensive or slow ILL requests (Alder, 2007; Imamoto, 2016, p. 372; Zopfi-Jordan, 2008). By purchasing, the library is certain the acquisition is meeting a real need in the collection development. The library is also certain the book will be lent at least once, which can't be said for other books selected for purchase through other collection development strategies (Ward, 2002, p. 96).

# Implementing POD-ILL, proposal and discussion

Purchase on Demand is a well-known concept in the library world ([Gee, 2014](), pp. 134-135) and is even included as a feature in larger ILL systems ([Foss, 2008](), p. 307). The University of Antwerp library's ILL system, Impala, which was developed in-house, does not, however, operate with an integrated POD module.[1] Furthermore, library collection budgets are provided and managed by the faculties themselves. All purchases on those budgets must be selected or approved by a Faculty Library Advisor, an academic staff member appointed by the faculty. A number of departments and faculties do invest sufficient time and budget in collection development, and standing orders often supplement the advisor's selection. Quick and easy purchase of items requested and needed by patrons is however not encouraged by this financing and selecting method.

Based on a short literature review the Acquisitions and Metadata team manager wrote a proposal to implement POD-ILL as an alternative to the print ILL and supplemental acquisition model. The proposal was discussed on a department head meeting in 2018 and further tweaked after addressing some concerns.

Expected positive effects that were agreed upon during the discussion were:
- Fast accessibility through purchase via online vendors such as Amazon.
- Inclusion of relevant books otherwise not obtained though classical collection development methods.
- Demand-driven acquisition, providing patrons with what they want and need quickly without depending on Library Faculty Advisors overdeveloping their own field of research.
- Avoiding possibly high international ILL delivery costs.
- Extended loan due to inclusion in the library print collection.

Some concerns were also raised and discussed, such as:
- Not submitting all eligible POD-ILL requests, resulting in disproportionate acquisition.
- Purchase of niche works or titles with low usage ([Tyler et al., 2010](), pp. 162-163).
- Academic staff catching on to this new strategy, and overusing ILL request as an alternative to the classical collection development strategy and financing.
- Purchase of E-books was initially included, but with high costs and specific licensing issues we have not yet included them in our parameters. We have also excluded purchase of articles.

---

1. Impala, Belgian ILL system: [https://anet.be/doc/impala/impalae/html/index.html](https://anet.be/doc/impala/impalae/html/index.html)

After discussion, all department heads agreed on implementing a pilot-project for POD-ILL. After one year an analysis of all POD-ILL requests would yield more information about the concerns raised earlier and would reveal whether the pilot could and should be continued in its agreed upon form.

The library provided a yearly POD-ILL acquisition budget of 10,000 euros. The initial parameters to determine if a book was eligible for purchase are listed below, but could be altered during the project.

- Print books, only accessible via international ILL.
- Requests by University of Antwerp academic staff.
- Books supplementing the collection profile.
- Recently published.
- Maximum cost of 60 euros.

## Workflow

There is no direct purchase module included in our ILL software Impala. To implement this, extensive development would be required, but the return on investment for this development is considered too limited. So to start POD-ILL, an additional administrative and acquisition workflow had to be set up and implemented throughout multiple library departments, from the ILL department to acquisition, through cataloguing back to ILL and from there to circulation. This workflow was meticulously documented on our online documentation platform.

### Incoming ILL request

An ILL request is submitted by a patron through our various catalogues or the ILL form and activated by the ILL staff in Impala. If the request is compliant with POD-ILL parameters it is sent to a lender string of two fictional ILL suppliers, all managed by the ILL staff.

- The first supplier is 'UA-POD'. All call numbers given to this supplier are 'acquisition + date'
- The second supplier is 'UA-X'. This supplier is used whenever we handle requests that are placed and hopefully fulfilled via ILL platforms other than Impala. The call number given to this supplier is '?'.

The submitted ILL form is printed as a pdf and immediately mailed to acquisitions general mail with mail subject 'POD-ILL'.

## Acquisition and cataloguing

If the request is in fact compliant with all parameters, the acquisitions team purchases the monograph online as soon as possible. A selection code 'POD-ILL' and the Impala request number is added as an internal message in the order form.

If the book is not available in retail or is not compliant to the parameters, the ILL staff member is informed by mail. The ILL staff will log in into Impala as the fictional supplier 'UA-POD' and refuse the request, causing the request to go to the next and final fictional lender 'UA-X' in the lender string. When the request pops up in our UA-X environment, we start the international ILL request route.

If, however, the requested title is available with an online vendor, the book is ordered as soon as possible. Upon arrival, a blue card with patron name, requesting campus library and Impala number of the original ILL request is added to the book. The book is then sent to our Metadata colleagues for cataloguing. By adding a sigillum UA-POD-IBL, the origins of this book as a POD-ILL book can always be traced. POD-ILL books are given a standard loan class, but an exception is made for the Law collection. This collection is only lent during weekends.

## ILL delivery to patron

After cataloguing, the book is handed over or sent to ILL staff at the correct campus library. Again we take on our fictional role as supplier to our own library, by logging into Impala as UA-POD library. The status of the pending ILL request is processed as 'success'.

In our own Impala library account, we see the outstanding request as successfully handled by the lending library. We await the arrival of the item and then confirm the delivery. The patron is sent the correct mail, informing them the item is available and ready for loan. Impala will charge the patron the normal ILL fee.

The item is sent to the shelved requested documents at the correct circulation desk. The blue card in the book marks it as a POD-ILL book. All library staff are instructed to clarify upon loan that this means the book has been purchased and is therefore available for loan for three months (UAntwerp staff) and not three weeks as with other ILL requests. Renewal is possible, which is rarely the case with ILL. The POD-ILL Law collection is an exception as it adheres to restricted loan terms and regulations. After loan or after pick-up time has expired, the book is added to the collection in open stack.

## Tweaking the parameters

After one year of establishing the workflow, the set parameters proved too strict. Publication date seemed less relevant, and the maximum cost of 60 euros per title was insuf-ficient. In addition, after discussion with the ILL department, we decided to expand the parameters to also include some requests placed by students. The parameters were adjusted as follows:

- Print books, only accessible via international ILL
- Requests by University of Antwerp academic staff
- Books supplementing the collection profile and no syllabi, tourist guides, schoolbooks, etc.
- Maximum cost per book is 150 euros. The maximum of 60 euros proved insufficient.
- Books requested by students, if in high demand or if mandatory/highly recommended literature and yet not purchased by the faculty or department.

## Analysis of four years POD-ILL purchases

### Submitted vs purchased requests

A total of 428 ILL requests were submitted for POD-ILL by our ILL staff, with totals of 121 in 2018, 84 in 2019, 119 in 2020 and 104 in 2021 (Fig. 1). The vast majority was submitted by staff at the Library of Humanities, Social and Design Sciences. Only five requests were submitted by staff at the Biomedical Library.

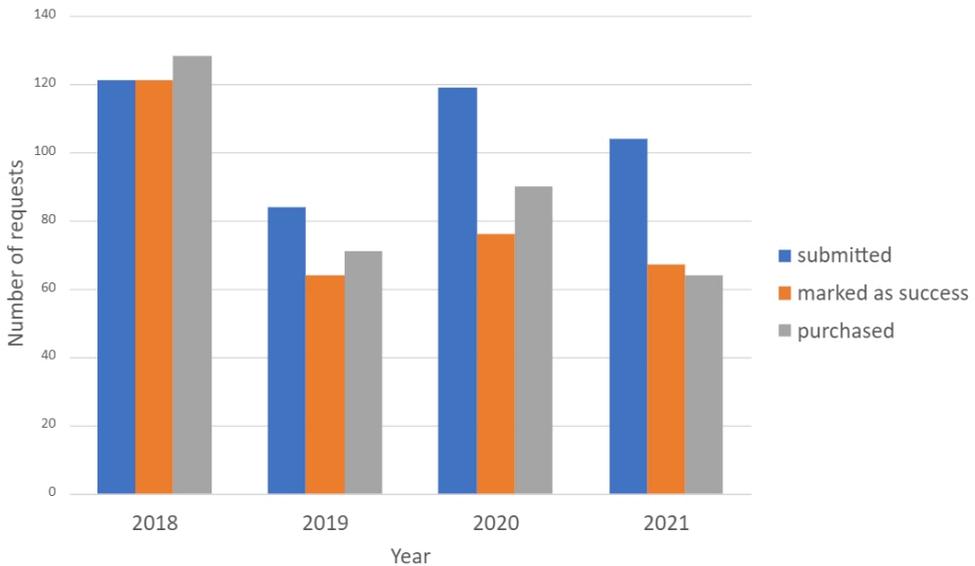

*Figure 1: Yearly overview of submitted vs purchases requests*

In 2018 all requests were marked in Impala as successfully delivered. For the following years, however, the number of failed POD-ILL requests has increased. From 20 failed requests in 2019 to 43 requests in 2020 and 37 in 2021. A total of 100 submitted requests were marked as failed.

It should be noted the successful requests in Impala do not match the number of orders placed and received via acquisitions. On analysing the data, between 2018 and 2021 a total of 40 requests could not be purchased and was therefore correctly marked unsuccessful.

The remaining 60 failed requests can solely be attributed to overdue delivery runtimes of ordered titles, even when purchased via online vendors. This means a part of the outstanding POD-ILL requests in Impala expired before the title was delivered to the library. The library staff had to resubmit these requests a second or even third time.

It is notable that each year, the number of purchased titles is higher than the successfully handled POD-ILL requests. A proportion of ILL requests marked as failed were in fact successful, but expired without resubmitting the request a second time. Given that the patron receives an automated e-mail concerning the successful or failed submission of their ILL request, ILL staff should be aware of the discrepancies in communications. In this regard, the workflow needs further tweaking.

Furthermore, we can see the number of submitted POD-ILL requests by library staff is generally on the low side. Further analysis is needed to determine if this is caused by an unwillingness or inattentiveness to implement POD-ILL or if there simply are few requests that fit the parameters.

## Requesting patrons

As Figure 2 shows, 148 unique patrons have placed at least one ILL request that resulted in a purchase through POD-ILL. All patrons so far have been university staff members. The parameter that facilitates purchases for students in some cases may need further promotion.

A little over 50% of individual patrons placed an ILL request that resulted in one delivered POD-ILL. Only 12 patrons have submitted ILL requests that resulted in five to nine purchases and four staff members have discovered the benefits of requesting via ILL and benefited from 10 or more purchases.

Apart from these 16 patrons, we see no large excesses in POD-ILL for certain patrons. Further analysis of the purchased titles for these 'big requesters' is needed, to determine if the titles are considered niche. If so, it might be better to decline the requests and advise them to purchase via their own research budget.

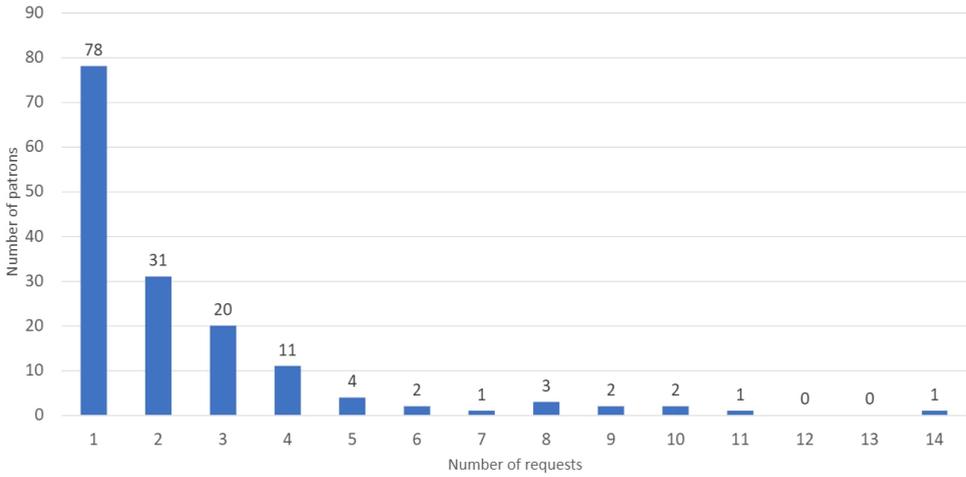

*Figure 2: Number of requests per patron*

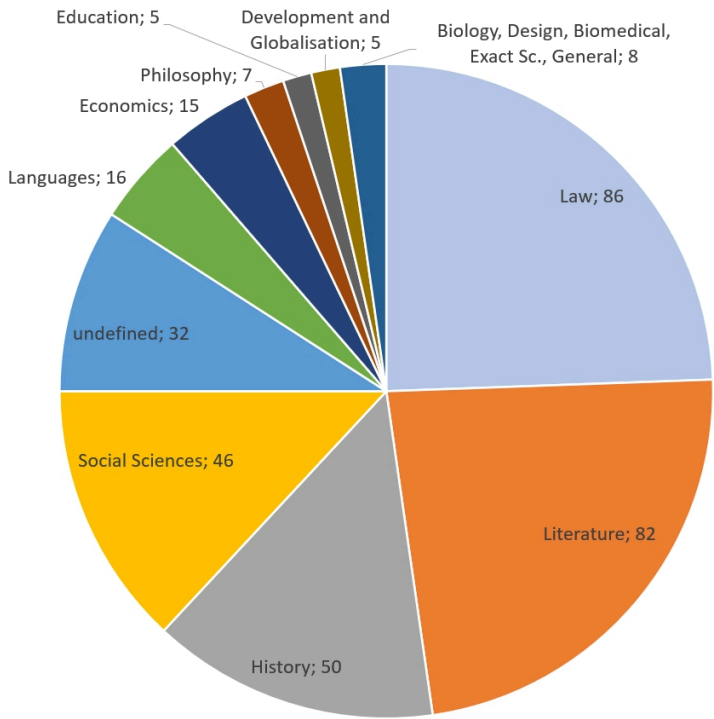

*Figure 3: Number of purchases per discipline*

As expected, the vast majority of purchases was made for the faculties or departments of Law, Literature, History and Social Sciences (Anderson et al., 2002, p. 3). Unfortunately, there is quite a large portion of 'undefined' requests that were not labelled with the appropriate cluster. A preliminary analysis of these titles however supports the overall conclusion stated above. Titles for Exact and Biomedical Sciences are rarely purchased. These findings are not surprising and are in line with our expectations. These research fields are more article oriented and online document delivery largely suffices for these disciplines (Fig. 3).

## Loan after purchase

From the 352 POD-ILL titles purchased about 7% was surprisingly never lent out. Our own ILL staff assesses the submitted ILL requests and checks if they comply with the parameters for purchase through POD-ILL. This might result in undesirable purchases. We did receive feedback from some patrons that in quite a few cases the delivery run-time was too high and the requester no longer required the title for research.

Luckily though, 93% of all POD-ILL purchases was lent out at least once and half of those were lent out from two up to 22 (!) times (Fig. 4). These loan numbers even exclude renewals, so loan could actually be even higher.

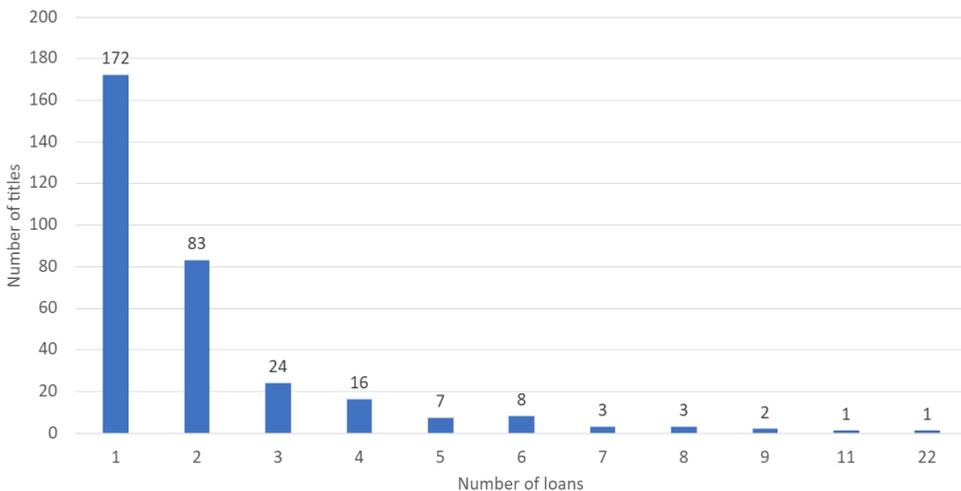

*Figure 4: Overview of loans per title*

## Vendors

In defining our pilot project as a means to quickly purchase and deliver ILL, we meant to acquire through online vendors such as Amazon and Bol.com. The library has a credit card for online purchases of publications. The financial department splits the amount of the monthly credit card bill between the different individual orders (and individual budgets).

No less than 70% of all purchases was in fact directly placed with Amazon (.fr, .de, .nl or .co.uk). We aimed to order within European borders as much as possible. All remaining orders delivered via other vendors were also purchased through Amazon. The largest suppliers after Amazon were The Book Depository, Bol.com and Dietmar Dreier. A myriad of small sellers delivered the remaining 49 requests (Fig. 5). With some exceptions, purchases are faster than ILL deliveries.

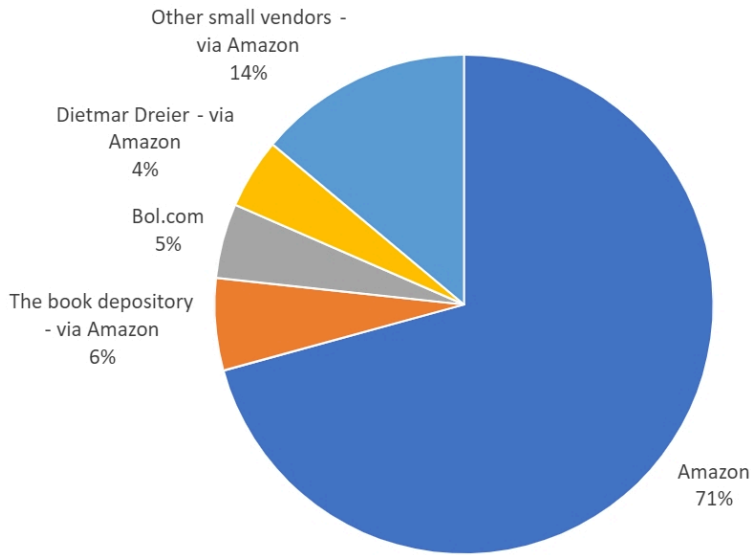

*Figure 5: Purchases per supplier*

# Evaluation

## Patron feedback

We have held no quantitative or qualitative survey, so feedback has mostly been garnered through e-mail or orally at the service desk. Patron feedback we have received has been positive to extremely positive. Only one or two patrons have complained about their ILL fee, seeing as the material was purchased and normally purchase is billed via the faculty acquisition budget. Our POD-ILL viewpoint concerning costs however still stands: the title would never have been purchased if it weren't for the ILL request. If we hadn't purchased the item, the patron would still have paid the due fee for a regular ILL or even more, if costs for delivery were exuberant. Thanks to POD-ILL however, the patron pays for swift delivery after purchase, with the added bonus of being able to lend the item for as long as needed.

## *Staff perception*

The staff perception is also largely positive, though some are hesitant to consider purchase because of a high amount of acquisitions for one or several departments or indeed faculty staff members. It would appear however that ILL staff are also hesitant to purchase due to prolonged delivery runtimes. This was not an issue we thought we would face. In some instances, it is still preferable to order international ILL in the Netherlands, Germany or German speaking Switzerland due to a short delivery runtime and relatively low ILL and postal costs. We have established clear terms of delivery through OCLC WorldShare ILL with Dutch research and university libraries. For our German-speaking neighbours, it's fairly easy and cost-efficient to order through Subito.

ILL staff are still however satisfied to be able to order titles that are more difficult or costly to obtain via ILL. Removing the parameter of only ordering for faculty staff was celebrated by our ILL staff, though we rarely see a purchase for students being submitted via POD-ILL.

## *Library managers' perception*

Our Library managers clearly see the benefits of POD-ILL. The proposed annual budget is more than sufficient to cover the cost of the purchases, and we can guarantee satisfied patrons while moderately supplementing the collection with required literature. We annually discuss the POD-ILL numbers, and in 2021 the library managers even discussed the continuation of the project as a whole. All were in favour of keeping the project on as an additional acquisition model that should not be called into question again, provided the yearly budget remains sufficient to cover the cost.

## Conclusion

As expected, most POD-ILL purchases are made for the disciplines Law, Literature, Social Sciences and History. These are the faculties and departments that already invest most heavily in collection development concerning monographs. These researchers generally place more ILL request, which in turn result in more POD-ILL. We have no indication that the POD-ILL system is being misused to divert purchasing and acquisition costs. The yearly allotted budget is more than sufficient to support POD-ILL. A few researchers will be made aware they are better served purchasing certain titles on their own research budget.

Staff members are thankful the requested title is included in the collection and therefore available for research without much restriction. Only a few patrons have contested the ILL fee, even though the purchase was made only due to their initial ILL request. Although we order through big online vendors to ensure quick delivery time, the delivery runtime is way below expectations.

The purchased titles are deemed appropriate and reflect the collection, though they do sometimes represent niche research. The orders mostly fit the preset parameters. These parameters can easily be adjusted, based on new emerging needs. As some faculties and departments are less preoccupied with collection development, we've expanded our parameters to include urgent ILL requests placed by students for mandatory reading. This expansion of the parameters hasn't really borne results yet, so staff must be made aware of this possibility again.

The administration concerning POD-ILL does not require much additional staffing time. The workflow could be improved with an automated integration between our ILL and acquisition software. This would require extensive development of Impala, which does not yield enough benefit. Our current workflow is a required work-around that serves its purpose nicely though.

In our experience POD-ILL is a perfectly adequate service for our university staff and students.

# Bibliography

Alder, N. L. (2008). Direct purchase as a function of interlibrary loan: Buying books versus borrowing. *Journal of Interlibrary Loan, Document Delivery & Electronic Reserve*, 18(1), 9–15. https://doi.org/10.1300/J474v18n01_03

Anderson, K. J., Freeman, R. S., Hérubel, J.-P. V. M., Mykytiuk, L. J., Nixon, J. M., & Ward, S. M. (2003). Buy, Don't borrow: Bibliographers' analysis of academic library collection development through interlibrary loan requests. *Collection Management*, 27(3–4), 1–11. https://doi.org/10.1300/J105v27n03_01

Chan, G. R. Y. C. (2004). Purchase instead of borrow: An international perspective. *Journal of Interlibrary Loan, Document Delivery & Information Supply*, 14(4), 23–37. https://doi.org/10.1300/J110v14n04_03

Foss, M. (2008). Books-on-demand pilot program: An innovative "patron-centric" approach to enhance the library collection. *Journal of Access Services*, 5(1–2), 305–315. https://doi.org/10.1080/15367960802199117

Gee, C. W. (2014). Book-buying through interlibrary loan: Analysis of the first eight years at a large public university library. *Journal of Interlibrary Loan, Document Delivery and Electronic Reserve*, 24(5), 133–145. https://doi.org/10.1080/1072303X.2015.1018473

Hostetler, M. (2010). Purchase-on-demand: An overview of the literature. *Against the Grain*, 22(2). https://doi.org/10.7771/2380-176X.5505

Imamoto, B., & Mackinder, L. (2016). Neither beg, borrow, nor steal: Purchasing interlibrary loan requests at an academic library. *Technical Services Quarterly*, 33(4), 371–385. https://doi.org/10.1080/07317131.2016.1203642

Tyler, D. C. (2011). Patron-driven purchase on demand programs for printed books and similar materials: A chronological review and summary of findings. *Library Philosophy and Practice (e-journal)*, (6). https://digitalcommons.unl.edu/libphilprac/635

Tyler, D. C., Melvin, J. C., Epp, M., & Kreps, A. M. (2014a). Don't fear the reader: Librarian versus interlibrary loan patron-driven acquisition of print books at an academic library by relative collecting level and by Library of Congress classes and subclasses. *College and Research Libraries*, 75(5), 684–704. https://doi.org/10.5860/crl.75.5.684

Tyler, D. C., Melvin, J. C., Epp, M., & Kreps, A. M. (2014b). Patron-driven acquisition and monopolistic use: Are patrons at academic libraries using library funds to effectively build private collections? *Library Philosophy and Practice (e-journal)*. http://digitalcommons.unl.edu/libphilprac/1149

Tyler, D. C., Melvin, J. C., Xu, Y., Epp, M., & Kreps, A. M. (2011). Effective selectors? Interlibrary loan patrons as monograph purchasers: A comparative examination of price and circulation-related performance. *Journal of Interlibrary Loan, Document Delivery and Electronic Reserve*, 21(1–2), 57–90. https://doi.org/10.1080/1072303X.2011.557322

Tyler, D. C., Xu, Y., Melvin, J. C., Epp, M., & Kreps, A. M. (2010). Just how right are the customers? An analysis of the relative performance of patron-initiated interlibrary loan monograph purchases. *Collection Management*, 35(3-4), 162–179. https://doi.org/10.1080/01462679.2010.487030

Ward, S. M., Wray, T., & Debus-López, K. E. (2003). Collection development based on patron requests: Collaboration between interlibrary loan and acquisitions. *Library Collections, Acquisition and Technical Services*, 27(2), 203–213. https://doi.org/10.1016/S1464-9055(03)00051-4

Ward, S. M. (2002). Books on demand: Just-in-time acquisitions. *The Acquisitions Librarian*, 14(27), 95-107. https://doi.org/10.1300/J101v14n27_12

Zopfi-Jordan, D. (2008). Purchasing or borrowing: Making interlibrary loan decisions that enhance patron satisfaction. *Journal of Interlibrary Loan, Document Delivery and Electronic Reserve*, 18(3), 387–394. https://doi.org/10.1080/10723030802186447

# About the Authors

## Jennifer Van den Avijle, Head of User Services
Universiteit Antwerpen (University of Antwerp)
https://orcid.org/0000-0001-8545-9168

Jennifer Van den Avijle (1987) is an archaeologist and librarian by training. After three years in the public library sector, Jennifer has been working as head of user services for the Biomedical Library and Library of Exact and Applied Sciences at the University of Antwerp since 2014. She manages the user services team, optimises loan and ILL services, oversees rebuilding and refurbishment projects, looks for new ways to expand our services and functions, all while trying to dissuade the team from adopting a library cat.

## Luca Maggiore, Librarian
Universiteit Antwerpen (University of Antwerp)
https://orcid.org/0000-0001-7827-080X

After high school in Velletri, Italy, Luca Maggiore (1968) graduated in Modern History at KU Leuven (Belgium) and later he obtained a postgraduate degree in Information and library science at the University of Antwerp. He currently works at the University of Antwerp as head of the Acquisition and Metadata Department (Library). He has published in the fields of library science and the history of Italy.

# Patron-Driven Acquisition et Evidence-Based Acquisition : Comment implémenter ces modèles afin d'étendre l'offre documentaire dans une bibliothèque ?

**Marion Favre and Manon Velasco**

## Résumé

Les modèles d'acquisition d'e-books basés sur l'utilisation comme le PDA et l'EBA sont aujourd'hui très répandus et permettent la mise à disposition de vastes collections que les bibliothèques n'ont pas réellement acquises. Un PDA d'Ebook Central a été mis en place en 2019 et un premier EBA de Wiley en 2021. La Bibliothèque de l'EPFL a testé ces modèles et a réalisé de nombreuses analyses pour en mesurer la pertinence après une ou plusieurs années, au regard des besoins de ses usager·ères. Les avantages et les limites de l'exploitation des statistiques d'usage dans le développement des collections ont également été mis en lumière dans cette étude. Ces deux modèles se sont révélés de bons compléments aux modèles d'acquisition dans lesquels les bibliothécaires de liaison effectuent la sélection. Tous deux allègent également considérablement ce travail de sélection, tout en augmentant les tâches de gestion et d'analyse du côté des bibliothécaires en charge des ressources électroniques. Le PDA a nécessité des ajustements au fil des années pour que les collections accessibles soient au plus près des domaines de l'EPFL. L'utilisation des titres acquis à travers ce modèle est globalement plus importante que celle des titres acquis par d'autres biais. En contrepartie le PDA présente un risque accru de générer des acquisitions à la popularité éphémère, malgré la mise en place d'un système de prêt avant l'achat définitif. Le programme d'EBA a révélé un intérêt des usager·ères pour des titres dont les sujets et les années de publication n'étaient pas forcément attendus. Ce modèle d'acquisition nécessite un gros investissement au niveau des analyses statistiques de la part des bibliothécaires chargé·es des ressources électroniques, mais les résultats de ces analyses permettent d'aller au-delà de la seule nécessité de sélection des titres au terme du programme d'EBA. Ces deux modèles nous ont amenés à nous interroger plus largement sur l'évolution du rôle de sélection des bibliothécaires, et sur la part de l'intervention humaine dans des modèles où les statistiques d'usage sont le matériau de départ pour la sélection.

# Abstract

Nowadays, usage-based acquisition models for e-books, such as PDA and EBA, are quite widespread and allow libraries to offer wider collections without acquiring them. A PDA with Ebook Central was implemented in 2019 and a first EBA with Wiley in 2021. The EPFL Library ran several analyses in order to evaluate their accuracy after one or more years, with respect to users' needs. The advantages and limitations of the use of usage statistics in collection development were also highlighted in this study. These two models have proven to be a good complement to the acquisition models based on liaison librarians' e-book selections. Both also greatly alleviate this selection work, while increasing the management and analysis tasks for the librarians in charge of electronic resources. The PDA model has required adjustments over the years, to ensure that accessible collection contents better match EPFL domains. The usage of the titles acquired through this model is generally higher than that of those acquired otherwise, nevertheless the PDA presents an increased risk of acquiring titles with an ephemeral popularity, despite the implementation of a loan system before the final purchase. The EBA program showed an interest among patrons in titles, related topics and publication years that were not necessarily expected. This acquisition model required a large investment in statistical analysis by the librarians responsible for electronic resources, but the results of these analyses made it possible to go beyond the sole need to select titles at the end of the EBA program. These two models prompted us to question more broadly the evolution of the role of selection made by librarians, and the part of human intervention in models where usage statistics are the starting point for the selection.

# Keywords

Patron-Driven Acquisition; PDA; Evidence-Based Acquisition; EBA; E-book; Livre électronique; Acquisition; Model; Modèle; Library; Bibliothèque; Collection development; Développement des collections; Usage statistics; Statistiques d'usage

# Article

## Introduction

La Bibliothèque de l'École polytechnique fédérale de Lausanne (EPFL) acquiert en moyenne 8 000 e-books par an, pour un budget d'environ 500 000 CHF. Les contenus sélectionnés couvrent de façon privilégiée les domaines d'études et de recherche de l'Ecole : mathématiques, physique, chimie, sciences du vivant, informatique et communications, sciences et techniques de l'ingénieur, ingénierie de l'environnement, architecture et urbanisme, génie civil, management de la technologie, technologie et politiques publiques et ingénierie financière. Les achats sont principalement réalisés en fonction des sélections des bibliothécaires de liaison et des propositions d'achat adressées par les usager·ères. Les demandes d'acquisition d'e-books sont traitées par l'équipe des ressources électroniques, depuis la demande de devis jusqu'à la mise à disposition dans notre outil de découverte (Ex Libris Primo), en passant par la négociation des tarifs et des conditions de licence. La Bibliothèque achète des e-books à la pièce, des collections, des bouquets annuels et/ou thématiques et souscrit également à des abonnements. Mais est-ce que ces acquisitions correspondent réellement aux besoins des usager·ères ? Chaque année, une part importante des titres acquis l'année précédente n'est, en effet, jamais consultée par les usager·ères.

Afin de mieux prendre en compte les besoins réels (et non supposés) des usager·ères, la Bibliothèque de l'EPFL a mis en place ces dernières années deux nouveaux modèles d'acquisitions pour les achats d'e-books : le *Patron-Driven Acquisition* (PDA) et l'*Evidence-Based Acquisition* (EBA). Ces modalités sont venues compléter les pratiques existantes (achats et abonnements), tout en donnant accès à des titres que la bibliothèque n'a pas réellement acquis. Les modèles PDA et EBA permettent la mise à disposition de vastes collections documentaires et dans lesquelles des acquisitions peuvent être faites en fonction des usages.

Dans cet article, une description, ainsi que les avantages et inconvénients de chaque modèle seront mis en avant, tout comme les détails du processus de mise en place et les modalités de sélection des titres. Finalement, plusieurs statistiques concernant les achats et usages, et les implications pour la Bibliothèque seront présentées.

## Patron-Driven Acquisition

### Description du modèle

Le *Patron-Driven Acquisition* (PDA) ou *Demand-Driven Acquisition* (DDA) est un modèle d'acquisition, où l'achat d'un titre est déclenché par une utilisation réelle. Les e-books sélectionnés dans un PDA sont visibles dans l'outil de découverte et consultables, avant d'avoir été acquis par la Bibliothèque. Ce modèle est proposé principalement par des agrégateurs d'e-books, comme EBSCO ou ProQuest Ebook Central par exemple, plutôt que par des éditeurs. Le fournisseur met à disposition des usager·ères un ensemble de documents (« pool ») défini selon des critères choisis par la Bibliothèque. Différents paramétrages pour la sélection des titres sont disponibles, tels que la langue, la date de publication, les thématiques, ou encore l'éditeur (ce qui permet d'exclure ceux avec lesquels la Bibliothèque travaille déjà en direct). Les fournisseurs proposent plusieurs options pour le déclenchement de l'achat : la location, la médiation avant achat ou l'achat direct. Ce déclenchement s'effectue lorsqu'un·e usager·ère consulte réellement le document (défini par un téléchargement ou un temps de lecture en ligne). Le fonctionnement est totalement invisible pour l'usager·ère qui, dans l'outil de découverte de la Bibliothèque ou sur la plateforme du fournisseur, ne voit aucune différence entre les e-books rendus disponibles par PDA et les e-books achetés de manière pérenne ou consultables via abonnement.

La réflexion derrière la mise en place de ce modèle d'acquisition dans une bibliothèque porte sur la pertinence des modes d'acquisition : vaut-il mieux acheter des documents sans besoin constaté des usager·ères, ou attendre la demande d'un·e usager·ère ? Walker et Arthur (2018) évoquent la différence « *Just-in-time* rather than *just-in-case* » qui illustre bien la logique du modèle de PDA.

## *Avantages et inconvénients du PDA*

Les avantages de ce modèle sont multiples. Tout d'abord, la certitude que la ressource sera utilisée, au moins une fois. Il permet ensuite de donner accès à des ressources nouvelles, ou auxquelles les bibliothécaires de liaison ne songent pas. Ce modèle permet enfin de réaliser un achat d'e-book en un temps très court, car l'acquisition ne nécessite pas l'intervention des bibliothécaires (de liaison et des gestionnaires de ressources électroniques), le processus étant automatisé (sauf si l'option de médiation avant achat est sélectionnée). L'intervention des bibliothécaires gestionnaires des ressources électroniques se situe en amont (paramétrage du pool) et en aval (analyse des titres achetés pour évaluer leur pertinence et leur utilisation).

Néanmoins, le PDA comporte également certains désavantages. Suivant la méthode de validation d'achat retenue, il n'est pas forcément possible de contrôler finement la pertinence des titres acquis : c'est le cas de la méthode d'achat sans médiation. Certaines collections thématiques sont larges et regroupent des ouvrages de niveau hétérogène et sur des sujets très variés, qui peuvent ne pas correspondre au profil de l'institution, même si la thématique générale semble pertinente. Dans le cas du PDA mis en place à l'EPFL, la thématique « géographie » avait semblé intéressante dans un premier temps : a posteriori il s'est avéré qu'elle contenait des e-books pertinents, mais également, après analyse des achats, des guides de voyage dont l'intérêt est moindre pour une école polytechnique.

Pour un accès illimité à un titre, le prix constaté des e-books disponibles via des agrégateurs est plus élevé que si l'achat est effectué directement chez l'éditeur, et les accès sont souvent plus restrictifs (DRM) : limites d'accès simultanés, de téléchargement, d'impression, etc. Il est également souvent nécessaire d'avoir un logiciel externe (ex. : Adobe Digital Editions) pour pouvoir télécharger le titre dans son intégralité.

## Test du PDA à la Bibliothèque de l'EPFL

La Bibliothèque de l'EPFL a commencé par tester ce modèle d'acquisition pendant une année, de septembre 2019 à septembre 2020. Trois fournisseurs ont été identifiés et évalués avant lancement : ProQuest, EBSCO et Dawsonera. Tous trois offraient des fonctionnalités de paramétrage du pool, d'ajustement de la somme consacrée aux acquisitions, ou encore de production de rapports. Le choix s'est finalement porté sur ProQuest Ebook Central, en raison de la taille et de la couverture éditoriale de son catalogue, ainsi que de son interface plus intuitive.

Lors de la période d'essai, plusieurs critères ont été retenus lors du paramétrage du pool :
- Date d'ajout des documents sur la plateforme, date de copyright et date de publication plus récente que 01.01.2018 ;
- Exclusion des éditeurs chez lesquels nous acquérons des e-books en direct (à la pièce, packages et abonnements) ;
- Prix d'achat par titre : entre 25 et 175 EUR ;
- Langue : français ou anglais ;
- Thématiques en rapport avec le profil d'enseignement/recherche de notre institution, choisies en collaboration entre les bibliothécaires gestionnaires des ressources électroniques et les bibliothécaires de liaison : 30 thèmes sélectionnés sur 54, exclusion des œuvres de fiction et de la littérature jeunesse.

Avec le fournisseur retenu, l'achat d'un titre se fait suivant l'option « 1U » (*One user*) avec un seul accès simultané. Pour éviter l'acquisition d'un titre qui ne sera peut-être utilisé qu'une seule fois, une option de location « short term loan » a été mise en place avant l'achat : cette option n'est pas disponible pour l'ensemble des titres du pool, mais lorsque la location coûte moins de 100 EUR ou moins de 50 % du prix d'achat du titre, une location d'un ou de sept jours se déclenche. L'achat définitif se fait après deux locations. Pour les titres sans option de location, l'achat se fait dès la première consultation.

Pour le PDA, la Bibliothèque de l'EPFL fixe une enveloppe budgétaire, d'où seront débités les montants des titres achetés. Ce dépôt est réapprovisionné au besoin. Certains fournisseurs proposent également de régler les achats avec une facture en fin de mois. Nous avons préféré le dépôt afin de contrôler et évaluer les dépenses plus facilement. Au début du test, un premier dépôt de 5 000 EUR a été effectué, puis un deuxième versement, pour la même somme, huit mois plus tard. Pour ne pas introduire de biais d'appré-ciation, l'équipe des ressources électroniques a décidé d'utiliser un budget dédié et de ne pas prélever cette somme sur les enveloppes affectées aux bibliothécaires de liaison.

À la fin de l'année de test, 105 titres avaient été achetés et 199 loués, pour un total de 224 titres uniques, certains ayant été loués puis achetés. Le coût moyen d'un e-book acheté dans ce modèle PDA est de 65,3 EUR et de 9,3 EUR pour une location. 8 381 EUR ont été dépensés sur les 10 000 EUR versés durant la phase de test.

Walters (2012) note qu'il est important de contrôler les achats de titres par PDA, que ce soit par médiation au moment de l'achat ou par analyse une fois l'achat réalisé. À la fin du test, nous avons pris le temps d'analyser les titres achetés pour identifier les dysfonctionnements éventuels et ajuster le paramétrage du pool.

Cette analyse nous a permis de réaliser que la sélection thématique initiale n'était pas suffisamment fine : certains thèmes étaient trop génériques et ne correspondaient pas au profil enseignement/recherche de notre institution. Nous avons trouvé certains titres non conformes à la politique documentaire de la Bibliothèque, qui n'auraient pas été achetés selon la voie traditionnelle, comme des guides de voyage ou des biographies d'hommes politiques par exemple. Après l'analyse des titres achetés, nous avons également réévalué les thèmes mis à disposition pour valider leur pertinence avec l'aide des bibliothécaires de liaison.

En parallèle de l'analyse thématique, la Bibliothèque a changé de résolveur de liens. Jusqu'en décembre 2020, la Bibliothèque de l'EPFL utilisait le résolveur de liens SFX (Ex Libris) pour signaler les ressources électroniques. Afin de ne pas le surcharger avec les notices provenant du PDA, le test a été réalisé avec un nombre limité de documents. Il y avait environ 20 000 titres au début du test et 30 000 à la fin. Désormais, avec Alma (Ex Libris) qui a remplacé SFX, il n'y a plus de contraintes liées à la limite du nombre de notices.

Grâce au changement de notre SIGB et du résolveur de liens, ainsi qu'aux analyses effectuées à la fin de la période de test, les critères de paramétrage du pool ont été révisés comme suit en mai 2021 :

- Date d'ajout des documents sur la plateforme, date de copyright et date de publication avec un *moving wall* des trois dernières années ;
- Pas de changement concernant l'exclusion des éditeurs chez lesquels nous achetons des e-books en direct (à la pièce, packages et abonnements) ;
- Prix d'achat maximum : 200 EUR ;
- Pas de changement sur la langue des documents : langue française ou anglaise ;
- Thématiques en rapport avec le profil d'enseignement/recherche de notre institution, choisies en collaboration entre les bibliothécaires des ressources électroniques et les bibliothécaires de liaison : 27 thèmes sélectionnés, exclusion des ouvrages de fiction et de la littérature jeunesse ; neuf thèmes ont été supprimés et six ont été ajoutés.

Après ces modifications, le nombre de documents dans le pool est passé à près de 30 000 titres. La taille n'a donc pas augmenté comme on aurait pu le penser, mais son contenu a été affiné, pour offrir des documents plus pertinents au regard de la politique documentaire. L'analyse des acquisitions survenues depuis la modification du pool montre que ces changements ont porté leurs fruits : il y a eu très peu d'e-books non pertinents et seules des erreurs d'indexation de la part du fournisseur peuvent ponctuellement afficher des titres inadaptés.

L'ajout et la maintenance des titres du PDA de ProQuest dans notre outil de découverte (Primo) ont été grandement facilités avec l'arrivée d'Alma. En effet, Ex Libris et ProQuest ont créé un profil d'importation quotidien, gérant automatiquement dans Alma les ajouts et suppressions des titres du pool. Les titres acquis changent également automatiquement le statut dans Alma de « Candidat PDA » en « Titre acquis ». Ainsi, il est facile de créer des rapports dans Alma Analytics, l'outil statistique d'Alma, pour connaître les titres faisant partie du PDA ou pour créer des analyses sur les acquisitions par PDA. Une fois paramétré, il n'y a plus besoin d'intervention de la part des bibliothécaires pour tenir à jour les titres provenant du PDA. Cela représente un important gain de temps par rapport aux autres modèles d'acquisition qui nécessitent une gestion manuelle dans le SIGB durant les différentes étapes de leur acquisition.

## Bilan après deux ans

Après plus de deux ans d'utilisation du PDA, nous disposons d'un recul suffisant pour évaluer sa pertinence et mesurer correctement ses bénéfices.

Notre outil de découverte s'est considérablement étoffé grâce au pool (plus de 30 000 titres), le temps de mise à disposition est réduit à zéro et logiquement, chaque document acquis à travers le modèle PDA a fait l'objet d'au moins un accès.

Nous avons surveillé si les titres du PDA suscitent des consultations sur le moyen terme. En 2021, 114 titres sur les 220 acquis depuis le lancement du projet du PDA ont été utilisés, ce qui correspond à 51,8 %. Concernant les titres acquis hors PDA sur Ebook Central, 75 titres sur 247 acquis[1] ont été utilisés, soit 30,3 %. Les documents provenant du PDA sont donc largement plus utilisés que ceux acquis par achat direct chez Ebook Central. A noter tout de même que 43 % des titres acquis via le PDA l'ont été pendant l'année 2021, ce qui leur garantit au moins un usage durant l'année.

Durant l'année 2021, il y a eu 221 locations pour un montant de 2 384,30 EUR. Le prix moyen de la location est de 10,8 EUR. Le coût des locations a donc légèrement augmenté par rapport à l'année de test. Sur les 178 titres uniques qui ont eu une ou deux locations, seulement 24 ont été finalement achetés. La demande pour des documents que nous n'avons pas à la Bibliothèque a donc été avérée.

---

1. Titres acquis durant la même période que le modèle PDA (septembre 2019- décembre 2021).

Les achats de l'année 2021 représentent 96 titres, pour une somme de 5 346,96 EUR. Le coût moyen d'un achat d'e-book est de 55,7 EUR, ce qui est plus bas que durant le test (65,3 EUR). C'est également plus bas qu'un e-book acheté chez Ebook Central hors du modèle PDA (120,6 EUR en moyenne).

En 2021, la Bibliothèque de l'EPFL a acquis 430 e-books à la pièce, tous fournisseurs confondus. Les achats par PDA constituent donc près de 20 % des acquisitions des e-books à la pièce (Figure 1). Le modèle d'acquisition par PDA offre ainsi un bon complément aux achats des bibliothécaires de liaison. Il est intéressant de se pencher sur la répartition par thématique des titres acquis par PDA (Tableau 1).

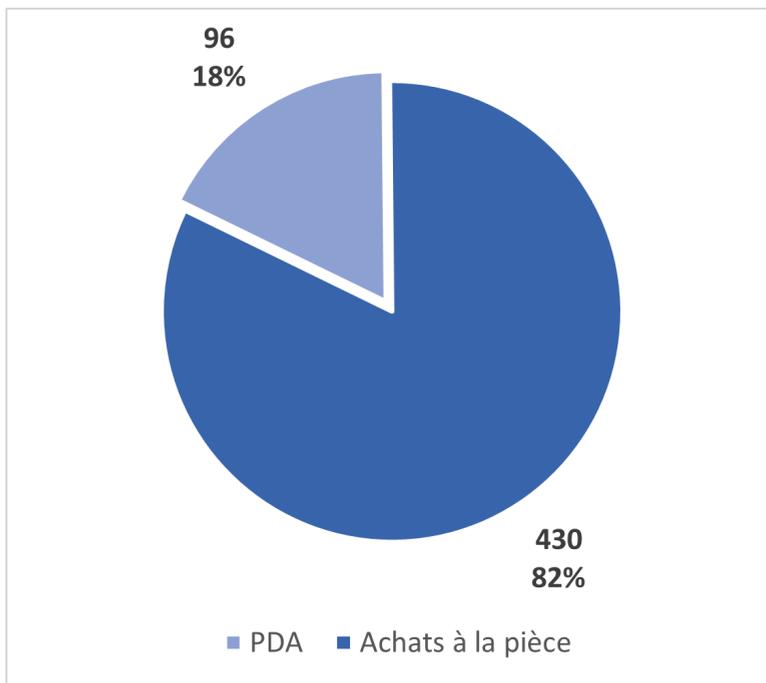

*Figure 1 : Comparaison des achats à la pièce et des achats PDA (2021)*

| Liaison | Titres PDA | Titres achat à la pièce |
|---|---|---|
| Informatique | 36 | 13 |
| Sciences humaines et sociales | 14 | 27 |
| Sciences et techniques de l'ingénieur | 14 | 13 |
| Architecture | 8 | 10 |
| Management | 7 | 16 |
| Sciences de la vie | 6 | 13 |
| Environnement | 4 | 60 |
| Mathématiques | 4 | 36 |
| Urbanisme | 1 | 14 |
| Chimie | 1 | 94 |
| Génie civil | 1 | 40 |
| Physique | 0 | 22 |
| Divers | 0 | 72[2] |

*Tableau 1 : Comparaison du nombre d'achats par liaison*

La répartition des achats PDA par domaine montre que les liaisons qui bénéficient du plus grand nombre d'achats par PDA sont des liaisons qui achètent principalement peu de titres à la pièce. Par exemple, la liaison « informatique » fait peu d'achats à la pièce, mais on remarque que la demande des usager·ères est élevée, que ce soit au niveau des achats (Tableau 1) ou des usages (Tableau 3).

Les achats du PDA peuvent ainsi orienter les bibliothécaires de liaison sur des sujets populaires auprès de nos usager·ères et actuellement peu couverts dans nos collections. La liste des achats par PDA, transmise chaque année aux bibliothécaires de liaison, peut les aider à étoffer leurs choix d'acquisition, tout en améliorant le paramétrage du pool.

---

2. Ce nombre correspond au rachat de titres suite à la faillite de Dawsonera en 2020. Ces titres n'ont pas été payés avec les budgets de liaison.

| Thèmes | Nombre de titres |
|---|---|
| Business/Management | 7 067 |
| Fine Arts (Architecture) | 6 348 |
| Medicine | 3 958 |
| Education | 3 439 |
| Economics | 3 152 |
| Engineering | 2 577 |
| Computer Science/IT | 2 310 |
| Science | 2 258 |
| Social Science | 1 800 |
| Agriculture | 1 504 |

*Tableau 2 : Top 10 des thèmes les plus fournis dans le pool PDA (2021)*

Concernant les titres utilisés, ils correspondent bien aux domaines d'étude et de recherche de notre institution. Le tableau suivant indique les cinq sujets les plus consultés dans le cadre du PDA, d'après une analyse de ProQuest Ebook Central (Tableau 3). Ils comprennent toutes les utilisations des titres du pool, locations et achats compris. Les sujets les plus consultés sont en adéquation avec notre politique documentaire et les thématiques que nous couvrons à la Bibliothèque.

| Sujets | Nombre de *unique title* |
|---|---|
| Computer Science/IT | 234 |
| Engineering | 83 |
| General Works/Reference; Science | 38 |
| Architecture | 36 |
| Mathematics | 32 |

*Tableau 3 : Top 5 des thèmes consultés (2021)*

Nous remarquons que le sujet « Computer Science/IT » est largement plus populaire que les autres sujets, que ce soit en nombre d'achats ou au niveau des usages. Si l'on met en relation les cinq sujets les plus consultés (Tableau 3) avec le total des documents disponibles dans le PDA (Tableau 2), le sujet « Computer Science/IT » est bien plus utilisé proportionnellement au nombre de titres disponibles (septième sujet le plus fourni) que les sujets les plus fournis comme « Business/Management » ou « Fine Arts » (ce qui est cohérent, puisque ces derniers ne font pas partie des facultés de l'EPFL). « Mathemat-ics » représente bien les changements effectués dans le pool en 2021. Ajouté en mai 2021, il est déjà dans le top cinq des sujets les plus consultés.

## *Enseignements du modèle PDA à la Bibliothèque de l'EPFL*

Bien que les titres provenant du PDA soient plus utilisés sur Ebook Central que les achats à la pièce, il serait malvenu de retenir ce seul modèle d'acquisition au détriment des autres. Pour Walters (2012), les usager·ères qui déclenchent l'acquisition de ces titres via le PDA ne sont pas formé·es aux acquisitions et n'ont pas une vision d'ensemble de notre collection, contrairement aux bibliothécaires. Certains des titres acquis sont très généralistes ou en rapport avec un sujet d'actualité, leur popularité sera sans doute éphémère, ce qui sera moins le cas des acquisitions réalisées par les bibliothécaires de liaison. Le PDA est un bon complément aux autres modèles d'acquisition, mais d'après Walters, il n'est ni possible ni souhaitable d'acquérir les e-books uniquement par ce biais.

À l'inverse, après un test du modèle PDA sur une période donnée, la bibliothèque de l'Université de l'Alabama a décidé d'utiliser le PDA comme modèle d'acquisition principal. D'après les auteur·rices, il ne faut justement pas se baser sur la qualité du contenu, mais uniquement sur ce qui est réellement utilisé par les usager·ères, quelle que soit la qualité ou la durée d'utilisation (Arthur & Fitzgerald, 2020).

Cela renvoie au questionnement essentiel, présent dans les bibliothèques depuis de nombreuses années : comment acquérir les meilleurs documents pour la bibliothèque et les usager·ères ? Quels sont les critères importants ? La pertinence d'un document doit-elle être définie par les bibliothécaires ou renvoie-t-elle à ce que les usager·ères utilisent réellement ? Comment acheter intelligemment lorsque l'on a un budget limité ? Est-ce pertinent d'acheter des titres qui ont un intérêt thématique pour la collection, mais sans certitude qu'ils seront consultés ou plutôt attendre que les usager·ères demandent un titre, quitte à ce qu'il ne soit plus consulté par la suite ? Faut-il acheter les titres ou les demander en prêt entre bibliothèques ailleurs ? Toutes ces questions sont légitimes et il n'existe actuellement pas une réponse universelle. Chaque bibliothèque décide ce qui est le plus adapté au profil du public ou de l'institution qu'elle dessert.

La Bibliothèque de l'EPFL a fait le choix du PDA en complément d'autres modes d'acquisition, après un test fructueux. Elle y voit la possibilité de donner à ses usager·ères l'opportunité d'intervenir dans ses acquisitions. Naturellement, les bibliothécaires gestionnaires des ressources électroniques continuent à suivre de près la pertinence et le coût de ces achats. L'impact organisationnel très faible de ce modèle, sur les bibliothécaires de liaison ainsi que les bibliothécaires de l'équipe des ressources électroniques, induit un grand gain de temps contrairement aux autres types d'acquisition. Le budget consacré annuellement à ce modèle représente une mince partie du budget total consacré aux acquisitions d'e-books (moins de 2 % du budget total). Le PDA s'inscrit ainsi dans une volonté de diversifier les modalités d'acquisition, chacune contribuant à un développement pertinent et raisonné des collections.

## Evidence-Based Acquisition

### *Description du modèle*

Le modèle d'acquisition dit « Evidence-Based Acquisition » (EBA) ou « Evidence-Based Selection » (EBS) se fonde sur les statistiques d'usage, et à la différence du PDA, les acquisitions ne sont pas immédiates, elles résultent de l'analyse des usages sur une longue période. Dans ce modèle, une somme de départ est payée par la bibliothèque cliente pour ouvrir l'accès pendant une période donnée, généralement une ou deux années, à tout ou partie de la collection d'e-books d'un éditeur. Selon l'éditeur, la bibliothèque dispose de plus ou moins de flexibilité dans la somme de départ et/ou dans le choix des collections accessibles. Les lecteur·rices peuvent ainsi consulter sans restriction un grand nombre de titres, souvent plusieurs milliers, pendant toute la durée du programme EBA, sans pour autant que ces titres aient tous été acquis. Au terme de la période de l'EBA, les statistiques d'usage des collections ouvertes sont fournies par l'éditeur et la bibliothèque peut sélectionner des titres, généralement au prix catalogue, dans les limites de la somme payée au départ. Les titres choisis en fin de période sont acquis de manière pérenne et resteront donc accessibles après la fin du programme.

De nombreux éditeurs proposent aujourd'hui ce modèle : on peut citer Wiley, Taylor & Francis, Oxford University Press, Cambridge University Press, JSTOR ou encore Elsevier. Springer a également fait une première incursion sur le terrain de l'EBA en 2021 avec son programme « Access & Select » essentiellement conçu pour les bibliothèques n'ayant que peu de contenus déjà acquis sur SpringerLink.

### *Avantages et inconvénients de l'EBA*

Le premier avantage de ce modèle est l'accès à un grand nombre de titres sans restriction d'utilisation de type DRM, dès leur parution pour les plus récents, pendant au moins toute une année sans les acquérir. Le second avantage est la réalisation d'acquisitions en fonction des besoins exprimés et non supposés des usager·ères en fin de période d'accès. Ces acquisitions sont en effet essentiellement réalisées sur la base des statistiques d'usage COUNTER, même si nous verrons que la contribution des bibliothécaires de liaison a son importance. Bien que ces statistiques demeurent un indicateur majeur de la pertinence des collections en bibliothèque, elles ne rendent compte que d'un usage brut qu'il convient d'analyser pour tenter de déterminer s'il exprime un besoin réel et/ou durable.

Ce modèle d'acquisition a déjà été très étudié dans les bibliothèques universitaires et les consortiums de bibliothèques qui l'utilisent parfois de manière intensive. L'étude de Abresch et al. (2017) expose clairement les objectifs des programmes d'EBA pour l'Orbis Cascade Alliance : prévisibilité des coûts, stabilité des titres et du signalement, titres récents sans DRM, large éventail de contenus répondant aux besoins des diverses institutions membres d'un consortium. Cette étude relève également quelques difficultés, en particulier la gestion et la communication aux bibliothécaires qui deviennent plus complexes à mesure que de nouveaux programmes d'EBA sont adoptés, chacun ayant ces spécificités en termes de coût, de collections accessibles ou encore de durée.

Un des principaux inconvénients de ce modèle est d'abord le prix des titres acquis qui ne bénéficient généralement pas de rabais, à la différence par exemple des achats dits « Pick&Mix » qui consistent en l'acquisition d'un grand nombre de titres dans une même commande chez un même éditeur. Un autre désavantage est une gestion plus complexe et nécessitant davantage de temps et de personnel pour les équipes en charge des ressources électroniques, contrairement à d'autres modalités d'acquisition des e-books comme l'achat de collections ou le PDA.

## L'expérience de l'EBA de Wiley à la Bibliothèque de l'EPFL

La Bibliothèque de l'EPFL a commencé au premier janvier 2021 deux programmes d'EBA d'un an, aux profils bien différents, avec les éditeurs Wiley et Wageningen Academic Publishing. La Bibliothèque travaillant déjà depuis de nombreuses années avec Wiley, avait déjà réalisé de nombreuses acquisitions pérennes et disposait aussi d'abonnements d'e-books, alors que Wageningen était un tout nouveau fournisseur. Seul le programme EBA de Wiley a fait l'objet d'analyses poussées, en raison de l'importance des collections et de l'ancienneté de la collaboration, et sera donc étudié en détail dans cet article.

Wiley proposait deux options à prix fixes : l'accès aux trois années de parution 2019-2021 tous domaines confondus ou l'accès aux dix années de parution 2011-2021 tous domaines confondus. Ces options, ainsi que leur prix, n'étaient pas flexibles, les années de parution et les thématiques ne pouvaient pas être adaptées, et les ouvrages de référence étaient exclus. D'autre part, la Bibliothèque de l'EPFL se trouvait confrontée à l'augmentation très importante du coût de la collection annuelle d'e-books en chimie de Wiley, qu'elle acquérait en principe chaque année et qui constituait l'essentiel de la dépense chez cet éditeur pour les e-books, le reste étant consacré aux achats à la pièce dans divers domaines. Sur la base des dépenses des années précédentes, bien qu'inférieures certaines années à la somme de l'EBA, la Bibliothèque a donc choisi la pre-mière option du programme EBA : l'accès aux trois années de parution 2019-2021 pour 25 000 USD.

Le rapport entre la valeur des collections accessibles (2019-2021), c'est-à-dire l'addition des prix à la pièce de tous les titres, et le prix de l'EBA était favorable : la valeur des collections accessibles représentait environ 14 fois le prix de l'EBA. Negrea (2019) note que la moyenne est généralement autour de 10 fois, mais cela peut parfois être plus, ou moins comme l'ont noté McCord et Hendrikx (2021) qui annoncent pour leurs 11 EBA une moyenne de sept fois le prix de la mise de départ.

Le contrat avec Wiley permettait à la Bibliothèque de l'EPFL d'opérer une première sélection après dix mois d'usage en novembre 2021, puis une seconde sélection affinée en janvier 2022 grâce aux statistiques des deux derniers mois de l'année 2021. Au début du mois de novembre 2021, les statistiques d'usage COUNTER étaient disponibles et ont été enrichies et analysées par la bibliothécaire chargée des acquisitions d'e-books, ce qui a apporté divers résultats dont la Bibliothèque a pu tirer des enseignements, notamment pour étayer la prise de décision. L'enrichissement des statistiques brutes a représenté un important travail d'ajout des données de thématiques et de prix, mais aussi de repérage des titres inclus dans le programme EBA, des titres déjà acquis les années précédentes, et des ouvrages de référence exclus du programme EBA. Il s'est également avéré que la colonne « Year of Publication » (YOP)  du rapport COUNTER TR_B1 ne correspondait pas nécessairement aux années de parution ou « Copyright year » des listes de titres disponibles sur le site de l'éditeur. Il n'a donc pas été possible de se fier uniquement à cette donnée pour isoler les titres 2019-2021 inclus dans notre programme EBA.

Environ 2 263 titres étaient inclus dans les collections accessibles dans notre programme EBA au 1er novembre 2021, et 10,5 % d'entre eux étaient présents dans les statistiques d'usage, soit 237 titres. On peut éventuellement ajouter 69 titres de 2019 et 2020, donc inclus dans les collections EBA, qui ont été utilisés mais avaient déjà été acquis précédemment. Depuis que la Bibliothèque de l'EPFL travaille avec le fournisseur Wiley, elle a acquis 1 781 e-books, et 525 d'entre eux ont été utilisés en 2021, soit 29,5 % (Tableau 6).

Au total, 2 075 titres ont été utilisés en 2021 sur les 10 mois (Figure 2), incluant des titres accessibles dans le cadre de l'EBA, des titres acquis les années précédentes (ouvrages de référence inclus) et des titres non acquis. Les 237 titres inclus dans le programme EBA représentent 11,4 % des titres utilisés. Le reste des titres utilisés sont pour 25,3 % des titres acquis précédemment et pour 63,3 % des titres qui ne sont ni acquis, ni inclus dans les collections EBA. Dans ce dernier cas, l'usage porte sur la consultation gratuite des premiers chapitres. Ce nombre important relativise la traduction en besoins des usages comptabilisés dans les rapports COUNTER, ces e-books non accessibles n'ont majoritairement pas fait l'objet d'un prêt de la version imprimée, de demande d'acquisition ou de demande de PEB.

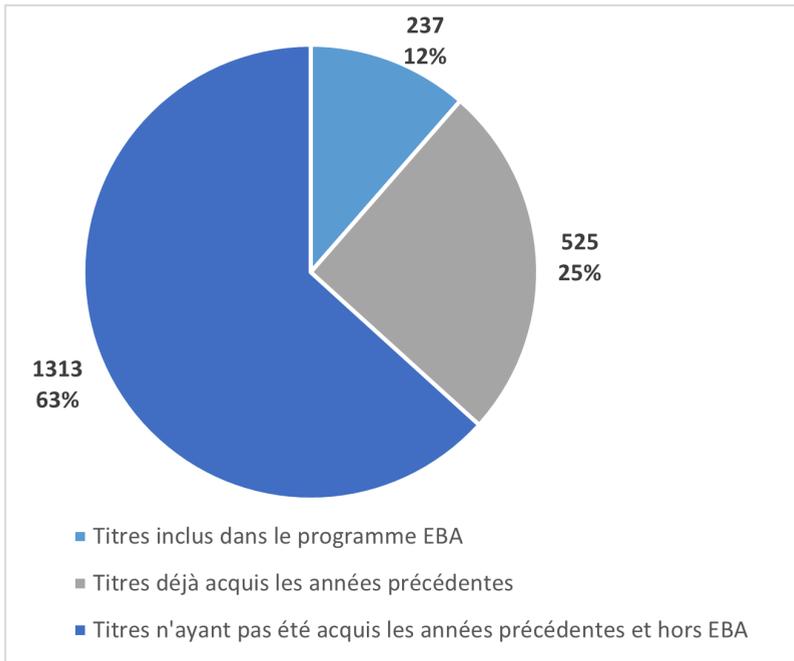

*Figure 2 : Distribution des titres utilisés en 2021*

Le total des usages « Unique Title Request » (COUNTER TR_B1) est de 5 768, et les titres inclus dans l'EBA représentent 445 usages, soit 7,7 %. Les autres usages sont pour 53,4 % des usages de titres acquis précédemment et pour 38,9 % des usages de titres non-acquis. Les titres acquis les années précédentes ont été utilisés en moyenne 5,8 fois chacun, ce qui reste dans la tendance des années précédentes, en revanche les titres inclus dans l'EBA ont été utilisés en moyenne 1,8 fois chacun (Tableau 4). En effet, seulement un tiers des 237 titres EBA utilisés l'ont été plus d'une fois (Tableau 5), et le titre le plus utilisé l'a été douze fois.

| Catégories | Nb moyen d'usages |
|---|---|
| Titres déjà acquis (525 titres utilisés 3 080 fois) | 5,87 |
| Titres EBA (237 titres utilisés 445 fois) | 1,88 |
| Titres non-acquis (1 313 titres utilisés 2 243 fois) | 1,71 |

*Tableau 4 : Usage moyen des titres en 2021*

| Usage des titres EBA | Nb de titres | % de titres |
|---|---|---|
| Utilisés plus de 4 fois | 18 | 7,6 % |
| Utilisés plus de 2 fois | 42 | 17,7 % |
| Utilisés plus de 1 fois | 82 | 34,6 % |
| Utilisés seulement 1 fois | 155 | 65,0 % |

*Tableau 5 : Répartition des usages des 237 titres EBA utilisés en 2021*

| | |
|---|---|
| Nombre de titres acquis jusqu'au 01.01.2021 (dont ouvrages de référence) | 1 781 |
| Nombre de titres acquis utilisés en 2021 (dont ouvrages de référence) | 525 |
| Pourcentage des titres acquis utilisés en 2021 | 29,5 % |

*Tableau 6 : Titres acquis utilisés en 2021*

La répartition thématique des titres des collections EBA utilisés a montré la prédominance de deux domaines, en lien avec la composition thématique des collections d'e-books de Wiley : les sciences de l'ingénieur pour 47,7 % et la chimie pour 18,1 % (Tableau 7). La répartition thématique des statistiques 2021 dans leur entièreté est en revanche nettement dominée par la chimie, ce qui est parfaitement logique, étant donné les acquisitions largement majoritaires dans ce domaine durant de nombreuses années (Tableau 8).

| Thèmes | Nb de titres | % de titres utilisés | Nb d'usages (UTR[3]) | % des usages |
|---|---|---|---|---|
| Physical Sciences & Engineering | 113 | 47,7 % | 241 | 54,16 % |
| Chemistry | 43 | 18,1 % | 91 | 20,45 % |
| Computer Science & Information Technology | 16 | 6,8 % | 24 | 5,39 % |
| Mathematics & Statistics | 16 | 6,8 % | 22 | 4,94 % |
| Earth, Space & Environmental Sciences | 10 | 4,2 % | 15 | 3,37 % |

*Tableau 7 : Répartition thématique des 237 titres utilisés en 2021 et inclus dans le programme EBA, top 5*

| Thèmes | Nb de titres | % de titres utilisés | Nb d'usages (UTR) | % des usages |
|---|---|---|---|---|
| Chemistry | 774 | 37,30 % | 2 955 | 51,23 % |
| Physical Sciences & Engineering | 711 | 34,27 % | 1 569 | 27,20 % |
| Life Sciences | 100 | 4,82 % | 334 | 5,79 % |
| Mathematics & Statistics | 149 | 7,18 % | 262 | 4,54 % |
| Humanities | 75 | 3,61 % | 90 | 1,56 % |

*Tableau 8 : Répartition thématique des 2 075 titres utilisés en 2021, top 5*

Les années de parution des titres consultés en 2021 sont très variées (Tableau 9), il ressort que seulement 24 % des titres consultés ont une année de parution entre 2017 et 2021, et 27 % entre 2012 et 2016.

| Année (YOP) | Nb de titres | % des titres |
|---|---|---|
| 2017-2021 | 506 | 24,4 % |
| 2012-2016 | 556 | 26,8 % |
| 2007-2011 | 446 | 21,5 % |
| 2002-2006 | 329 | 15,9 % |
| Antérieures | 238 | 11,5 % |

*Tableau 9 : Répartition par année de publication des titres utilisés en 2021*

Il en va de même des titres présents dans les statistiques d'accès refusé (COUNTER TR_B2) : 13 % des titres ont une année de parution entre 2017 et 2021, et 28 % entre 2012 et 2016.

Les statistiques d'accès refusé de 2021, mais aussi des deux années précédentes, ont également été analysées de manière plus approfondie. Nous avons ainsi pu noter dans 22,4 % des cas un recoupement des 237 titres EBA utilisés en 2021 avec les statistiques d'accès refusé des deux années précédentes. Ces statistiques d'accès refusé, ainsi que les statistiques d'usage des titres non-acquis en 2021 (pour rappel 63 % des titres consultés), nous ont montré que de nombreux titres non couverts par notre programme d'EBA semblent susciter de l'intérêt de nos usager·ères, en particulier des ouvrages de référence. En effet, le rapport TR_B2 indique que le titre avec le plus grand nombre de tentatives d'accès dispose de 178 tentatives.

Ces quelques chiffres sont riches en enseignements. Contrairement à nos attentes, les statistiques d'usage des titres inclus dans notre programme EBA n'ont pas été très élevées, et la chimie n'a pas été le domaine le plus utilisé, les titres en sciences de l'ingénieur ont été davantage consultés. Le fait que la Bibliothèque de l'EPFL ait réalisé des achats chez Wiley jusqu'à l'année précédant la mise en place de l'EBA a obligatoirement été à l'origine de quelques recoupements avec les collections incluses dans l'EBA. Et les usages des titres déjà acquis parus en 2019 et 2020 ont été assez bons (pour rappel, 29,3 % des titres acquis avec année de parution 2019 ou 2020 ont été consultés en 2021). Les autres acquisitions réalisées par les bibliothécaires de liaison les années précédentes peuvent aussi être jugées globalement pertinentes car bien utilisées, en nombre de titres (29,5 %) et en nombre d'usages (en moyenne 5,8 fois chacun). Globalement les statistiques d'usage de 2021 ont montré que les titres les plus utilisés ne sont pas forcément les plus récents, d'où la relative faiblesse des statistiques d'usage des titres du programme EBA tel qu'il était en place en 2021. Cela se vérifie aussi au niveau des statistiques d'accès refusé, les titres les plus plébiscités ne sont pas récents. McCord et Hendrikx (2021) ont également noté dans leurs programmes d'EBA que les titres récents ne sont pas plus utilisés que les titres anciens. Le test de l'EPFL confirme qu'il faut souvent plusieurs années après sa parution à un titre pour trouver son public ou le chemin vers la bibliographie des professeur·es. Si besoin était de le prouver encore, le règne de l'immédiateté des journaux ne s'applique pas aux monographies.

## La sélection finale des e-books

La Bibliothèque de l'EPFL avait décidé de ne pas imputer directement la somme sur les budgets des bibliothécaires de liaison afin de se focaliser en priorité sur les statistiques d'usage. Une répartition initiale du montant de l'EBA entre différent·es bibliothécaires aurait pu favoriser une sélection des titres par domaine, plutôt que selon les statistiques, pour correspondre à la somme dépensée par chacun au départ. Cependant il avait été établi que la liaison de chimie bénéficierait de la primauté dans la sélection finale des titres, dans la mesure où la frontlist de chimie représentait la plus grande part des dépenses des années précédentes.

Wiley a permis à la Bibliothèque de réaliser des sélections en dehors des trois années de parution accessibles dans le programme EBA. Ainsi la chimie et les autres liaisons ont pu réellement acquérir des titres qui suscitaient de l'intérêt sans se limiter à trois années de parution. La sélection du bibliothécaire de liaison en chimie en novembre 2021 a été plus faible qu'initialement envisagée, du fait du nombre de titres réellement utilisés et pas encore acquis dans ce domaine. Elle a été utilement complétée par des titres hors EBA qui figuraient dans les statistiques d'usage 2021 et aussi dans les statistiques d'accès refusé. Quelques titres inclus dans notre programme EBA mais absents des statistiques d'usage ont aussi été « repêchés » dans la sélection du bibliothécaire de liaison en chimie. Les autres domaines ont pu bénéficier d'une place plus grande qu'initialement prévue dans la sélection finale.

La sélection finale des titres, incluant la sélection de chimie, a été réalisée en janvier 2022 par la bibliothécaire chargée des acquisitions d'e-books et a été soumise aux bibliothécaires de liaison afin d'être affinée grâce à leur connaissance de leurs domaines et des pratiques de leurs communautés. Cette sélection a été réalisée selon plusieurs critères :

- Sélection du bibliothécaire de liaison de chimie ;
- Titres inclus dans les collections de l'EBA ayant été utilisés au moins une fois et ayant été signalés comme intéressants en cours d'année 2021 par certain·es bibliothécaires de liaison ;
- Nombre d'usages en 2021 et bonne répartition sur plusieurs mois de ces usages. La distribution sur plusieurs mois des usages indique en effet la constance de l'intérêt pour un titre, tandis que la concentration des usages sur une période très courte indique un intérêt éphémère ;
- Année de publication ;
- Nombre d'accès refusés en 2021 et bonne répartition sur plusieurs mois de ces tentatives d'accès ;
- Nombre d'accès refusés en 2019-2020 et bonne répartition sur plusieurs mois de ces tentatives d'accès.

Seule la bibliothécaire de liaison en environnement a souhaité remplacer des titres sélectionnés dans son domaine sur la base de signalements en cours d'année 2021 par d'autres titres utilisés un plus grand nombre de fois. Le reste de la sélection finale a été conservée telle quelle, mais tou·tes les spécialistes disciplinaires ont examiné chaque titre afin de valider sa pertinence dans la sélection. Il faut également souligner l'importance de l'analyse fine des statistiques d'usage, notamment la répartition des consultations sur plusieurs mois de l'année et l'exploitation des accès refusés. Grâce à la combinaison de ces facteurs, une partie des données brutes d'usage a ainsi pu être interprétée comme l'expression d'un besoin. La sélection soumise à Wiley à la fin du mois de janvier 2022 comptait en tout 121 titres, dont 48 pour la liaison de chimie, 45 pour les sciences de l'ingénieur, sept pour les sciences de la vie, sept pour l'environnement et 14 dans d'autres domaines. Parmi ces titres, 73 étaient inclus dans notre programme d'EBA, les autres e-books ont été sélectionnés parmi les titres avec des accès refusés et parmi les 63 % de titres consultés et non accessibles (Figure 2).

D'autres expériences montrent que la sélection ne s'effectue pas seulement sur la base des statistiques d'usage, mais qu'elle nécessite une collaboration étroite avec les spécialistes disciplinaires. À la Western University, McCord et Hendrikx (2021) notent que l'usage des monographies ne peut pas être interprété de la même manière que celui des journaux, ils remarquent aussi que les usages minimums déclenchant l'achat peuvent être affinés selon les disciplines. Certaines facultés peuvent en effet avoir été plus actives pendant la durée de l'EBA au détriment d'autres, indépendamment de leur taille, et l'implication des bibliothécaires de liaison permet de réduire le biais de la discipline privilégiée : des titres avec des statistiques légèrement trop basses par rapport au seuil peuvent être « repêchés ». D'autre part, lors de l'arrêt du programme EBA, certains titres qui ne sont plus accessibles peuvent être demandés par les usager·ères et être acquis ou intégrés dans un programme de PDA. Solomon et Gray (2018) soulignent également l'importance pour la Case Western Reserve University de la collaboration entre les bibliothécaires spécialisé·es et le personnel chargé des acquisitions lors des phases de décision. Le modèle EBA présente enfin selon eux un avantage en termes de temps pour les bibliothécaires spécialisé·es, car la mise en œuvre du programme est essentiellement gérée par les équipes en charge des ressources électroniques.

Lors de cette expérience du programme EBA de Wiley au cours de l'année 2021, la Bibliothèque de l'EPFL a rencontré quelques difficultés de signalement dans son outil de découverte. Tout d'abord, la collection des e-books Wiley 2021 dans la zone communautaire d'Alma[4] n'a été disponible et conforme aux attentes en termes de couverture des titres que plusieurs mois après le début de l'année 2021, le signalement de ces titres dans l'outil de découverte de la bibliothèque a donc pris du retard. Il s'est avéré également que les années de parution ou « Copyright year » telles qu'utilisées par Wiley dans les listes de titres disponibles sur son site web et les années de publication des titres présents dans les collections d'e-books 2019, 2020 et 2021 d'Alma ne correspondaient pas toujours, ainsi il a pu arriver que des titres exclus du programme EBA soient signalés dans notre outil de découverte. Enfin, nous avons également rencontré le cas de quelques titres retirés de la vente en cours d'année et pourtant encore référencés dans les collections d'Alma activées pour l'EBA.

L'étude de Abresch et al. (2017) souligne en effet que le signalement est rendu parfois difficile du fait de l'absence de certains titres dans la zone communautaire d'Alma. Robbeloth et al. (2017) ont étudié le programme EBA de Wiley adopté par le consortium de bibliothèques , l'Orbis Cascade Alliance,  et relèvent eux aussi les défis du signalement des titres inclus dans le programme d'EBA.

## Enseignements tirés du programme d'EBA de Wiley en 2021

En ne prenant en compte que les statistiques d'usage comme mesure des besoins, il apparaît clairement que les seules statistiques des titres du programme EBA ne suffisent pas. Il faut aussi prendre en considération les statistiques d'usage globales de 2021, notamment celles des titres non-acquis (premiers chapitres consultés gratuitement), ainsi que les statistiques des accès refusés.

---

4. La zone communautaire est un catalogue de données bibliographiques, ainsi que d'informations administratives et
   d'accès pour les ressources électroniques (partagées et enrichies par les bibliothèques et ExLibris)

Dans le cadre d'un programme EBA, il n'est en principe pas possible d'inclure d'autres titres que ceux accessibles via l'EBA dans la sélection finale, mais heureusement Wiley a permis à la Bibliothèque de l'EPFL d'effectuer ses sélections au-delà des collections incluses dans le programme EBA. Comme nous l'avons vu, les deux tiers des titres consultés inclus dans l'EBA ne l'ont été qu'une fois et, en principe, cela ne devrait pas nécessairement justifier un achat, c'est pourquoi il a été très utile de pouvoir sélectionner aussi d'autres titres. La sélection finale de la Bibliothèque a pu être finalisée et atteindre le montant initialement dépensé. Les propositions des bibliothécaires de liaison ont complété la sélection de titres préparée par la bibliothécaire chargée des acquisitions d'e-books.

L'offre pour le renouvellement du programme EBA en 2022, ainsi que l'offre pour la collection annuelle de chimie 2022, ont été reçues entre septembre et octobre 2021. La Bibliothèque a alors pris la décision de renouveler le programme EBA de Wiley en 2022, mais avec davantage d'années de parution (dix années). Cette décision a été prise d'une part sur la base des informations fournies par les statistiques d'usage sur les années de parution des titres plébiscités, et d'autre part en fonction du prix de la collection annuelle de chimie et de l'usage des titres acquis les années précédentes. Les statistiques ont en effet montré qu'environ un tiers des collections acquises était utilisé (Tableau 8), ce qui laisse penser que le coût par usage des collections annuelles et achats à la pièce est peut-être trop important. Cette décision sera évidemment questionnée à la fin de l'année 2022 et les stratégies d'acquisition chez Wiley pourront alors être revues.

Les monographies ne suscitent pas nécessairement l'intérêt des usager·ères dès leur parution et la sélection des titres conservés dans le cadre des deux années du programme d'EBA de Wiley permettront de combler, sur la base des usages, les lacunes des collections acquises avant 2021. La Bibliothèque de l'EPFL considère ainsi pour le moment ce modèle d'acquisition EBA comme une solution pertinente pour une période donnée d'une ou deux années, avant de revenir à des modes d'acquisitions plus classiques fondés sur les propositions d'achat et les statistiques d'usage (dont les accès refusés), ou d'envisager éventuellement de nouveau un programme d'EBA.

Sur le plan des ressources humaines, l'EBA représente une charge de travail plus importante que les autres modes d'acquisition pour l'équipe des ressources électroniques, notamment pour l'analyse et l'enrichissement des statistiques COUNTER brutes et pour la sélection finale. Conduire plusieurs programmes d'EBA simultanés avec des éditeurs importants ne serait pas gérable pour une équipe comme celle de la Bibliothèque de l'EPFL. En revanche, pour les bibliothécaires de liaison, le temps consacré au cours de l'année aux sélections, au traitement des propositions d'achat et au suivi budgétaire disparaît presque complètement, au profit d'un temps réduit consacré en fin d'année à la prise de décisions, sur la base de statistiques retravaillées par la bibliothécaire spécialiste des acquisitions e-books. Le rapport entre le bibliothécaire de liaison et l'acquéreur se trouve ainsi modifié par cette redistribution des responsabilités.

À titre comparatif, la Western University avait 11 programmes d'EBA en cours durant l'année 2021, McCord et Hendrikx (2021) ont présenté les enjeux et les défis de ce modèle. Ils évoquent entre autres les variations importantes d'un éditeur à l'autre en termes de collections et de prix, et la complexité des négociations chaque année : dépenses précédentes chez l'éditeur, collections ouvertes, mise de départ, fourniture des statistiques, rabais, etc. Les capacités de la bibliothèque de la Western Universityétant limitées, les opérations de gestion des programmes d'EBA doivent aussi être réalisables pour les équipes en charge des acquisitions. Ces opérations incluent la gestion du budget, le signalement pendant et après la période de l'EBA, la gestion des statistiques, et le dialogue avec les bibliothécaires spécialisé·es ou de liaison. Torbert (2020) de l'University of Mississippi remarque également qu'à la différence des modèles de PDA via un agrégateur, les bibliothèques dans un programme d'EBA travaillent directement avec l'éditeur et doivent assumer une plus grande partie de la charge administrative.

Davantage de recul sera nécessaire pour évaluer correctement l'impact du programme EBA de Wiley à la Bibliothèque de l'EPFL. Dans les prochaines années, il sera intéressant de savoir si le coût par usage des titres acquis via ce programme est plus faible que celui des autres titres acquis suivant d'autres modèles plus classiques chez le même éditeur. Le principal critère de réussite d'un programme d'EBA à la Case Western Reserve University (Solomon et Gray, 2018) est l'obtention d'un coût moyen par usage inférieur à celui des titres acquis selon d'autres modèles. D'après leur analyse et selon le profil des acquisitions globales dans leur institution, le programme EBA d'Elsevier a par exemple su remplir ce critère, tandis que celui de CRC Press a été jugé inadapté. C'est également l'approche adoptée par Tran et Guo (2021) dans les bibliothèques de l'Université de Stony Brook pour répondre aux besoins croissants en ressources électroniques avec un budget stable ou en baisse. Pour cette université, il est ainsi ressorti que les titres acquis via un programme d'EBA ont été un bon complément à la collection générale de livres électroniques, et tous les modèles d'acquisition ont continué de coexister.

Durant l'année 2021, les usager·ères de la Bibliothèque de l'EPFL ont eu accès à environ 2 500 titres, et durant l'année 2022 ce seront près de 12 000 titres accessibles avec des années de parution plus larges. Les coûts de ce modèle pour la Bibliothèque sont légèrement plus élevés que les acquisitions réalisées les années précédentes, la collection annuelle de chimie et les achats à la pièce représentaient entre 20 000 et 23 000 USD par année. Mais les collections accessibles sont quatorze fois plus importantes et les titres qui resteront accessibles en 2023 auront été choisis en fonction des besoins constatés à travers l'analyse des usages et la collaboration avec les spécialistes disciplinaires.

En plus du renouvellement du programme EBA de Wiley, la Bibliothèque de l'EPFL a également commencé un autre EBA avec l'éditeur IOP (Institute of Physics) en 2022. La différence majeure avec le programme EBA de Wiley, au-delà du coût plus faible, est avant tout le fait que la Bibliothèque n'a réalisé aucune acquisition d'e-books chez cet éditeur durant les trois dernières années, du fait de l'impossibilité de réaliser des achats à la pièce et du prix élevé des collections annuelles par rapport aux besoins estimés par le bibliothécaire de liaison chargé de la physique. Ce programme d'EBA semble donc une solution intéressante pour compléter nos collections chez cet éditeur en accord avec les besoins réels dans ce domaine.

Le panorama des offres d'EBA ne permet que peu de flexibilité aux bibliothèques clientes en termes de coûts et de collections accessibles, certains fournisseurs, indépendamment de la variété de leurs catalogues, proposent des offres trop rigides pour être adaptées aux besoins de toutes les institutions. Une offre idéale se rapprocherait davantage du PDA, elle permettrait entre autres à la bibliothèque cliente plus de souplesse dans le choix des années de parution et des thématiques accessibles. Cette offre donnerait le choix entre un accord d'une ou plusieurs années, notamment pour laisser le temps à certains titres récents de trouver leur public et pour répartir la charge de travail, et enfin elle devrait proposerait davantage de paliers dans la somme initiale.

Les offres de programmes d'EBA pourraient également être davantage pensées dans la durée, et permettre de redéfinir les sommes et les collections accessibles lors du premier renouvellement. La Bibliothèque de l'EPFL, comme de nombreuses autres, continuera de privilégier les programmes d'EBA les plus flexibles et adaptés au profil de l'institution ainsi qu'au budget disponible, et réinterrogera aussi chaque année la pertinence de ce modèle par rapport aux autres modalités de développement des collections de livres électroniques.

## Conclusion

Les acquisitions d'une bibliothèque universitaire se doivent d'être variées et aussi pertinentes que possible pour les thématiques correspondant à l'institution liée. Il est difficile pour un·e bibliothécaire de liaison de connaître et d'acquérir l'ensemble des titres pertinents parus durant l'année. Permettre une acquisition selon d'autres systèmes offre la possibilité d'enrichir la collection de la bibliothèque. Le PDA comme l'EBA sont basés sur les demandes des usager·ères, mais s'appuient très largement sur l'expertise des bibliothécaires gestionnaires de ressources électroniques et des bibliothécaires de liaison, lors de l'élaboration du pool PDA et de la sélection finale des titres à acquérir dans le cadre des EBA.

Ces modèles permettent d'acquérir un contenu qui trouve sa légitimité dans une forte demande des usager·ères, soit à travers des consultations qui déclenchent un achat soit à travers les statistiques d'usage. C'est sur la durée que ces acquisitions peuvent conserver cette légitimité en continuant d'être utilisées, ou la perdre si le besoin ne s'inscrivait pas dans le temps, ce qu'il est malheureusement difficile d'anticiper. De nombreuses autres modalités d'acquisition comme les demandes reçues en prêt entre bibliothèques, les bibliographies des professeur·es et surtout les achats des bibliothécaires de liaison, grâce à leur expertise et leur relation avec leur section restent ainsi privilégiées dans la constitution de collections d'e-books.

Il n'est néanmoins pas interdit de se questionner sur l'utilité d'une sélection humaine dans notre contexte, ou en tout cas questionner le moment de l'intervention humaine. Actuellement le processus de sélection des e-books à la Bibliothèque de l'EPFL commence par une intervention humaine, puis finit par l'analyse des données statistiques pour contrôler l'usage des titres acquis. On pourrait envisager davantage de programmes d'EBA pour inverser cette tendance comme on l'a vu avec l'expérience de Wiley : partir des usages pour effectuer une partie des sélections, puis faire intervenir l'humain. L'expérience nous a en effet montré que l'exploitation des statistiques d'usage peut se suffire à elle-même pour les titres les plus consultés, puis elle devient plus délicate quand les chiffres de consultation baissent, dans ce cas les données laissent la place à l'intervention humaine pour finaliser la sélection.

Il est important d'analyser les données à disposition pour contrôler la pertinence des acquisitions, quelles que soient leurs modalités. Les statistiques d'usage COUNTER doivent être travaillées, enrichies, analysées, et il ne faut leur faire dire que ce qu'elles peuvent dire, et se rappeler aussi qu'elles ne sont qu'un des moyens possibles pour évaluer les besoins. Il faut expérimenter, ajuster, embarquer les bibliothécaires de liaison, considérer qu'en termes d'usage, les tendances ne sont pas forcément définitives. Il faut souvent du recul pour dire si un mode d'acquisition des e-books est pertinent, il faut du temps pour savoir si un titre acquis sera utilisé, si son prix est justifié.

L'ajout de ces deux modèles d'acquisition dans notre institution a permis de mettre à disposition plus de 32 500 titres pour les usager·ères en 2021, et plus de 40 000 en 2022, sans qu'ils soient préalablement acquis par la Bibliothèque. Au début de l'année 2023, seulement une petite partie de ces titres mis à disposition sera acquise de manière pérenne par la Bibliothèque. Ces acquisitions, réalisées sur la base des besoins exprimés à l'instant T pour le PDA et des usages sur une année pour l'EBA, présentent l'avantage de limiter par ailleurs les demandes d'acquisition classiques dont le processus d'achat prend plus de temps pour aboutir à l'ouverture de l'accès. Ces méthodes d'acquisition réduisent aussi potentiellement les conséquences des besoins non couverts (car passés hors des radars de sélection des bibliothécaires de liaison) sur d'autres services de la Bibliothèque comme le prêt en réseau ou la fourniture de copies.

Pour la Bibliothèque de l'EPFL, les modèles d'EBA et de PDA sont de bons compléments aux modèles déjà utilisés, malgré leurs limites, mais ne sauraient les remplacer. Tant que la demande et les usages resteront élevés et le prix raisonnable, ces modèles continueront à faire partie des acquisitions de la Bibliothèque.

Dans un contexte de restrictions budgétaires, il est compréhensible de vouloir ouvrir de vastes collections à moindre coût, de favoriser la pertinence immédiate des acquisitions, avec des dépenses rapidement justifiables. Pour autant la mesure des besoins réels est complexe et le développement des collections d'e-books mérite l'expertise de plusieurs professionnels différents, les bibliothécaires de liaison et les bibliothécaires gestionnaires de ressources électroniques. L'expérience de ces deux nouveaux modèles d'acquisition des e-books à la Bibliothèque de l'EPFL a permis à chacun de développer ses compétences, d'apprendre à aborder avec finesse les données d'expression des besoins, les consultations et les statistiques COUNTER, de découvrir (ou redécouvrir) la complexité de ce marché du livre électronique pour les bibliothèques. Les bibliothécaires ont pu voir les limites et les possibilités des offres existantes et ont appris à composer avec elles pour accomplir au mieux notre mission de mise à disposition de l'information.

## Bibliographie

Abresch, J., Pascual, L., & Langhurst Eickholt, A. (2017). EBA in practice: Facilitating evidence-driven e-book programs in both consortium and individual library settings. *Proceedings of the Charleston Library Conference.* https://docs.lib.purdue.edu/charleston/2017/collectiondevelopment/12

Arthur, M. A., & Fitzgerald, S. R. (2020). Rethinking collection development: improving access and increasing efficiency through demand driven acquisition. *The Journal of Academic Librarianship*, 46(1). https://doi.org/10.1016/j.acalib.2019.03.005

Cramer, C. J. (2013). All about demand-driven acquisition. *The Serials Librarian*, 65(1), 87-97. https://doi.org/10.1080/0361526X.2013.800631

McCord, D., & Hendrikx, S. (2021, March 8-11). *Push and pull: Negotiating EBA programs in a pandemic.* [Conference session]. ER&L virtual event. https://pheedloop.com/erl202/virtual/?page=sessions§ion=SESHG93H2S4LUA4Z4

Negrea, S. (2019, June 18). *Acquiring e-books for college and university libraries.* University Business. https://universitybusiness.com/acquiring-e-books-for-college-and-university-libraries/

Robbeloth, H., Ragucci, M., & DeShazo, K. (2017). Evidence-based acquisition: A real life account of managing the program within the orbis cascade alliance. *The Serials Librarian*, 73(3-4), 240-247. https://doi.org/10.1080/0361526X.2017.1388331

Solomon, D., & Gray, B. C. (2018, June 24-27). *Applicability of evidence-based acquisition model to collection development in engineering subjects.* [Conference session]. ASEE Annual Conference & Exposition. Salt Lake City, UT, United States. https://doi.org/10.18260/1-2-29805

Torbert, C. (2020). Management of concurrent DDA and EBA e-book programs: Or, adventures in deduplication for acquisitions. *Theology Cataloging Bulletin*, 28(3), 17-19. https://doi.org/10.31046/tcb.v28i3.1823

Tran, C. Y., & Guo, J. X. (2021). Developing user-centered collections at a research library: An evidence-based acquisition (EBA) pilot in STEM. *The Journal of Academic Librarianship*, 47(5). https://doi.org/10.1016/j.acalib.2021.102434

Walker, K. W., & Arthur, M. A. (2018). Judging the need for and value of DDA in an academic research library setting. *The Journal of Academic Librarianship*, 44(5), 650-662. https://doi.org/10.1016/j.acalib.2018.07.011

Walters, W. H. (2012). Patron-driven acquisition and the educational mission of the academic library. *Library Resources & Technical Services*, 56(3), 199-213. https://doi.org/10.5860/lrts.56n3.199

## About the Authors

### Marion Favre, E-resources Librarian

École Polytechnique Fédérale de Lausanne (Swiss Federal Institute of Technology in Lausanne)

https://orcid.org/0000-0001-9756-8249

Marion Favre (1990) is an e-resources librarian, working mainly with ebooks. She oversees Alma for her team and also manages the library's video game collection. Marion also likes to dig in COUNTER reports and to create analyses with Analytics.

### Manon Velasco, Librarian

École Polytechnique Fédérale de Lausanne (Swiss Federal Institute of Technology in Lausanne)

https://orcid.org/0000-0002-6775-7433

Manon Velasco is a multipurpose librarian at EPFL where she is currently in charge of e-book collections. She is also involved in the EPFL's institutional repository and she is a member of the copyright team. Her professional interests include revolutionizing the relationship with publishers, negotiating the best possible agreements with e-book providers, supporting researchers in opening their research and playful learning applied to copyright literacy education.

# INTERLIBRARY LOAN WORKFLOWS AND EXPERIENCES

# How library systems influence the interlibrary loan workflow: A comparison between Alma and Koha

**Martina Kassler and Veronica Fors**

## Abstract

The interlibrary loan (ILL) workflow is influenced by several outside and inside factors, which can variously affect its efficacy in providing items to library users. This paper will show differences and similarities in the ILL workflow, from customer order to delivery or return of the item, when using two different library systems. The ILL workflow includes both internal and external document supply because the libraries offer and provide services both to their own users and to other libraries. Some supplied items are returnable (for example books) and others are non-returnable (for example copies of articles). This workflow chain involves a series of steps and starts with a request made by a user. This is followed by the entry of necessary data into both the receiving and lending libraries' ILL systems to administrate the request and continues with the delivery to the end user. For returnable items, the final step is returning them to the provider. This paper will compare how two Swedish libraries (Örebro University Library and Karlstad University Library) established their ILL workflows and how decisions made in the ILL process are based on the different library systems used (Alma and Koha). It concludes with a discussion of what can be learned from the comparison, the benefits and challenges presented by each system and any improvements that can be made.

## Keywords

# Article

## Introduction

Libraries' limited finances and resources mean that their collections cannot contain all the material that their users request. ILLs are therefore an important part of a library's information supply.

The concept of ILL existed as far back as the 8th century in Western Europe. During the Middle Ages, borrowing and lending occurred between monasteries and later, during the Renaissance, libraries in Italy, France and England flourished and an informal practice of ILL was used by the scholars of the time. Since then, the extent of ILL has varied between times and places. In the first half of the 20th century, the international interlending practice increased which led to the creation of the International Federation of Library Association (IFLA). This organization established international rules and guidelines for lending which made international borrowing and lending more formalized and efficient. Today's librarians and library patrons see ILL as a viable, economical and natural option ([Miguel, 2007](#)).

There are several different aspects that have affected the current ILL workflow. The workflow should both facilitate good service for the library's patrons and be efficient for the staff working with ILL. Other elements such as economical and technical factors have, however, also had to be taken into consideration when trying to streamline the workflow.

Both Örebro University Library and Karlstad University Library have relatively recently implemented new library systems. Örebro acquired the proprietary system Alma, which is provided via subscription from Ex Libris, in 2017. Karlstad, on the other hand, acquired Koha, which is an open-source system, in 2020. It is therefore interesting to investigate whether and how the workflow for ILL management is affected by working in two different systems that are also provided through different business models.

The ILL process at the two libraries consists of management of ILL for the library's patrons (ILL borrowing) and management of loans to other libraries (ILL lending). In this paper we focus on the management of the ILL borrowing process.

For books, the main source for both borrowing and lending is LIBRIS Fjärrlån, the ILL module in the Swedish union catalogue LIBRIS. Most journal articles ordered for library patrons are ordered via LIBRIS and the vendor Subito. Articles sent to other libraries are mainly ordered by other Swedish libraries via LIBRIS.

The paper starts with background information about the two library systems Alma and Koha and the implementation process at the libraries. There is also a description of their current ILL workflows and then a comparison of the systems and workflows. This is followed by a section describing the workflow adjustments which have been made depend-ing on the systems' features. Finally, the discussion looks ahead to the improvements which can be made to increase the efficacy of the workflows and contribute to developing in services for the library patrons.

## Örebro University and Örebro University Library

Örebro University is a comprehensive university with programs, courses, and research distributed across three faculties, eight schools, 85 degree programs and 920 single courses. The university has about 16,000 students 1,700 members of staff and 490 doctoral students. The university is located in the city of Örebro in the middle of Sweden with two campus areas in Örebro and another in Grythyttan situated one hour north of Örebro. One of the campuses in Örebro constitutes the main campus area where most of the faculty and administration can be found while the other one is located at the University Hospital.

The University Library consists of three staffed physical libraries, one at each campus area, and one unstaffed Sheet Music Library, also located at the main campus. The Medical Library located at the University Hospital campus does not only serve the university, but also provides library services, including ILLs, for medical staff in the region (Örebro University, 2021).

In 2020, the library's collections consisted of 247,770 printed books, 232,061 electronic books, 215 print journals and 14,500 electronic journals; 773 study places are provided and there are around 350,000 visitors a year (based on a 5-year period before the pandemic) (National Library of Sweden, 2021).

The Library has a staff of 38. Five staff members at the main library work part-time with ILL (equal to two full-time positions), one staff member at the Medical Library also works part-time with their ILLs, as does one staff member at the library located in Grythyttan.

In 2020, the library borrowed 553 physical materials (mostly books) from other libraries and lent 2,996 materials to other libraries. During the same period, 361 non-returnable materials (mostly journal articles) were ordered from other libraries (and vendors) and 153 non-returnable materials were sent to other libraries.

## The Alma library system

Alma was first released 2012 and is a cloud-based library system maintained and developed by Ex Libris, a Clarivate company, based in Israel. Ex Libris is one of the largest companies in the library technology sector. They have specialized in products for acade-mic, research, national libraries, and consortia, and Alma was created to meet the needs of those libraries (Breeding, 2015b).

Alma is a unified next generation system which integrates disparate features for managing print, electronic and digital resources. The system includes the whole suite of library operations such as selection, acquisition, print management, electronic management, metadata management, link resolution, digitization, user management, fulfillment, and discovery. The sharing of data and collaborating of services facilitate more efficient workflows (Branch, 2014).

Ex Libris provides Alma as a proprietary system which customers can access through subscription. Like many of the other next generation systems, Alma is provided by a multi-tenant/cloud solution. Every customer accesses the same cloud-based version of the software by the SaaS (Software as a Service) method instead of installing it. The vendor is responsible for hosting and supporting the system. Each library can have customized features, but the underlying code is the same for everyone. When the software updates all the customers get it at the same time (Machovec, 2014).

The staff at a library using Alma does not have access to the entire system, instead each staff-member is assigned various so-called roles that give them access to different sections of the system where they can perform their tasks. These sections consist of in-built workflows in the system (Ex Libris, n.d.-b). For example, if you work with ILLs you are assigned the role Fulfillment Services Manager and are then given access to the workflow for Resource Sharing.

Alma is a widely used library system. In 2021, it was used at libraries in 41 countries (Breeding, 2021). In Sweden it is used at about 20 libraries, including Umeå University Library, the Royal Institute of Technology, Mälardalen University and the Swedish Film Institute Library.

## From Voyager to Alma at Örebro University Library

Before acquiring Alma, Örebro University Library used the Voyager ILS (also from Ex Libris), but had a separate ILL system. SAGA, the ILL system used until the end of 2013, was a system developed at the Karolinska Institute University Library and was used by several Swedish university libraries.

Both Voyager and SAGA had been acquired as part of a consortium called GSLG. The consortium consists of Örebro University Library, Linnaeus University Library and Borås University Library. At the end of 2013, SAGA was discontinued and the solution for the GSLG consortia was to develop a new ILL system in-house. The new system was called GSLG ILL or simply GILL and developed at Borås University. However, it had already been decided that the system to replace SAGA would only be a temporary solution since the procurement of a new library system was planned for the near future.

In 2014, an inventory was made where each library in the consortium produced a decision document that described the systems that the individual libraries had and what needs were judged to exist in the future. During 2016-2017, Örebro and the members of the consortium then decided to purchase and implement a new ILS-system instead of Voyager. There were several reasons for changing the system: the development and new features for Voyager had started to decline, the system was hosted locally on a consortium-owned server at Örebro University and the server was approaching its end-of-life, and Voyager also lacked any real support for e-resource management (Johansson & Wiman, internal document, 2014 and Nordström, personal communication, 2021).

The new library system was acquired through a public tender where an added value model, taking both price and quality into consideration, was used for evaluating the offers that were received. Within the consortium it was also considered that an important feature when procuring a new system or suite of systems was that all functionalities should be covered, including the acquisition, description, management, searching and circulation of physical and electronic resources, with integrated workflows for all resources. Support for the entire lifecycle of library resources was also required. The new system was not only to replace the current ILS, but also the discovery service, knowledge base and Link Resolver. The new system should also include Electronic Resource Management System (ERMS) functionality and eliminate the need to enter duplicate information into multiple, stand-alone systems.

Even if ILL management was not the main focus in the procurement there were some requirements, like integration with LIBRIS Fjärrlån. Integration of the ILL workflow into the main library system was also considered an improvement. ILLs were almost not managed at all in Voyager – any system use for this was limited to lending material to other libraries. The ILLs from other libraries for the library's users were managed in systems that were separate from Voyager, as mentioned above. In order to display the ILLs among other loans to the library patrons, an in-house developed OPAC was needed. Using a unified system would eliminate the need to invest library resources in in-house solutions.

The consortium finally decided to accept the tender from Ex Libris which received the lowest total cost, calculated using the added value model, with the Alma/Primo system suite, where Primo is the Discovery interface. Alma/Primo then went live at Örebro University Library in 2017 (Nordström, personal communication, 2021).

# ILL workflow at Örebro University Library

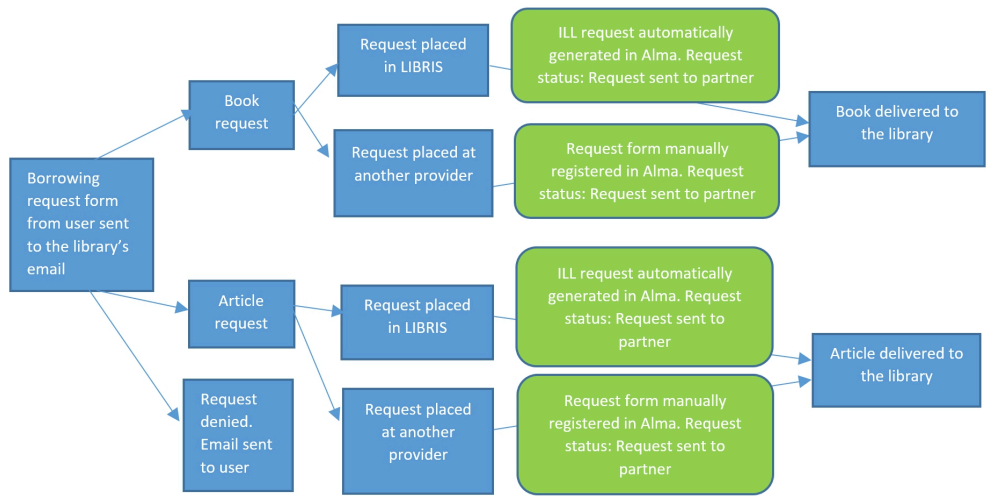

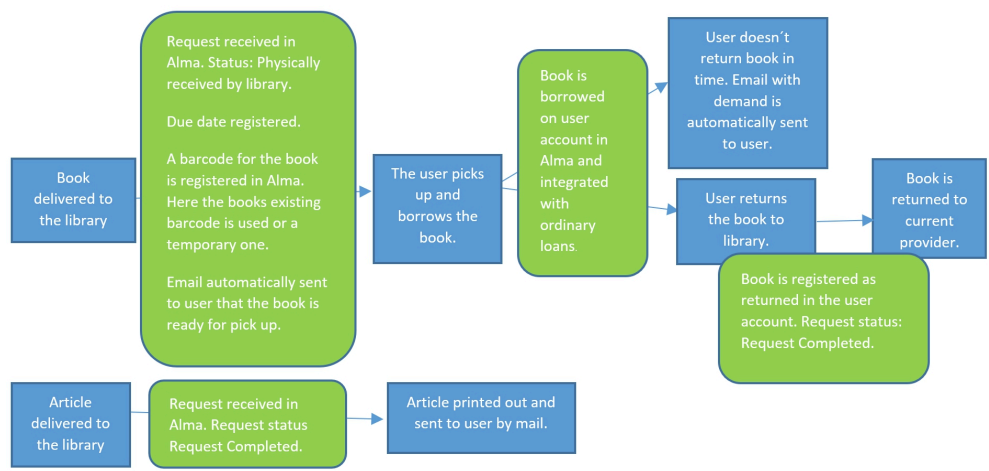

*Figures 1a and 1b: ILL workflow at Örebro University Library*

The ILL process at Örebro University Library starts with an ILL request from a library user on the library's website. There is one request form for books and another for articles/chapters. The ILL form is automatically sent to a functional email for ILL at the library. The ILL staff checks the email and verifies whether the request is for an item in the library's own collections. If this is the case, the request is denied. An email is sent to the user with information that the book is available at the library or that the book can be reserved. If the book or article is not available at the library, a search for suitable providers is performed.

There is no charge for staff and students at Örebro University to make ILL requests for books which can be ordered from the Nordic countries. For ILL requests for books from other countries, a fee of 175 SEK (1 SEK is approximately 0,09 EUR) per book is charged for both staff and students. Requests for articles through ILL are free of charge for staff and for the moment also for students. There are no fees for the staff at Region Örebro County (which is a user group at the Medical Library, which is part of Örebro University Library) because of an agreement with Region Örebro County.

The ILL staff normally makes ILL book requests from LIBRIS Fjärrlån, the Swedish union catalogue. If the request cannot be made in LIBRIS, attempts are made to order from other providers, such as libraries in other Nordic countries, Subito or OCLC. Article requests are normally made from Subito. If they are not available for delivery, attempts are made with other providers, such as LIBRIS, then Nordic Countries Libraries, OCLC, Get it Now or Reprints Desk.

Requests made in LIBRIS are imported once a day using an integration profile in Alma (the integration profile uses LIBRIS's API). The import only takes place once a day due to a limitation in Alma. It is only possible to make four imports at fixed times daily and because Örebro University Library consists of four separate libraries, one import time slot is used for each library.

If the request is made elsewhere, it is added manually in Alma. Manual registration is helped by an in-house plug-in where the DOI and PMID number can be entered and then the bibliographical records are registered automatically in Alma (the same functionality is now built into Alma but the plug-in is still used since it provides some extra flexibility). Records about the user, owning library and patron (provider) are also added and the status of the request is marked as *Request sent to partner*. Both the staff at the library and the user who has requested the item can see the request status: the staff in Alma and the user via My Account in Primo.

The chosen providers then, in turn, receive the request and decide if they can fulfill it. If they cannot, the ILL staff mark the status of the record in Alma as *Cannot be fulfilled* and the record is then removed in the system. A search for other possible providers is made. If no provider is found, the user is notified that the request is denied.

If the provider can fulfill the request, the book or article is delivered to the library. When the item arrives, the ILL staff register it in Alma as *Received*. If the ILL is a book, a temporary record is created. A lending period is registered on the record, a temporary barcode is assigned to the item (here the existing barcode on the book is used or, if there is no existing barcode, a barcode is added on the ILL-slip which comes with the slip from the library) and an email is automatically sent to the user notifying them that the item is ready for pick up. The status now becomes *Physically received by library*. The item is placed on the library's hold shelf at the information desk, ready for the user to pick up. If the received request is a non-returnable, such as an article, it is printed out and sent to the user. The status is then marked as *Request complete*.

When the user picks up the item at the information desk the temporary barcode on the item is scanned in and registered in Alma and the status is now *Loaned item to patron.* The loan is considered a standard loan in the system and integrated with a user's other loans on their account. When the loan is due, a due date notice is automatically sent to the user by Alma.

Finally, the user returns the loan to the library's information desk and its status is marked as *Request complete* and returned to the current provider. If the book is lost or not returned at all a fee of 500 SEK is placed on the user's account.

## Karlstad University and Karlstad University Library

The 2021 statistics show that Karlstad University offers teaching in about 75 programs, 750 courses and in over 50 subjects. There are about 19,000 undergraduate students, about 265 PhD students and the staff number is 1,334 (Karlstad University, 2021).

The university has two campuses, one in Karlstad and another in Arvika, where the Ingesund School of Music trains music teachers, musicians and music producers.

In 2021, Karlstad University Library's collection consisted of 139,099 printed books, 397,776 electronic books, 195 print journals and 13,607 electronic journals. The library has 1,488 study places and around 187,745 visitors a year. In 2021 Karlstad University Library's ILL department lent 2,279 books and articles and borrowed 1,088 (Karlstad University, 2021). Three staff members work with ILL part time, the equivalent of one full-time position.

## The Koha library system

Koha is an open-source system. This library system is used worldwide by all kinds of different libraries like public libraries, university libraries, private libraries, school libraries, research libraries and more.

Koha is a based on an SQL database, and is web based and therefore very flexible and can be adapted to different needs, as testified to on the website of the Koha library software community.[1] Koha is a Maori word that means "present" as a hint that joining Koha is free of charge. The library system was invented in 1999 in New Zealand.

There is a Swedish Koha user group[2] networking to help each other to develop the sys-tem. They provide information on their website about news, discussions, a wiki and the announcements of meetings for exchanges and discussions.

---

1. https://koha-community.org/about
2. https://koha.se

Some of the university and college libraries working with the Koha system in Sweden are: Stockholm University Library, Mid Sweden University Library, Luleå University Library, Lund University Library, Gothenburg University Library, and the libraries of the following colleges: Kristianstad University, Blekinge Institute of Technology, Dalarna University, and the University of Gävle. And last, but not least Karlstad University Library, which changed from the Sierra library system to Koha in January 2021.

## From Sierra to Koha at Karlstad University Library

From the early 2000s, Karlstad University Library had the integrated library system Millenium owned by the American company Innovative. The system later changed name from Millenium to Sierra. Karlstad University Library's contract was about to expire in 2020 and the management decided to have a closer look at all possible alternatives. During 2019, pre-studies were carried out by a university project manager through interviews with the employees working the most with the system. The options were to continue with Sierra, change to the Alma library system, or start with the open-source system Koha. Considerations involved not only the costs but also how well the system could communicate with other systems, which is relevant for the work at the library and how it could contribute to making day-today work as smooth as possible.

The decision was in Koha's favor because it seemed to be the most effective system for the work at the library and in the long run it would also be the cheapest alternative. A big advantage is that Karlstad University Library would be able to influence and make changes which are perfectly suited to the library's way of working and collaborate with other Swedish libraries to develop the system. In December 2020, Karlstad University Library went live with Koha according to Nilsson (internal document, 2019).

Sierra was not the best tool for ILL, and the process involved finding ways inside the system to make sure that loans were booked and the stage of their delivery process visible. One advantage was that a webform to order ILL was automatically integrated into the system. The biggest disadvantage was that Sierra and the Swedish national ILL system LIBRIS Fjärrlån were not connected to each other.

The Swedish Koha group had, together with a consultant, already done work on developing the ILL module before Karlstad's University Library joined Koha.

# ILL workflow at Karlstad University Library

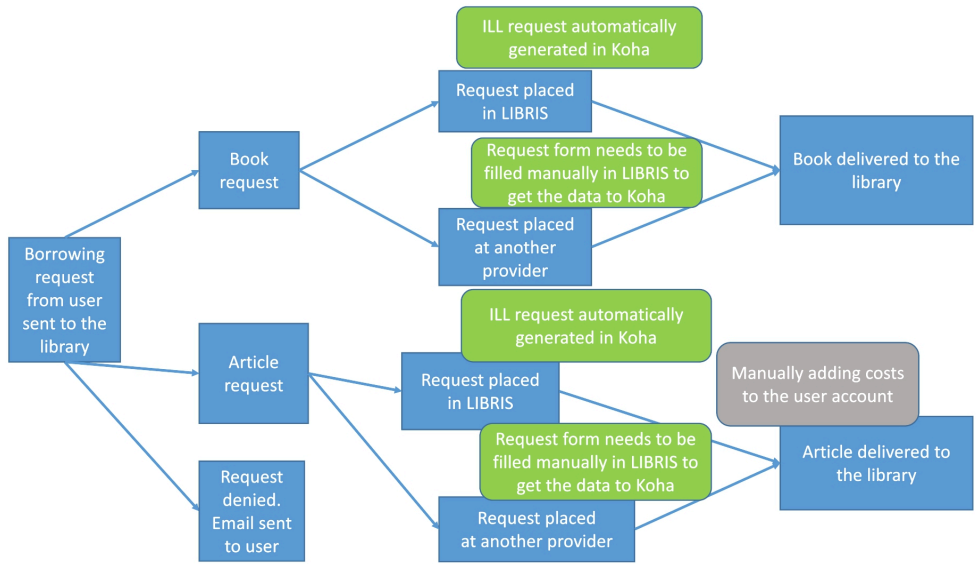

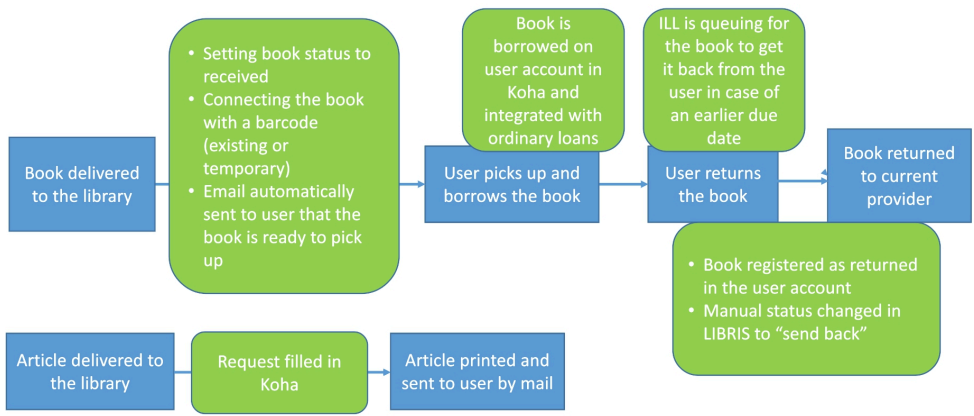

*Figures 2a and 2b: ILL workflow at Karlstad University Library*

The ILL staff looks up the references sent by the user. A first manual check is done to determine if the book or article is owned by the library or if the article can be reached via open access. In these cases, the user is informed on how to access the material and the ordering process is stopped. When the book or article is not available, the Swedish ILL system LIBRIS Fjärrlån is checked to see if it is possible to place an order with a Swedish library. If this is the case, the order for the book or article is placed through LIBRIS because the Koha library system is connected to LIBRIS and all orders made in LIBRIS are automatically implemented into Koha with an update every 15 minutes.

If it is not possible to order through LIBRIS other providers are checked. For books, it is the Nordic libraries first and then Subito and OCLC. For articles, Subito, Get it now, Nordic libraries and OCLC are checked in this order.

If the order is sent to providers other than LIBRIS, the ILL staff must place a manual order in LIBRIS through a so-called "empty form" order to import the data into the Koha library system.

Ordering books through ILL from Nordic countries is free of charge at Karlstad University Library. Books ordered from outside the Nordic countries cost 200 SEK and articles 50 SEK each. Those fees must be added manually to the user's account in Koha if it is a student or other user. For university staff, costs are charged to the faculties once a year.

As soon as the ordered material is delivered, the status is set to *Received* in Koha. If it is an article, the printed version is sent to the user by post and the order is fulfilled. If it is a book, it is set as *Received* in Koha, the returning dates (guaranteed time and possible duration of loan) are filled in and a barcode is associated to the book so that the user is able to borrow it at the self-service machines. An email is sent automatically to the user notifying them that the book is now ready for pick up from the reserved books shelf. As soon as the user borrows the book at the library, it is added to the user's account and can be kept until an email is sent informing the user to return the book. This will usually happen at the end of the longest possible borrowing period. If the book needs to be returned earlier because the lending library claims it back, the book is reserved for the interlibrary loan department in Koha. That generates an email to the user to return the book. As Karlstad University Library has 14 days as an automatic re-borrowing period, the return date is set to the end of that ongoing period. As soon as the user returns the book, it is removed from their account. The ILL staff will change the status in LIBRIS Fjärrlån to *Returned* and will send the book back by post.

## Alma-Koha Comparison

Both Örebro and Karlstad University Libraries stated that the cost was an important issue in the procurement, and both claimed that they accepted the tender that was the most economically advantageous given the current requirements. The different models have different costs at different times in the operation process. For an open-source system, like Koha, the software is free, but services such as system implementation, data migration, training materials, onsite training, server and ongoing maintenance, come at a fee ([Jost, 2016](#)). For a proprietary system like Alma, on the other hand, the library pays more for the software subscription, but saves on service fees which are included in the subscription (Ex Libris, n.d.-a).

Subscribing to a proprietary system means that the library itself does not have to manage system development because this is managed by the company which provides the system. If a library subscribes to Alma, for example, it can use the system and automatically receives updated versions of it through the Ex Libris cloud solution. Support is also provided by the company. This saves time and resources which suited Örebro University Library. A disadvantage is that the library cannot influence the system to any great extent and must instead adapt its workflows to the system. The library is dependent on the vendor for updates of the system and its functions, solutions to any systemic problems and service enhancements (Breeding, 2017).

With an open source-system, the library has an opportunity to cooperate with other Koha users and to influence the system. Karlstad University considered this an advantage in their procurement. The system can to some extent be adapted to the needs of the library. However, potential problems may occur if the situation arises that not everyone is willing to share their source code at all or to the same extent (Askey, 2008). Cooperation may then be tilted. Another disadvantage is, as stated, that the library does not have access to the technical support that a providing company can offer (Helling, 2010). Here the challenges are a question of resources and the need for relevant skills and competence (Thacker & Knutson, 2015). These skills and competence can come from in-house staff or a consultant.

An overall comparison of Örebro and Karlstad's University Libraries' experiences of Alma and Koha seems to indicate few differences in functionality and usability between the systems. The libraries both have a total solution, and ILL functionality is integrated in their systems which makes work easier and more effective both for the staff and users. Previous investigations of the degree to which open-source systems have the same functionality as proprietary systems have shown that they are roughly equivalent. Koha, in particularly, seems to be the equal of any proprietary system in terms of functionality (Breeding, 2015a; Pruett & Choi, 2013). In the past, Koha lacked a module for ILL (Pruett & Choi, 2013), but now the Koha network has integrated one. It appears that some functions were not as well developed in Koha as in Alma, for example manual registration of records in the system (see below).

There follows a more detailed comparison between the libraries' workflows which shows the similarities and differences in functionality.

At the beginning of the ILL process, users at Örebro University Library complete one of two separate web-based forms (one for books and one for articles/chapters) generating an email to the ILL administration. It should be noted that Primo has the functionality to include an ILL request form which is, when it is completed, automatically imported into Alma. The library has not yet started to use this functionality, but there are discussions about doing so. Karlstad University Library does not currently have any web-based forms and instead users contact the administration by email. Koha has a function for integrating an acquisition form in the system. Recently this and other possibilities for ILL acquisition forms have been investigated to decide how to proceed. A similar process to the current one of checking if the ILL requests should be actual requests with regard to the current terms will then be implemented.

If the request proceeds, both libraries use LIBRIS to investigate whether requests for books are available at Swedish Libraries. As both Alma and Koha are integrated with LIBRIS, they both import records to their systems when the requests are made. Karlstad seems to receive imports more frequently than Örebro: Koha is updated every 15 minutes (this can be adjusted to any suitable interval) and Örebro University Library can get one import of records per day to Alma due to a limitation in the system which permits a limited number of imports per day for system users.

Regarding requests to other suppliers, such as Subito, Get it Now, Reprints Desk and OCLC, the systems differ in terms of functionality for manual registration of records. In Alma it is possible to register a record manually directly within the system. In Koha, on the other hand, this is not possible. Instead, the staff at Karlstad University Library places a manual request in LIBRIS through the so-called *empty form* and the record's data are then imported to Koha.

When the book arrives to the libraries, the circulation-process can be managed in one system for both Alma and Koha. The request status changes when different steps are taken in the workflow. The books are received, lent, and returned in the system. It is searchable and both library staff and users can see the status of the request. The request is registered on the user's account with their ordinary loans, which makes it possible to get an overview of all the user's loan types.

Communication with users takes place through the systems. Emails are sent automatically by the systems when, for example, the book is ready for pick up, is delayed or must be returned. When the book is delayed, Karlstad University Library must place a reservation for the ILL department to be able to stop the automatic renewal.

## Adjustments in the ILL workflow connected to the ILL systems limits

The choice of provider is very closely connected to the ILL system. As all orders set in LIBRIS are automatically implemented into Alma and Koha, it is the libraries' first choice for books and at Karlstad University Library also for articles. With regard to books, it is very clear that they should be ordered within Sweden if possible. For articles, there are several factors in play.

At Karlstad University Library, articles are ordered in LIBRIS as much as possible as the workflow is the easiest and most time efficient. The normal range of costs for an article is 80 to 100 SEK in these cases. Delivery time can range from a few hours to a few days. If an article request is very urgent, it is ordered from document delivery services like Subito or Get it Now. This is also done if the article is unavailable through LIBRIS. In such cases orders are made from the providers' websites and additionally a proforma order is placed in LIBRIS to import the request into Koha. The pricing can differ much more in these cases, ranging from six to 54 EUR per article for Subito and 24 EUR for Get it now. In these cases, the availability, pricing, and delivery format are taken into consideration. Get it Now delivers via email directly to the user, while Subito delivers to the library in digital form via email. The article then must be printed and sent to the user as a paper copy.

At Örebro University Library most articles are ordered from Subito. This provider is considered to have relatively low prices and delivers the articles quickly (see above). Even if the records are not imported directly to Alma like the records from LIBRIS, it is possible to register requests manually in Alma and this is facilitated with the in-house plug-in which imports bibliographical data to the record if there is a DOI or PMID number. As a second alternative, articles are ordered from LIBRIS. Although the price is in the normal range, as mentioned above, delivery time can vary, often taking longer than articles from Subito.

## Discussion and future work

One of the challenges regarding the ILL workflow is to get the requests into the system and make the workflow as effective as possible. At Örebro University Library, the request from library patrons is first placed through web-based forms and then comes to the library's ILL email. Then it must be registered in Alma, either via import or manually. The process of activating the ILL form function available in Primo is ongoing. Through this, ILL requests with their bibliographical data would come directly into Alma which will streamline the workflow. Far fewer records will then have to be created manually in the system. An additional advantage is that the request will be visible directly at the user's Alma account. Currently, users may be uncertain whether their orders have been processed because there is a delay between the request being made and its registration in Alma.

The Koha development group for ILL has the option to import data from Subito directly on their list of future developments. Several libraries are using this document delivery provider and face the same problem. If data can be directly imported from Subito to Koha, the ILL workflow will probably change with Subito again being Karlstad University Library's first choice for ordering articles. Once Subito is connected to Koha maybe other providers can be connected as well.

RapidILL is another alternative to LIBRIS that is directly connected to Alma for sourcing articles. Alma and RapidILL are both Ex Libris products. The system is more fixed to the company's solutions, but the solutions provided are easy to implement and use. RapidILL also has an automated design for ILLs, making it possible to obtain the loans and non-returnables fast and at a low cost. It therefore seems like a solution that would make the entire ILL workflow more efficient. However, no Swedish libraries appear to have started using RapidILL permanently and Örebro University Library is also still waiting with implementation. However, it should be interesting to follow developments and see if it could be something for Örebro University Library in the future.

Another area under development is the billing of articles. At the moment, Karlstad University Library charges a fee per article on student accounts. The faculties pay for articles requested by Karlstad University employees, which means that the ILL department manually registers, outside the library system, how many articles per faculty are ordered. At the end of the year internal bills are sent to the faculties. The future goal is to automatically generate the fee on the students account as soon as an article is ordered. At present there is an in-between solution where the fees are prefilled beside the order number and can be connected to the account with one click. For the employees, it should be possible to create accounts for the faculties or to have a statistics part in which orders can be counted by faculty. That would make the work easier as all the information would be registered in one system.

Another area of future development are claims. At the moment books need to be reserved for the ILL department as soon as they are claimed by the lending library. As Karlstad University Library has automatic renewals of 14 days, in the worst case scenario 14 days are needed to get the book back, and there are better solutions already in progress to claim ILLs immediately.

There are also features in Alma which can facilitate billing. But because almost all ILL requests are free of charge for the library's users, for example, students can currently make ILL requests for articles for free, the library does not have much invoicing and activating these features is not a priority right now.

## Conclusion

As can be seen in the workflow presentations, the Alma and Koha systems are not that different. Both provide the libraries with the possibility to take the basic steps from requesting material to returning it. The differences are in the details: for example, both systems connect to LIBRIS, but not to Subito, Get it now or RapidILL, and manual orders can be made directly in the system (Alma) or have to be entered in LIBRIS to be implemented (Koha).

At both these University Libraries, our aim is to achieve a balance of convenience, time efficiency and economical workflow for ILL. We consider the best possible solution for our customers, as well as for the ILL employees and the library's finances.

# Bibliography

Askey, D. (2008). We love open source software: No, you can't have our code. *The Code4Lib Journal*, (5). https://journal.code4lib.org/articles/527

Branch, D. (2014). Alma in the cloud: Implementation through the eyes of acquisitions. In B. R. Bernhardt & L. H. Hinds (Eds.), *Too much is not enough! Charleston Conference proceedings, 2013* (pp. 545–550). https://doi.org/10.5703/1288284315322

Breeding, M. (2015a). Adoption patterns of proprietary and open source ILS in U.S. libraries. *Computers in Libraries*, 35(8), 17-18, 20.

Breeding, M. (2015b). Introduction and concepts. *Library Technology Reports*, 51(4), 5–19. https://journals.ala.org/index.php/ltr/article/view/5686/7062

Breeding, M. (2017, November 1). Open source software. *American Libraries Magazine*. https://americanlibrariesmagazine.org/2017/11/01/open-source-software/

Breeding, M. (2021, May 3). 2021 library systems report: Advancing library technologies in challenging times, *American Libraries*. https://americanlibrariesmagazine.org/2021/05/03/2021-library-systems-report/

Ex Libris. (n.d.-a). *Support services*. Retrieved October 13, 2021, from https://exlibrisgroup.com/services/support-services/

Ex Libris. (n.d.-b). *Managing user roles*. Ex Libris Knowledge Center. Retrieved October 13, 2021, from https://knowledge.exlibrisgroup.com/Alma/Product_Documentation/010Alma_Online_Help_(English)/050Administration/030User_Management/060Managing_User_Roles

Helling, J. (2010). Cutting the proprietary cord: A case study of one library's decision to migrate to an open source ILS. *Library Review*, 59(9), 702–707. https://doi.org/10.1108/00242531011087024

Jost, R. M. (2016). *Selecting and implementing an integrated library system: The most important decision you will ever make*. Chandos Publishing.

Karlstad University. (2021). *Facts and figures*. Retrieved April 29, 2022, from https://www.kau.se/bibliotek/om-biblioteket/fakta-om-biblioteket/fakta-om-biblioteket/biblioteket-i-siffror

Machovec, G. (2014). Consortia and next generation integrated library systems. *Journal of Library Administration*, 54(5), 435–443. https://doi.org/10.1080/01930826.2014.946789

Miguel, T. M. (2007). Exchanging books in Western Europe: A brief history of inter-

national interlibrary loan. *International Journal of Legal Information*, 35(3), 499–513. https://doi.org/10.1017/S073112650000247X

National Library of Sweden. (2021). *Biblioteksstatistik – Rapporter*. Retrieved November 10, 2021, from https://bibstat.kb.se/reports

Örebro University. (2021). *Facts about Örebro University*. Retrieved November 10, 2021, from https://www.oru.se/english/about-us/facts-about-orebro-university/

Pruett, J., & Choi, N. (2013). A comparison between select open source and proprietary integrated library systems, *Library Hi Tech, 31*(3), 435–454. https://doi.org/10.1108/lht-01-2013-0003

Thacker, C., & Knutson, C. (2015). Barriers to initiation of open source software projects in libraries. *The Code4Lib Journal*, (29). https://journal.code4lib.org/articles/10665

## About the Authors

### Martina Kassler, Interlibrary Loan Librarian
Karlstads universitetsbiblioteket (Karlstad University Library)

https://orcid.org/0000-0002-1557-1397

Martina Kassler is a current interlibrary loan librarian at Karlstad University Library. After completing a bachelor in economics and a master in ethnology and sociology and another one in library and information science, she took up a position as an interlibrary loan librarian. She is also interested in the topic of interculture.

### Veronica Fors, Librarian
Örebro universitetsbibliotek (Örebro University Library)

Veronica Fors (1976) has a master's degree in library and information science and is currently librarian at Örebro University Library at the unit Library Acquisitions and Scholarly Publishing. She is responsible for the library's interlibrary loans. She also works with the library's publication database DiVA.

# International loans in the Main Library of AGH UST: The technical library as a witness of change

Karolina Forma

## Abstract

Workflows and methods used in the Main Library of AGH UST changed with the challenging history of Poland and rapid changes in sourcing information for the Library's patrons and STEM professionals. The article describes this journey specifically from the point of view of the Interlibrary Loans Department. Since 1952 it has served hundreds of thousands of users domestically and outside of Poland, maintaining continuing relationships with its partners. The most important milestone in document delivery was in 2001, when the library started using IFLA Vouchers, which made managing loans much easier. Nowadays, in times of digitization and open access ILL is still important – as its core role is providing access to materials which often can seem unobtainable.

## Keywords

International library loans; Interlibrary loan; AGH UST (Krakow); AGH UST Library; Polish libraries; Data analysis

## Article

## History and context of the Main Library of AGH UST

The Main Library of the Academy of Mining and Metallurgy in Krakow (*Akademia Górniczo–Hutnicza im. Stanisława Staszica w Krakowie*)[1] opened in the academic year 1919–20, as a vital part of a new university in Krakow. The need for a technical university in the 20[th] century in southern Poland was urgent. So in 1912, when Krakow was still a part of the Austro-Hungarian Empire, the Ministry of Public Works in Vienna agreed to the founding of the Mining Academy. After the restoration of Poland's sovereignty in 1918, the University was officially opened, and a library followed, in 1920. The collection was and still is very specialised: the Library holds mainly scientific works about mining, machine building, foundry, technology, and the environment – historically the most important disciplines for the University. Additionally, there are mathematics, management, and social sciences collections.

---

1. In abbreviation "AGH UST", "AGH" stands for "Akademia Górniczo–Hutnicza" and "UST" for "University of Science and Technology".

The library network consists of the Main Library and thirteen faculty libraries, two institute libraries and three libraries of non-faculty units.[2] The Library collection mirrors the needs of the Academy and has always been based on various sources and regular international exchanges. The library holds books in 50 languages.

From its very first days the Academy and the Library connected different cultures and languages, as it was founded just after the First World War, when Poland became an independent state again. Its core collection was founded by Polish professors, who were often trained abroad, speaking various languages. Furthermore, Marie Curie-Skłodowska founded the Library's subscription of French journals, specifically *Revue Scientifique*, *Revue Rose*, and *Comptes Rendus Hebdomadaires des Séances de l'Académie des Sciences* ([Krawczyk, 2009](#)).

Up until 1939, the University and the Library were connected to their Austro-Hungarian roots and Upper Silesia, both receiving and lending materials – partially in German. After the Second World War Poland and all its administration were heavily influenced by the Soviet Union. As the connections to "the West" were limited, University professors had to build contacts within Poland and in the Eastern Bloc (Fig. 1).

The most important way of informing others about what the Library was holding was printed catalogues sent out to libraries in the Eastern Bloc or personal connections.

The Library was always ready for international connections, but did not always document internal affairs – so the earliest data about ILL in the Library can be found in 1951, when polyglot Zofia Żebrowska was hired to lead the Interlibrary Loans Office ([Krawczyk, 2009](#)). Internal statistical documents give us some information about the circulation of prints. In 1952, 77 publications were lent and 12 borrowed by the Library, in 1953 – 93 lent, 27 borrowed.

Circulation got larger each year, but only to and from Polish institutions. Currently, the Interlibrary Loans Office still works primarily with Polish institutions. Sources requested by our patrons can be accessed only in the scenic Main Reading Room. Services are available to the employees, doctoral students, and full-time and extramural students of AGH UST.

---

2. Numbers applicable to May 2022.

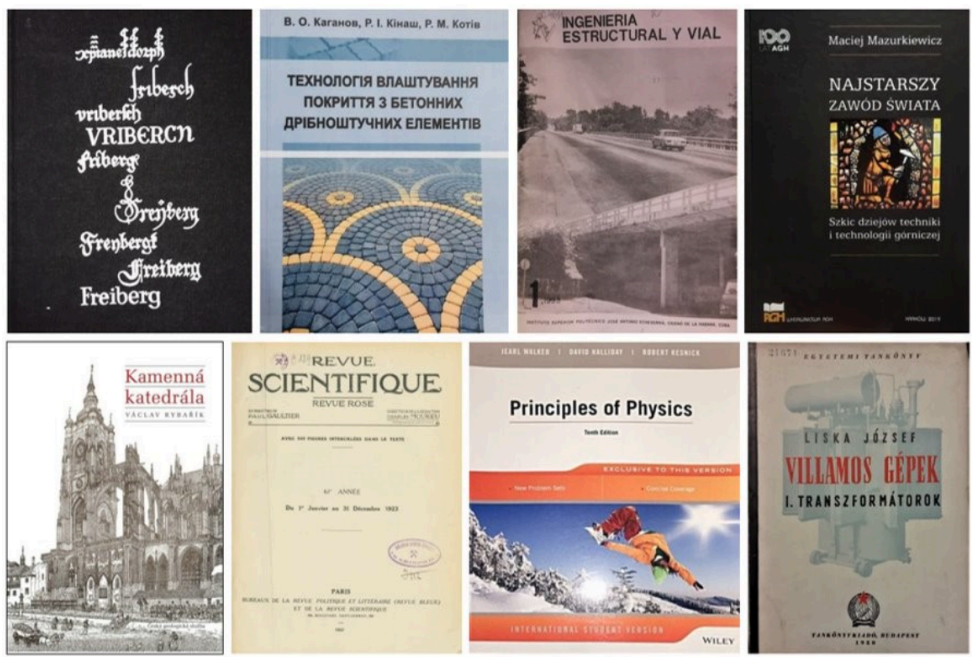

*Figure 1: Prints from AGH Main Library collection in various languages (German, Ukrainian, Spanish, Polish, Czech, French, English, and Hungarian)*

The first international loan took place in 1972. A professor requested a conference paper from Bergbau-Bücherei in Bochum, Western Germany, one of the oldest libraries in the Ruhr area. It was managed without payment, as the libraries were operating exclusively on mutual agreements. To this day, German libraries remain our closest partners, both lending and receiving materials.

In the 1970s, 56 international borrowings were documented. The most important partners for the AGH UST Library were European units, mostly from the Eastern Bloc although the University was developing a vast web of connections – from Japan and India to South America, due to scientific research in mining and metallurgy.

Statistics in the 1970s showed a proportion of 13:1 (when one book was lent from outside Poland, 133 were lent internally). At the end of 1979 the Library had an agreement on Interlibrary Loans with 24 libraries, some of them in our sister institutions – Technical University in Ostrava or Freiberg University.

- Kungliga biblioteket – Stockholm, Sweden
- Utrikespolitiska institutet – Stockholm, Sweden
- DTU Bibliotek – Lyngby, Sweden
- Central Scientific Library of the USSR Academy of Sciences – Kiev, then Ukrainian Soviet Socialist Republic
- Lviv Polytechnic Library – Lviv, then Ukrainian Soviet Socialist Republic
- Univerzitná Knižnica – Bratislava, then Czechoslovakia

- National Széchényi Library – Budapest, Hungary
- Österreichische Nationalbibliothek – Vienna, Austria
- Central Scientific and Technical Library – Sofia, then People's Republic of Bulgaria
- Ústřední knihovna:Vysoká škola báňská – Ostrava, then Czechoslovakia
- Eindhoven University of Technology – Eindhoven, The Netherlands
- Universitetsbiblioteket Norges teknisk-naturvitenskapelige universitet – Trondheim, Norway
- Universitätsbibliothek – Freie Universität Berlin, then Federal Republic of Germany
- Universitätsbibliotheken TU Berlin. then Federal Republic of Germany
- Universitätsbibliothek – Technische Universität Braunschweig, then Federal Republic of Germany
- Universitätsbibliothek – TU Clausthal. then Federal Republic of Germany
- Württembergische Landesbibliothek Stuttgart, then Federal Republic of Germany
- Universitätsbibliothek Stuttgart, then Federal Republic of Germany
- Zentralbibliothek – Universität Hohenheim – Stuttgart, then Federal Republic of Germany
- Bergbau-Bücherei in Essen, then Federal Republic of Germany
- Bibliothek des Karlsruher Instituts für Technologie – Karlsruhe, then Federal Republic of Germany
- Bayerische Staatsbibliothek – Munich, then Federal Republic of Germany
- Universitätsbibliothek der Technischen Universität München – Munich, then Federal Republic of Germany
- Universitätsbibliothek TU Bergakademie Freiberg, then German Democratic Republic

The last two decades of the 20th century were characterized by the rise of electronic sources and methods. More and more requests were sent by fax and information about the Library's holdings could increasingly be found on the internet. Librarians stopped receiving random library slips sent out by readers who needed a book but were uncertain if the Library had it. In the 1980s, requests were received from France and Soviet Union. The Library of Congress was also an important partner, lending twenty-five books, one copy of an article and one journal. The eighties also brought noteworthy political changes; changes in University administration, and consequently, changes in the Library and its international co-operations. After a dramatic period of strike-breaking and Martial Law, Poland regained its independence from soviet influence in 1989 and the transformation started. The borders opened and let in many more academics, independent education, and new sources (Fig. 2).

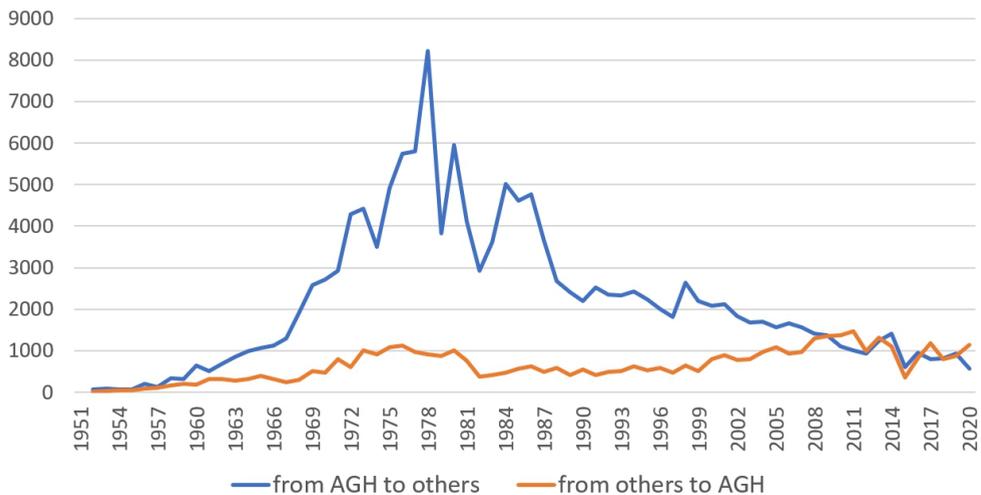

*Figure 2: Interlibrary loans internal and external – summarized*

The 1990s was a prime time for developing further cooperation with the US – University of Illinois, The New York State Library, University of California, among many others. With the Iron Curtain falling in 1989, new possibilities arose and more contacts were managed – from Portugal, Australia, Brazil, China and South Africa. Nowadays, we are partnered with 265 libraries in 31 countries.

## COVID-19 response

The Library was closed to patrons from March 16[th] to May 15[th], 2020, due to the COVID-19 pandemic. As the world tried to manage these "unprecedented times" requests still had to be fulfilled and books had to be sent. ILL Office employees had to physically be in the workplace at least a few hours a day throughout those two months – as they were still a part of the Circulation Department which was still working, partially offline. All the safety precautions were taken, patrons weren't allowed to enter premises and the employees tried to work in isolation.

The most important task of the Office was providing information – as many patrons contacted the Library asking about all the changes that were then happening. The Office got lots of questions by social media or bibliochat, in addition to multiple telephone calls a day.

As internal work was managed, external communication had to change. March 2020 was when online forms started to be used for ILL in the Main Library (Fig. 3). Requests subsided a lot as the library started promoting online services. Ordering online was managed via a system made for the library by Academic Computer Centre CYFRONET AGH.

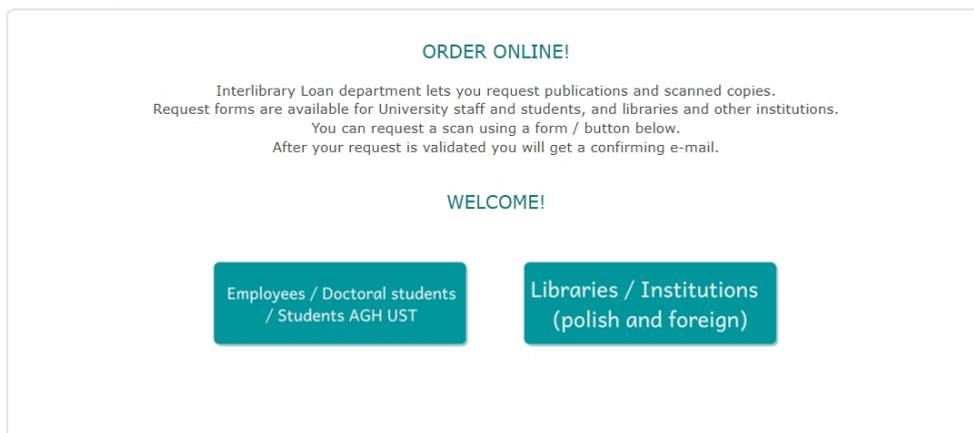

*Figure 3: Screen of an ILL request form*

All the requested printed sources were scanned. If a patron requested a book, a scan of the table of contents was sent with an inquiry about which part of the book was needed, and then only the requested part was scanned and sent. Full books are not digitized in the Interlibrary Loans Office due to copyright laws. Furthermore, patrons do not express a need for complete books, only specific chapters.

It was a trying time for loaning as many people turned to searching for resources online out of pure convenience. The pandemic brought about a serious decrease in requests – so the changes were very much needed.

## The change in the Library

From its foundation up to 2021, the Interlibrary Loans Office was a part of the Circulation Department. Nowadays the Office is part of the International Cooperation and Inventory Control Division. With the new administration some methods changed – mostly in terms of communication. Online request forms were developed and the Library joined the Polish interlibrary loan system ACADEMICA.

Circulation is done online. Currently, the Library is using Virtua, but the migration to a new ILS will happen in the future. Books can be borrowed for a month, theses for two weeks, and journal issues for one week. The full process, from taking the request to completion takes a mere hour.

Electronic request forms are used for communicating online for: registering new patrons, ordering scans of journal articles or book excerpts, which are not available for loan, or ordering printouts of Polish and foreign standards, ordering magazines from the storeroom for use in the Main Reading Room, managing Interlibrary Loan Service, requesting the purchase of books for the AGH Library's collection as well as for the Fiction Collection.

The Office has also opened to the public in a different way: Agnieszka Zych, from the Interlibrary Loans Office, is a Polish Sign Language expert and helps tremendously with the visibility of the Library and its social media.

On the other hand, we decided to continue logging the international lends on handwritten cards and yearly reports, mostly for aesthetic and historical purposes (Fig. 4).

*Figure 4: Handwritten lending card for National Library of Russia. Columns are: referential number of the loan, date of arrival, author and title, lending library's signature number, number of volumes, name of the borrower, date of return, signature of the librarian from ILL Office, notes.*

## The Library and the University

The University is home to 2,184 teachers, including sport instructors and language teachers, and 20,562 students, including doctoral students and post-grads[3]. 95% of users of the Interlibrary Loans Office are professors. The core audience is STEM professionals but the Library carries a sizeable portion of sources for the Faculty of Humanities and the Faculty of Management.

The Library doesn't participate in many projects – as we work closely with German libraries, TIB in Hannover being one of them, we joined TIBORDER in 2000 but decided against joining SUBITO. The year after, we joined the IFLA Voucher programme. In 2015, the Library joined the *Bibliotheksverbund Bayern* (BVB). In 2021, we joined ACADEMICA – Interlibrary loan system of books and scientific publications, a library project by the

3. Numbers applicable for 31 December 2021.

Polish National Library in cooperation with the Foundation for Polish Science and the Research and Academic Computer Network (Academica, n.d.). The aim of the project was to introduce a new level of quality to the interlibrary loan system by replacing its traditional form of sending paper copies by post with lending publications in digital form. The project was financed by the European Union from the European Regional Development Fund under the Innovative Economy Operational Programme. The project gives its users access to more than 3.7 million publications in all fields of knowledge, including the latest, protected by copyright.

In times of digitalization and open access, ILL is still very important—as its core role is lending materials, quite often those which seem unreachable. The central idea we all agree on is reciprocity. If a partnered library offers its services for free, the Library also lends materials free of charge.

Being more open also means communicating more. So the Office took some steps to make itself more visible and accessible. Besides social media campaigns, ILL was promoted through various Open Day events in May 2022. These promotion activities revealed just how many users simply didn't know they could use the interlibrary loan service. Many of the young people questioned said that they just didn't want to bother the Library and preferred finding sources online.

In summary, in 50 years since its beginning, the Interlibrary Loans Office has undergone many changes due to the political climate, work methods, and circumstances. It has always served a specific clientele – professionals in need of very particular sources of information, so specific that they were unobtainable in Kraków (the city is known for its scientific prowess and creative potential). The Office changed with the Digital Revolution by scanning and sharing the unique works of the University professors and by constantly connecting with other libraries, creating new kinships.

The Library is a central part of the University, holding the vision of knowledge, passion, and affinities; everchanging, but always there.

# Bibliography

*Academica. Cyfrowa wypożyczalnia międzybiblioteczna książek i czasopism naukowych.* (n.d.). Retrieved October 21, 2022, from https://academica.edu.pl

Akademia Górniczo-Hutnicza im. Stanisława Staszica w Krakowie. (2022). Fakty i liczby. Retrieved October 21, 2022, from https://www.agh.edu.pl/o-agh/fakty-i-liczby

Biblioteka Główna AGH. (2014). *Maria Skłodowska Curie w 100-lecie otrzymania Nagrody Nobla.* Retrieved October 21, 2022, from http://www.bg.agh.edu.pl/MSC/msc.php?page=01rok

Fulara, I., Zych, A., & Imiołek-Stachura, K. (2017). Międzynarodowa współpraca w zakresie wypożyczeń międzybibliotecznych – na przykładzie Wypożyczalni Międzybib-liotecznej Biblioteki Głównej Akademii Górniczo-Hutniczej. *Biblioteka i Edukacja*, (12), 1-18. https://bazhum.pl/bib/article/578365

Krawczyk, J., Janczak, B., & Dudziak-Kowalska, M. (2009). *Bibliotekarze w dziewięćdziesięcioleciu Akademii Górniczo-Hutniczej*. Uczelniane Wydaw. Nauk.-Dydakt. AGH im. S. Staszica.

## About the Author

Karolina Forma, Librarian

Biblioteka Główna Akademii Górniczo-Hutniczej (AGH University of Science and Technology Main Library)

https://orcid.org/0000-0003-1876-1951

Karolina Forma (1990) started her work in AGH University of Science and Technology Main Library in 2018, joining the Branch Libraries and Inventory Control Division. She takes an active part in ongoing financial and quantitative inventories of various collections. In 2019, she became a co-manager of the library's social media and took a creative role in the library's marketing team. Since 2021, she has been expanding her ILL knowledge by helping in the Interlibrary Loans Office. Her work interests include 20th century illustrations, history of advertisement, and social media.

# The gentle art of giving and receiving: Resource sharing services in a museum library

**Florian Preiß**

## Abstract

Deutsches Museum library is one of the largest collections of literature on the history of science and technology. It shares its resources to a broad audience by interlibrary loan and document delivery. The latter is carried out as part of the Subito library network. Also, to share its collection online, the museum library takes part in the platform Deutsches Museum Digital.

## Keywords

## Article

The Deutsches Museum, one of the world's most distinguished museums for science and technology houses a library that is the largest museum library in Germany and one of the world's leading research libraries on the history of science and technology.

Oskar von Miller, an engineer and pioneer of early electrification, founded the Deutsches Museum in 1903 to promote interest in scientific and technological advances among a broad audience. The museum's exhibition building, situated on an island in the river Isar (Fig. 1), opened its doors in 1925 and replaced a provisional exhibition in Maximilianstraße ([Füßl, 2010](#)). The object collection grew fast and over the following decades the museum became a prominent place for science and technology education – and one of the major tourist attractions in the Bavarian capital. Before the COVID-19 pandemic broke out in Europe in 2020, the museum attracted well over one million visitors a year.

The Deutsches Museum perceives itself as a research museum and operates its own research institute focusing on the history of science and technology and maintains close ties with the relevant departments of the three large Munich universities: Ludwig-Maximilians-Universität (LMU), Technische Universität (TUM) and Universität der Bundeswehr (UniBW).

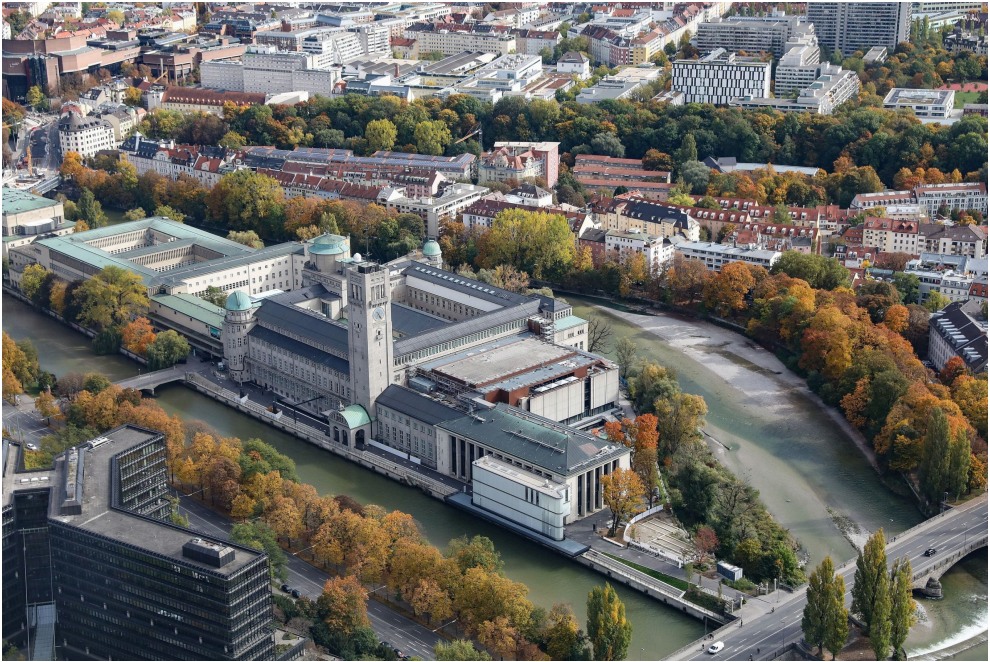

*Figure 1: Deutsches Museum (Photo: Bernd Wackerbauer)*

## The Deutsches Museum library and its collection

The library has been an integral part of the museum from the very beginning. It relocated multiple times to several different locations in Munich until the present building was inaugurated in 1932 (Hilz, 2017; Hilz, 2021). Since then, the library building has housed the reading rooms, the offices and media storage in the immediate vicinity of the exhibition building (Fig. 2).

The library collection comprises more than 950,000 volumes, including a spectacular rare book collection of more than 15,000 books published prior to 1800. An extensive stock of historical journals, grey literature, patent documents, telephone directories, railway timetables, address books and unique resources also account for the library's importance as the place to be for anyone conducting research on the history of science and technology in Germany. This is one of the reasons why the library, in cooperation with Bayerische Staatsbibliothek (Bavarian State Library, BSB) has been approved a "Fachinformationsdienst" (specialized information service) by the Deutsche Forschungs-gemeinschaft (German Research Foundation) since 2016, which entitles to additional acquisition and personnel funds ([Hilz, 2021](#); [Winkler, 2020](#)).

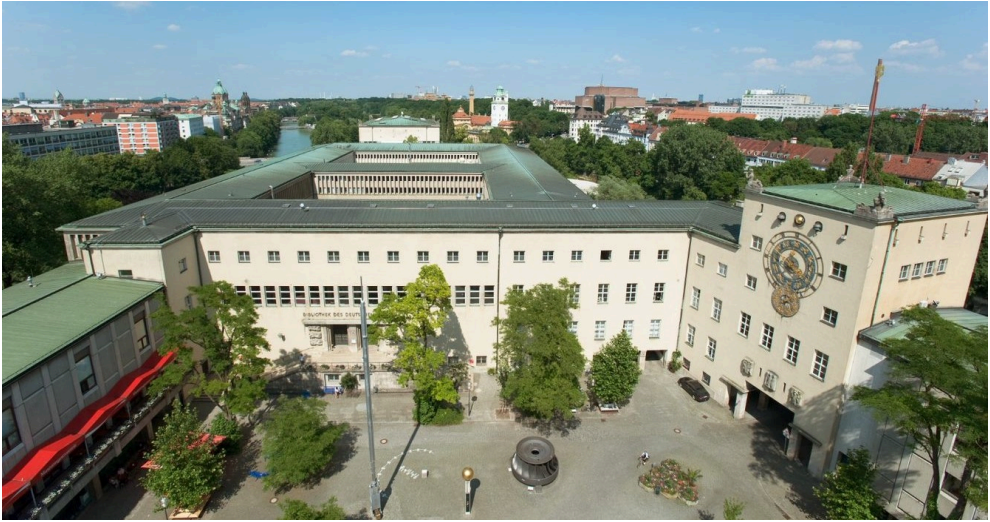

*Figure 2: Deutsches Museum library (Photo: Dt. Museum Archiv CD 62745)*

## Library services

While this Fachinformationsdienst aims to assist state-of-the-art research in the field of history of science and technology, the 25 library staff members also serve the everyday needs of museum staff, both academic and non-academic. All 586 employees and 184 volunteers (end-of-year 2021) may benefit from the full range of the library's services including free interlibrary loan orders and they are entitled to request staff member library passes to borrow books. The same favourable conditions apply to those visiting scholars who are granted a scholarship by the museum's research institute. The reading room with about 130 seats is not only open to staff and scholars, but also to the general public (Fig. 3). However, these external patrons are limited to a basic service without the possibility of borrowing books or placing interlibrary loan orders.

The number of external visitors reached the 50,000 mark in 2018 and just missed this peak in 2019 (Fig. 4). Due to the restrictions imposed by the authorities because of the COVID-19 pandemic, the number of visitors dropped sharply in 2020, reached a low in 2021 and have gradually started to recuperate in 2022. This dramatic decline in visitors had little impact on the statistics for inter-lending services: interlibrary loan and document delivery services remained in demand, and even increased in some parts. While many academic libraries in Germany had to shut their doors from mid-March 2020 for nearly two months and interrupted their inter-lending services, the Deutsches Museum library could continue operation despite the fact that some of the staff were required to work remotely ([Hilz & Winkler, 2020](#)).

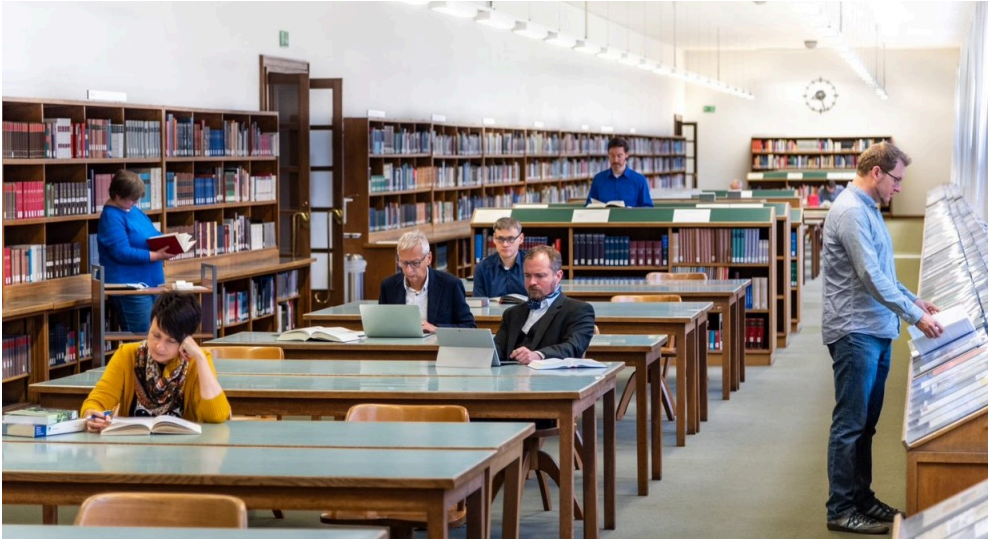

*Figure 3: Reading room (Photo: Dt. Museum Archiv CD L 7596)*

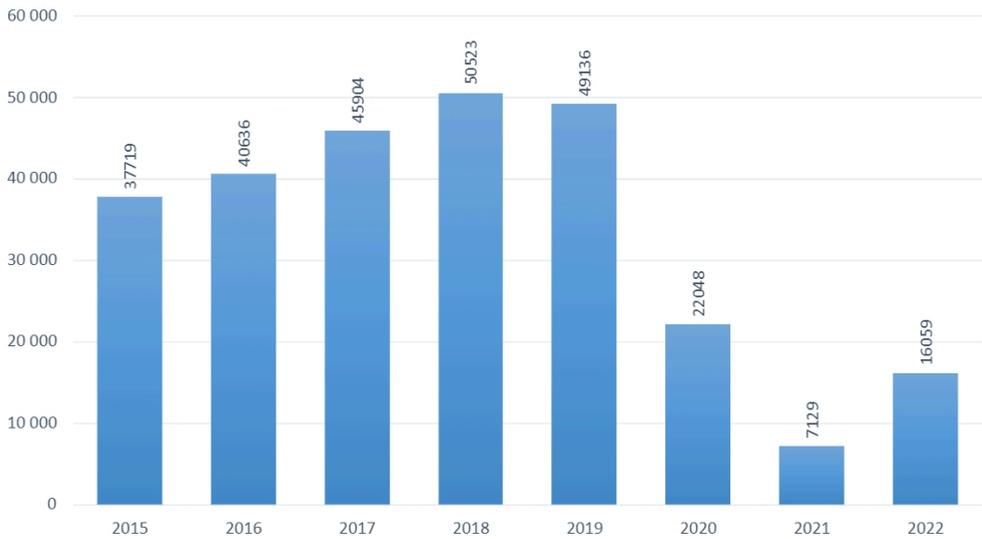

*Figure 4: Number of library visitors*

# Deutsche Museum library as a giving partner for sharing resources

The Deutsches Museum library is both a giving partner in the conventional interlibrary loan system as well as in the Subito network. In conventional ILL, it gives books on loan nationally and internationally and delivers paper copies to requesting libraries.

In 2019, the Deutsches Museum library became a full member of the library network Subito (Homann, 2019). This registered association is a non-profit organization of about 40 cooperating libraries, mostly from Germany with few members from Switzerland and Austria. These libraries started to cooperate in the mid-1990s to promote delivery of documents and to establish new (digital) methods of delivery. For a small fee, libraries worldwide as well as individuals and companies from Germany, Austria, Switzerland and Liechtenstein can open a user account and order via a simple webform. These orders come with a guarantee that they will be fulfilled within a maximum of 72 hours (24 hours for urgent orders, subject to a surcharge).

Digital deliveries for ILL and Subito however face various legal hurdles. Clasen (2019) recently published a comprehensive overview of legal issues and challenges in German interlibrary loan and document delivery. For example, the legal restrictions limit orders of copies from monographs to a maximum of only 10% of the content. And it gets most complicated when (paper or digital) copies from newspapers or newsstand magazines are needed: The strong position of publishers makes it virtually impossible to deliver these via ILL or Subito (unless the tangible originals are sent – but this is rarely the case, as huge bound volumes or fragile paper make it difficult).

In recent years, the numbers of conventional ILL orders at Deutsches Museum library slowly descended from a peak of 454 delivered items in 2018 to a low of 261 in 2021. Subito orders on the other hand showed a huge growth in the first four years that the library has been participating: The amount more than tripled from 422 in 2019 to 1,599 in 2022 (Fig. 5). Although it looks as if this significant increase in orders resulted in an unreasonable additional workload, this is not the case. The workflow for processing Subito orders could be kept very lean and the easy delivery options via the Subito web application helped to minimize the workload. The centrally processed invoicing by the Berlin-based Subito head office ensures that the library staff is not burdened with this task in any way. In addition, the head office and the "Zeitschriftendatenbank" database (the German union catalogue for journals) as underlying metadata are responsible for checking deliverability from a legal point of view – so the library staff does not have to check whether incoming orders comply with rights management and licence contracts. All necessary licence fees that might potentially be requested by rights holders are cleared by the head office. In summary, the decision to participate in Subito meant, with minimal effort, increasing the visibility of the library's collection and at the same time making it more available worldwide. This is clearly reflected in the origin of the orders received: While conventional international interlibrary loan orders regularly come in from barely a dozen countries, the Museum library has received Subito orders from six continents and 40 countries so far (Fig. 6).

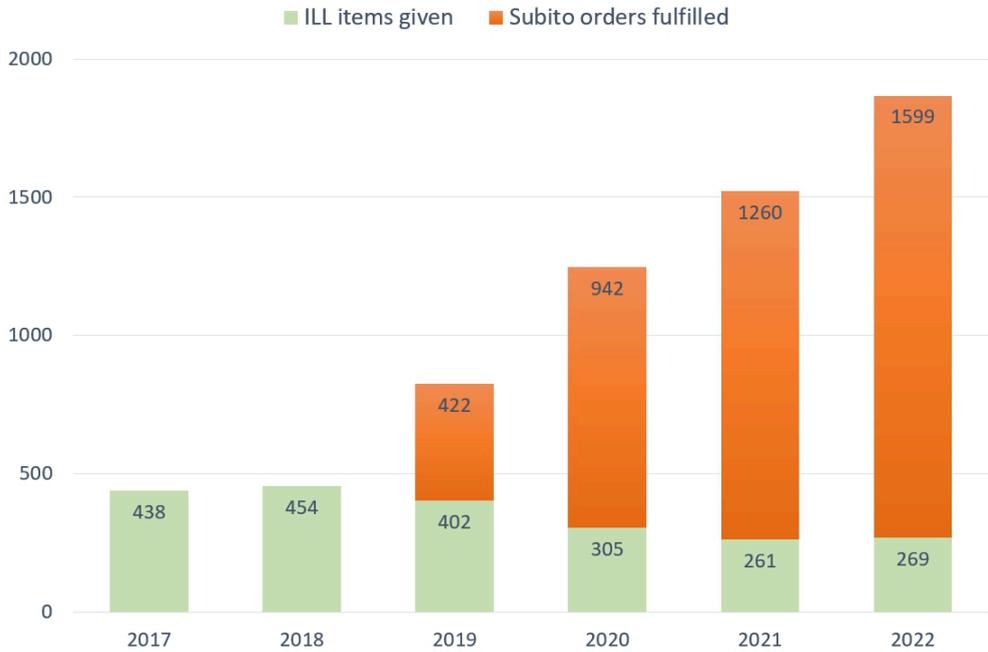

*Figure 5: Document delivery (giving)*

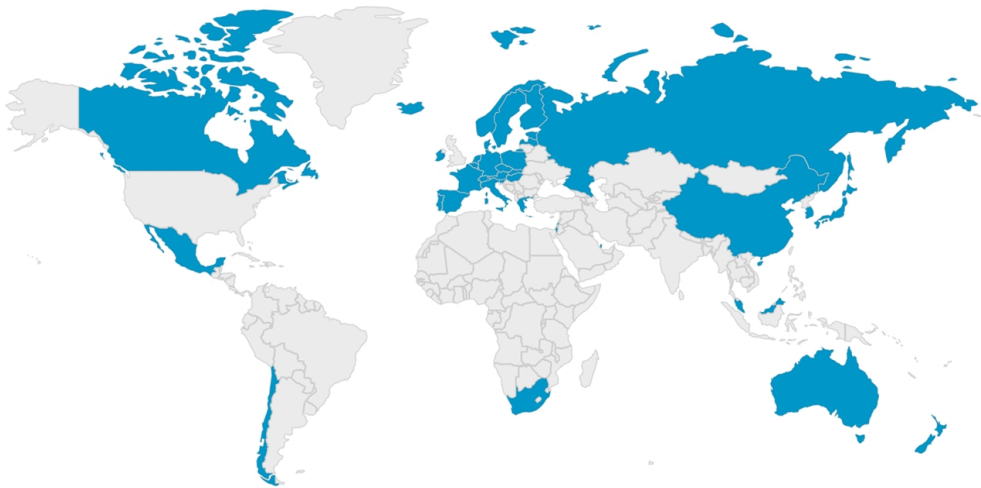

*Figure 6: Countries, from which Deutsches Museum's direct delivery service customers originate from (as of end of year 2022)*

# Deutsches Museum Digital: sharing resources online

Deutsches Museum Digital[1] is the place where all the digitization efforts of the museum departments (archives, library and object collection) are brought together (Huguenin, 2019). In the future, this platform is thought to act as a one-stop shop to provide users with important research material from archival sources, library material and museum objects.

So, when considering the giving part of its resource sharing activities, it should not be forgotten, that the museum library's digitization projects also play an essential role as they contribute to the online availability of relevant literature. To do this, the library is pursuing two strategies – "custom-made" vs. "mass business", but in the end, the results from both processes will be hosted on Deutsches Museum Digital (Bunge, 2019). While the "custom-made" digitized books, on the one side, are created by the library staff themselves and focus on small, coherent collections of predominantly rare books, the "mass business", on the other side, is an outsourcing cooperation in partnership with Google Books. Within the scope of this partnership, to date more than 50,000 non-copyrighted volumes (both journals and monographs, approx. 5% of the collection) have been digitized. Google Books is presenting these volumes on its own platform and hands over a library copy of the digitized material, which the museum successively provides on Deutsches Museum Digital.

# The receiving part

For its internal users (i.e. staff and visiting scholars), the library procures virtually any book available worldwide that is needed. The importance of this procurement service is highlighted by the results of an anonymous survey conducted among library users in May 2022 (Heller et al., 2022): Among the 100 museum staff participants in the survey (among which 56 academic and 44 non-academic), only 16% always find all the literature they need in the museum library's collection. This means that in most cases external library material is essential for professional and scientific work.

Most of the orders (regularly three fifths in the past years) can be fulfilled as a loan service from state library BSB and the two large university libraries LMU and TUM (Fig. 7). This service is called "Ortsleihe" (city loan). Therefore, a courier visits the respective libraries on a weekly basis and if needed more frequently to pick up the media that has been ordered by the museum's ILL staff. The courier will then bring the media to the museum library and notify the users that their ordered documents are ready for collec-tion at the circulation desk. The courier will copy or scan those media which are only for exclusive use on the premises of the respective libraries. As a result, thanks to the Ortsleihe service, users save themselves long walks or subway and tram trips through the city. They also do not have to worry about reminder fees, as the museum library's accounts at BSB, LMU and TUM enjoy preferential conditions.

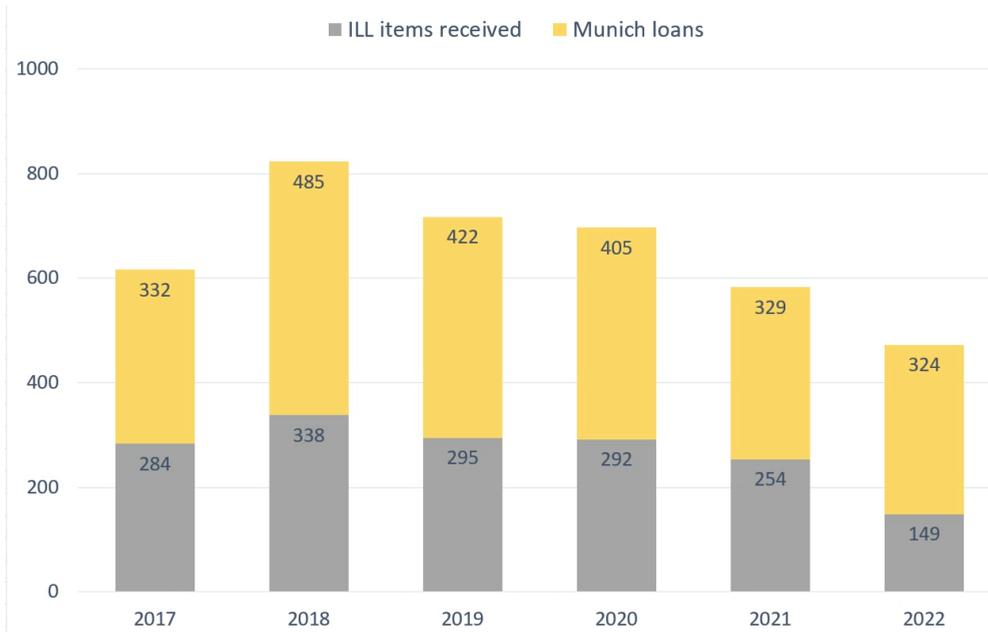

*Figure 7: Document delivery (receiving)*

All documents that cannot be ordered within Munich will be ordered via ILL, first via the regional Bavarian Interlending network and, if necessary, Germany-wide. For ordering items from libraries in Bavaria and Germany, the library uses the ZFL tool ("Zentraler Fernleihserver" – a centralized ILL server application maintained by the German library networks). This tool can also be used for international orders, but the lending library in this case receives all the order details via a standardized e-mail.

Virtually all orders can be kept free of charge for the museum's library patrons: Monographs in German interlibrary loans are shared on a reciprocal basis between libraries. Costs arise only for packing and shipping, but these are not charged to the user. The same applies to the occasional fees charged for copies/scans. However, as higher fees are often incurred in international interlibrary loans, it is not always guaranteed that these costs can be covered by the library. To pay for orders from foreign libraries, the Deutsches Museum library participates in the IFLA voucher scheme and also accepts payment vouchers from libraries abroad. In practice, a considerable surplus of incoming IFLA vouchers can keep all orders placed at foreign libraries free of charge for museum staff.

For urgent orders of papers needed by museum staff, the Deutsches Museum library may also act as an ordering party itself in Subito. In this case, it will be charged the regular fees by Subito. However, since one result of the anonymous survey (Heller et al., 2022) was that only about a quarter of internal users use these interlibrary and document delivery services, and almost a third are not even aware of it, there is still potential for increasing its visibility.

## Allowing users to focus on their research

All these library services in their entirety form an essential part of the research infrastructure at the Deutsches Museum. They are highly appreciated by Deutsches Museum researchers, as they do not have to concern themselves with how and where to get literature and scientific information. They simply address their own institution's library, which defines the supply of literature as one of its core tasks. On a national and international level, the Deutsches Museum library contributes to a wider network of resource sharing and inter-lending services, so that external users also may benefit from the Museum's extensive historical sources.

## Bibliography

Bunge, E. (2019, September 4-6). *Boutique-Digitalisierung vs. Massengeschäft: ein Erfahrungsbericht aus der Bibliothek des Deutschen Museums* [Conference session]. ASpB-Tagung 2019, Frankfurt a. M., Germany. https://opus4.kobv.de/opus4-bib-info/files/16695/Bunge_Digitalisierung_DNB.pdf

Clasen, N. (2019). Digital possibilities in international interlibrary lending – with or despite German copyright law. In P. D. Collins, S. Krueger & S. Skenderija (Eds.), *Proceedings of the 16th IFLA ILDS conference: Beyond the paywall – Resource sharing in a disruptive ecosystem* (pp. 92-100). National Library of Technology. http://www.nusl.cz/ntk/nusl-407836

Füßl, W. (2010). The Deutsches Museum and its history. In W. M. Heckl (Ed.), *Technology in a changing world: The collections of the Deutsches Museum* (pp. XIV–XVI). Deutsches Museum.

Heller, L., Morich, M., & Pietsch, A. (2022). *Umfrage zur Nutzung der Bibliothek des Deutschen Museums.* Unpublished survey conducted as part of a team course achievement at Hochschule für den öffentlichen Dienst, Fachbereich Archiv- und Bibliothekswesen München.

Hilz, H. (2017). *Die Bibliothek des Deutschen Museums: Geschichte – Sammlung – Bücherschätze*. Deutsches Museum.

Hilz, H. (2021). Bibliothek auf der Isarinsel: Geschichte und Gegenwart der Bibliothek des Deutschen Museums. *AKMB-news*, 27(1), 51-58. https://doi.org/10.11588/akmb.2021.1.89932

Hilz, H., & Winkler, C. (2020). Corona, subito und die Bibliothek. *Kultur & Technik*, 44(3), 50.

Homann, M. (2019). subito Document Delivery Zurück in die Zukunft in Zeiten des Wandels. *b.i.t. online*, 22(5), 422-425. https://www.b-i-t-online.de/heft/2019-05/nachrichtenbeitrag-homann.pdf

Huguenin, F. (2019). Deutsches Museum Digital: Online-Portal von Archiv, Bibliothek und Objektsammlung. *AKMB-news*, 25(2), 3-11. https://doi.org/10.11588/akmb.2019.2.72577

Winkler, C. (2020). Googeln Sie noch oder finden Sie schon? *Kultur & Technik*, 44(3), 48-53.

## About the Author

Florian Preiß, Librarian

Deutsches Museum (The Deutsches Museum)

https://orcid.org/0009-0007-0649-192X

Florian Preiß studied medieval and postmedieval archaeology in Bamberg and library sciences in Munich. Since 2016, he has been librarian and head of the reading room department at Deutsches Museum library in Munich. His main responsibilities include the interlibrary loan and document delivery activities of the Museum library.

# Four years of ILL service at the Faculty Library of Arts and Philosophy (Ghent University): Findings and challenges

**Joris Baeyens**

## Abstract

This paper aims to give an insight into the working of ILL services at the Faculty Library of Arts and Philosophy of the Ghent University. Additionally, we take a closer look at two challenges in the field of ILL: e-book lending and controlled digital lending (CDL). Putting these challenges into practice can give a new incentive to ILL. At this moment, however, it's still work in progress.

## Keywords

Interlibrary loan; ILL; E-book lending; Controlled digital lending; CDL; Digital rights management; DRM; Licences; Copyright

## Article

## Introduction

In June 2018 the Faculty Library of Arts and Philosophy (Ghent University) started its own ILL services. Previously ILL requests were handled by Ghent University's central library (also known as the Book Tower).

During this four-year period we have gained an insight in the world of ILL. This paper is about these insights, but also about the challenges ILL faces more broadly.

These challenges mainly concern the transformation of print ILL into digital ILL. At the present moment, ILL staff still have to deal frequently with processing print books, the so called 'returnables'. This may change from the moment a legal framework is put into practice for the cases outlined below:

- making e-book licences lendable for ILL;
- making a digital copy of a print book lendable through CDL.

Both these challenges will be explored further on.

This paper is derived from the presentation I gave at the European Staff Week in Liège, June 2022.[1] The enriching presentations and discussions with colleagues gave extra input to this paper.

## ILL platforms as primary vehicles

When started in 2018 our ILL Service had access to four ILL platforms (or providers): Impala (Belgium), Subito and Gateway Bayern (Germany) and Sudoc (France). Later on, we joined OCLC WorldShare ILL (June 2019) and RapidILL (May 2020). It was clear from the beginning that these platforms are essential vehicles for ILL work. To underline this: in 2021, 96% of all ILL requests were processed via ILL platforms. As a consequence, sending an email directly to libraries to obtain an ILL has rather become an act of last resort ([Posner, 2019](#)).

Figure 1 shows the fulfilment rate of ILL requests in 2021 per provider. The two main providers, Impala and RapidILL, are explained in more detail below.

Having access to a range of providers significantly increases the chances of a positive ILL outcome. Which provider to choose first is often made on the basis of a WorldCat search as this gives a good indication of which libraries hold the item. Besides availability of the item, other criteria are also taken into account, such as rate, sending costs, import duties, loan period. Overall, one can say that an assessment of the above elements determine which provider and consequently which library or series of libraries are addressed.

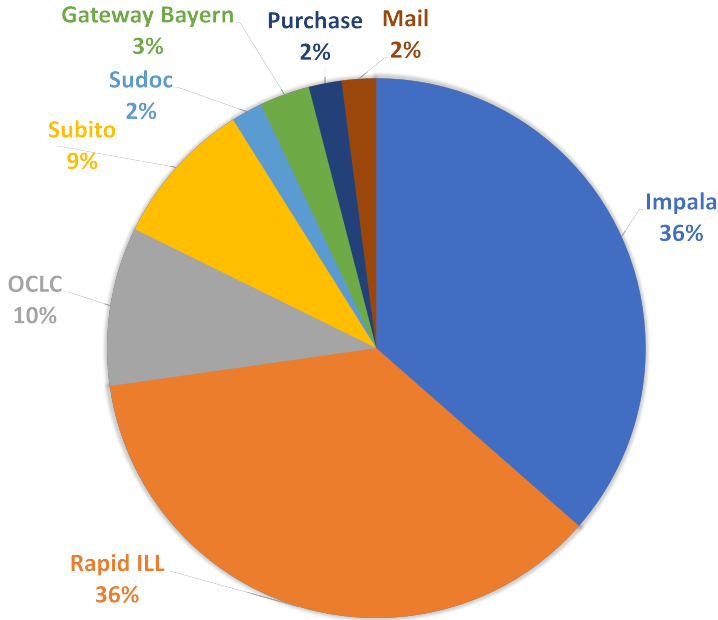

**2021**

*Figure 1: Faculty library of Arts and Philosophy (UGent): ILL requests per supplier (2021)*

## Impala

Impala—developed by Antwerp University—is Belgium's national ILL provider since 1992. It has a large pool of members ranging from Belgian university libraries to public and subject libraries. In 2021, Impala represented 36% of our ILL transactions, mainly for print books. Twice a week a car shuttle transports ILL requested print books to and from Belgian universities. Being centrally located, the campus of Vrije Universiteit Brussel serves as exchange hub. Given that Belgium is a small country this shuttle system has proven to be cost-effective and efficient. Therefore Impala is our first choice for print material.

## RapidILL

In terms of ILL requests, RapidILL is on a par with Impala: both represented an equal share of 36% of ILL requests in 2021. Thanks to its high fulfillment rate and short delivery time RapidILL has 'rapidly' become our premium provider of non-returnables (articles and book chapters).

The foundations of this platform were developed by ILL staff at Colorado State Uni-versity Libraries in 2001, then taken over in 2019 by Ex Libris/ProQuest (Ex Libris, 2019) that further expanded and commercialized the platform.

To unravel the success of Rapid ILL it may be interesting to have a closer look at its main characteristics (Ex Libris, n.d.):

- Related libraries are organized around 'pods'. These pods are created to support peer or consortium resource sharing. A library can take part in several pods. A group of libraries can ask ProQuest to create a new Pod (California Digital Library, 2022).
- The use of algorithms that utilize load-leveling and time zone awareness.
- A holdings database tailored to resource sharing needs. The database matches requests down to the year level. Holdings are updated on a six-month cycle.
- A library's local holdings are checked in terms of availability prior to sending the ILL request.
- Built around a community of reciprocal users.

## *Recent initiatives*

The "HERMES Project" is a library driven initiative, a European project connected with the interlibrary loan and document delivery services, that came into existence during the COVID period (Lomba et al., 2023):

> The whole objective of HERMES Project is to support effective access to knowledge for the academic community. To achieve this, HERMES plans to strengthen skills in searching and getting quality academic texts. In addition, and to provide easier access to scientific literature, librarians are the other target of HERMES' objective: the improvement of librarians' competencies in terms of resource sharing (RS), and the creation of a more effective international system to share electronic resources.

Early in 2022 the 'International ILL Toolkit' was set up by librarians Brian Miller (Ohio State University Libraries) and Lapis Cohen (University of Pennsylvania) in partnership with OCLC. It's a freely available crowd-sourced database, easily accessible on a Google spreadsheet. The core of the toolkit is a listing of lenders from around the world with contact, delivery, and payment information. It also provides valuable tips & tricks and ILL request templates in different languages (OCLC SHARES, n.d.).

These two initiatives can be applauded for their efforts to make resource sharing freely accessible for all libraries worldwide, especially as membership with most platforms tends to come with a high price tag.

Another new service came from German provider Subito. After 15 years in existence, it has expanded its ILL service to include the delivery of scanned book chapters.

Purchase is considered when ILL is not possible or too expensive. In these cases a short assessment is carried out before switching to purchase. Most often priority is given to students as they cannot rely on funds to order books with our acquisition service. Also publications on subjects that do not fit into our collection are prone to be purchased. ILL requests for recent publications that are not yet available for ILL are redirected to our acquisitions department.

A specific category are (unpublished) PhD theses. Making theses available in 'open access' repositories is a fairly recent trend. And older dissertations are still often excluded from ILL, either by the defender's copyright choice or by a 'no loan' policy of the issuing university. Over the years ProQuest Dissertations & Theses database (PQDT) has built a vast collection of PhD theses. When a dissertation cannot be obtained through ILL but is available at PQDT a purchase from this provider is made.

## Print versus digital

An interesting trend that occurred during 2018–2021 was a shift from 'print' to 'digital' document delivery at our faculty library. By 'print' we mean traditional paper books whereas 'digital' refers to either an article or a chapter delivered electronically.

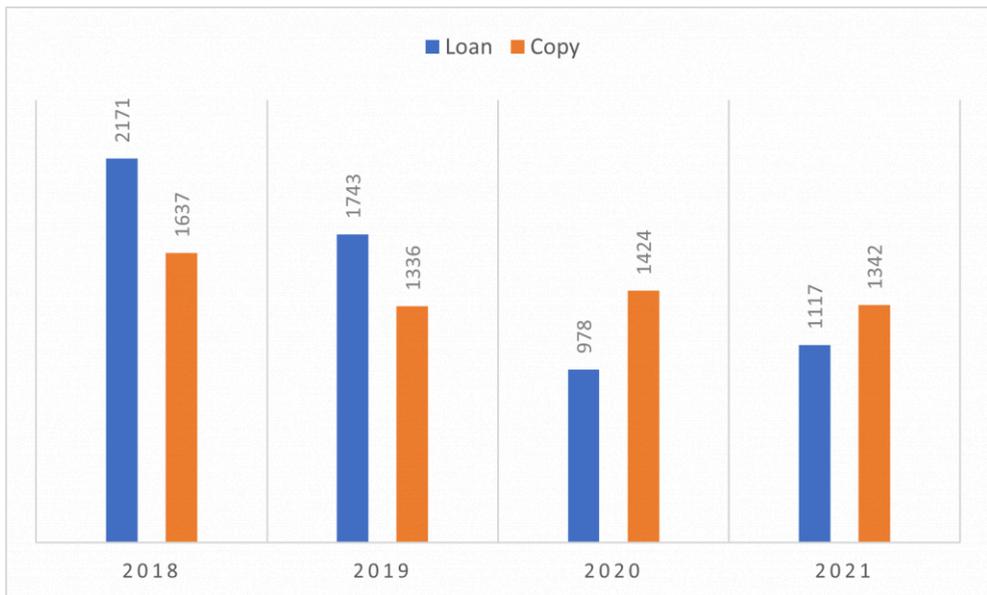

*Figure 2: Faculty library of Arts and Philosophy (UGent): loan vs digital supply (= copy)*

This shift from print to digital occurred in the midst of the COVID period (2019–2020) and continued in 2021 (Figure 2). This evolution towards (more) digital document delivery was undoubtedly accelerated by the physical library closure during the pandemic. Figures for 2022 indicate that this trend more or less persists: we processed 1,019 print requests versus 1,040 digital requests.

## Challenges

### E-books and ILL

Libraries are evolving towards hybrid providers offering both a paper and digital collection. Within our library, e-book purchase represented 10% of overall acquisition in 2022. In the meantime, several initiatives were taken to increase e-book acquisition such as a financed digital-first collection policy and five EBA's with major publishers, while two DDA's have been launched since the beginning of 2023.

A well-known concern amongst librarians is the issue of e-books not being eligible for ILL. This concern is fully legitimate: as libraries' e-book collections grow less material will become shareable through ILL.

That being said, our ILL service has not yet had to deal with 'unfulfilled' requests as a result of "e-book only" availability. To date, there has always been a library that had a print copy for loan.

But it goes without saying that ILL e-book lending would be a big step forward. Several initiatives have already been made to address this issue. Occam's Reader (USA), a university driven initiative is one of them. With a technical infrastructure fully in place it proves that ILL e-book lending can be successful. The scale of this project is however limited to a number of affiliated research libraries (Greater Western Library Alliance) and one publisher: Springer (OCCAM'S READER, 2021).

A project aiming at a wider scale is ProQuest's "Whole e-book ILL pilot" (Roll et al., 2022). It was launched in November 2021, with expected outcomes by the end of 2022. This project involves six major publishers and four libraries. The key concept behind this project is making e-book licences lendable between libraries. In a nutshell: when circulating an e-book through ILL, the lending library loses one licensed access during the E-ILL period. The patron of the borrowing library receives an E-ILL URL and has access to the e-book for two weeks via the ProQuest E-book Central platform. The borrowing patron can then make use of the e-book in accordance with the DRM restrictions of the e-book licence. These DRM restrictions can however put a damper on the patron's user experience. Fortunately, the industry shift towards DRM-free e-books has accelerated in recent years (Rodriguez, 2021, p. 17), which gives libraries the possibility to build a DRM-free e-book collection.

The lack of "lendability" of e-books is the most common issue that has recently been put forward ([Weston, 2015](), p. 53). If this pilot turns out to be successful and scalable to a wide range of university libraries, it might turn out to be a breakthrough for e-book ILL.

## *Print books and CDL*

As mentioned earlier, print ILL is gradually losing ground to digital ILL. In 2021 however, 45% of our patrons' ILL requests were still for print books. Almost half of these print book requests were fulfilled through the Belgian ILL Impala network and the other 55% of print books came from abroad.

There are however some significant downsides relating to print book ILL, such as:
- risk of loss or damage during sending,
- cumbersome import formalities for non-EU parcels,
- sending costs and import duties,
- delivery time,
- ecological foot print,
- antiquarian books: too precious for sending.

Another interesting finding is that in 2021 46% of our print ILL requests were for books published before 2000. Apart from some exceptions, it is unlikely that pre-2000 publications will ever appear as an e-book. This is where CDL or Controlled Digital Lending can offer a durable, cost-effective and efficient alternative. Controlled Digital Lending is a model by which libraries digitize their print materials with purpose of making the digital version available for lending.

Controlled Digital Lending has three important requirements ([SPARC, n.d.]()):
- A library must own a legal copy of the physical book.
- A library must maintain an "owned to loaned" ratio: simultaneously lending no more copies than it legally owns.
- A library must take technical measures to ensure that the digital file cannot be copied or redistributed (= implementation of DRM).

Many libraries applied CDL on a local scale during the pandemic and the resulting closure of libraries. They achieved this by making their print materials available in digital format on the basis of user requests. On a broader scale, CDL can be a mechanism for ILL, and certainly one that can be a solution to offset the downsides outlined above.

US and Canadian libraries as well as organizations such as ProQuest and OCLC are already anticipating on the use of CDL as a modern method of ILL. There is however a legal issue preventing CDL from being widely implemented. At the time of writing the Internet Archive, a pioneer in CDL, is involved in a lawsuit with four major publishers on the use of CDL.

This lawsuit and the long-lasting international debate on copyright explain why enabling Controlled Digital Lending is a long-term process. Let's hope for a positive outcome in favor of CDL. An outcome that gives legal ground to CDL could surely act as a catalyst for wider implementation (International Federation of Library Associations and Institutions, 2021).

## Concluding thought

Referring to a statement by the organizers of this staff week: "No library can buy or hold everything its users need. At a certain point, librarians need to pool their resources and collaborate to provide access to what they don't have." The presentations held during the staff week at University of Liège demonstrated the interesting initiatives taken by libraries on how to deal with this challenge in various ways.

The challenges that are dealt with in this paper concern two domains: "e-book ILL" and "ILL-CDL". Both these domains involve legal and technical requirements that need to be met. The continuing efforts made by libraries, library associations, advocacy organizations and aggregators are smoothing the way. Once these requirements are successfully fulfilled, these domains may, in my humble opinion, well become foundations of a future Interlibrary Loan/Resource Sharing landscape.

## Bibliography

Baldwin, P. (2014). *The copyright wars: Three centuries of Trans-Atlantic battle*. Princeton University Press.

California Digital Library. (2022, April 18). *Announcing the WEST RapidILL pod*. https://cdlib.org/cdlinfo/2022/04/18/announcing-the-west-rapidill-pod/

Ex Libris. (2019, June 20). *Ex Libris acquires RapidILL, provider of leading resource-sharing solutions*. https://exlibrisgroup.com/press-release/ex-libris-acquires-rapidill-provider-of-leading-resource-sharing-solutions/

Ex Libris. (n.d.). *About Rapid*. Retrieved December 21, 2022, from https://rapid.exlibris-group.com/Public/About

International Federation of Library Associations and Institutions. (2021, June). *IFLA position on controlled digital lending*. https://repository.ifla.org/bitstream/123456789/1835/1/ifla_position_-_en-_controlled_digital_lending.pdf

Lomba, C., Marzocchi, S., & Mazza, D. (2023). HERMES, an international project on free digital resource sharing. In F. Renaville & F. Prosmans (Eds.), *Beyond the Library Col-* lections: *Proceedings of the 2022 Erasmus Staff Training Week at ULiège Library*. ULiège Library. https://doi.org/10.25518/978-2-87019-313-6

OCCAM'S READER. (2021). *About*. Retrieved December 21, 2022, from https://occam-sreader.lib.ttu.edu/about/index.php

OCLC SHARES. (n.d.). *International ILL toolkit*. Retrieved January 30, 2023, from https://docs.google.com/spreadsheets/d/1bL94PplzLTXYyxWNRn-wQj4cn6YhCo62V4kFHhW0S15E/edit#gid=0

Posner, B. (2019). Insights from library information and resource sharing for the future of academic library collections. *Collection Management*, 44 (2–4), 146–153. https://doi.org/10.1080/01462679.2019.1593277

Rodriguez, M. (2021). Licensing and interlibrary lending of whole eBooks. *Against the Grain*, 34(4), 17–18. https://issuu.com/against-the-grain/docs/atg_v33-4/s/13538491

Roll, A., Lee, C., Fraenkel, J., & Murphy, W. (2022, February 23). *The key to modern resource sharing: whole ebook lending and more* [Video]. YouTube. https://www.youtube.com/watch?v=TCbppWTh3wg&t=11s

Roncevic, M. (2020). Digital rights management and books. *Library Technology Reports*, 56(1). https://doi.org/10.5860/ltr.56n1

SPARC. (n.d.). *Controlled digital lending*. Retrieved December 21, 2022, from https://sparcopen.org/our-work/controlled-digital-lending/

Weston, W. (2015). Revisiting the Tao of resource sharing. *The Serials Librarian*, 69(1), 47–56. https://doi.org/10.1080/0361526X.2015.1021988

Wu, M. M. (2019). Shared collection development, digitization, and owned digital collections. *Collection Management*, 44(2–4), 131–145. http://dx.doi.org/10.1080/01462679.2019.1566107

## About the Author

Joris Baeyens, Acquisitions and ILL Librarian

Universiteit Gent, Faculteitsbibliotheek Letteren en Wijsbegeerte (Ghent University, Faculty Library of Arts & Philosophy)

https://orcid.org/0000-0001-5612-561X

Joris Baeyens (1969) obtained an associate degree in Information Management: Library and Archives. He started working in 2012 at Faculty Library of Arts & Philosophy as acquisitions librarian and joined the library's ILL team in 2018.

# Changing tools, changing habits, changing workflows: Recent evolutions of the interlibrary loan service at ULiège Library

**Fabienne Prosmans and François Renaville**

## Abstract

Over the last few years, the interlibrary loan (ILL) service of the University of Liège Library has evolved considerably, both in terms of habits and workflows. In this article, we will explain the main stages of this evolution: (1) first reduction in the number of ILL units (from eight to five) and involved operators (from 15 to 10) within the homemade ILL solution (2015); (2) use of the resource sharing (RS) functionality in the new Alma library management system (2015); (3) second reduction in the number of ILL units (from five to only one) and in involved operators (from 10 to six) (2018); (4) subscription to an international broker ILL system (RapidILL) for electronic and digital materials and its integration with Alma (2020); (5) project of peer-to-peer resource sharing for print materials between Alma instances of university and research libraries in Belgium (2022), and (temporary) free ILL service to all University users (2020–2022). The aim of these changes is to harmonise the practices of ILL operators, reduce the quantity of manual and administrative operations and tasks devoted to ILL and supply materials that do not belong to the library collections in a fairer, faster and more fluid way. While all of these changes have been implemented gradually over the years, not all of them have been deployed in a concerted manner among all stakeholders.

## Keywords

# Article

# Introduction

## *About ULiège Library*

The University of Liège (ULiège) was founded in 1817 by William I of the Netherlands, then King of the United Kingdom of the Netherlands, as it existed between 1815 and 1839. For a long time, the *Bibliothèque Générale* ('General Library') was the only real official university library. However, from the end of the 19th century, due to the expansion of university institutes throughout the city and the growing specialisation of fields of knowledge, certain departments and faculties started to develop their own collections and libraries ([Cuvelier et al., 2008](#)).

In 1954, the decision was taken to locate the whole university outside of the city, at the Sart Tilman hill. This led to a radical change in its documentary landscape in 1956, with the creation of specialised *Unités de Documentation* ('Documentation Units') that were close to users and therefore related to faculties or departments. All these units were characterised by a high degree of autonomy and independence. Very few services, such as cataloguing, were common and shared.

In the early 2000s, with the rise of digital technology, the university made the decision to modernise its libraries and documentation units, to reorganise them in depth and to group them into four, and subsequently five, major entities: Philosophy and Literature Library; Law, Economics, Management and Social Sciences Library; Science and Technology Library; Life Sciences Library; and Agronomy Library. "In addition to the much-needed modernisation and development of the electronic library, this restructuring aimed to improve and extend the service to users and to coordinate the acquisitions policies in a more efficient manner" ([ULiège Library, 2022](#); [Cuvelier et al., 2008](#)).

In January 2014, the members of the Library Board organised a 'greening' in order to define the main lines of reorganisation of the library and to review the library strategy in the context of the forthcoming bicentenary. They laid the foundations for various elements of reflection aimed at reviewing the strategy and missions, increasing collaboration between branches and services and reflecting on the necessary reconversion from a 'traditional collection library' to a new model where services are at the centre ('Library as a Service').

## ILL at ULiège Library, 2005–2015

From 2005 to 2015, eight library branches and 15 operators managed ILL services. We mainly used Impala and Subito[1] as partners for our resource sharing requests. Impala is used for both borrowing and lending requests, while Subito is only used for borrowing requests. It is important to note that the 15 operators were not working full-time on ILL services. Most worked less than 0.5 FTE, while some even less than 0.1 FTE.

Our ILS was Aleph, but we did not use its ILL module. Our interlibrary loan was managed within an in-house solution called MyDelivery, which was developed with APIs and integrated within the library website and the Primo discovery tool. A blank ILL request form was accessible in Primo MyAccount.

Interlibrary loan operators received an email informing them that a new request required processing. They then manually created the request in Impala or Subito and followed the evolution of the request in these brokers. MyDelivery was used for internal use only, e.g. for the creation, cancellation and management of requests, for invoicing and for some tracking features.

## Changes in ILL services

### 1) Reduction in the number of ILL units and operators

There were several reasons for which a reduction in the number of units and ILL operators was needed:
- The eight independent ILL units available were too many, and the smallest ones received very few requests. We wanted to avoid operators in these libraries having to process ILL requests only occasionally;
- Some libraries are not particularly far from each other. We wanted to free these libraries from the burden of providing an ILL service of their own if a close collaboration with another library located nearby could be established;
- We also wanted to avoid the risks of interruption of the ILL service for libraries with only one or two ILL operators. Even when libraries experienced no understaffing issues, some ILL units used to be closed for several weeks during holiday periods, for example;
- We migrated to Alma in early 2015. By reducing the number of ILL operators before the migration, we could consolidate the existing workflows between operators and reduce the number of operators to train in resource sharing (RS) in Alma.

---

1. Impala is the Belgian ILL platform for resource sharing while Subito is the German platform.

We decided to keep five ILL units and 10 operators (Fig. 1). At ULiège Library, interlibrary loan tasks do not occupy operators full-time since we estimated that the total ILL tasks represent 1.5 FTE.

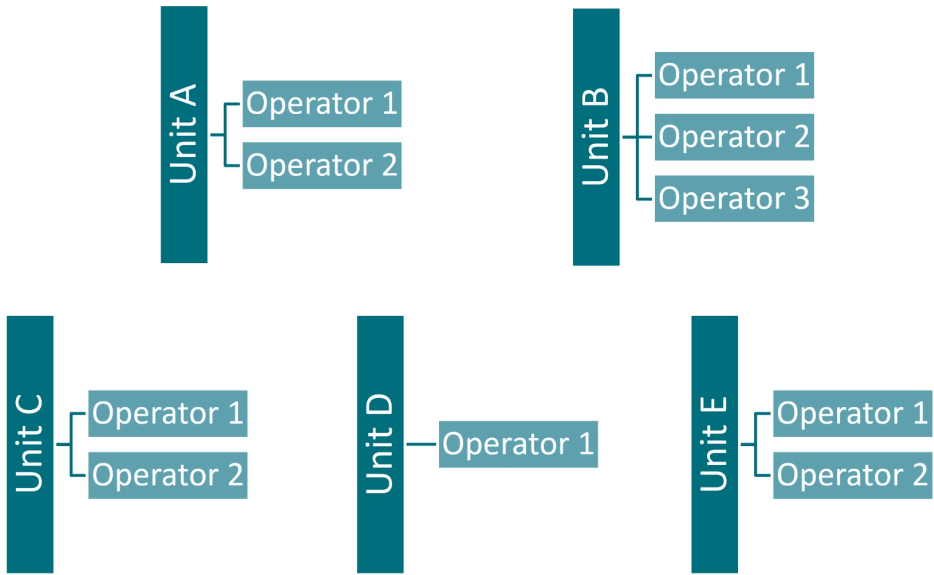

*Figure 1: New organisation after the first reduction (2015)*

After a few weeks, we noticed that this first reduction allowed us to simplify some tasks. Additionally, no negative impact on service quality was reported, and we did not receive any negative feedback from users.

Working with five ILL units allowed us to maintain significant freedom of organisation for each unit. However, we noticed that each library saw itself as an independent unit; there were very few interactions between operators and very little communication between units themselves. At that stage, the service was still strongly reliant on the library type: the unit associated with the Health Library used to only serve medicine students and members of the Faculty of Medicine and the University Hospital Centre, the unit associated with the Science Library limited its service to students and researchers in chemistry, physics, and astrophysics, etc.

This first reduction was accepted easily because of the migration to Alma: many organisational changes were expected, not only for ILL (Bracke, 2012; Renaville, 2018; Brisbin et al., 2020). Additionally, some ILL staff members saw this as a good opportunity to leave the service and to redirect their professional career within ULiège Library instead of changing their habits or following training sessions on resource sharing in Alma.

## 2) Moving to the Resource Sharing functionality in Alma

Context

During the Alma implementation project, we decided to stop working with our in-house solution and to start fully using the resource sharing functionality in Alma. Our goal was to make the most of the new library system and to take advantage of Alma's various features.

Concretely, for borrowing requests, a book provided by a partner is temporarily located in one of the five resource sharing libraries and receives a temporary barcode. This allows the patron to loan the book. Letters such as the On Hold Shelf Letter, Courtesy Letter or Overdue Notice Letter are automatically managed and sent by the system, as these items are owned by the library. It is possible to define circulation rules (overdue fine, maximal renewal period, etc.) and to add resource sharing fees to the user account.

For lending requests, it is easier to know when a book is sent to a partner for resource sharing purposes. Previously, with the in-house solution, operators had to lend the book in their own name. Moreover, with Alma, the borrowing partner receives the Borrower Overdue Email Letter when the due date is reached.

Finally, it is easy to retrieve statistics from Alma Analytics such as the number of borrowing requests by user group or the number of borrowing or lending requests by partner or material type.

This allowed us to stop using our home-made solution MyDelivery, but we still had to use our main partners (Impala and Subito) for delivery.

What we did at ULiège Library

Each of our five ILL units was associated in Alma to a library for which the option 'Is resource sharing library' had been checked. In Alma, libraries "within an institution or campus may be configured to have relationships where they enable patrons to check in or check out resources at another location, send items back and forth, or acquire (purchase) items on behalf of each other. If a library is configured to do this for libraries at other institutions (and not only within the institution), it is known as a *resource sharing library*" ([Ex Libris, n.d.-a](#)).

We assigned the most appropriate resource sharing library to each of our 40,000 users (students and faculty and staff members) based on their field of research or studies.

Then, we created two specific locations for each Resource Sharing Library in Alma. The first location was for borrowing requests. A physical document that comes from a partner is temporarily assigned to this location; as mentioned above, the document receives a temporary barcode which allows the user to loan the document. The other location was for lending requests; documents that go to a partner 'leave' their original location and are temporarily located in this new location.

We also created two circulation rules by library, one for borrowing and one for lending requests. Libraries did not always define the same rules. Some libraries allowed readers to borrow materials from a partner, while others only accepted consultation in the reading room.

Finally, we had to create our partners for resource sharing in Alma. At this stage, all our partners were of the email profile type.

There are three ways to create borrowing requests in Alma at ULiège Library (Fig. 3):

1. First, for a resource not available at the University of Liège, the patron can use the View It or Get It tab in Primo or a database. In this case, we used the Ex Libris's resource sharing form where metadata are automatically taken from the bibliographic record. Once the form is submitted, a borrowing request is created in Alma, and interlibrary loan operators can find it in their task list;

2. Users can use a blank form for their ILL requests. Since we wanted a more customised request form than the one proposed by Ex Libris by default, we set up a new blank form (Fig. 2) and integrated it with the Alma APIs (De Groof, 2017). Once the form is submitted by the patron, the Alma API creates a borrowing request in the system, and ILL operators can start working on it;

3. Patrons can send an email to the ILL unit with their request details. Operators then manually create the request in Alma and start the fulfilment process.

*Figure 2: In-house blank form*

Compared to the Ex Libris default resource sharing blank form used at the time, we found that ours had a lot of advantages:

- It can be used for borrowing requests created by registered and non-registered

users such as university services or third-party services. It can also be used for lending requests created by other libraries or partners;
- The blank form is interactive and adapts according to the selections made (requested material, etc.), and only the necessary fields are displayed;
- If the user enters the DOI or PMID, the fields for bibliographic details are automatically filled in;[2]
- The journal title field is interfaced with a locally managed journal database containing over 45,000 journals, which makes it possible to obtain the standardised journal title and the ISSN in the ILL requests;
- The 'delivery and payment' fields are fully configurable according to the status or user group of the requester;
- In case of physical delivery, it is mandatory to select a pickup location from the drop-down list. No pickup location field is displayed for digital delivery.[3]

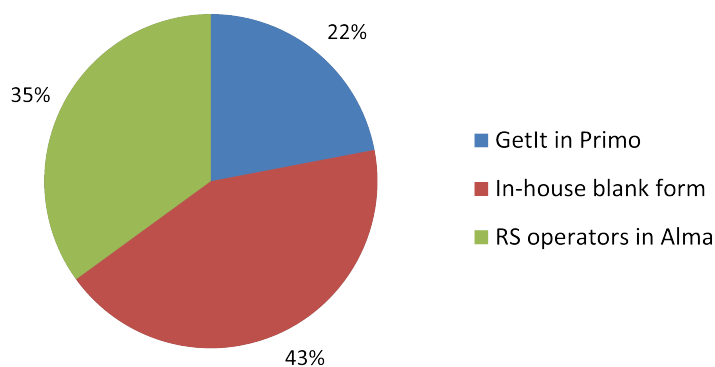

*Figure 3: Origin of borrowing requests (2017)*

In February 2015, when ULiège Library went live with Alma, we stopped using MyDelivery and exclusively worked with our five resource sharing libraries in Alma.

### 3) Simplification of the ILL backend

Working with five ILL libraries and 10 operators represented an improvement, but that number was still too high. Therefore, in autumn 2017, the Library Board decided to reor-ganise and centralise the interlibrary loan service and to create a new single entity. From January 2018, the new entity would virtually group all ILL operators together, who would work as a team and serve all library branches and patrons.

---

2. This feature is now also available in the Primo default form.
3. Currently, the pickup location is still proposed in the Alma/Primo blank form. See: Hide Pickup Location on Blank ILL Article/Book Chapter Request Form https://ideas.exlibrisgroup.com/forums/308173-alma/suggestions/46394707-hide-pickup-location-on-blank-ill-article-book-cha.

We deactivated the five existing resource sharing libraries and decided to use the default resource sharing library in Alma. This library is only used for resource sharing.

Due to the new organisation and the pooling of human resources, it was no longer necessary to spread the work across 10 librarians, so we also reduced the number of operators to six (Fig. 4).

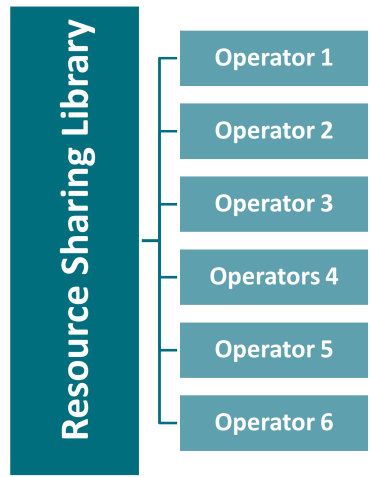

*Figure 4: Virtual Resource Sharing Library – New organisation after the second reduction (2018)*

Although ILL operators do not physically work at the same place, the resource sharing library has a physical address; this is important for physical items provided by partners. Thanks to a shared calendar, operators complete their resource sharing tasks in turn. They only spend a few hours per week on these tasks, and they complete these tasks in the library where they are employed.

Thus far, we have experienced various advantages of using only one resource sharing library:

- The same resource sharing library is now automatically allocated to all users. Previously, the resource sharing library allocated to users was based on each user's field of research or studies. Not all users had the most appropriate resource sharing library assigned to them in their user record, and some users had no assigned resource sharing library at all.
- This centralisation has reduced the number of transits of physical documents for borrowing requests.

- ILL operators collaborate more closely now as they are all members of the same virtual team.
- There is more fluency in ILL delivery. All ILL operators complete their RS tasks in turn. Additionally, in the past, there was the possibility of all ILL operators being out of the office at the same time (because of sickness, holiday, etc.), thus interrupting the ILL service in a library. Now, the probability that all six operators are absent at the same time is close to zero, and a substitute can easily be found if needed.
- The practices of the resource sharing operators have been strongly harmonised. Previously, there were various differences in workflows. For example, some RS libraries allowed the loan of books (provided by partners) to local researchers, while others only allowed consultation in the reading room. Now, all operators follow the same procedure, and there is one single circulation rule for borrowing requests and one single rule for lending requests.

However, we encountered some difficulties during this transition from five resource sharing libraries to one. The interlibrary loan service was the preserve of some operators. The transition involved more transparency across ULiège Library and the pooling of human resources, and costs and profits became shared at the top level. The change was not easily accepted by some colleagues: some operators had concerns regarding having more work, while others regarding having too little work. Some operators worried about having to process requests related to less familiar disciplines. Some operators also regretted the loss of conviviality due to the automation of certain tasks, such as automatic emails sent from Alma. Finally, since the Library Board decided to reduce the number of libraries from five to one at the end of November 2017, we only had a few weeks to change the workflows to be ready for early 2018. This was not easily accepted by some operators, as several thought the change had been applied too fast (Prosmans & Renaville, 2018).

## 4) Moving forward with RapidILL for digital requests

For almost 30 years, ULiège Library had used the Impala system as its main ILL platform for managing its outgoing and incoming requests. Developed by the University of Antwerp Library in 1990, Impala was chosen in 1992 as the document delivery platform for interlibrary loan by the national conference of chief librarians of Belgian universities (Corthouts et al., 2011). Impala is accessible to all types of libraries (academic, research and public) and can handle different types of documents. Journal articles and physical books are mainly provided; Impala does not handle the provision of e-books. Of the 177,458 requests processed in 2000, the year with the highest number of requests, 134,556 requests came from university libraries (23,660 for books and 110,896 for articles) – which, however, only represent about 10% of Impala's membership (Corthouts et al., 2011). This shows the important use of the platform among Belgian university libraries.

However, despite the reliability of the tool and the fact that the Impala solution was still evolving (Saerens, 2014), ULiège Library representatives disliked the fact that there was no integration between the Impala and Alma systems. Until early 2015, with the use of the in-house MyDelivery application, incoming requests from Impala partners were managed exclusively within Impala. With the switch to Alma and in order to work in the most integrated way possible with the new system, it emerged that ILL staff often had to enter data twice: requests via Impala (lending requests) were entered manually into Alma, whereas direct requests from users (borrowing requests) using the Primo form or the blank form had to be re-entered by the operators (copy and paste) into Impala in order to be fulfilled. This was particularly inefficient, and there was a need for simplification. This situation was also present between Alma and Subito, but with the latter being a more occasional supplier compared with Impala, the cumbersome nature of the need for double data entry was less noticeable.

In spring 2020, the COVID-19 pandemic had a strong impact on library services and accessibility, and many libraries tried to reinvent themselves or innovate where possible (Whitfield et al., 2020; Ashiq et al., 2022; Skalski, 2023). In March, Ex Libris, which had acquired the RapidILL solution in 2019 (Ex Libris, 2019; Baeyens, 2023), offered its customers the opportunity to join the new COVID-19 pod to test the resource sharing solution and participate in the collective effort to address the pandemic (Veinstein, 2020). Interested libraries could simply ask to join the COVID-19 pod to submit requests for articles and book chapters (digital delivery), with the RapidILL client community fulfilling their requests as much as possible.

ULiège Library saw this as an excellent opportunity to test a new tool at a time when it was needed:

- Due to the lockdown that started on 18 March 2020 in Belgium, the library had to close its doors and limit its services and activities to anything that could be executed remotely. The ILL service was therefore heavily impacted.
- Most of the libraries we worked with in the Impala network also temporarily ceased their activities, and it was difficult, if not impossible, for the physical provision to meet the ILL requests of our users.

Participation in the COVID-19 pod was therefore timely for us, and ULiège Library joined the pod in a trial. The results of this trial proved to be conclusive for our library.

Between 23 April and 31 August 2020, approximately 200 requests sent to RapidILL were satisfied (a satisfaction rate of 95%). Delivery time ranged from a few hours to, exceptionally, a few days – e.g. when requests were sent on a Friday at the end of the day.

Regarding the learning curve, the ILL team members expressed their satisfaction with the use of RapidILL. They found that this new solution provided added value, especially when requests for items were more difficult to meet using Impala.

In addition, unlike Impala, RapidILL can interface with Alma to avoid double data entry, which saves time considerably for staff. Although this integration was not tested in the trial, it was of great interest to the staff. This Alma/RapidILL integration not only allows users' requests to be transferred directly to RapidILL but also, when these requests are fulfilled by a RapidILL partner, to mark the request as completed in Alma and to send, from Alma, a delivery notification letter with a download link directly to the user. The automation process is such that a student or researcher can now submit a request for an article via the form and receive the requested copy a few hours later, without any intervention from ILL staff.

RapidILL's business model is based on a fixed annual fee set out in a contract. The number of requests processed by RapidILL is therefore irrelevant – unlike in Impala or Subito, where there is a cost per use. However, there is an obligation of means to be provided since RapidILL's clients are both borrowers and lenders (Delaney & Richins, 2012).

Finally, the predictable cost of a RapidILL subscription also allows libraries to consider certain issues and services differently. For example, it may become easier to forego certain non-essential journal subscriptions if their articles can quickly and easily be supplied through RapidILL. Similarly, offering a free ILL service to students and researchers becomes an easier option to consider (Renaville & Prosmans, 2023).

In view of these positive elements following the trial and analysis of the possibilities, ULiège Library therefore decided to subscribe to the RapidILL solution in September 2020. While it had initially seemed interesting to us to integrate a French-speaking pod or a pod with other Belgian libraries which would have also subscribed to RapidILL, this option quickly proved pointless because the default pods which we had joined were sufficient for our needs. Moreover, one must keep in mind that the more pods a library joins, the more likely they are going to be asked to supply other pod members with materials (Baeyens, 2023). A balance must therefore be struck between the two.

Although no French or Belgian pods were created, we were able to benefit from the arrival of several important new partners for us in the RapidILL community:

- Belgian universities: KU Leuven, Ghent University, Free University of Brussels (ULB);
- French universities: Université Clermont Auvergne;
- Swiss universities: BCU Lausanne, SLSP consortium (via Rapido).

While RapidILL was initially an additional supplier for article and book chapter requests, it became the preferred partner (by default) from November 2020 onwards thanks to the activation of processing automation between Alma and RapidILL (Fig. 5).

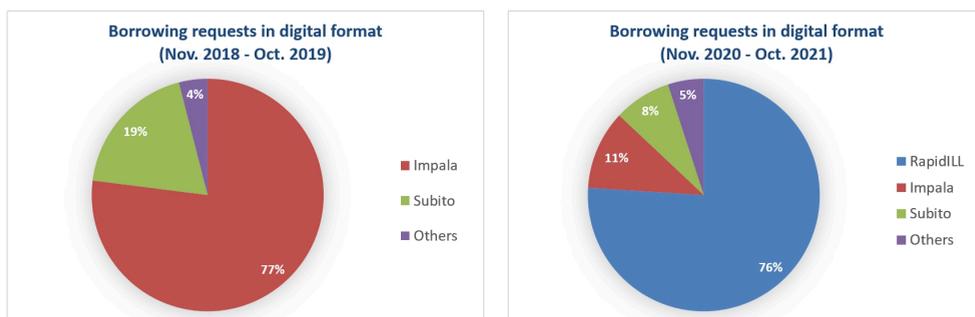

*Figure 5: Distribution by partner for digital delivery*

After a few months of use, we were able to establish the following advantages of working with RapidILL:

- Full and easy integration with Alma: Our ILL operators only sign in to RapidILL to access statistics or check potentially problematic cases;
- Fast delivery of requested documents: 10 hours on average for 2021 and 12 hours for 2022 (Table 1);
- Known annual cost (high borrowing usage has no financial impact);
- The growth of the Rapido solution (fully integrated in Alma itself), which benefits the RapidILL customer community as every new Rapido customer also joins the general COSMO pod (Ex Libris, n.d.-c);
- The fact that it is a global community and, therefore, that RapidILL libraries' collections are richer than Impala's historical partners (Table 2);
- The fact that it is not necessary to be an Alma customer to use RapidILL. For example, in 2021, the Internet Archive became a provider within RapidILL (Kahle & Coelho, 2021).

| | 2021 | 2022 |
|---|---|---|
| ULiège Borrowing Requests | 2,267 | 2,391 |
| System Average Borrowing Requests | 2,947 | 2,898 |
| ULiège Borrowing Filled | 2,009 | 2,187 |
| ULiège Borrowing Unfilled | 199 | 146 |
| ULiège % Filled | 89% | 91% |
| System Average % Filled | 95% | 95% |
| ULiège Average Filled TAT (Hours) | 10 | 12 |
| System Average Filled TAT (Hours) | 13 | 13 |

*Table 1: RapidILL borrowing statistics for 2021 and 2022*

| Lending Institution | Number of Requests |
|---|---|
| Internet Archive | 283 |
| Tufts Univ., Hirsh Health Sciences Lib. | 139 |
| Université de Clermont-Ferrand | 102 |
| Université Libre de Bruxelles | 97 |
| Rowan University | 90 |
| University of Pennsylvania | 86 |
| Olivet Nazarene University | 83 |
| Univ. of MD Baltimore Health Sciences Library | 83 |
| Ghent University Library | 69 |
| University of Chicago | 64 |
| Binghamton University | 60 |
| Oregon Health and Science University | 53 |
| Indiana Univ./Purdue Univ. Indianapolis | 51 |
| University of Arkansas Libraries | 49 |
| KU Leuven LIBIS | 48 |

*Table 2: Top 15 of our RapidILL lending suppliers for digital delivery in 2021 and 2022*

On the negative side, there is currently no priority for electronic holdings over print holdings of a lender library in RapidILL. Presently, format is not a parameter in the algorithm used to select the most appropriate partner. If it were possible for electronic holdings to take priority over physical holdings, it would help institutions to avoid digitising print materials (a procedure that requires time and human resources) if others have an electronic version that can be provided more easily and quickly. This is, however, a request for improvement that has been introduced and is supported by the customer community.[4]

## 5) Peer-to-peer resource sharing between Alma instances for physical items

While digital delivery of articles and book chapters was now mainly conducted via RapidILL, Impala remained the preferred solution for physical delivery (almost exclu-

4. Electronic holdings should take priority over physical holdings https://ideas.exlibrisgroup.com/forums/935109-rapidill/suggestions/43633749-electronic-holdings-should-take-priority-over-phys.

sively books). Here, too, however, there was a need for interfacing: ULiège Library's ILL staff still had to manually re-enter borrowing requests from our users into Impala.

In 2021, Ex Libris (n.d.-b) introduced the Resource Sharing Directory:

> The directory is a central place with up-to-date information about resource sharing libraries in Alma. Libraries opt-in and define themselves in the directory with a few simple steps. Being in the directory will allow other libraries to find your institution and create a peer-to-peer resource sharing relationship with you simply and with less back and forth communication over technical connection details.

Libraries in the Alma Resource Sharing Directory are grouped into regional pods. In this way, they can easily create peer-to-peer relationship for physical supply. ULiège Library joined the Resource Sharing Directory by the end of 2021. In Belgium, the Free University of Brussels (ULB) and the European Commission Library also joined the Directory.

Partnering within the Directory brings several advantages:
- It helps to avoid filing encoding and tracking borrowing requests for physical items in the Belgian ILL platform Impala.
- Several steps of the workflow can be made automatic in Alma:
  - Sending a borrowing request from Alma to the partners of the Directory can be automatic;
  - With rules created by the library, partners from the Directory can easily be automatically assigned to requests in physical format;
- It allows a reduction in the cost of the ILL service: each request made via Impala is charged at the broker system level (cost per use). Using the Resource Sharing Directory is free of charge and no borrowing costs are expected between partners. Additionally, if the costs of the service decrease, this represents an additional argument in favour of a free ILL service for the benefit of our students and researchers.

Working with the Free University of Brussels (ULB) Library has been successful so far. In the first two months of 2023, out of a total of 78 successful physical borrowing requests, 28 were delivered by the Free University of Brussels and 34 via the Impala network. Previously, all requests completed by ULB would have been processed within Impala.

We hope that we can now expand this P2P experience to other partners in Belgium, such as KU Leuven or libraries from European institutions.

## 6) Free ILL service for our patrons

In 2018, the members of the Library Board decided to conduct a feasibility study on the possibility of offering a free ILL service to the university community. The study started in 2019 and was not completed until 2022, partly due to the pandemic. However, the pandemic offered some interesting insights into the study. To mitigate the impact of the national lockdown in March 2020 on library services, it was decided that the ILL service would become free for students, faculty and staff during the pandemic. As a consequence, and unexpectedly, the lockdown allowed for a real pilot phase for the free service, with consistent usage data, in parallel to the study itself (Renaville, 2020; Renaville & Prosmans, 2022; Renaville & Prosmans, 2023).

## Feasibility study

In order to avoid rushing into this project and repeating the mistakes made a few years ago when the five ILL units were merged too fast into one service (see above), it was decided to take time for analysis. The study paid particular attention to:
- The internal context;
- The external context;
- The stakeholder analysis;
- A strengths, weaknesses, opportunities and threats (SWOT) analysis.

### Internal context

This part of the study focused on the current organisation of the ILL service, the history of the service and the various recent changes (see above), including the tools used so far, and several statistical data going back to 2018 (outgoing requests by format and faculty, profiles of requesters, suppliers, costs and revenues).

### External context

In addition to a literature review, attention was also paid to the national context, and a comparative analysis with other Belgian universities was conducted: Did other universities offer a free service? If so, under what conditions and for whom? Finally, two conjunctural elements were added: the COVID-19 crisis which closed libraries and thus temporarily prevented any physical supply, increasing in parallel the digital demands, and the trial of the RapidILL solution, becoming in itself an internal influencing factor.

*Stakeholder analysis*

The objective of this section was to identify the different stakeholders impacted by the project:
- Primary stakeholders: most affected, either positively or negatively;
- Secondary stakeholders: indirectly affected;
- Tertiary stakeholders: least impacted.

A power-interest grid was also helpful to identify, and eventually contain, the stakeholders' interests, any potential risks or misunderstandings ('me issues'), any mechanisms that positively influence other stakeholders and the project and, finally, any negative stakeholders as well as their adverse effects on the project. An important aspect of the work consisted in individual interviews with eight key stakeholders, i.e. all ILL staff members and those responsible for billing the service. Surprisingly, the stakeholder analysis showed that all ILL operators were in favour of a completely free ILL service for students, and the only ones (slightly) against a free service were was among the sponsor themselves (the Library Board, in 2021).

*SWOT Analysis*

All these elements from the internal and external factor analyses and the stakeholder analysis were then synthesised in a SWOT analysis, which, as the acronym suggests, was a compilation of the project's strengths, weaknesses, opportunities and threats

## Pilot phase during COVID-19

The pilot phase was conducted from spring 2020 (first lockdown) to spring 2022, which is an exceptional length of time to have solid statistics and usage data, but conversely, so long that reverting to a paid service would have been difficult and would have required evidence in the form of consistent data and arguments. In addition to the statistics, the ILL operators also had two years' experience in a free context.

*Usage data*

While the provision of physical documents decreased between 2019 and 2020 due to the temporary closure of some ILL services, there was a very significant increase in digital requests from 2020 onwards, mainly due to the free service provision (Fig. 6). Unsurpris-ingly, it was mostly university staff members (x2 between 2020 and 2021) and graduate students (x3 between 2020 and 2021) who benefitted the most from the service. Although undergraduates are moderate users of the service in absolute terms, they used it five times more between 2020 and 2021 (Fig. 7).

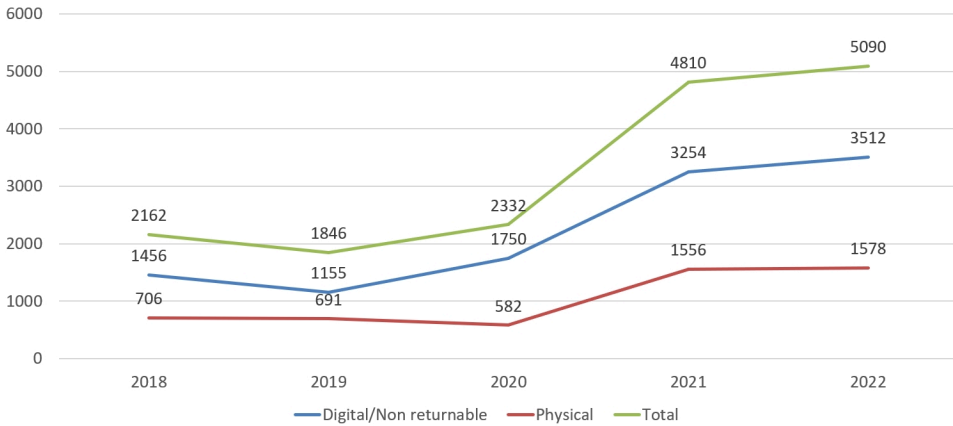

*Figure 6: Completed borrowing requests (2018-2022)*

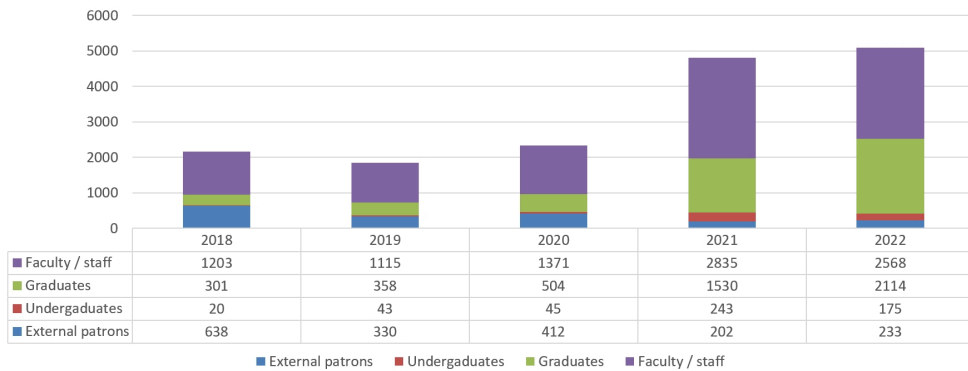

*Figure 7: Completed borrowing requests by patron type (2018-2022)*

## First reactions among users and ILL staff

Not surprisingly, feedback from students on the free service was positive, and the increase in requests suggested that a need had been met.

Among ILL operators, it was appreciated that there was no longer any administrative burden (ILL fees to be charged to students' accounts, invoices to be managed, follow-up of payments, etc.).

There were some rare abuses from users in physical requests, but most striking was probably the fact that for the supply of books outside Belgium, postal charges were now borne exclusively by the library, which increased the cost of the service. However, due to the subscription to RapidILL and its integration with Alma (see above), the significant increase in the number of requests did not impact the library's finances or the workload of the staff.

## Recommendations to the Library Board Committee

The feasibility study, consolidated by a two-year pilot phase, recommended to the Library Board to consider that free ILL should be a priority within the Library as a Service model and that it should be maintained for the ULiège community:

- With no limit for digital articles and chapters delivered via RapidILL;
- With an increase budget for postal charges;
- With no symbolic contribution to the costs.

In parallel to these measures, Alma P2P for physical delivery should be encouraged. The study also recommended some practical safeguards to prevent abuses and avoid uncontrolled costs, including purchasing a print copy or an e-book version when necessary (Van den Avijle & Maggiore, 2023), and to raise awareness among patrons that there are always hidden costs for the university and the library.

## Conclusion

The ILL service of the University of Liège Library has undergone major changes for seven years: changes in habits with the reduction of the number involved operators and ILL units and the reorganization of the team, changes of tools (from a homemade solution to Alma/RapidILL), and changes in workflows for more integration and automation.

Thanks to these changes, it was observed that the professionalisation and involvement of the staff have increased and that processes have become more highly standardised. A smaller number of people does not imply a lower-quality service. On the contrary, delivery is even more efficient as the service is rarely closed and supply is even faster. ILL operators were not necessarily against the changes, but they must be engaged in the evolution of the service because they need to understand what is taking place and must want to be involved in the thinking, decisions and changes in their work. Because change management takes time and preparation, it is important that managers carefully consider their priorities in implementing changes and take it step by step. From an institutional point of view, peer-to-peer resource sharing has led to new collaborations with partners. Finally, if costs remain under control, a free ILL service is possible.

# Bibliography

Ashiq, M., Jabeen, F., & Mahmood, K. (2022). Transformation of libraries during Covid-19 pandemic: a systematic review. *The Journal of Academic Librarianship*, 48(4). https://doi.org/10.1016/j.acalib.2022.102534

Baeyens, J. (2023). Four years of ILL service at the Faculty Library of Arts and Philosophy (Ghent University): findings and challenges. In F. Renaville & F. Prosmans (Eds.), *Beyond the Library Collections: Proceedings of the 2022 Erasmus Staff Training Week at ULiège Library*. ULiège Library. https://doi.org/10.25518/978-2-87019-313-6

Bracke, P. J. (2012). Alma at Purdue: the development partnership experience. *Information Standards Quarterly*, 24(4), 16–20. https://doi.org/10.3789/isqv24n4.2012.03

Brisbin, K., Storova, H., & Enoch, T. (2020). Out with the old, in with the new: revising ERM workflows in a time of change. *Serials Librarian*, 78(1–4), 74–78. https://doi.org/10.1080/0361526X.2020.1715762

Cuvelier, C., Vanhoorne, F., & Thirion, P. (2008). Le Réseau des Bibliothèques de l'ULg. *Lectures*, (158) (Novembre-décembre), 18–25. https://hdl.handle.net/2268/642

Corthouts, J., Van Borm, J., & Van den Eynde, M. (2011). Impala 1991–2011: 20 years of ILL in Belgium, *Interlending & Document Supply*, 39(2), 101–110. https://doi.org/10.1108/02641611111138905

De Groof, R. (2017, July 3). Blank form for resource sharing requests. *Ex Libris Developer Network*. https://developers.exlibrisgroup.com/blog/Configuration-and-main-Perl-scripts-of-the-resource-sharing-request-blank-form-at-ULg/

Delaney, T. G., & Richins, M. (2012). RapidILL: an enhanced, low cost and low impact solution to interlending, *Interlending & Document Supply*, 40(1), 12–18. https://doi.org/10.1108/02641611211214233

Ex Libris. (2019, June 20). *Ex Libris acquires RapidILL, provider of leading resource-sharing solutions*. https://exlibrisgroup.com/press-release/ex-libris-acquires-rapidill-provider-of-leading-resource-sharing-solutions/

Ex Libris. (n.d.-a). *Configuring parameters of a resource sharing library*. Ex Libris Knowledge Center. Retrieved February 28, 2023, from https://knowledge.exlibris-group.com/Alma/Product_Documentation/010Alma_Online_Help_(English)/030Ful-fillment/080Configuring_Fulfillment/060Resource_Sharing/030Configuring_Parameters_of_a_Resource_Sharing_Library

Ex Libris. (n.d.-b). *Contributing to the Resource Sharing Directory*. Ex Libris Knowledge Center. Retrieved February 28, 2023, from https://knowledge.exlibrisgroup.com/Alma/Product_Documentation/010Alma_Online_Help_(English)/030Fulfillment/050Resource_Sharing/040Contributing_to_the_Resource_Sharing_Directory

Ex Libris. (n.d.-c). *Rapid pods*. Ex Libris Knowledge Center. Retrieved February 28, 2023, from https://knowledge.exlibrisgroup.com/RapidILL/Product_Documentation/RapidILL_Tools/Rapid_Pods

Kahle, B., & Coelho, J. (2021, August 24). *Interlibrary loan with the Internet Archive.* [Conference session]. IGeLU 2021 Digital Conference. https://drive.google.com/file/d/1zHF0ZDRyqDwtZtoFdA2fYczCw1iKwAFs/view?usp=sharing

Prosmans, F., & Renaville, F. (2018, August 22). *One resource sharing library to rule them all* [Conference session]. 13th IGeLU Conference, Prague, Czech Republic. https://hdl.handle.net/2268/227174

Renaville, F. (2018). Un SGB comme catalyseur de modernisation. L'expérience à l'Université de Liège. *Arabesques*, (89), 22–23. https://doi.org/10.35562/arabesques.251

Renaville, F. (2020). *Passage à un service de prêt interbibliothèques gratuit au regard de la gestion des changements : le cas de ULiège Library* [Unpublished executive master thesis]. University of Liège.

Renaville, F., & Prosmans, F. (2022). *Gratuité du service du prêt interbibliothèques à ULiège Library – étude de faisabilité.* ULiège Library. https://hdl.handle.net/2268/296151

Renaville, F., & Prosmans, F. (2023, February 8). *ILL at the University of Liège: from fee to free* [Conference session]. FIL Online 2023. United Kingdom. https://hdl.handle.net/2268/299864

Saerens, R. (2014). Impala: Uitdaging en vernieuwingen. *Cahiers de la documentation = Bladen voor Documentatie*, (4), 22–27. https://hdl.handle.net/10067/1237080151162165141

Skalski, F. (2023). COVID-19. New challenges and new solutions at the University of Warsaw Library. In F. Renaville & F. Prosmans (Eds.), *Beyond the Library Collections: Proceedings of the 2022 Erasmus Staff Training Week at ULiège Library.* ULiège Library. https://doi.org/10.25518/978-2-87019-313-6

ULiège Library. (2022, June 11). *Background.* Retrieved February 28, 2023, from https://lib.uliege.be/en/uliege-library/about/background

Van den Avijle, J., & Maggiore, L. (2023). Purchase on demand via interlibrary loan: analysis of the first four years at University of Antwerp Library. In F. Renaville & F. Prosmans (Eds.), *Beyond the Library Collections: Proceedings of the 2022 Erasmus Staff Training Week at ULiège Library.* ULiège Library. https://doi.org/10.25518/978-2-87019-313-6

Veinstein, B. (2020, March 30). 5 digital initiatives to improve library services during COVID-19. *Ex Libris Blog.* https://exlibrisgroup.com/blog/covid-19-new-initiatives-support-ex-libris-community/

Whitfield, S., Bergmark, M., Dawson, P., Doganiero, D., & Hilgar, R. (2020). Rider University: keeping resources available through the pandemic & construction, *Journal of Interlibrary Loan, Document Delivery & Electronic Reserve*, 29(3–5), 97–104. https://doi.org/10.1080/1072303X.2021.1939222

# About the Authors

Fabienne Prosmans, Fulfilment and ILL Manager
Université de Liège (University of Liège)
https://orcid.org/0000-0002-1408-5207

Fabienne Prosmans (1972) studied mathematics at the University of Liège and obtained a PhD at the University of Paris Nord. Since 2005, she has been working at ULiège Library as a subject librarian in the fields of mathematics and applied sciences. She has been the fulfilment and ILL manager at ULiège Library since 2015 and is now coordinating the "User Services" unit.

François Renaville, Head of Library Systems
Université de Liège (University of Liège)
https://orcid.org/0000-0003-1453-1040

François Renaville (1976) studied linguistics, literature, and translation. After a two-year experience as a teacher in Finland, he joined the University of Liège Library as a subject librarian. He has been working on library systems since 2005. Since 2022, he has been member of the IGeLU (International Group of Ex Libris Users) Steering Committee and has been coordinating the "Systems & Data" unit at ULiège Library. He is interested in discoverability, integrations, delivery and user services. On a private level, he is a great coffee, chocolate and penguin lover.

PART IV

# ENHANCED ILL SERVICES THROUGH COLLABORATIONS AND NETWORKS

# From fee to free: How to reduce expenses by eliminating revenue

**Léa Bouillet**

## Abstract

Interlibrary loan (ILL) service has always been considered as a special service within the library offer. As such, it is customary to find that ILL has its own cost policy and is not a free service for registered users. At UT2J university, ILL is free, both for end users and for other ILL services that request materials from our collections. Indeed, by providing free ILL for other libraries, we are able to provide free ILL for our staff and students. The fact of eliminating the revenue generated by lending materials also eliminates the expenses related to borrowing requests thanks to the principle of reciprocity. This economic model is virtuous, guarantees equal access to the documentation to all users, and allows us to promote our collections at a national level.

## Keywords

Interlibrary loan; Free ILL; Reciprocity; Equality of access; Positive communication

## Article

## Introduction

Interlibrary loan (ILL) services are nothing new in academic libraries in France ([Nortier, 1965a](#), [1965b](#)), or the rest of the world, but our new environment, strongly based on digital communication, instantaneousness and free access (either actual or imagined), has had quite an impact on the way ILL is organised and provided as a service to users. One of the main goals of ILL at University Toulouse Jean Jaurès has always been to maintain easy accessibility to documentation for everyone, students and staff. As a provincial university, it is especially important for researchers to have access to the same level of resources as researchers in Paris.

As a consequence, the cost policy of the service is one of the main focuses, and has been greatly challenged in the last few decades by local and national decisions. The principle of free ILL for students and staff, and of reciprocal free ILL with other libraries, is an economic model to which the library is particularly attached, but that has to be defended as it is not the most obvious choice when it comes to saving public money.

## Background at UT2J

So how is eliminating revenue leading to a reduction in costs? Both the local and national contexts of the university Jean Jaurès are key to understanding this policy but some simple calculations can easily show how this has become an obvious choice for our ILL.

University Toulouse Jean Jaurès (UT2J) is part of a network comprising four universities and 17 engineering or specialist schools, with about 130,000 students in all fields, from literature to civil aviation. The University's 20 libraries (Bibliothèques de l'UT2J, n.d.) specialise in human and social sciences and are part of a network of 54 academic libraries in Toulouse and its region, employing about 400 staff members. This library network is coordinated by an inter-institutional service for documentary cooperation (Service Inter-Établissements de Coopération Documentaire, n.d.), whose mission it is to provide and manage services that the network needs, such as its library system Alma, the remote Ask a Librarian service, its shared catalogue Archipel, staff training, etc.

This network implies that each student, irrespective of which university or school they are registered in, has access to every library and can borrow documents without having to register again or pay a fee. A common library card, using the student ID system, allows for easy access to this service. As such, no ILL is provided between the libraries of this local network, except between the libraries located in Toulouse itself and those located outside the city. Since the three main universities were once one single entity (before 1970 and the Faure law following the events of May 1968) (Alcouffe et al., 2019, pp. 118–119), this state of affairs is historic and self-evident to everyone. Services such as a shuttle service for borrowing and returning documents between the libraries have been implemented in the last few years to develop this cooperation, free of cost to users.

## National organization of ILL in France

On a national level, things are much more complicated. French academic libraries share a national catalogue, the SUDOC, managed by the French national agency for academic libraries (ABES). Interlibrary loans are based on this catalogue and its back-office interface, SUPEB. This software, which uses an in-house software solution, enables the ILL service to manage their borrowing requests and to receive lending requests and process them, though digital documents cannot be sent through this interface and no billing system exists.

Until the 1990s, French libraries could rely on a postal franchise allowing French institutions to send each other letters and packages at no cost. Each library had its own policy regarding the cost for the end user, some deciding on free ILL and others preferring to bill their users for the time it took to locate and obtain the document for them; but in both cases, no shipping costs were incurred by the libraries and consequently no shipping costs were passed on to library users. However, at the end of the 1980s, the postal franchise on which the academic libraries relied was deemed to be illegal and a misuse of an agreement between the postal services and the French government. Academic libraries were in fact benefiting from the laxity of controls on the use of the postal franchise, but were not allowed, legally, to use it to send documents for ILL (Deguilly, 1987).

In 1996, the postal franchise was no longer available to French academic libraries, which greatly changed the way ILL costs were dealt with. ILL operations witnessed a significant increase in costs, and each university had to decide on how to manage these new costs. Some libraries decided to introduce a fixed price for any book they lent to other libraries, to cover postal charges. Others suggested that another economic model would be more cost efficient: a reciprocity system, based on an equilibrium of requests and loans, to avoid billing and potential added costs. The University Jean Jaurès quickly decided that this solution was the easiest one to implement. A librarian at the University of Saint Etienne took it upon herself to create a list of libraries implementing this reciprocity system, to provide visibility to all ILL services operating in this way (SCD de l'Université de Saint Etienne, n.d.). This list now comprises over 400 libraries which have agreed to share their documents for free with other French academic libraries.

In view of this initiative, the ABES was called upon in 2006 to study the possibility of an integrated exchange and billing system for ILL services (Baraggioli, 2018, pp. 53–54). The billing platform was designed on the basis of the reciprocity system: every document, whether it be an original sent by post or a copy sent by email, would be sent for free between academic libraries. At the end of the year, the balance would be zero, in surplus or in deficit. In the event of deficit, a financial compensation would be provided directly from the Ministry of higher education.

Several difficulties have prevented the completion of this project. Firstly, only libraries that are part of the SUPEB network, almost exclusively French academic libraries (Agence bibliographique de l'enseignement supérieur, 2020), would have had access to this central billing system, which excluded municipal and foreign libraries, among others. At UT2J, for example, 32% of the requests made by students and staff cannot be completed through SUPEB. ILL would have had to maintain another billing system for those requests, which would have been time consuming and complex. Secondly, each university is allowed to choose its own rates, and no consensus was ever found on the cost that would apply in the event of deficit. Finally, those universities whose collections were in particularly high demand argued that this system would result in the loss of significant revenues, despite the financial compensation. As a result, this project was abandoned and to this day the ILL reciprocity system is still based on local decisions and an informal list maintained by one of the participants.

## Organization of free ILL at UT2J

When the ILL service at UT2J was confronted with the added cost of shipping in 1996, a decision had to be made as to whether that new cost should be passed on to service users and requesting libraries. Since no one wanted to take the controversial decision of having students pay for the ILL requests, it was decided that the service would remain free of charge for all service users and that the ILL service would join the reciprocity list of libraries providing documents for free.

For users, participating in the reciprocity system guarantees that books and theses will be provided for free, as long as they can be found in a French library. The ILL service subsidizes the cost of shipping the document back to the lending library and, if the only library able to provide the document has a fee-based ILL service, it also subsidizes that cost. As for copies, for a long time it was common practice for service users to pay for the price of scanning and printing the material. A standard cost of €3 for a copy of 10 pages, and €1.70 for every 10 pages hereafter was applied, except if the lending library practiced reciprocity: in this case, the article was provided for free to the user. Since this happened in 90% of the cases, it was decided in July 2022 to provide all articles for free, for an added cost to the service of about €90 per year. The only documents users have to pay for are documents that can only be obtained from abroad, because the shipping cost is much higher and there is concern that subsidizing these requests would lead to an explosion of requests for documents that are unavailable in France, adding too great a cost to the service.

With a view to simple and universal access, there is no need to be a student at the University to enjoy these conditions. All users registered in the library network can access these rates, within the limit of three requests per week. It should be noted that only one of the other two universities in Toulouse provides its users with free ILL access, and only since 2019.

## How much does it cost and save?

Our ILL service has been working this way at UT2J for many years now, and despite budget cuts and great attention on how public money is spent, free access to the ILL service and reciprocity between us and other libraries has never been abolished because having no revenues also means we spend very little money.

| | With reciprocity | | Without reciprocity | |
|---|---|---|---|---|
| **Costs** | | **€36,055** | | **€55,233** |
| | Documents provided by libraries that don't practice reciprocity | €1,790 | Documents provided by other libraries | €20,968 |
| | Shipping back documents | €19,185 | Shipping back documents | €19,185 |
| | Shipping documents to borrowing libraries | €15,080 | Shipping documents to borrowing libraries | €15,080 |
| **Revenues** | | **€1,154** | | **€17,258** |
| | Invoices to users (copies of articles or documents from abroad) | €778 | Invoices to users (copies of articles or documents from abroad) | €778 |
| | Documents sent to libraries that don't practice reciprocity | €376 | Documents sent to other libraries | €16,480 |
| **Total cost of service** | | **€34,901** | | **€37,975** |

*Table 1 – Cost comparison With reciprocity vs Without reciprocity*

When it comes to simple calculations, this system is the most cost-efficient for our ILL service. Not all costs are easy to identify, but even just taking into account the most obvious ones, such as shipping, the balance is clearly in favour of reciprocity.

Using the 2019 data for borrowing and lending requests,[1] it is possible to make a financial projection of costs for the scenario in which reciprocity is not the norm, and compare it with actual costs for ILL that year (Table 1, Figure 1).[2] This projection doesn't take into account a change of fares for users (hence possible revenues), since it is not an acceptable option at UT2J to begin charging students for the service.

---

1. Statistics from 2019 were preferred to those of 2020 or 2021, since the activity of ILL was greatly impacted by COVID restrictions.

2. The average cost practiced by libraries that are not part of the reciprocity network is 8€ for an original document. This simulation was made based on this cost.

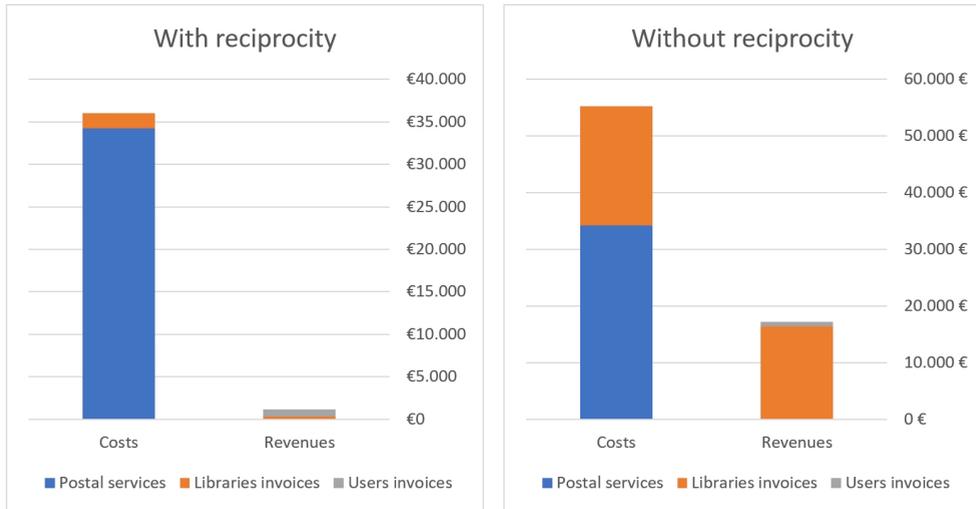

*Figure 1: Cost comparison With reciprocity vs Without reciprocity*

Reciprocity saves us 8.8% of the total cost. With 3,001 patron requests received (and 2,621 sent once documents available in Toulouse or online are deducted) and 2,060 documents lent, the average cost of sending or receiving a document is €7.45, instead of €8.11 without reciprocity.

The balance is positive mostly because our ILL service borrows more documents than it lends, but it is necessary to add the working time saved on preparing invoices, tracking payments, checking bills received and paying them to this €3,000 saving, which might seem slight considering the overall cost. Indeed, in practical terms, practicing reciprocity means that no invoice needs to be prepared nor bill to be paid. The ILL takes great care to send borrowing requests to libraries that practice reciprocity, unless the document in question cannot be found in these libraries. This way, the ILL staff doesn't spend its time preparing invoices and paying bills, and the university accounting department doesn't have to be involved in ILL proceedings, all of which are hidden costs when providing a fee-based service.

The ILL also has to be on the look-out for unnecessary spending:

- No document available in another library in Toulouse can be requested through ILL. The staff is quite vigilant in checking the Archipel catalogue, as well as the local public libraries catalogue and some very specialist libraries before sending a request for a document. If need be, the patron is required to go to the local library that owns the document instead of requesting it through ILL.
- No document available online can be provided in physical form, though ILL staff make it a point to provide this online document to the service user.

- The staff strive to ensure that all requested documents are picked up. Service users that don't come to pick up the requested documents can be excluded from the service if it happens several times.

## Overall benefits

Besides the economic considerations, the library finds many reasons to maintain this policy, from the user standpoint, as well as from that of the professional back office. Whilst UT2J may only be the 24th biggest university in France (in number of students), its ILL service was the 5th busiest in French academic libraries in 2019 ([Ministère, n.d.](#)). The number of ILL service users has been stable since 2014, with about 740 people using the service every year, at a time when a lot of ILL have seen a decrease in their activity, due in part to online resources. Libraries now have to compete with SciHub or social academic networks like ResearchGate, which allow free, if illegal, access to articles. It is essential that libraries are able to propose a legal alternative to these platforms, with the same easy and free access. Unfortunately, copyright laws in France don't allow for exactly the same ease of use, as it is not permitted to send an electronic copy of an article directly to the service user, except when software is used that prevents the recipient from viewing the entire document online and requires them to print the article to read it. Free ILL is a good start, however: it is easy to communicate about the service; it unifies the library offer, where there is no fee-based service; it plays a role in providing a positive image of the library and what it has to offer, by placing the user in the centre of library policy. Finally, it is a good way to raise awareness of copyright laws and also, in a roundabout way, to promote open access. As it is, free access to the ILL service is key to explaining that the ILL service at UT2J has been able to maintain this level of activity. 48% of service users are master degree students, mostly from the field of history. These students would have to pay for all their requests themselves if the service was not subsidized by the library, unlike most PhD students (12% of service users) and academic staff (13% of service users), who would most likely charge the costs associated with their ILL requests to their research programme. Finally, free ILL is a lever to ensuring equality between all users, and to guarantee that research is possible to students outside Paris, where a lot of research material is located. One student once said to the head of our ILL service: "Thanks to you, it is possible to write a quality thesis outside Paris".

From a professional standpoint, reciprocity allows our collections to be more visible and more accessible to other researchers outside Toulouse. In particular, the UT2J library houses a national collection of excellence on Spain and Portugal called "CollEx-Persée études ibériques" ([CollEx Persée, n.d.](#)). This collection represents 20% of the borrowing requests the ILL service receives. For other libraries, being able to request these documents for free guarantees that this national collection of excellence, for which the university receives special funds, is accessible to everyone in the country. It is a way to promote our collections and to be able to rely on the national network of academic libraries to develop our acquisition and conservation policy.

## Conclusion

The UT2J ILL service can rely on several decades of experience to justify the need for free ILL with other libraries and for library users. This economic model allows for ease of access, ease of communication with patrons, better visibility of our collections on a national level, and saving public money, whether they be direct savings or in terms of human resources. If practicing reciprocity between libraries is quite widespread and develops to include international sharing (HERMES project (HERMES, n.d.; Lomba et al., 2023) or NILDE network (Guerra, 2023; NILDE network, 2017) for example), access to a free ILL service for users is less common. However, several French university libraries have moved to this economic model in the last few years, such as the University Toulouse III Paul Sabatier in 2019 (Bibliothèques de l'UT3, 2019), the University Lyon 2 in 2020 (Bibliothèques de l'Université Lumière Lyon 2, n.d.), or the University Savoie Mont Blanc in 2022 (Bibliothèques de l'USMB, n.d.), hence proving the viability of this model.

## Bibliography

Agence bibliographique de l'enseignement supérieur. (2020). *Liste des bibliothèques participant au réseau SUPEB.* https://abes.fr/wp-content/uploads/2020/07/liste-bibliotheques-supeb.pdf

Alcouffe, A., Auger, F., Barrera, C., Bousigue, J.-Y., Cantier, J., Devaux, O., Espagno, D., Gounelle, A., Grosclaude, L., Grossetti, M., Jalaudin, C., Jolivet, A.-C., Lacoue-Labarthe, I., Lamy, J., Meynen, N., Pech, R., Périé, G., Barrera, C., & Ferté, P. (2019). *Histoire de l'Université de Toulouse. Volume III. L'époque contemporaine XIXe-XXIe siècle.* Editions midi-pyrénéennes. Université fédérale Toulouse Midi-Pyrénées.

Baraggioli, J.-L. (2018). *Étude sur la modernisation du prêt entre bibliothèques et la fourniture de documents à distance dans le cadre du GIS – CollEx-Persée.* https://www.collex-persee.eu/1664-2/

Bibliothèques de l'Université Lumière Lyon 2. (n.d.). *Le Prêt entre Bibliothèques devient gratuit.* Bibliothèque Universitaire; Webmestre Université Lumière Lyon 2. Retrieved October 5, 2022, from https://bu.univ-lyon2.fr/pret-entre-bibliotheques/le-pret-entre-bibliotheques-devient-gratuit

Bibliothèques de l'USMB. (n.d.). *Nouveau: Gratuité du PEB (prêt entre bibliothèques).*

Bibliothèques Universitaires. Retrieved October 5, 2022, from https://www.univ-smb.fr/bu/2022/04/26/nouveau-gratuite-du-peb-pret-entre-bibliotheques/

Bibliothèques de l'UT2J. (n.d.). *Liste des bibliothèques de l'UT2J*. UT2J – Bibliothèques; Valerie Peyrou. Retrieved September 22, 2022, from https://bibliotheques.univ-tlse2.fr/accueil-bibliotheques/bibliotheques

Bibliothèques de l'UT3. (2019, September 3). *Le service du PEB est désormais gratuit!* Bibliothèques de l'UT3. https://bibliotheques.univ-tlse3.fr/toutes-les-actualites/le-service-du-peb-est-desormais-gratuit

CollEx Persée. (n.d.). *Le réseau CollEx Persée*. Retrieved October 5, 2022, from https://www.collexpersee.eu/le-reseau/

Deguilly, F. (1987). Mariage forcé: La Poste et le prêt inter. *Bulletin des bibliothèques de France*, (2), 163-169. https://bbf.enssib.fr/consulter/bbf-1987-02-0163-002

Guerra, E. (2023). The NILDE network and document delivery in Verona University Libraries. In F. Renaville & F. Prosmans (Eds.), *Beyond the Library Collections: Proceedings of the 2022 Erasmus Staff Training Week at ULiège Library*. ULiège Library. https://doi.org/10.25518/978-2-87019-313-6

HERMES. (n.d.). *HERMES project*. Retrieved October 5, 2022, from https://www.hermes-eplus.eu/

Lomba, C., Marzocchi, S., & Mazza, D. (2023). HERMES, an international project on free digital resource sharing. In F. Renaville & F. Prosmans (Eds.), *Beyond the Library Collections: Proceedings of the 2022 Erasmus Staff Training Week at ULiège Library*. ULiège Library. https://doi.org/10.25518/978-2-87019-313-6

Ministère de l'enseignement supérieur, de la recherche et de l'innovation. (n.d.). *ESGBU, données 2019*. Retrieved October 5, 2022, from https://esgbu.esr.gouv.fr/broadcast/database-export

NILDE network. (2017, January 3). *Rules and Regulation of NILDE*. NILDE World. https://nildeworld.bo.cnr.it/en/content/rules_and_regulations

Nortier, M. (1965a). Le prêt entre bibliothèques en France. *Bulletin des bibliothèques de France*, (4), 119-131. https://bbf.enssib.fr/consulter/bbf-1965-04-0119-002

Nortier, M. (1965b). Le prêt entre bibliothèques en France (fin). *Bulletin des bibliothèques de France*, (5), 155-168. https://bbf.enssib.fr/consulter/bbf-1965-05-0155-001

SCD de l'Université de Saint Etienne. (n.d.). *Liste de réciprocité pour le PEB*. Retrieved August 19, 2022, from https://scdnum.univ-st-etienne.fr/peb/liste_reciprocite.php

*Service Inter-établissements de Coopération Documentaire*. (n.d.). Retrieved August 19, 2022, from https://bibliotheques.univ-toulouse.fr/bibliotheques/trouver

# About the Author

## Léa Bouillet, Head of Public Services Department

Université Toulouse Jean Jaurès (University Toulouse Jean Jaurès)

https://orcid.org/0000-0002-9927-070X

Léa Bouillet (1989) has a master's degree in book history and passed the competitive examination to become a library curator in 2011. After one and a half years of training at the National School of Library and Information Technology (ENSSIB) in Lyon (France), she began her career in 2013 as head of public services at the University Health Library of the Paul Sabatier University in Toulouse (France). After 4 years, she was appointed director of public services in the libraries of the University Jean Jaurès in Toulouse. She is primarily responsible for the public services policy, projects related to handicaps, training in information skills and ILL. In parallel, she provides training in preparation for library examinations with the Occitanie library careers training centre (Mediad'Oc).

# HERMES, an international project on free digital resource sharing

**Carmen Lomba; Stefania Marzocchi; and Debora Mazza**

## Abstract

In 2020, the COVID-19 situation forced academic libraries to move exclusively to the Web. Library closure led to key issues for the scientific community: the unavailability of physical collections, the lack of digitization of paper assets, the inadequacy of catalogues, the absence of information on e-book and electronic resource availability, problems in delivering digital documents to users, and a general lack of competencies in searching and retrieving digital documents. To respond to the COVID-19 emergency in the library field, in April 2020 a group of volunteer librarians under the aegis of IFLA created "Resource Sharing during COVID-19" (RSCVD): the first experiment in free digital resource sharing worldwide. After the success of the RSCVD Project in the face of the COVID emergency, the Erasmus Plus Programme, through the HERMES Project, provided the opportunity to put in place a multifaceted action which can have long-lasting impacts.

## Keywords

## Article

## Introduction

The HERMES Project is a European project closely connected with the interlibrary loan and document delivery services. The title of the project is "HERMES: Strengthening dig-ital resource sharing during COVID and beyond".[1] This project was approved in February 2021 by the European Commission and is financed by the Erasmus Plus sub-Programme "Partnerships for Digital Education Readiness". The duration of the project was 18 months, from May 1st 2021 to October 31st 2022, but it has been extended for six months, to April 2023. The budget is €193,320.

The leader of the project is the Italian Consiglio Nationale delle Ricerche (CNR), and the partners are the International Federation of Library Associations (IFLA), the University of Balamand (Lebanon), the University of Cantabria (Spain) and the MEF University (Turkey). There are also associated partners in every partner country, in addition the RSCVD community, that we will mention later.

The Scientific Committee of HERMES is formed by members of the IFLA Document Delivery and Resource Sharing Section (hereafter DDRS), and by experts on librarianship, open science, copyright, and resource sharing from the partner countries and United States, United Kingdom and Qatar.

The direct beneficiaries are librarians and information professionals, researchers, and university students.

The whole objective of HERMES Project is to support effective access to knowledge for the academic community. To achieve this, HERMES plans to strengthen skills in searching and getting quality academic texts. In addition, and to provide easier access to scientific literature, librarians are the other target of HERMES' objective: the improvement of librarians' competencies in terms of resource sharing (RS), and the creation of a more effective international system to share electronic resources.

HERMES grew out of the RSCVD initiative, that was created precisely to facilitate the access to scientific literature during COVID.

## RSCVD

The idea of RSCVD "Received" was born during the COVID lockdown of 2020. The global pandemic resulted in the closure of most libraries and services were moved online. Therefore, resource sharing became either impossible or very difficult to perform. But at the same time, users' information needs remained the same or even increased.

It was then that members of the IFLA DDRS Committee, led by Peter Bae and Silvana Mangiaracina, created and launched the initiative called Resource Sharing during COVID-19 (RSCVD) (IFLA Document Delivery, n.d.) to facilitate electronic document delivery during the lockdown. They prepared this initiative in record time, with the help of InstantILL: a website with a very simple form to request materials. Peter Bae invited librarians all over the world to facilitate the inter-lending of their electronic holdings. Through RSCVD, the librarians could make requests for articles and chapters that were not available through their usual networks.

The requests were visible in a spreadsheet, where the volunteering librarians found them. Then the volunteers checked request by request, one by one, their digital collections. In the event of availability, they provided the material through a safe electronic transmission mechanism, like the OCLC Article Exchange for example. Delivery was electronic, and the service was free, and of course compliant with the international standards of inter-lending and document supply. These requests were marked as "Done" in the spreadsheet.

Soon the spreadsheet was improved with important changes: for example, borrowers could add a due date, so that requests had a validity period, after which the requests disappeared from the spreadsheet.

The DDRS Committee members, and the librarians, also verified the affiliation of the borrowers, to make sure that the requests were coming from libraries, not from individuals.

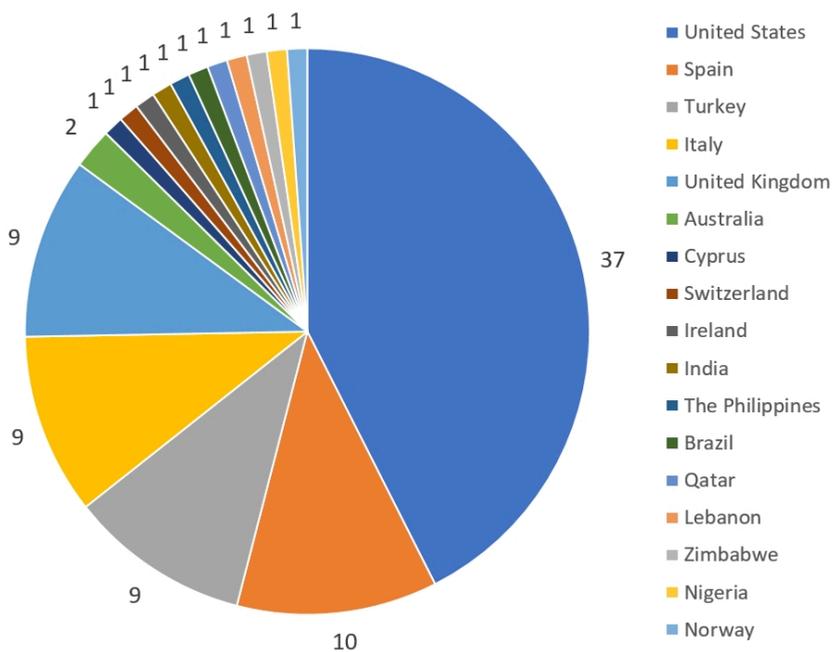

*Figure 1: List of countries with volunteering libraries (as of June 2022)*

The RSCVD initiative was designed very quickly during the COVID lockdown: it was put in service in almost one month. Since 21 April 2020, when the first request from an Italian library was sent, they have received 19,387 requests from libraries in 26 different countries. And 60% of those requests were fulfilled by volunteers from over 80 libraries (Fig. 1).

And, more importantly, a community of volunteering librarians was created. Peter Bae kept things together through messages and online meetings.

On seeing how well the initiative was received, the promoters realized that there was a need for an international communication channel that could fill the gaps between well-established national resource sharing networks.

So even though RSCVD had been planned as a short-term initiative (the plan was to finish in August 2020), in fact this initiative is still running today, and one of the main HERMES Project objectives is to ensure the sustainability, and long-lasting impact of RSCVD. In fact, the old initiative is being developed now through HERMES.

# Outputs

The HERMES Project combines the goal of developing the old RSCVD initiative with the goal of making it easier for the academic community to access quality scientific documents. To achieve these goals the HERMES team works on three outputs:

- A publication that wants to be a reflection of the meaning and practices of resource sharing today.
- A piece of software that will replace the RSCVD spreadsheet. It will be a new open source software for resource sharing management, called TALARIA, Hermes' wings.
- And the third output is a number of training activities and tools, aimed at librarians and the academic community.

## Output 1: Publication

Partners of HERMES together with IFLA DDRS Committee members have written a manual entitled "Document Delivery and Resource Sharing: Global perspectives". This manual, in addition to reflecting on the meaning and practices of RS nowadays, wants to foster the emergence of a common European and international perspective.

The manual starts with a Manifesto, that aims to be an update of the Manifesto of the Rethinking Resource Sharing initiative of 2007 (Rethinking Resource Sharing Initiative, 2007). The new Manifesto emphasizes the idea of reducing boundaries: the libraries are gateways to the whole world's cultural heritage. Researchers from anywhere in the world are our users.

This manual also wants to explore the current state of the European and international resource sharing systems, to understand how they work, and the challenges they face, as well as to stimulate the communication of best practices and technical advances.

Another objective is understanding the impact of copyright laws and regulations on our service, along with the impact of the terms of the licences. The publication also presents practical issues relating to management and content access to help librarians and other stakeholders improve skills to locate and obtain documents, with a particular focus on Open Access.

So, the expected impacts of the publication would be:

- wider knowledge of resource sharing
- better understanding of existing systems
- guidelines for legal issues that affect ILL including copyright and licensing
- development of a shared vision for cooperation between national systems

This manual is now under peer-review at IFLA. It will be published as an IFLA professional report in IFLA series, under Creative Commons licence.

## Output 2: TALARIA software

Due to the COVID pandemic, it has become impossible for many libraries to complete their activities. However, users' information needs have increased as a consequence of this situation. With the help of IFLA's DDRS Committee members and volunteer librarians from all over the world, it was possible to create the initiative "Resource Sharing during COVID-19" – RSCVD (or "Received"), a service that enables libraries around the world to request materials needed by their users from other libraries that can provide them.

Currently, the website RSCVD.org has an interface based on a web form powered by Open Access Button (built by OA.Works) where users can submit their requests, and the back-end is managed by volunteers through a Google Sheet to manage requests and delivered through OAB's InstantILL.

The service is efficient, but the handling of requests is done manually. In this context, HERMES comes into play. In Greek mythology, Hermes is the messenger of the gods. He wears a pair of sandals that allow him to fly as fast as the wind to deliver messages that were assigned to him. Output Two of the HERMES Project aim is to build, or deliver, a new resource sharing platform called TALARIA, open-source software that will enhance the existing RSCVD system, improve the document-sharing tools already used worldwide and ensure independence from commercial platforms (Mangiaracina et al., 2022). Just like Hermes' sandals, TALARIA will deliver resources reliably and quickly between libraries, and the RSCVD international community will be the first to use this new software.

## TALARIA's technical features

TALARIA development is led by CNR Dario Nobili Library Bologna, which is also the developer of the Italian NILDE software for document delivery ([Guerra, 2023](#)). The software will be released as open source software, it will use only free open source third-party software solutions, such as plugins, API and external services. TALARIA is also designed to support library communities (or consortia) in their RS management, for instance: the RSCVD international community, the NILDE community (Italy), and any other national, regional, consortium community willing to use it for their RS management needs since it is designed to be a flexible platform. For this reason, TALARIA offers the possibility to customize configurations. Most of the pillars of the TALARIA software that are already been implemented have customizable features, first of all, the visual identity (logo, TALARIA colour scheme), in order to respect the graphic identity of each RS community. The library communities can customize their Community policies on retrieving the ILL costs: document exchange among libraries can be set up to be free of charge, free of charge with an imbalance threshold or with a fixed unit cost. Another customization regards the Library RS profiles, depending on the community to which the library belongs, they can choose between a "Basic profile" to request only and "Full profile", where every library is borrower and lender and can request and provide resources.

The user registration and authentication systems available include traditional access with e-mail and a password; Federated Access, a simplified way to access online services, made possible by agreements between content providers and federations of research institutions and organizations based on SAML (Security Assertion Markup Language) and through Social login, which enables users to authenticate themselves through their accounts on social networks such as Facebook and Google.

TALARIA introduces the user roles: the Library Operators are responsible for managing the RS workflow. The role of the Library Manager is acquired when a user registers a new library or when their account is associated to a library with that role. The Library Manager has complete control over all operations (borrowing, lending) and is the one managing their library, or multiple libraries, inside the system. The permitted actions are to invite, add, remove and delete the other Library Operators of their own library. The other user roles are the Borrowing Operator and the Lending Operator, which only handle respectively the submissions and the fulfilment of requests.

TALARIA also has an environment dedicated to end-user management, where patrons can collect their bibliographic references and make a request from a library they are associated with. This feature enables user requests that each library can decide to set up or not.

In order to use TALARIA, libraries have to be registered and the system asks them to provide the GPS coordinates (latitude and longitude) of the library's location. They can be set manually, if known, or they can be automatically detected if the librarian is inside the building.

The Open Access Search and bibliographic citation import are the major features. Through the TALARIA homepage it is possible to immediately search and import bibliographic references in the search box in different ways, by unique identifiers such as DOI, PMID, ISBN or by the title of the resource and automatically import the bibliographic metadata. A new reference can also be imported via the OpenURL protocol. Alternatively, it is always possible to fill in the form manually by "Go to form" option. The search activates the Open Access button API, which can check whether a resource is freely available and whether an archived version exists in an institutional repository (Fig. 2). If the resource is available, the URL to the resource or the resource itself is immediately available through the TALARIA user interface, as in the current RSCVD software version.

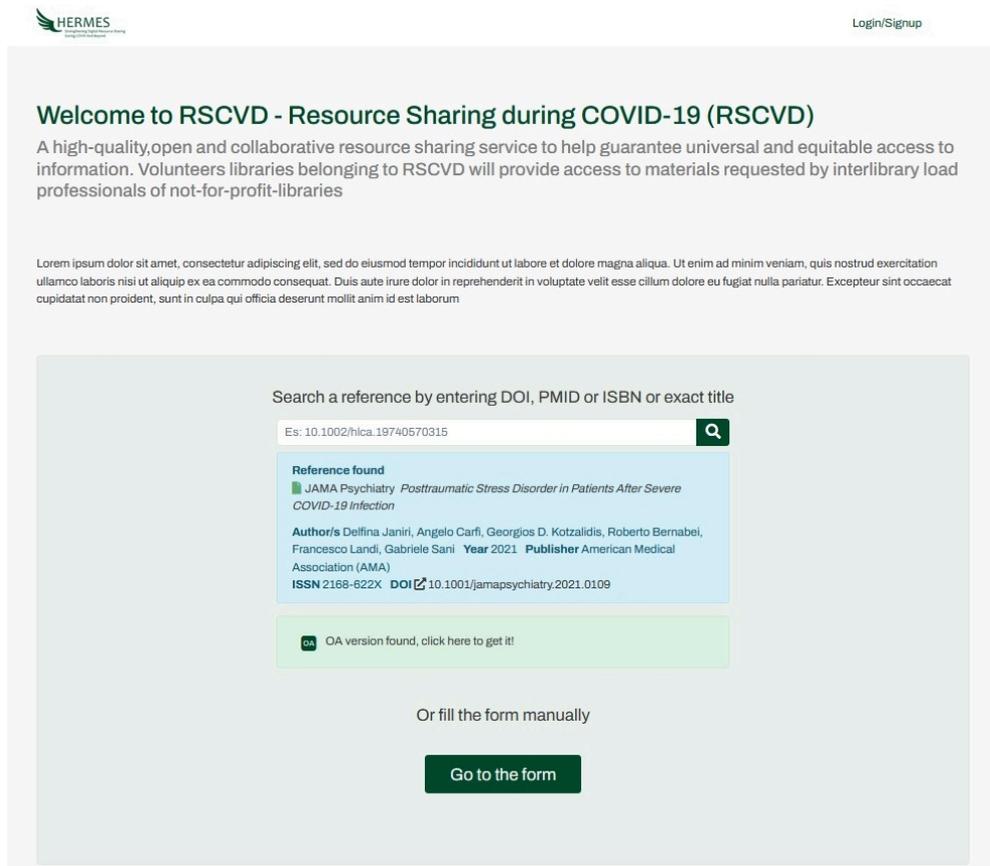

*Figure 2: TALARIA homepage showing an Open Access reference that has been found*

In TALARIA, librarians can manage requests in two environments: Borrowing and Lending. Each request has an ID, the status of the request, the name of the operator creating the request and the date of entry. The bibliographic reference is provided in a short form. Since TALARIA is a very articulate piece of software, icons have been used for the interface: each icon corresponds to an action that the librarian is permitted to do.

The Borrowing environment for the request submission is composed in "New request", "Pending requests" and "Archive". A new request can be opened to view the detail and the full bibliographic reference, it can be sent to a library or it can be cancelled. It is possible to modify the bibliographic reference, search for the OA version (this action will be possible until the request is sent to a Lender library) or retrieve the OA version of the resource and check own holdings if a link resolver has been set up. In the event of unfulfilment, the actions permitted are the reiteration of the request to another Library or archiving it as unfulfilled. Furthermore, creating one or more tags for each request is possible. A request can be sent to a single library or to "All libraries" in the system. This second option is suggested when the search in the catalogues does not generate any results or the request has been unfulfilled several times. The Archive panel allows the storage of the concluded borrowing requests.

Request processing and fulfilment are managed in the Lending environment. Besides the Archive, there are two relevant panels: Pending requests and Orphaned requests.

Pending requests displays all the requests that are made to the user's own library. In order to manage the request, it is always necessary to accept to fulfil it. The fulfilment supports a set of delivery methods: Secure E-Delivery (SED) embedded into the system, Article Exchange and via URL. The idea behind the URL came from RSCVD: volunteer librarians noted that some of the requested resources were available in public digital archives in the form of digitized documents. This makes it possible to send the resource directly through a direct URL. Otherwise, a library cannot fulfil for a variety of reasons. Reasons for non-fulfilment are:

1. not available for ILL
2. not held
3. not on shelf
4. wrong reference
5. ILL not permitted by licence or copyright law
6. order exceeding the maximum number of requests

A borrowing library may ask for the cancellation of a request once the request has already been forwarded to the lending library. In this case, the lender library can accept or not the cancellation request.

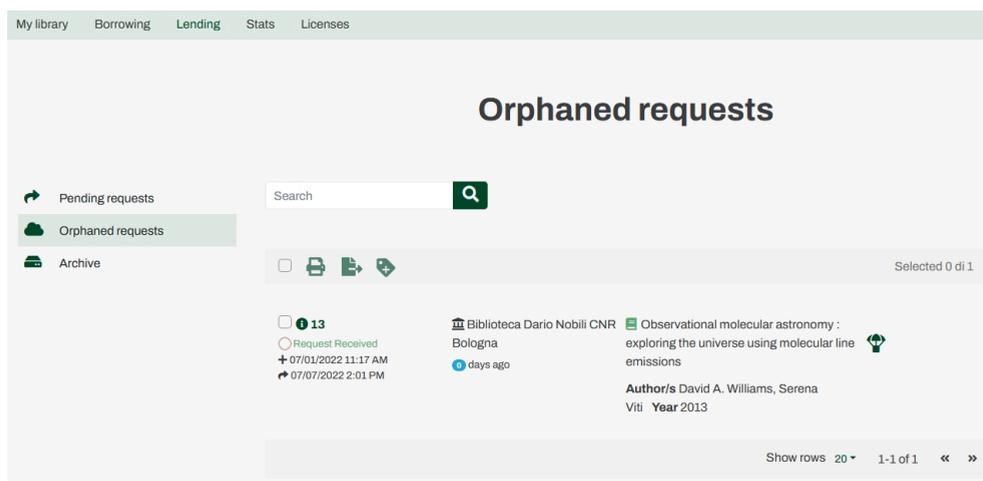

*Figure 3: Lending environment in "Orphaned requests" panel*

The requests sent to "All libraries" are displayed in the "Orphaned requests" panel (Fig. 3), a secondary mode of requests. Orphan requests must be accepted to be managed. At this point, the request disappears from the panel "Orphaned requests" for all libraries and the request will be only in the accepting library's "Pending requests" panel.

## TALARIA state of development

The RSCVD community will start using TALARIA in 2023, after the release of TALARIA 1.0, but the development is not complete.

Discussion with RSCVD volunteers and librarians is fundamental to the development process in order to gather feedback, detect bugs and make improvements. They have been involved in software testing several times since the beginning of the project. The TALARIA demo of the first prototype was presented in December 2021 by the software team to the CBN (NILDE Libraries Committee), which is the guarantor of the functioning of the NILDE community, and involved 20 librarians. In February 2022, the software was presented to a group of 30 RSCVD librarians. In summer 2022 TALARIA vers.Alpha was tested by 12 partner librarians and HERMES Scientific Committee. The users were assigned tasks for checking the proper functioning of the software with the support of the draft software user manual. At the end of the tests, a meeting of the whole group was arranged with the developers to discuss the results of the feedback.

For the TALARIA 2.0 release, work is planned on the supporting the connection with other RS systems, based on ISO-18626 protocol, licence management, catalogue search, payment management and statistic reports. Connecting to other RS systems is in fact one of the most difficult challenges. The ISO 18626 Interlibrary Loan Transactions stan-dard is the protocol that will be used to ensure interoperability among the different ILL systems. ISO 18626 is based on XML and Web services. The aim is to make uniform/ standardize management communication between the different ILL software used by libraries all over the world and thus fill the gaps when documents cannot be found in national RS communities. In addition, the first version of the software user manual will be released in 2023.

## Output 3: Training activities

Another output of HERMES are the training activities and tools. These activities are designed by a HERMES Project team led by the MEF University, and they are directed at the two target groups that the project is intended for: the librarians, especially the staff of interlibrary loan and document delivery services, and the academic community in general: students, teachers, and researches.

For the librarians, the HERMES team has planned to do webinars and tests about the TALARIA software, in order to present it to the librarian community, and to evaluate it with the librarians' feedback. These webinars will be launched when the beta version of the software is ready. Along with this presentation, some other webinars about topics covered by the publication will also be launched. Topics include resource sharing, and practical knowledge and instruments for a more effective service. The goal is to share professional knowledge between librarians and to learn about the new "Received" software and service.

The intention is to launch these live webinars to volunteer librarians pertaining to the HERMES associated partners, and to the RSCVD community. Eventually a final version of the webinars will be recorded and kept on the IFLA website, subtitled, open and free.

The planned training activities and tools for the academic community consist of a set of six live webinars. The webinars are about:
- Introduction to Open Access
- Open Access ethic matters. Legal, national, and EU mandates. Initiatives. Trends
- Copyright and licences
- Resource discovery. Use cases in the academic field
- How to get the information. Sources and Tools
- Future directions

A training activity has already been undertaken with the students in Librarianship at the University of Hacettepe (Turkey) in March 2022. Despite the difficulty stemming from the fact that the webinars were in English, the sessions were highly valued by the students.

Another training activity was held during the IFLA DDRS 2022 Meeting in Qatar,[2] where the professionals who attended it had the opportunity to attend training sessions about HERMES and TALARIA, resource discovery, Open Access and copyright and licences.

## Conclusion

The HERMES Project will end on 30 April 2023, with the 5th Transnational Project Meeting held in The Hague (The Netherlands), the headquarters of IFLA.

The new training activities will be focused on TALARIA software. In January 2023, a training session was held in Istanbul for RSCVD volunteers in Turkey. HERMES' YouTube channel has been updated with the new webinars for librarians and the academic community.[3]

The project is ambitious and challenging, as are the expected results. HERMES Project and RSCVD initiative want to build a community of libraries willing to share their resources internationally, crossing national and cultural borders and solving problems related to the connection between the different ILL systems. To educate librarians and academics about the principles and practices of resource sharing, especially with the support of online free training resources.

## Bibliography

Guerra, E. (2023). The NILDE network and document delivery in Verona University Libraries. In F. Renaville & F. Prosmans (Eds.), *Beyond the Library Collections: Proceedings of the 2022 Erasmus Staff Training Week at ULiège Library*. ULiège Library. https://doi.org/10.25518/978-2-87019-313-6

IFLA Document Delivery and Resource Sharing Section Standing Committee. (n.d.). *Resource Sharing during COVID-19 (RSCVD)*. https://rscvd.org

Mangiaracina, S., Tugnoli, A., Mazza, D., & Kahaleh, R. (2022). *Designing TALARIA – A new software to support resource sharing of international communities*. International Federation of Library Associations and Institutions (IFLA). https://repository.ifla.org/handle/123456789/2379

Rethinking Resource Sharing Initiative. (2007). *A Manifesto for rethinking resource sharing.* https://rethinkingresourcesharing.org/wp-content/uploads/2013/05/Manifesto-english.pdf

## About the Authors

### Carmen Lomba, Interlibrary Supply Manager

Universidad de Cantabria (University of Cantabria)

https://orcid.org/0000-0003-2479-3669

Carmen Lomba (1964) has a degree in History of Art (University of Valladolid). She has been working as librarian at the Library of Cantabria University since 1989 and has been responsible for the Interlibrary Loan-Document Delivery Service (ILL-DD) since 1995. She is currently part of the Working Group on ILL-DD for REBIUN (Network of Spanish University Libraries) and has been a member of the NILDE Internationalization Working Group since 2019. She participates as a volunteer in RSCVD initiatives and works in the European project "HERMES –Strengthening digital resource sharing during COVID and beyond" ERASMUS PLUS PROGRAMME, led by the Italian CNR (*Consiglio Nazionale delle Ricerche*) 2021-2023. She is a member of the IFLA Document Delivery and Resource Sharing (DDRS) Standing Committee (2021-2025).

### Stefania Marzocchi, Research Fellow

Consiglio Nazionale delle Ricerche (National Research Council)

https://orcid.org/0000-0001-8777-9720

Stefania Marzocchi (1969) is a librarian and project manager. Her collaboration with the Italian CNR (*Consiglio Nazionale delle Ricerche*) started in 2018 and is related to the implementation of CARONTE, BRAIN@WORK, and HERMES project activities, all funded by the European Commission. She is experienced in the design, management, and evaluation of international technical assistance projects with specific reference to the field of methodological approaches to research activity.

### Debora Mazza, Research Fellow

Biblioteca Dario Nobili - Consiglio Nazionale delle Ricerche (Dario Nobili Library - National Research Council)

https://orcid.org/0000-0001-7842-2922

Debora Mazza (1996) graduated in History and Oriental Studies from the University of Bologna. She has been collaborating with Dario Nobili CNR Library in Bologna since 2019.

She is in charge of the library's communication activities (graphic design, social media, website management) and she is involved in front-office, reference services, and document delivery with NILDE. She also develops the TALARIA software as part of the European HERMES Project.

# The NILDE network and document delivery in Verona University Libraries

**Elisabetta Guerra**

## Abstract

NILDE is a document supply system and a library network used by more than 900 libraries, based on reciprocity. This article describes how this system works and how NILDE libraries cooperate to get reciprocal advantages. The paper also presents the example of Verona University Libraries, which participate in NILDE: the data show that through NILDE the ILL Service of Verona University Libraries can meet almost 90% of the document delivery requests of internal users, and almost all requests of external libraries. In conclusion, it is hoped that the experiences of reciprocity among libraries, even from different countries, can increase to ensure access to information for all users. In our opinion, this service also contributes to the achievement of the objectives of the UN 2030 Agenda for Sustainable Development.

## Keywords

Alma; Document delivery; Interlibrary loan; Library cooperation; NILDE; University of Verona

## Article

## NILDE: a model of cooperation among libraries

NILDE (Network Inter-Library Document Exchange)[1] is both web-based software for document delivery (DD) and a community of libraries sharing bibliographic resources in the spirit of reciprocal collaboration. The NILDE project was implemented in Italy by Biblioteca Dario Nobili, the Bologna National Research Council Library, in 2001. Since then,

---

1. https://nilde.bo.cnr.it

NILDE has been constantly updated and improved (Mangiaracina et al., 2012). Thanks to the integration with the Italian Catalogue of Periodicals (ACNP)[2] (Brunetti et al., 2015) and with the Collective Catalogue of the Italian National Library Service (OPAC SBN),[3] it is possible to check holdings of many Italian libraries directly in the NILDE software. In 2012, after the integration with the software GTBib-SOD used by Spanish Libraries for document delivery, 33 Spanish libraries joined the NILDE community.

Moreover, it is now possible to integrate NILDE with the main bibliographic databases using the standard protocol ANSI/NISO Z.39.88-2004 OpenURL Framework for Context-Sensitive Services. This protocol makes it possible to connect the most widely used bibliographic databases to NILDE, so that the user gets the automatic compilation of bibliographic citations in the NILDE request form (Mangiaracina et al., 2008). OpenURL also ensures the integration with link resolvers, such as LinkSource and SFX, that can be configured by the library to check immediately if the requested document is in its collections.

NILDE libraries take an active part in the community and subscribe to NILDE rules and regulations. Rules and Regulations define Representing Bodies of NILDE and the commitments of members.[4] The representing bodies of NILDE are the Administrator (National Research Council, Library Dario Nobili), the Assembly (composed of all participating libraries), and the Committee of NILDE Libraries (elected by the Assembly).

NILDE subscription is free for the first year, and requires a very low participation cost in the succeeding years.[5]

The commitments of each library towards the other members are:

- To supply DD services reciprocally.
- To participate in one of the national collective catalogues (ACNP, SBN or REBIUN for Spanish libraries).
- To fulfil document requests in as short a time as possible (average two working days/ maximum five working days).
- To supply documents at no charge, unless there is a strong imbalance between the number of documents requested and supplied.
- To distribute its own requests fairly between all the libraries.

NILDE's policy, expressed by Rules and Regulations, provides all participating libraries with significant advantages:

- The policy protects the interest of the entire community: by joining national cata-

---

[2]. https://acnpsearch.unibo.it

[3]. https://opac.sbn.it/web/opacsbn

[4]. https://nildeworld.bo.cnr.it/en/content/rules_and_regulations

[5]. https://nilde.bo.cnr.it/subscriptions.php?inc=economic_conditions

logues, libraries commit themselves to supplying a quality service, rather than just using it.

- The policy leads to balance across the entire system.
- The mechanism of requests encourages self-regulation and avoids overloading a small number of libraries.

NILDE libraries pledge themselves to comply with the Copyright Laws in force (Law no. 633 of the 22/4/1941 and successive amendments and supplements)[6] and, where prevailing, to the existing contractual clauses.

NILDE provides a useful tool to check licences before fulfilling a document delivery request for electronic documents: the ALPE licence archive. ALPE is the NILDE "E-journals Licenses Database" and contains the terms related to the document delivery service, extracted from the electronic licensing agreements (negotiated or individual agreements) signed by libraries participating in NILDE (Bessone et al., 2017). Before sending an electronic document, the librarian can interrogate the ALPE licence archive to check if the publisher allows delivery and which are the accepted delivery methods. Many publishers accept delivery through SED, Secure Electronic Delivery. NILDE grants SED through NILDE Digital HardCopy: this is a secure electronic delivery method that permits the sending of documents in pdfs or image formats (jpg, jpeg, gif, tiff, tif, bmp, png) directly from NILDE requests. With NILDE Digital HardCopy every page of the document is converted into an image (png) and the resolution is lowered to 200 dpi. In addition, the document contains a disclaimer that informs the receiving library of legal requirements and of the obligation to deliver a paper copy to the final user (Mangia-racina, 2013).

The NILDE community is now composed of 918 active libraries, of which 71.2% are university libraries, 7.3% research libraries, 8.0% libraries of the National Health Service (*Servizio Sanitario Nazionale*), 8.7% other type of public libraries and 4.7% not for profit private libraries. Most of them are Italian (877), but there are also 34 Spanish libraries and libraries from Australia, Brazil, Croatia, Luxembourg, Sweden, Switzerland, and Norway.

During 2022, 136,288 documents were exchanged, with a success rate of 84% and an average response time of 0.5 days. The total number of exchanged documents since the creation of NILDE has been of more than 3,000,000.

The NILDE workflow can be described briefly as follows:

- The user makes a request (through "NILDE Users" module or other system).
- The library receives the request (by mail, online form, NILDE Users module or other means, depending on the method the library is using), checks holdings and sends the request to the supplying library (requests and requests statuses are visible in NILDE in the "Borrowing" tab).

- The supplying library receives an alert via email, sees the new request in the "Lending" tab and sends the document through the preferred delivery method: directly in NILDE in "hard copy" format, or other methods.
- The requesting library receives the document: if it is a hard copy, the document is available for seven days in the NILDE server and it must be cancelled just after print.

In accordance with Italian copyright law, only a paper copy can be delivered to the user.

## Verona University library system: the advent of Alma and the role of NILDE in interlibrary services

In 2019, within the context of a complete reorganization of Verona University Libraries, the interlibrary services were unified. In the same year, the University library management system switched from Aleph to Alma (Ex Libris).

It was a big change for our libraries. Before Alma, resource sharing was managed separately in the two Central Libraries (Biblioteca Arturo Frinzi for Humanities, Economics and Law areas and Biblioteca Egidio Meneghetti for Medicine, Science, and Technology areas). Resource sharing requests were managed by the two services with different methods and workflows. Both libraries had single subscriptions for NILDE, ACNP and ILL SBN (the ILL service of the Italian National Library Service).[7] Meneghetti library developed an internal program for the management of ILL requests: requests were made by users completing an online form and then librarians put requests in NILDE or sent them via email to partners. Frinzi library users sent DD requests through the "NILDE users" module and ILL requests from the online catalogue (OPAC) of the University libraries.

Thanks to Alma, it is now possible to manage interlibrary requests from a single virtual desk and users can make requests directly in UNIVERSE, the Verona University Libraries discovery tool.[8]

In 2021, Frinzi and Meneghetti merged their profiles in NILDE and ACNP, and they are now identified as "Sistema Bibliotecario Università di Verona" with the ACNP code VR009.

Despite big changes within the organization of the interlibrary services, NILDE has been a point of continuity with the past: Verona University Libraries continue to participate in NILDE, and NILDE is still the major document delivery tool in use.

---

7. https://www.iccu.sbn.it/en/interlibrary-loan-and-document-delivery-ill-sbn/index.html

8. https://universe.univr.it

## Some figures: borrowing

In 2022, the Resource Sharing Service of Verona University library system received 5,718 document delivery requests submitted by patrons.

Users make borrowing requests from the portal and discovery tool (UNIVERSE) and the operator finds them in Alma. Then, the operator chooses a partner and sends the requests through NILDE. The request is only sent outside NILDE when the requested material is not available in NILDE libraries.

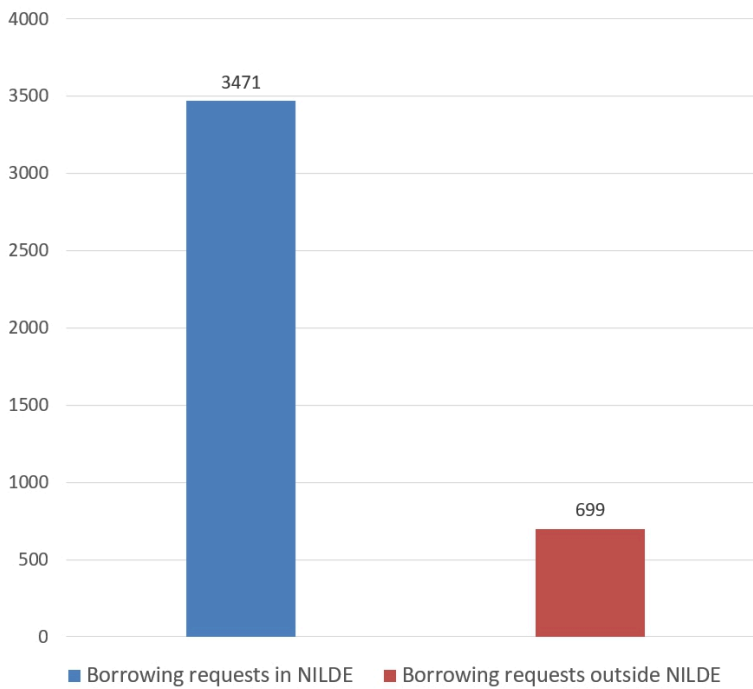

*Figure 1: Borrowing requests at Verona University Libraries (2022)*

If we exclude requests related to material owned by University libraries (and therefore that have been cancelled) or requests that for various reasons could not be fulfilled, 4,170 requests were sent to other libraries: 83.2% of them were sent through the NILDE network, and only 16.8% outside NILDE (Fig. 1), that is via email to other Italian or foreign libraries or through other systems (RapidILL, Subito, RSCVD, acnpDoc, ILL SBN)[9]

9. RapidILL is an Ex Libris interlibrary loan solution that was tested within a trial in 2021/2022. Subito is the ILL service of a network of 40 academic libraries with its headquarters in Berlin. RSCVD is the IFLA resource sharing project developed during COVID-19 (Lomba et al., 2023). acnpDoc is a document delivery service available for ACNP subscribers.

(Table 1). The success rate in NILDE was 85.3% and the average response time was 0.79 days.

| Supplier | Number | Comment |
| --- | ---: | --- |
| NILDE | 3,471 | |
| Unfulfilled | 517 | Cancelled by users, unobtainable document, etc. |
| Italian libraries outside NILDE | 439 | |
| Available online in Verona University | 299 | |
| GIRARTICOLO | 289 | Delivery of documents owned by Verona University Libraries, for internal users |
| Available in Verona University Libraries | 246 | Physical items |
| Open access articles | 193 | |
| Foreign libraries | 122 | |
| RapidILL | 72 | |
| Subito | 47 | |
| acnpDoc | 9 | Document delivery service for ACNP subscribers only |
| RSCVD | 4 | |
| Available in Verona Public Libraries | 4 | |
| Other | 6 | |
| **Total document delivery requests in 2022** | **5,718** | |

*Table 1: Distribution of borrowing requests by supplier (2022)*

## Some figures: lending

The Resource Sharing Service of Verona University library system uses NILDE for nearly all lending requests for articles, book parts and digitized material. The requests are processed separately in Alma and NILDE.[10]

---

10. The integration between NILDE and Alma is not yet available, but it is an ongoing project. The resource sharing working group within ITALE, the Italian association of Ex Libris users, is collaborating with the NILDE team to find solutions for a possible integration of the two systems. See: https://itale.igelu.org/gruppo-di-lavoro-resource-sharing. In March 2023 however, Luisella Consumi, Nazareno Bedini, and Francesca Mocchi from the ITALE

In order to save time, it was decided to process lending requests coming from NILDE directly in NILDE, without registering them in Alma. NILDE statistics are constantly available, so it is not necessary to duplicate requests. Only lending requests that come outside NILDE are created in Alma.

Verona University Libraries holdings are searchable in ACNP. At the end of 2021, we added our e-journals collection in the ACNP catalogue. The result was an immediate increase of requests for electronic documents. Moreover, our catalogue is searchable through Z39-50 protocol, in the "other catalogues" section of OPAC SBN.[II]

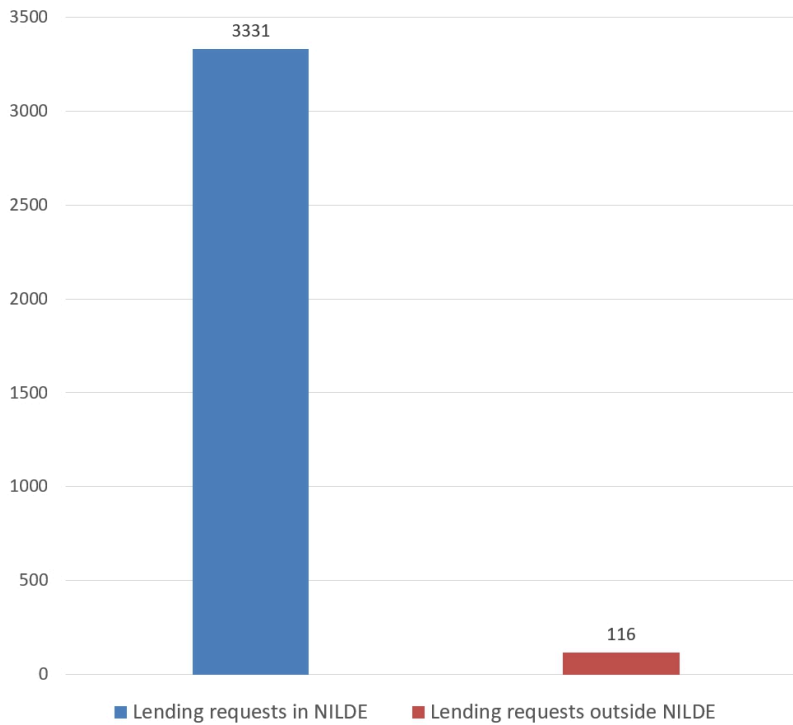

*Figure 2: Lending requests at Verona University Libraries (2022)*

In 2022, Verona University Libraries received 3,331 lending requests from NILDE partner libraries; 59 of them were cancelled by requesting libraries before fulfilment, 622 went unfulfilled and 2,650 of them were fulfilled (Fig. 2). The success rate was 81% and the average response time was 0.3 days. Among the reasons for unfulfilment were the working group launched an app called *Nilde Openurl* within the Alma Cloud App centre. This app makes it possible to send an Alma RS borrowing request to NILDE via openurl. See: https://developers.exlibrisgroup.com/app-center/nilde-openurl/.

II. https://opac.sbn.it/cataloghi-z39.50-in-rete

unavailability of paper/electronic documents, but also the impossibility of sending the document due to electronic licensing restrictions. Only 116 lending requests were processed outside NILDE: 93 in Italy and 23 abroad.

All in all, as the 2022 figures show (Fig. 3), there is a balance in the total number of borrowing and lending requests processed in NILDE at Verona University Libraries.[12]

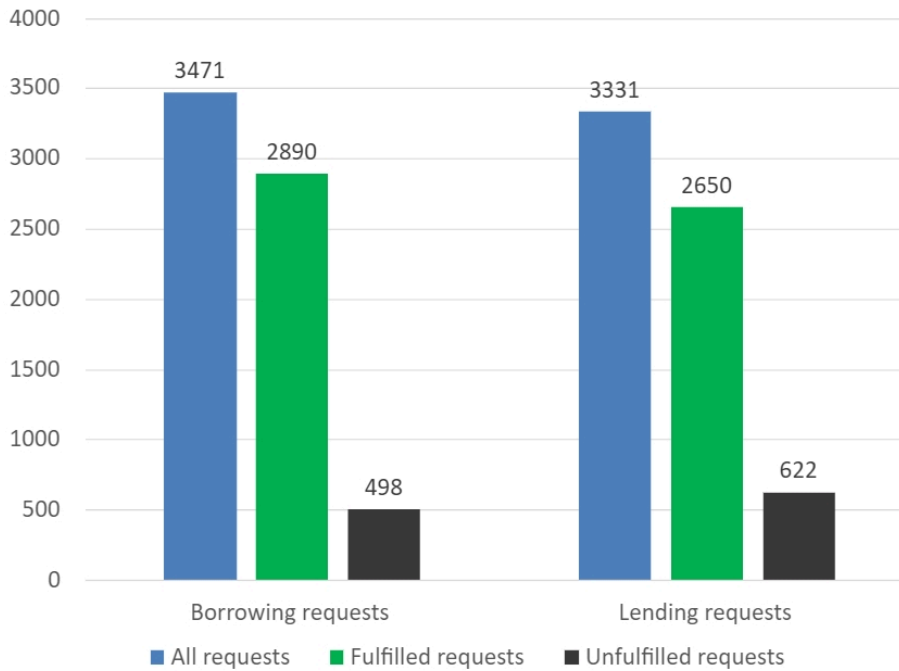

*Figure 3: Borrowing and lending requests in NILDE at Verona University Libraries (2022)*

## A few thoughts in conclusion

To conclude, we would like to share some reflections on resource sharing and, in particular, on the great opportunities arising from collaboration among libraries in the framework of mutual exchange and reciprocity. NILDE is an example of how the joint, cooperative and not for profit engagement of libraries can lead to excellent results. We believe that resource sharing is, by definition, an important tool for guaranteeing access to information. It is one of the pillars on which we, as libraries, can build the foundations for achieving UN 2030 Agenda for Sustainable Development Goals.

---

12. The total number of requests ("All requests" in the chart) includes also requests that were cancelled by requesting libraries before the fulfilment process starts.

Access to quality information can play fundamental role in achieving the 17 goals of the Agenda. According to the IFLA Statement on Evidence for Sustainable Development (International Federation of Library Associations and Institutions, 2022):

> Libraries represent an essential link in the chain between those who are gathering or producing evidence, and those who should be using it to take decisions. They fulfil a number of roles, including supporting research, and managing data and dissemination (including through supporting the development of technical possibilities for this), enabling the sharing of and access to information, the effective packaging of information for decision-makers, teaching the skills needed for information access and use, and long-term preservation in the service of future research and accountability.

Resource Sharing is one of the key activities of Libraries in this process, as it ensures "sharing of and access to information", when this information is not available in the user's library. Sharing resources in the form of free reciprocity allows us to achieve these objectives at low cost, and provides mutual help in critical moments, such as in the case of the RSCVD project in 2020 during the COVID-19 pandemic.

The hope is that reciprocity will increasingly cross national borders, particularly among European libraries. The wish is that copyright regulations will take into account the need to disseminate information in the fields of education and research and will be updated as new technologies become available.

Particular attention should be paid to licences for the use of electronic resources which too often do not allow document delivery outside the country of origin. The impossibility of sharing resources caused by licence restrictions is particularly limiting when it involves higher education and research institutions, even when they are involved in common research projects or educational programs.

## Bibliography

Bernardini, E., Colombo, G., Lomba, C., Mangiaracina, S. Merlini, F., & Secinaro, E. (2017). A NILDE survey on International ILL Exchanges: results and considerations. *ILDS – 15th Interlending and Document Supply Conference. No Library Left Behind: Cross-Border Resource Sharing (Paris, France, 04-06 October 2017). Proceedings, IFLA Document Delivery and Resource Sharing Section.* https://nildeworld.bo.cnr.it/it/node/488

Bessone, F., Colombo, G., De Carolis, E., Filippucci, G., Garbolino, L., Gasbarro, E., Mangiaracina, S., Russo, O., Tamburini, E., & Tugnoli, A. (2017). To lend or not to lend? With ALPE it is easier! An Italian cooperative system for checking ILL permitted uses in e-resource licenses. *ILDS – 15th Interlending and Document Supply Conference. No Library*

Left Behind: Cross-Border Resource Sharing (Paris, France, 04-06 October 2017). Proceedings, IFLA Document Delivery and Resource Sharing Section. https://nildeworld.bo.cnr.it/it/node/489

Brunetti, F., Bonora, O., & Filippucci, G. (2015). ACNP and NILDE: Essential tools for access to scientific research. In A. Holl, S. Lesteven, D. Dietrich. & A. Gasperini (Eds.), *Library and Information Services in Astronomy VII: Open Science at the Frontiers of Librarianship* (pp. 275-283). Astronomical Society of the Pacific. http://www.asp-books.org/a/volumes//?book_id=560

IFLA Document Delivery and Resource Sharing Section. (2012). *Guidelines for Best Practice in Interlibrary Loan and Document Delivery*. International Federation of Library Associations and Institutions. https://repository.ifla.org/handle/123456789/705

International Federation of Library Associations and Institutions. (2022). *IFLA Statement on Evidence for Sustainable Development.* https://repository.ifla.org/handle/123456789/2191

Lomba, C., Marzocchi, S., & Mazza, D. (2023). HERMES, an international project on free digital resource sharing. In F. Renaville & F. Prosmans (Eds.), *Beyond the Library Collections: Proceedings of the 2022 Erasmus Staff Training Week at ULiège Library*. ULiège Library. https://doi.org/10.25518/978-2-87019-313-6

Mangiaracina, S. (2013). *NILDE technical description: Secure Electronic Document Delivery and Digital Hard-Copy.* https://nildeworld.bo.cnr.it/en/pub/nilde-technical-description-secure-electronic-document-delivery-and-digital-hard-copy

Mangiaracina, S., & Tugnoli, A. (2012). NILDE reloaded: a new system open to international interlibrary loan. *Interlending & Document Supply*, 40(2), 88–92. https://doi.org/10.1108/02641611211239551

Mangiaracina, S., Zaetta, M., De Matteis, D., Tugnoli, A., Beghelli, E., & Tenaglia, G. (2008). NILDE: developing a new generation tool for document delivery in Italy. *Interlending & Document Supply*, 36(3), 167-177. https://doi.org/10.1108/02641610810897908

## About the Author

Elisabetta Guerra, User Services Manager
Università di Verona (University of Verona)
https://orcid.org/0000-0002-8880-3646

Elisabetta Guerra (1972) graduated in Foreign Languages and Literatures at Verona University and has been librarian at Verona University Libraries since 2006. Since the beginning of her activity, Elisabetta has been working in the interlibrary loan and document delivery services. She is also engaged in information literacy programs for library users.

Since 2022, she has been the User Services Manager for the Libraries of the Medicine, Science and Technology Campus. She works in the central library "Egidio Meneghetti", where she manages the loan and consultation services.

PART V

# IMPACT OF COVID-19 ON LIBRARY ACCESSIBILITY AND SERVICES

# COVID-19. New challenges and new solutions at the University of Warsaw Library

**Franciszek Skalski**

## Abstract

This article aims to address how the COVID-19 pandemic influenced the University of Warsaw Library's work from 2020 to 2021. Analyses of BUW's situation in 2020 and 2021 based on published reports by the heads of the library's organisational units as well as the author's own observations and experiences reveal that the COVID-19 pandemic forced the BUW to organise its work and services in unusual ways. The results are evident in the use of new technologies and electronic resources. Furthermore, the experience gained during the pandemic helped BUW introduce new solutions, processes, and workflows which could be valuable for future use in similar situations and for the patrons' benefit.

## Keywords

## Article

## Introduction

No one suspected that 2020 would be a year of great change and disaster. The global COVID-19 pandemic affected various spheres of society. Universities and academic libraries also had to find solutions to new problems, continue their work, and adjust to the new reality imposed by the pandemic.

# Beginning of the pandemic and its initial problems

On 10 March 2020, the rector of the University of Warsaw (UW) decided to shut down the University of Warsaw Library (*Biblioteka Uniwersytetu Warszawskiego* – BUW). It was the beginning of a long series of changes (Łukaszewska, 2020). We had to decide on the library's closure, the duration of this closure, how resources would be delivered to patrons during that period, and whether such deliveries would be possible at all. Furthermore, we also had to determine what to do with the due dates of the borrowed books and other materials and the penalties for overdue items. Finally, we needed to decide how to inform BUW's patrons that the library would be closed. BUW used to be very flexible. It was crucial for the library to provide its patrons with instant access to information. Thus, we immediately started an information campaign addressed to the patrons through our social media channels and website. The first piece of information concerned the increased number of books that could be borrowed at a time, up to 20. Moreover, we extended the period for returning the borrowed books to 14 April 2020 (later, this date was changed several times during the first lockdown until the final date, 30 May 2020). We also announced that overdue fines would not be charged for keeping books borrowed beyond the set deadline. Finally, BUW's authorities decided that the library would be accessible until 10 pm on 10 March 2020 and that the entire library would be operational until the last patron had left. Ultimately, this meant that the library was closed at midnight. BUW was very crowded that day; 8,137 books were borrowed (including 2,331 in self-check kiosks) and 3,255 were returned to the library (Fig. 1). In comparison, before the pandemic, about 1,450 books were borrowed in one day. These decisions did not solve all the problems that the pandemic brought about later. However, we hoped that the extended limits would help our patrons, especially UW students and employees. In BUW, we also had to cancel the scheduled staff meetings, training, and travel for the following weeks. Furthermore, we had to stop registering new patrons. Although we did not forget about the second important group of our patrons, those from outside the UW, we could not do much to help them. External users had access to BUW, but they were not allowed to borrow books; instead, they could use the library's collections onsite. However, because of the pandemic situation, they lost admission to the library building. To compensate for our inability to offer them access to traditional or electronic collections, we extended their accounts until the time they could enter the library again because of health restrictions (Wołodko, 2020, p. 62).

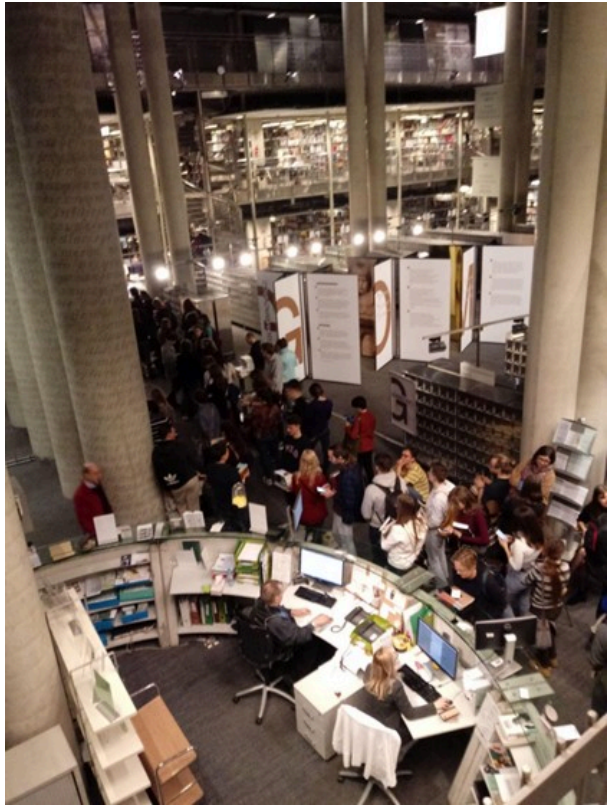

*Figure 1: Crowds of students waiting in line at the circulation desk and looking for information at the information desk on 10 March 2020*

## First lockdown in Poland

On 11 March 2020, a national lockdown was enforced in Poland. However, the lockdown did not mean that BUW staff automatically switched to teleworking. On the first day, we had to prepare books ordered on the previous days—they had not been borrowed because the library had to be closed. Another problem appeared here as to what should be done with those books. Previously, patrons had one week to come and collect requested books, but now adhering to this collection period had become impossible. The solution appeared simple: we extended that period to the week after the end of the first lockdown. No one could predict when we would be able to reopen the library. It turned out that we had to apply this solution several times. A parallel, more complicated problem was with the books from the interlibrary loan. First, we had to contact each library from where we had requested books and ask whether we could keep them until the sit-uation would allow us to deliver them to our patrons or return them immediately to the libraries in Poland and abroad.

Another key challenge brought about by the COVID-19 pandemic was the reorganisation of the library departments' work, which previously focused primarily on contact with the patrons. Even though academic lectures were immediately organised online, students and employees still needed books and other library materials. During the first lockdown, these students and employees had to exclusively rely on electronic resources provided by BUW. We decreased the price for scanning documents to 1 PLN (approximately 0.21 EUR) for the A3 page format for all our patrons. Moreover, we prepared a special offer for the UW students and employees, according to which they could issue a free scanning order of up to 50 pages weekly. Although this was not a significant amount, we could not offer more services due to the large number of BUW patrons. In the beginning, we hoped that the lockdown would end shortly. The Information Services and Training Department communicated with our patrons through phone calls, emails, and social media.

A week later, from 18 March 2020 onwards, all university staff had to switch to working from home in line with the decision of the rector of the UW. This decision caused new challenges as it was a completely new situation for the library staff. We had no experience with remote work, yet we had to manage and organise the tasks to fit the situation. We could no longer scan and send documents to our patrons, nor could we deliver books and other materials. Information services could not respond to the phone calls, and the only information channels that remained operational were our social media and website. The other major problem was setting up secure remote home offices. For this purpose, our IT Department set up and provided 80 new VPN accounts, which were necessary for all staff whose work was related to the BUW intranet and Virtua/VTLS integrated library system. Furthermore, the library laptops were specially configured and lent to the staff members who needed compatible hardware (Kamińska et al., 2020).

The circulation department staff were organised into two groups, and each group was assigned a different task. Furthermore, the end date of the lockdown was postponed several times. First, the date was shifted to 14 April 2020. However, on 20 March 2020, it was extended indefinitely until further notice by the decision of the Minister of Health of the Republic of Poland (Ministra Zdrowia, 2020).

Other key tasks during this first lockdown and home office period were to analyse how to prepare the library for reopening under new conditions and determine how to minimise contact with patrons to prevent the spread of the virus. BUW's work was based on the following four different but related topics:

- the organisation of the new location of the circulation desk;
- the organisation of work at the circulation desk;
- an analysis of the devices and alternative ways that could help with circulating the materials between the patrons and the library to minimise the contact between the library staff and patrons; and
- a summary of the ways to protect the library staff from the virus.

To ensure we adhered to the safety protocols, we followed the decisions of the Ministry of Health of the Republic of Poland. We also followed the recommendations provided by the National Library of Poland (Biblioteka Narodowa, 2020a; 2020b). Due to the lack of information regarding the spread of the virus, BUW decided to extend the quarantine period for returning books and other library materials. Moreover, to minimise direct contact with patrons, BUW bought and installed a book return box outside its main entrance (Fig. 2). We introduced a procedure to deal with possibly contaminated books from the return box. From 22 April 2020, two librarians from the circulation department came to the library to empty the return box every Wednesday. The procedure was simple. One librarian emptied the return box and delivered the books to a dedicated place at the circulation desk. The second librarian removed the books from the patron's account using the laser reader to minimise contact with possibly contaminated materials. Finally, the first librarian put the books away on a designated bookshelf for a five-day quarantine.

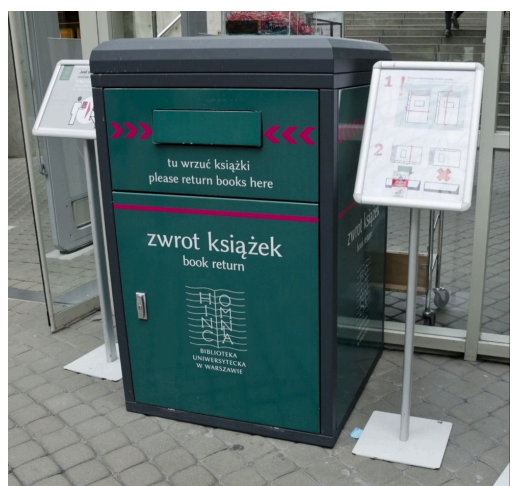

*Figure 2: Book return box located outside the library*

## Temporary circulation desk

As mentioned before, our tasks while working from home included preparing a plan for reopening the library under the pandemic restrictions, protecting the staff, and introducing certain restrictions and procedures to guarantee safety in BUW. Based on the recommendations of the National Library of Poland (Biblioteka Narodowa, 2020a), BUW decided that the library staff would have to work with gloves and masks when the lockdown was lifted. The other recommendations included maintaining the libraries' focus on borrowing and returning books as soon as this was possible and keeping reading rooms closed until the pandemic situation was under control. Moreover, the BN recommended organising the circulation of books outside the libraries, which created another problem, namely ensuring the safety of the place where the circulation desk was installed. We decided to organise the temporary circulation desk in one of the empty office spaces, which was a former restaurant, near the entrance of the library. According to the recommendations, the space was organised as follows:

- There were two entrances.
- Two librarians were appointed from the circulation department and seated at a distance of 1.5 metres from each other.
- One librarian was dedicated to borrowing, while the other was in charge of the returned books and other materials.
- A transparent plexiglass screen separated librarians from the patrons to avoid spreading the virus.
- Bookshelves with the ordered books were placed behind the librarians.
- Bookshelves also separated the circulation desk from the place where the returned books were quarantined.
- There was a separate exit to prevent contact between the patrons entering the room and those leaving the temporary circulation desk.

The temporary circulation desk was opened on 18 May 2020, after two days of technical preparation (Figs. 3–5). For the first time since the beginning of the lockdown, patrons could come to the library to borrow and return books and other materials. The circulation desk was open from Monday to Friday, between 9 am and 6 pm. We further introduced changes to allow borrowing and returning the books and materials under the new special restrictions. For instance, only two people other than the library staff were allowed in the room at a time. Everyone had to wear a mask and sanitise their hands before entering. There was a security desk in front of the entrance, where security staff checked each patron's body temperature. The restrictions were strictly enforced to prevent the spread of the virus, so no one with a higher temperature was allowed to enter. Two librarians issued and collected books, and two others put the returned books away on the specially designated bookshelves for a five-day quarantine. The special status 'on quarantine' was created in our integrated library system (ILS) to prevent books from being requested before the end of the quarantine period. BUW had to introduce this solution since previously there was a possibility to reserve books that had already been borrowed by someone else. When a requested book was returned to the library, the ILS automatically activated the request for that book and sent a notification to the requester.

After that, the patron had five days to come and pick up the item. Through these new conditions, the status 'on quarantine' blocked the book in the holding process.

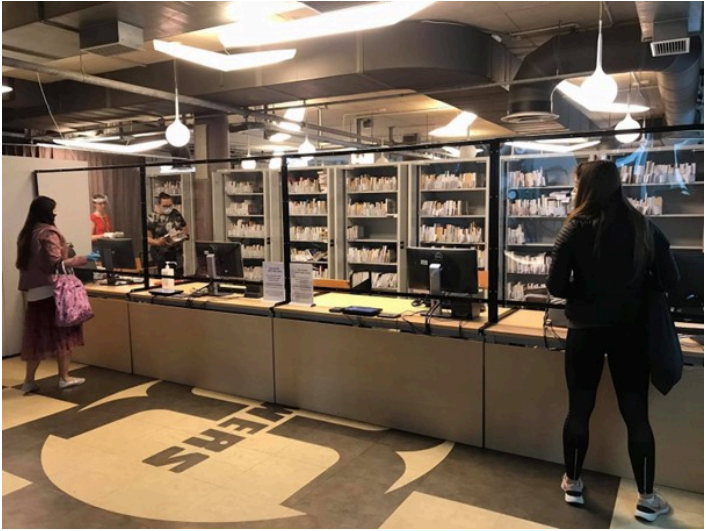

*Figure 3. Temporary circulation desk. A view from the patron's side*

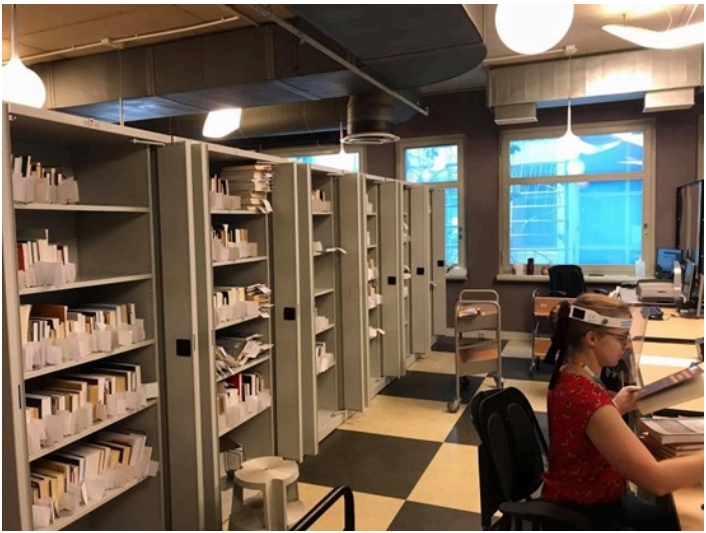

*Figure 4. Temporary circulation desk. A view from the staff's side*

While two librarians worked at the temporary circulation desk, others looked for and put on shelf the requested books located in the open stack area. When the requested books were found and prepared, the librarians delivered them to the temporary circulation desk. Due to numerous requests, BUW had to introduce a change in the delivery time. Usually, the time between the request and delivery of the book was one hour. Because patrons could not take the books from the open stack area by themselves and had to similarly request books from the closed stacks, we extended the delivery time to one day. Practically, it meant that a requested book would be ready for pick-up the following day.

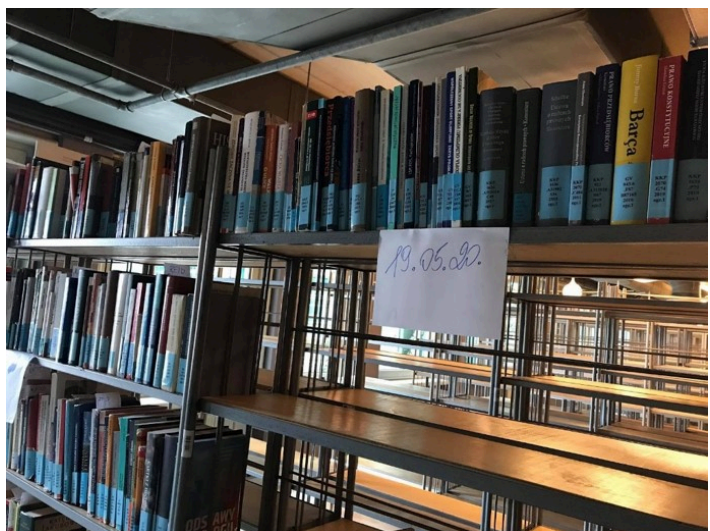

*Figure 5: Temporary circulation desk (Quarantine zone)*

## New borrowing limits

We changed the limits on borrowed books and other materials. On 10 March 2020, we harmonized and increased the limits on 20 items to 60 days for all patron groups. However, this was a temporary solution dictated by the specific conditions of the COVID-19 pandemic. When the temporary circulation desk was opened, BUW increased the pre-pandemic limits on borrowing books to 10 books for the UW's bachelor's and master's students and 20 books for the doctoral students, employees, and professors.

## Free of charge scanning service

To provide patrons with access to the library's resources which could not be borrowed, BUW offered the BW's students and employees a new service, namely scanning up to 50 pages weekly free of charge. We understood that it could not satisfy all the needs of our patrons. People needed a wide variety of content to work and study, especially during the global pandemic ([Anderson, 2020](#)). We did the best we could, but the 50-page limit was driven by the abilities of our Imaging Services staff, the high number of patrons, and the copyright regulations.

## Interlibrary Loan

Changes were introduced in the Interlibrary Loan Services. According to an internal decision taken by BUW authorities, we permitted patrons to borrow home books from the interlibrary loan, which were previously only accessible onsite. This decision was limited to the books from the libraries that allowed us to provide such a service. We also introduced another solution for the UW's students and employees who could not pick up books requested in Interlibrary Loan (ILL). They could request the partial digitisation of items received by ILL under the same conditions as BUW materials (50 pages weekly, free of charge). This solution was restricted to print books requested by BUW from the ILL services. Previously, our students and employees could only request articles within the Subito portal.

## Rotational work

On 29 June 2020, the circulation department introduced work on a rotating basis to prevent the spread of the COVID-19 virus among the library staff. A large number of infections could lead to the library shutting down. Hence, the librarians were organised into two groups. One group worked from the library, while the second one worked from home. After one week, the groups switched their work locations. In that case, if someone was infected, only one group would go into quarantine, and the second group could still work in the library to keep BUW's services open. Dividing the staff into two groups meant that only half of the employees were in place; therefore, shift work was impossible. We had to shorten the circulation desk opening hours to seven hours per day. The temporary circulation desk operated under those conditions for three and a half months, until 14 October 2020.

# Reopening of the Library

On 15 October 2020, after six months of work, the temporary circulation desk was removed and the BUW building was reopened to the public. Reopening resulted in new challenges and restrictions that had to be introduced to prevent the spread of the virus. We had to limit the number of people who could be in the library building at one time to 300. Previously, no such restrictions existed. Everyone entering the library had to check their body temperature and sanitise their hands. We also had to limit the number of people in the reopened main reading room to 10.

We relocated the registration desk outside the library to prevent new, unregistered patrons from entering the library. The presence of unregistered patrons could have exceeded the limit of people allowed in the library simultaneously. We also included an information desk alongside the new registration desk, where the newly registered patrons could seek help from the IT Department's librarian.

The Interlibrary Loan Services returned to the pre-pandemic loan policy. Thus, the books requested from other libraries could be read onsite only.

We cancelled the 'on quarantine' status of the returned books in the ILS. This decision was made in line with the BN's new recommendations (Biblioteka Narodowa, 2020c). Moreover, it would have been redundant to keep the returned books 'on quarantine' when the open stack area had been reopened, allowing people contact with books and periodicals located there. We hoped that the sanitary restrictions and the limited number of people would prevent the spreading of the virus. Due to the increasing number of sick and hospitalised staff, BUW returned to the rotational work system on 19 October 2020. The library was open from 2 pm to 9 pm on Monday and Tuesday and between 9 am and 4 pm from Wednesday to Friday. In the following days, the pandemic situation consistently worsened.

## Second lockdown in Poland

On 9 November 2020, due to a large number of COVID-19 cases and according to the decisions of the Ministry of Health (Rady Ministrów, 2020), the UW authorities announced a second lockdown and closed the library once again. However, this time, the decision to shut down the library was made rapidly and formalised in the evening, after the library's working hours. BUW could not apply solutions similar to those made for the first lockdown. We could only inform our patrons via social media channels and our website that from the following day onwards, the library would be closed until further notice. We repeated the good practices from the first lockdown, extending the validity of the library accounts and the period for returning the borrowed books. Once again, the library staff had to work remotely.

## Second reopening of BUW

On 29 November 2020, BUW reopened for the second time. As the situation was unstable, the library retained the rotational work system with one significant change. Previously, the library's opening hours were from 2 pm to 9 pm on Monday and Tuesday and between 9 am and 4 pm from Wednesday to Friday. This time, BUW decided to extend the opening hours by introducing the self-service mode from 9 am to 2 pm on Monday and Tuesday and between 4 pm and 9 pm from Wednesday to Friday. During the self-service mode hours, the patrons could enter the library building, read books, and borrow them from the open stack area in our self-check kiosks. This solution not only extended the time of access to the library collection but also minimised contact between the library staff and patrons. For example, if the patrons needed a book from the open stack area, they did not have to wait until the library staff's working hours. Instead, the patrons could independently borrow the required items from self-check kiosks without contacting anyone or waiting for the opening of the circulation desk. BUW worked under these conditions throughout the remainder of 2020 and beyond, until the third lockdown in April 2021.

In January 2021, we introduced another change, namely self-service mode hours on weekends. The patrons could access the library on Saturdays from 9 am to 4 pm and on Sundays from 3 pm to 8 pm.

## Third lockdown in Poland

On 15 March 2021, the pandemic situation forced the Ministry of Health to issue another lockdown. However, this time the library remained open. According to the decision of the rector of the UW ([Rektora UW, 2021](#)), BUW had to change the previous conditions of working during self-service hours to avoid contact between patrons and library staff. The rotational work system remained but was modified to fit the new requirements. Furthermore, the registration desk, main reading room, and Interlibrary Loan Services had to be closed. We introduced new procedures to allow our patrons to borrow books not only from the open stack area but also from closed stacks. From 7 am to 12 am, the circulation department staff had to bring in the books from the return box, check the books returned in the self-check kiosks, and pick from shelf requested books. The requested books were then placed on the bookshelves in a specially designated area near the circulation desk. At noon, the library building was opened to patrons. They could find the requested books on the bookshelves and borrow them from self-check kiosks. Moreover, they could also return books if needed. This solution allowed BUW to maintain the flow of materials without contact between library staff and patrons. We did our utmost to inform our patrons about the new restrictions, and two days before the introduction of the new working conditions, we published the relevant information on the library's social media channels and website.

## Third reopening of BUW

On 14 May 2021, BUW reopened for the third time under the same conditions as the second reopening. We retained the rotational work system with self-service mode hours during the week and on weekends. After 30 May 2021, BUW stopped working on a rotat-ing basis as it was no longer necessary. However, opening and self-service mode hours remained the same until October 2021. The free scanning service changed, and only UW employees could request 50 scans free of charge (including items from the ILL service).

## Comparative statistics

A comparison of the statistics presented in BUW's annual report for 2020 (Wołodko et al., 2021) and data from previous years indicates how the global COVID-19 pandemic impacted the library's work. According to this report, the use of electronic resources increased by 87%. The main impact was seen in the circulation of books, which decreased by 44%, and the number of onsite visits to the library, which decreased by 72%. In 2020, there were 8,751 new registrations, 5,486 of which were made online. In 2019, the library registered 13,046 patrons; of these, 5,275 patrons registered themselves online. The total number of registered BUW patrons dropped slightly by 10%, and the number of new registrations decreased by 33%. Moreover, the number of active borrowers decreased by 33%. BUW noted a sharp 55% increase in communication with patrons facilitated through the 'ask the librarian' chat service and a further increase of 224% through social media messages. There was also a significant increase in digitisation requests (166%) and in the general number of requested items (32%) and scans (34%). Most requests were issued by UW students and employees, of which 96% were successful. The COVID-19 pandemic also impacted the Interlibrary Loan Services. The number of requests from other libraries decreased by 43%. The rate of unsuccessful requests was 30% in 2020 compared to 25% in 2019. The number of items delivered via the Subito platform decreased by 51%, and the number of scans delivered in other ways decreased by 24% (Wołodko et al., 2021).

Those numbers show how the pandemic situation affected the library. We observed an increase in the usage of electronic resources and a significant decrease in the number of borrowed books and onsite visits. These findings are unsurprising given the pandemic conditions since, for a long time, only electronic channels were open for delivering materials and information. A comparison of the statistics from 2020 and 2021 (Wołodko at al., 2021), indicates a further increase in the usage of electronic resources (33%) and a further decrease in the number of active borrowers (9%). We can also observe an increase in onsite visits (38%), circulation of books (7%), and new registrations (22%). The total number of registered BUW patrons increased by 7% from 2020 to 2021. However, the use of the 'ask a librarian' feature and communication through social media channels dropped significantly by 40% and 46%, respectively.

The number of digitisation requests decreased by 22%, while the number of requested items and the general number of scans increased by 25% and 13%, respectively. ILL services statistics indicate that the number of requests from other libraries increased by 17% compared to 2020 but the number of unsuccessful requests was higher than in previous years (34%). The number of items delivered via the Subito platform increased by 25%, and the number of scans delivered through other means increased by 20%. Moreover, the number of requests for onsite scanning per the new solution introduced in 2020 was almost at the same level in 2020 and 2021 (Wołodko et al. 2022). These numbers indicate that the pandemic situation still affected the library in 2021 but had a lesser impact compared to 2020.

| | 2019 | 2020 | 2021 |
|---|---|---|---|
| Visits to the library | 719,809 | 201,682 (↓72%) | 277,573 (↑38%) |
| Borrowed books | 844,105 | 472,749 (↓44%) | 504,573 (↑7%) |
| New registrations | 13,046 | 8,751 (↓33%) | 10,700 (↑22%) |
| Total number of registered patrons | 112,826 | 102,091 (↓10%) | 109,515 (↑7%) |
| Number of active borrowers | 25,086 | 16,833 (↓33%) | 15,375 (↓9%) |
| Use of electronic resources | 2,003,679 | 3,746,964 (↑87%) | 4,975,796 (↑33%) |
| Communication transmitted through "Ask the librarian" | 352 | 1,772 (↑403%) | 1056 (↓40%) |
| Communication transmitted through social media channels | 123 | 399 (↑224%) | 216 (↓46%) |

*Table 1: Summary of comparative statistics*

| | 2019 | 2020 | 2021 |
|---|---|---|---|
| Total number of requests | 1,167 | 3,104 (↑166%) | 2,430 (↓22%) |
| Total number of requested items | 3,908 | 5,177 (↑32%) | 6,468 (↑25%) |
| Requests from BUW users (paid requests) | 349 | 467 | 374 |
| Requests from UW faculties and BUW staff | 818 | 595 | 418 |
| Free requests from UW staff and students | N/A | 2,042 | 1,638 |
| Free requests completed | N/A | 1,961 (96%) | 1,596 (↑97%) |
| Total number of scanned pages | 175,677 | 235,329 (↑34%) | 266,585 (↑13%) |

*Table 2: Digitization requests*

Tables 1 and 2 present a numerical summary of the comparative statistics discussed above. Values in brackets show the percentage decrease and increase compared to the previous year. Tables 3 and 4 present the numerical summary of comparative statistics from ILL services.

|  |  | 2019 | 2020 | 2021 |
|---|---|---|---|---|
| Requests from other libraries | General | 1174 | 669 (↓43%) | 780 (↑17%) |
|  | Polish libraries | 1080 | 585 | 712 |
|  | Foreign libraries | 94 | 84 | 68 |
| Requests not completed | General | 293 (25%) | 204 (30%) | 268 (34%) |
|  | Polish libraries | 257 | 140 | 231 |
|  | Foreign libraries | 58 | 64 | 37 |
| Requests completed | General | 881 (75%) | 465 (70%) | 512 (66%) |
|  | Polish libraries | 823 | 445 | 481 |
|  | Foreign libraries | 36 | 20 | 31 |

*Table 3: ILL requests*

|  |  | 2019 | 2020 | 2021 |
|---|---|---|---|---|
| Requests delivered via Subito | General | 57 | 28 (↓51%) | 35 (↑25%) |
|  | Polish libraries | 0 | 0 | 0 |
|  | Foreign libraries | 57 | 28 | 35 |
| Scans sent by libraries | General | 58 | 44 (↓24%) | 53 (↑20%) |
|  | Polish libraries | 31 | 36 | 32 |
|  | Foreign libraries | 27 | 8 | 21 |
| Requests for onsite scanning | Nr of requests | N/A | 35 | 32 |
|  | Nr of items | N/A | 45 | 34 |
|  | Nr of scans | N/A | 7,077 | 9,542 |

*Table 4: Digital ILL requests*

## Conclusions

The COVID-19 global pandemic caused major changes in the libraries' work. Lockdowns and health restrictions forced BUW to reorganise their work in unusual ways, such as by creating a temporary circulation desk or introducing rotational work. Despite those solutions, the statistics from 2020 showed a significant decrease in the traditional aspects of library work, such as borrowing books, registrations, and visits. It is apparent that the pandemic also disrupted the work of the ILL services, especially in the circulation of physical materials, which is unsurprising because the library was shut down for an extended period. The lockdown forced librarians to use new technologies. Moreover, VPN use became increasingly necessary to work from home during lockdowns and rotational work. The increased use of technology was particularly evident in communication. We noted a large volume of information transmitted through the 'ask the librarian' chat feature and an increase in the use of social media channels. The statistics also revealed an increase in the use of digital materials and electronic resources. A significant increase was noted in digitisation requests, which were also applied to the ILL services. Whether this trend will continue after the end of the pandemic remains to be seen.

# Bibliography

Anderson, J. (2020, July 17). *Research under pressure: impact of Covid-19 through a librarian's eyes.* LIBER. https://libereurope.eu/article/research-under-pressure-impact-of-covid-19-through-a-librarians-eyes/

Biblioteka Narodowa. (2020a, March 23). *Bezpieczeństwo Epidemiczne w Bibliotekach.* https://bn.org.pl/aktualnosci/3938-bezpieczenstwo-epidemiczne-w bib-liotekach.html

Biblioteka Narodowa. (2020b, April 16). *Rekomendacje Biblioteki Narodowej dla bibliotek po zniesieniu zakazu prowadzenia działalności bibliotecznej.* https://www.bn.org.pl/aktualnosci/3961-rekomendacje-biblioteki-narodowej-dla-bibliotek-po-zniesieniu-zakazu-prowadzenia-dzialalnosci-bibliotecznej.html

Biblioteka Narodowa. (2020c, August 8). *Rekomendacje Biblioteki Narodowej dotyczące funkcjonowania bibliotek podczas epidemii, stan na 24 października 2020 roku.* https://bn.org.pl/aktualnosci/4035-rekomendacje-bn-dotyczace-funkcjonowania-bib-liotek-podczas-epidemii%2C-stan-na-24-pazdziernika-2020-roku.html

Kamińska, A., Książczak-Gronowska, A., & Wiorogórska, Z. (2020). The use of information and communication technologies in academic libraries in a crisis situation. *Zagadnienia Informacji Naukowej*, 58(2A), 44–60. https://doi.org/10.36702/zin.704

Łukaszewska, K. (2020). Wiosna, która przejdzie do historii. *Pismo Uczelni UW*, 2, 6-7. https://uw.edu.pl/wp-content/uploads/2020/06/pismo_uw_2_95_2020.pdf

Ministra Zdrowia. (2020). *Rozporządzenie Ministra Zdrowia z dnia 20 marca 2020 r. w sprawie ogłoszenia na obszarze Rzeczypospolitej Polskiej stanu epidemii. Dz. U. Poz. 491.* https://dziennikustaw.gov.pl/D2020000049101.pdf

Rady Ministrów. (2020). *Rozporządzenie Rady Ministrów z dnia 6 listopada 2020 r. zmieniające rozporządzenie w sprawie ustanowienia określonych ograniczeń, nakazów* *i zakazów w związku z wystąpieniem stanu epidemii.* Dz. U. Poz. 1972. https://dziennikustaw.gov.pl/DU/rok/2020/pozycja/1972

Rektora Uniwersytetu Warszawskiego. (2021). *Zarządzenie nr 47 Rektora Uniwersytetu Warszawskiego z dnia 19 marca 2021 w sprawie zmiany zarządzenia nr 217 Rektora Uniwersytetu Warszawskiego z dnia 29 września 2020 r. w sprawie funkcjonowania Uniwersytetu Warszawskiego w okresie stanu epidemii ogłoszonego w związku z zakażeniami wirusem SARS-CoV-2.* Poz. 64. https://monitor.uw.edu.pl/Lists/Uchway/Attachments/5803/M.2021.64.Zarz.47.pdf

Wołodko, A. (2020). Academic libraries in unusual situations. *Przegląd Biblioteczny*, 88, 57–69. https://doi.org/10.36702/pb.772

Wołodko, A., Rowińska, M., & Ślaska, K. (2021). *Biblioteka Uniwersytecka w Warszawie oraz biblioteki wydziałowe Uniwersytetu Warszawskiego w roku 2020.* Biblioteka Uniwersytecka w Warszawie. https://www.buw.uw.edu.pl/wp-content/uploads/2021/07/BUW_sprawozdanie_2020_na-www.pdf

Wołodko, A., Rowińska, M., & Ślaska, K. (2022). *Biblioteka Uniwersytecka w Warszawie oraz biblioteki wydziałowe Uniwersytetu Warszawskiego w roku 2021.* Biblioteka Uniwersytecka w Warszawie. https://www.buw.uw.edu.pl/wp-content/uploads/2022/05/Sprawozdanie-BUW-za-rok-2021.pdf

## About the Author

Franciszek Skalski, Junior Librarian

Biblioteka Uniwersytetu Warszawskiego (University of Warsaw Library)

https://orcid.org/0009-0008-5168-0647

Franciszek Skalski (1995) has been employed at the University of Warsaw Library since 2020. Between 2013-2018, he studied archaeology at the University of Warsaw.

# How COVID has changed our world. New opportunities for the future? A case study for UPM Libraries

**Reyes Albo**

## Abstract

Delivery services, loan policies, formats, workflows, etc., all of them have changed a lot due to the pandemic situation. The University has adapted itself and has adapted the new situation to students, teachers and researchers. A new way to understand library life has come to stay and old procedures have started to disappear. Only time will tell which of all these measures will stay for the future, but a new view and shape of libraries is here to stay.

## Keywords

## Article

## Introduction. Facts and figures at UPM

In 2022, the population in Spain is 47 million, of which more than six million people live in the Madrid region and we have 1.6 million students in 75 universities (50 public and the rest, privately owned). Fifteen of these universities are in the Madrid region. Universidad Politécnica de Madrid (UPM) is the only exclusively technical university in Madrid and the largest and oldest technical university in Spain.

UPM was founded in 1971 (but the earliest School has its origin in 1777), the University is an umbrella for 16 Schools and one Faculty, with a budget in 2021 of 409.3 (€million) and public funding 263.3 (€million).

Human resources include 2,900 professors, 1,700 University staff, of which 300 are library staff members, and the number of students in 2022 is of 39,500 (29,200 undergraduates + 10,300 postgraduates), making a total of 44,100 library users.

The University academic offer covers a wide catalogue of course formats, such as accredited bachelor and master degree programs, accredited PhD programs, double degree program agreements with international universities, international exchanges by over 1,800 UPM students every year, cooperation for development groups, and a complete set of sports clubs and associations.

In the field of research, which is one of the main strengths of the University, there are 202 research groups, five research institutes, research centres, with funding of over 40 million euros/year from regional, national and international projects. Developing close partnerships with industry, enterprises, leading companies and promoting start-ups are the core goals of the University. UPM is a partner of the EIT Digital, Health, Raw Materials and Climate Knowledge and Innovation communities.

UPM places the Sustainable Development Goals, SDGs for the UN 2030 Agenda (2015), at the heart of its key fields of action, mainly in research and in academic works.

## Short presentation of UPM Libraries and environment

The University library has 16 library branches with 300 librarians, there is no central library, and the ratio is 8.4 students per seat (2021), with a total of 4,700 seats.

The collection has 1,000,000 printed books, 200,000 e-books, 200 databases, 67,000 subscribed e-journals, and a total of 800 laptops available for loan with a budget of €2,500,000.

Regarding transaction data, borrowing transactions are over 52,000/year. If we have a closer look at figures over the last four years, we can see that, in the case of interlibrary loans and e-resources, there is a decrease in numbers for these two services. On the other hand, if we look at electronic resources, such as those relating to downloads and e-journal subscriptions, there is a significant increase in figures, specially relating to 2020 numbers.

## Trends

So, we can say there is a strong relationship between interlibrary loan requests (Fig. 1) and electronic resources of any kind: downloads from database contents, general search (including catalogue) and e-journal subscriptions (Figs. 2 and 3).

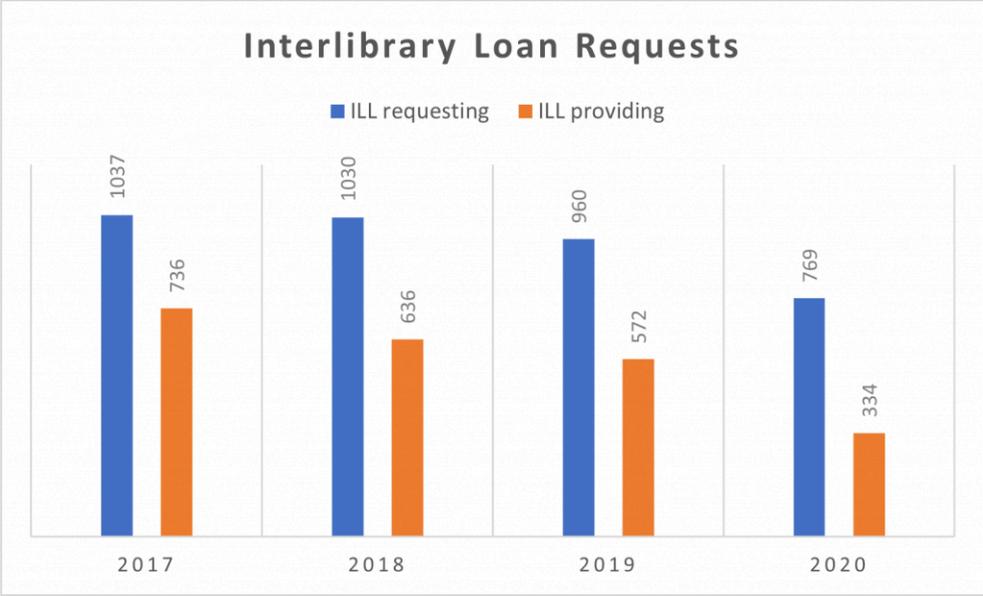

*Figure 1: Trends in interlibrary loan at UPM library from 2017–2020. Decreasing data*

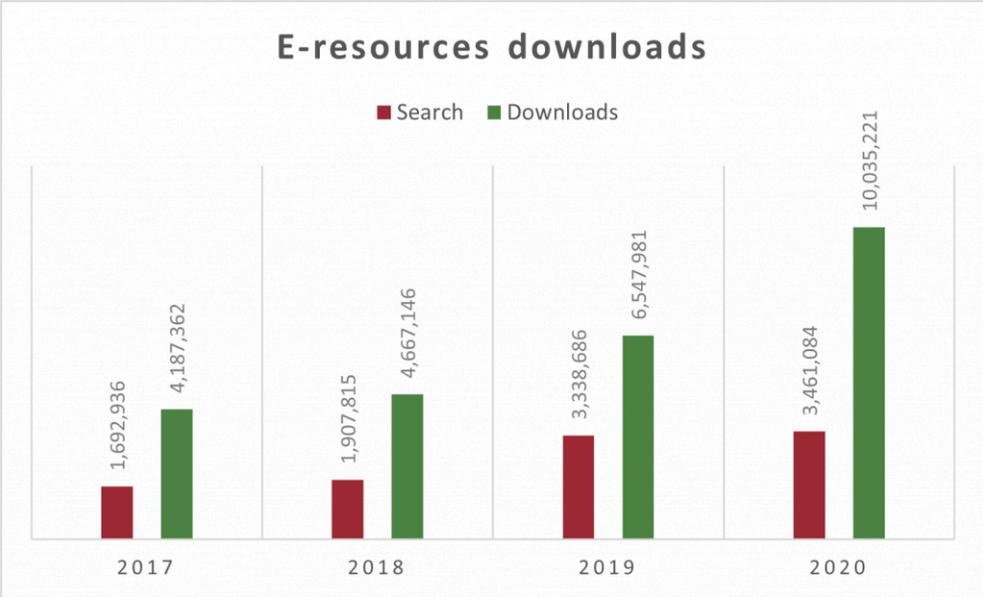

*Figure 2: Trends in e-resources downloads at UPM library from 2017–2020. Increasing data*

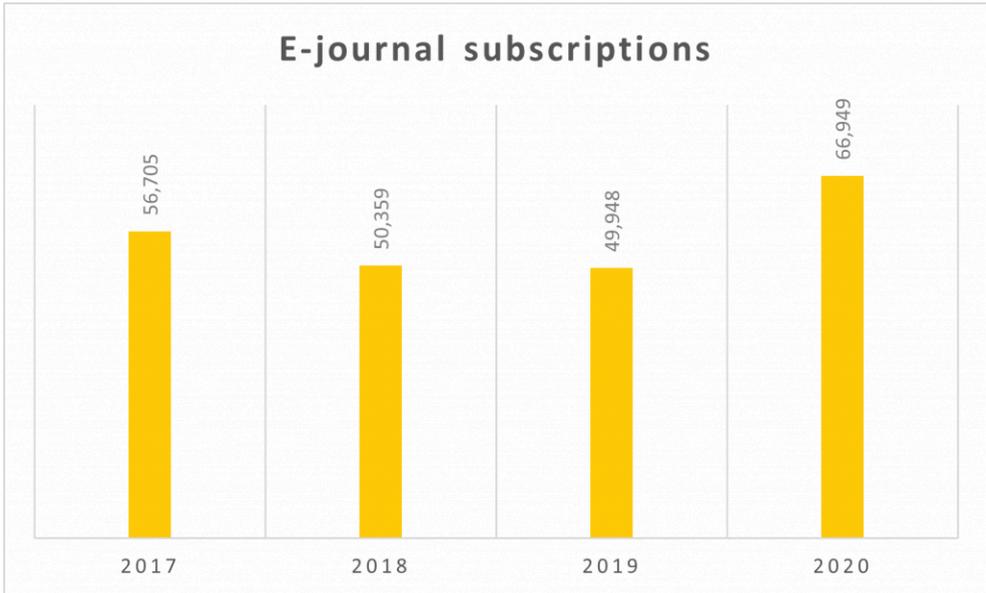

*Figure 3: Trends in e-journals subscriptions at UPM library from 2017–2020*

If we look at the latest figures (2021 and 2022), still not officially fully processed today, we can also see that this trend is here to stay among users and that there is a difference between users before COVID-19 and users after COVID-19 (Fig. 4). Also, librarians have changed because society has changed.

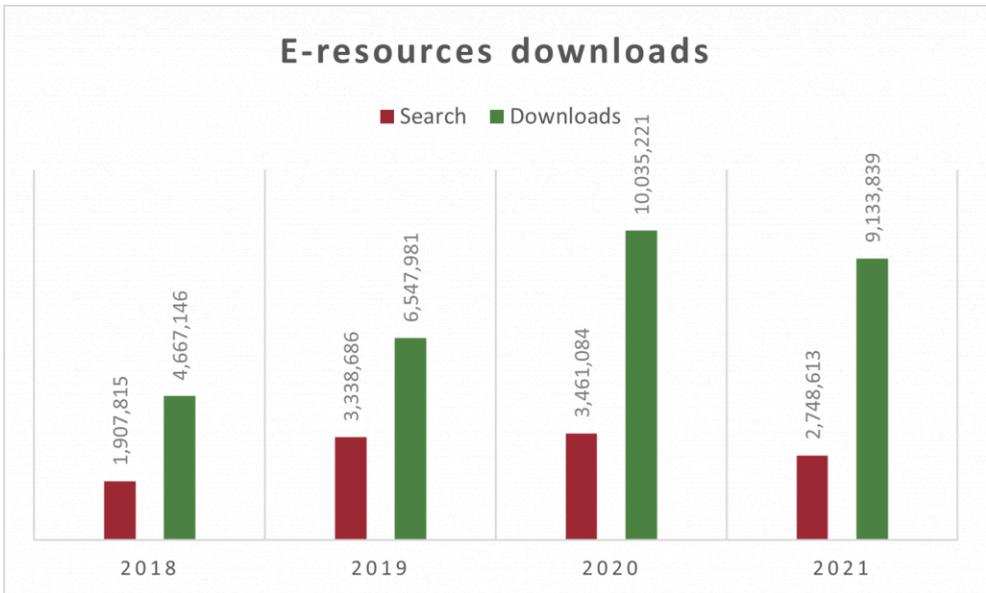

*Figure 4: Trends in downloads and search from 2018 to 2021. Electronic resources have increased with the pandemic and seems to have stabilized in the last year*

# How COVID-19 has changed our daily life and its impact on our libraries

Digitization in organizations has changed our lives (Fig. 5), we live in a changing world where the speed of change in daily life is almost imperceptible and yet we deal every day with new scenarios in different environments, and our libraries are part of this changing world.

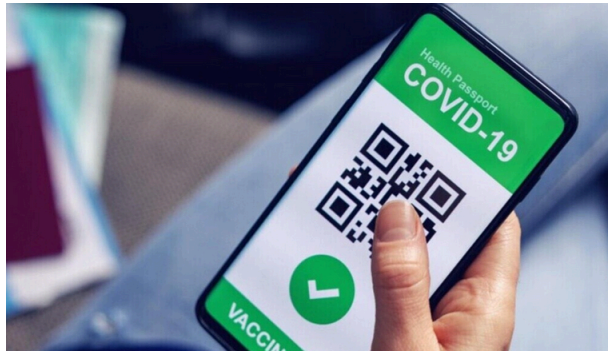

*Figure 5: A digital scenario has come to stay*

Examples of this include the way people look for new jobs, using online platforms and social media apps (Infojobs, Job finder, Online job, LinkedIn), and extend to include other aspects of life such as:

- new ways of communication (social media, TikTok, Facebook, YouTube, WhatsApp, Instagram, Twitter, etc.);
- new ways of working (home office) (Zoom, Teams, Skype, video apps, etc.);
- new ways of buying (eBay, Amazon, online shopping);
- new ways of travelling (online checking, applications to design your own trip, flight comparators, virtual trips, street views);
- new ways of medicine (E-health, online appointment, online doctor, 3D devices for medicine, distance surgery, etc.);
- new ways of searching (in a connected world, data science, analytics, the world in our hand with smart devices that rule our close environment);
- new ways of learning (using MOOCs, Moodle and online platforms);
- new ways of enjoying oneself (realistic videogames, tv platforms like Netflix, or even the Metaverse which can have wider and wider uses).

So, the truth is that before COVID-19 we were already a long way down the path towards integrating these changes and the foundations for the great change that came with COVID were already laid, even if we didn't realized it.

The COVID-19 pandemic took place at the same time everywhere in the world and all of us had to face the same problems simultaneously, we all had to rethink our present and our future at the same time, and this made us work in an even more synchronous way.

Libraries did their best, starting with simple needs like using NO PAPER (Figs. 6 and 7). With this kind of decision, and the success of this change, a kind of collective concern related to sustainability was foremost in our minds.

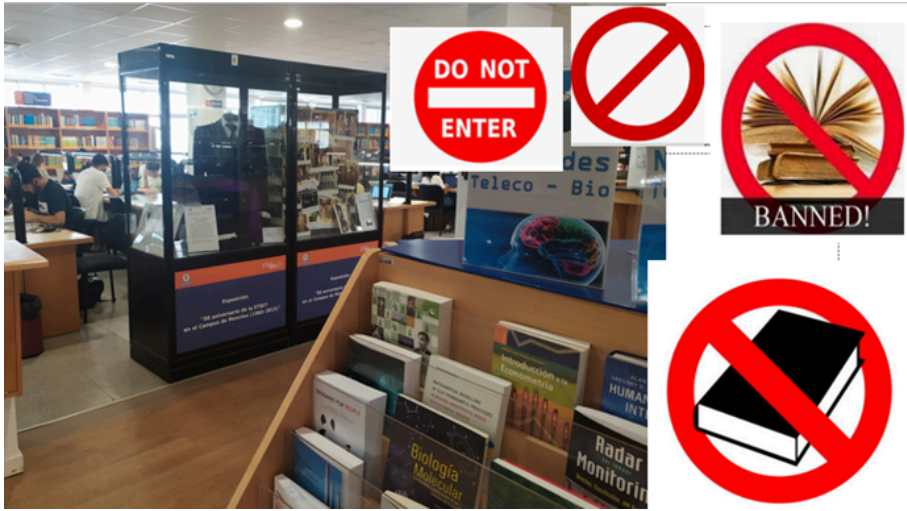

*Figure 6: We had to make up a new reality and new solutions*

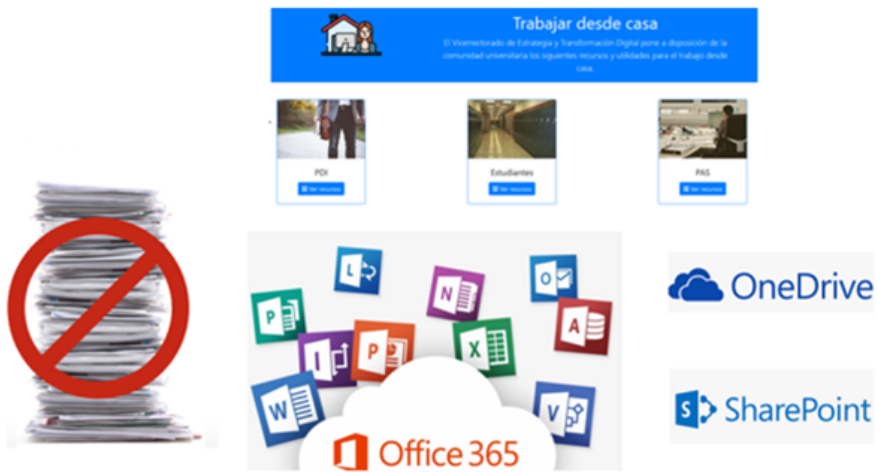

*Figure 7: Starting from simple needs, we had to get used to work in a virtual environment*

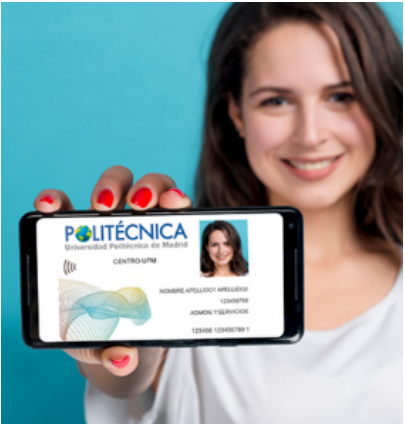 New facilities like the new digital UPM userID (mobile)

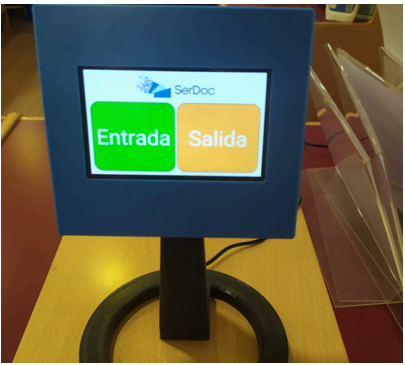 New devices to register the library capacity in real time

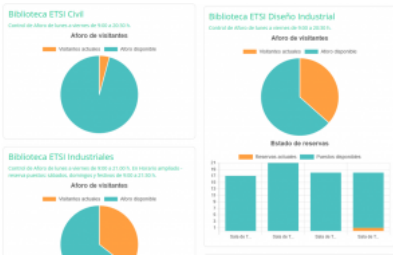 New online statistics to share use of the library

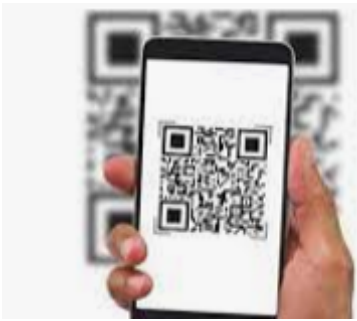 New reservations online for the use of group study rooms, for example

*Table 1: Impact of COVID-19 on library services*

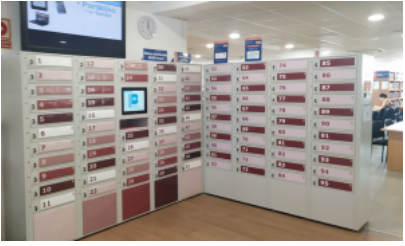

Self borrowing machines for laptops

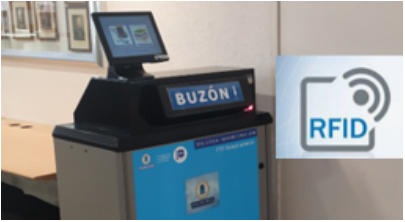

New drop boxes with RFID systems 24 hours

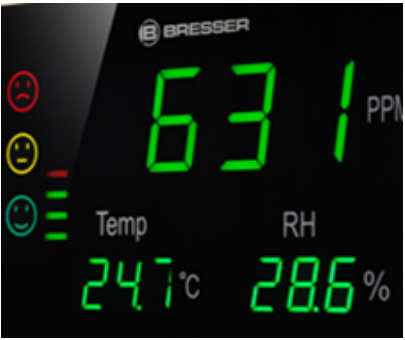

New devices to measure C02, Humidity and Temperature at the reading room

In our case, we got some decisive changes that will stay for the future (Table 1):

- new digital UPM userID (mobile), a long sought-after goal for us and long requested by our authorities. At last, it took actually very little time to create this useful environment for the students;
- new accesses for the library with new security gates;
- new devices like the check-in and check-out machine, where you can register yourself and where we register the library capacity in real time and can be seen online on the website. Same device with different uses: capacity, reservations, check in and check out;
- new devices that can read QR codes, barcodes and other information like logs (in-out);
- new use for laptop lending (long term lending) and home working;
- new self-lending machines and also a 24 hour dropbox with RFID systems;
- new devices to measure CO2 concentration, humidity and temperature.

Sharing resources has become a must for library services, reinforcing links between consortia and services and all sorts of collaborative and sharing resources in university library networks.

## Conclusions

So, the question could be: What should the library of the future be? To get towards the answer, we might remember the following saying, wrongly attributed to Charles Darwin: "It is not the most intellectual of the species that survives; it is not the strongest that survives; but the species that survives is the one that is able best to adapt and adjust to the changing environment in which it finds itself."

And the answer to the previous question is that the library of the future will be a space for the user in constant movement and adapting its reality to new environments and a place where librarians and users move at the same speed.

## Bibliography

Consorcio Madroño. (n.d.). *Acerca de*. Retrieved March 15, 2022, from http://www.consorciomadrono.es/acerca-de/

Instituto Nacional de Estadística de España. (n.d.). *Population figures. Latest data*. Retrieved March 15, 2022, from https://www.ine.es/dyngs/INEbase/en/operacion.htm?c=Estadistica_C&cid=1254736176951&menu=ultiDatos&idp=1254735572981

Red de Bibliotecas Universitarias Española. (n.d.). ¿*Quiénes Somos?* Retrieved March 15, 2022, from https://www.rebiun.org/quienes-somos/rebiun

Universidad Politécnica de Madrid. (2022). UPM, *facts and figures*. https://www.upm.es/sfs/Rectorado/Gabinete%20del%20Rector/UPM%20en%20Cifras/brochure2022english.pdf

## About the Author

Reyes Albo, Library Director
Universidad Politécnica de Madrid (Technical University of Madrid)
https://orcid.org/0000-0002-4772-7405

Reyes Albo (1965) is library director at the School of Telecommunications Engineering of the Polytechnic University of Madrid and coordinator of the University libraries that belong to the TIC Area. She has experience in budget management, human resources and coordination of working groups at University level, library infrastructure management, collection management, electronic resources, and experience in digital content management for digital environment. She is an expert in advanced training in information resources for librarians, academics and students on topics such as bibliographic reference management, information literacy, plagiarism and teaching students how to create academic works for their curricula.

# Book Contributors

## Reyes Albo, Library Director

Universidad Politécnica de Madrid (Technical University of Madrid)

https://orcid.org/0000-0002-4772-7405

Reyes Albo (1965) is library director at the School of Telecommunications Engineering of the Polytechnic University of Madrid and coordinator of the University libraries that belong to the TIC Area. She has experience in budget management, human resources and coordination of working groups at University level, library infrastructure management, collection management, electronic resources, and experience in digital content management for digital environment. She is an expert in advanced training in information resources for librarians, academics and students on topics such as bibliographic reference management, information literacy, plagiarism and teaching students how to create academic works for their curricula.

## Joris Baeyens, Acquisitions and ILL Librarian

Universiteit Gent, Faculteitsbibliotheek Letteren en Wijsbegeerte (Ghent University, Faculty Library of Arts & Philosophy)

https://orcid.org/0000-0001-5612-561X

Joris Baeyens (1969) obtained an associate degree in Information Management: Library and Archives. He started working in 2012 at Faculty Library of Arts & Philosophy as acquisitions librarian and joined the library's ILL team in 2018.

## Tina Baich, 2021-2023 Chair of IFLA Document Delivery & Resource Sharing Section Standing Committee

Indiana University–Purdue University Indianapolis

https://orcid.org/0000-0002-8046-2461

Tina Baich is the Senior Associate Dean for Scholarly Communication & Content Strategies at IUPUI University Library, Indianapolis, Indiana, USA. Before becoming a library administrator, she led the library's resource sharing efforts for 12 years. Tina is currently SPARC's Visiting Program Officer for the U.S. Repository Network and Chair of IFLA's Document Delivery & Resource Sharing Section Standing Committee.

### Léa Bouillet, Head of Public Services Department
Université Toulouse Jean Jaurès (University Toulouse Jean Jaurès)
https://orcid.org/0000-0002-9927-070X

Léa Bouillet (1989) has a master's degree in book history and passed the competitive examination to become a library curator in 2011. After one and a half years of training at the National School of Library and Information Technology (ENSSIB) in Lyon (France), she began her career in 2013 as head of public services at the University Health Library of the Paul Sabatier University in Toulouse (France). After 4 years, she was appointed director of public services in the libraries of the University Jean Jaurès in Toulouse. She is primarily responsible for the public services policy, projects related to handicaps, training in information skills and ILL. In parallel, she provides training in preparation for library examinations with the Occitanie library careers training centre (Mediad'Oc).

### Karin Byström, Librarian
Uppsala universitetsbibliotek (Uppsala University Library)
https://orcid.org/0000-0002-1912-106X

Karin Byström (1973) is a librarian at Uppsala University Library in Sweden where she works as a coordinator at the Media division. She has extensive experience of working with many parts of collection development – cataloguing and acquisition, e-resources and open access, digitization and shared print. Karin has an interest in library collaboration and she has been involved in many Swedish library collaboration projects. She is currently the project manager of the collaboration for Digitization of Swedish print and part of the working group for the Swedish shared print initiative. Since 2019 Karin has been a member of the standing committee of the IFLA Acquisition and Collection Development section, where she currently holds the position as secretary.

### Camille Carette, Library Systems Manager
École nationale des chartes (The École nationale des chartes)
https://orcid.org/0000-0001-7123-1584

After studying history at the Panthéon-Sorbonne University, Camille Carette (1994) trained in the use of digital technologies in heritage and research. Since 2019, she has worked at the library of the École nationale des chartes, first exclusively on the digitization of the theses of archivists-paleographers, then as library systems manager from 2020. Among other things, she is in charge of the ILS, the discovery tool and the digital library. In 2023, she joined the organizers of AUFO, the association of French-speaking users of Omeka.

### Marion Favre, E-resources Librarian

École Polytechnique Fédérale de Lausanne (Swiss Federal Institute of Technology in Lausanne)

https://orcid.org/0000-0001-9756-8249

Marion Favre (1990) is an e-resources librarian, working mainly with ebooks. She oversees Alma for her team and also manages the library's video game collection. Marion also likes to dig in COUNTER reports and to create analyses with Analytics.

### Karolina Forma, Librarian

Biblioteka Główna Akademii Górniczo-Hutniczej (AGH University of Science and Technology Main Library)

https://orcid.org/0000-0003-1876-1951

Karolina Forma (1990) started her work in AGH University of Science and Technology Main Library in 2018, joining the Branch Libraries and Inventory Control Division. She takes an active part in ongoing financial and quantitative inventories of various collections. In 2019, she became a co-manager of the library's social media and took a creative role in the library's marketing team. Since 2021, she has been expanding her ILL knowledge by helping in the Interlibrary Loans Office. Her work interests include 20th century illustrations, history of advertisement, and social media.

### Veronica Fors, Librarian

Örebro universitetsbibliotek (Örebro University Library)

Veronica Fors (1976) has a master's degree in library and information science and is currently librarian at Örebro University Library at the unit Library Acquisitions and Scholarly Publishing. She is responsible for the library's interlibrary loans. She also works with the library's publication database DiVA.

### Elisabetta Guerra, User Services Manager

Università di Verona (University of Verona)

https://orcid.org/0000-0002-8880-3646

Elisabetta Guerra (1972) graduated in Foreign Languages and Literatures at Verona University and has been librarian at Verona University Libraries since 2006. Since the beginning of her activity, Elisabetta has been working in the interlibrary loan and document delivery services. She is also engaged in information literacy programs for library users. Since 2022, she has been the User Services Manager for the Libraries of the Medicine,

Science and Technology Campus. She works in the central library "Egidio Meneghetti", where she manages the loan and consultation services.

### Martina Kassler, Interlibrary Loan Librarian

Karlstads universitetsbiblioteket (Karlstad University Library)
https://orcid.org/0000-0002-1557-1397

Martina Kassler is a current interlibrary loan librarian at Karlstad University Library. After completing a bachelor in economics and a master in ethnology and sociology and another one in library and information science, she took up a position as an interlibrary loan librarian. She is also interested in the topic of interculture.

### Susanne Kirchmair, Head of Library Services

MCI | Die Unternehmerische Hochschule (MCI | The Entrepreneurial School)
https://orcid.org/0000-0002-7339-5192

Susanne Kirchmair (1975) has an academic background in International Business & Economics from the University of Innsbruck (A), Aarhus Business School (DK), and University of Groningen (NL), as well as Library & Information Studies from the University of Vienna (A). She started her career as a librarian at the University and State Library of Tyrol. She worked in circulation, acquisition, and training in the Social and Economic Sciences faculty library. Since 2011, she has been responsible for Library Services at MCI, implementing resources, tools, and services to enhance the user experience and provide high-quality content for studying and research. She also teaches in various study programs and continuing staff education to support the academic success of students and faculty.

### Carmen Lomba, Interlibrary Supply Manager

Universidad de Cantabria (University of Cantabria)
https://orcid.org/0000-0003-2479-3669

Carmen Lomba (1964) has a degree in History of Art (University of Valladolid). She has been working as librarian at the Library of Cantabria University since 1989 and has been responsible for the Interlibrary Loan-Document Delivery Service (ILL-DD) since 1995. She is currently part of the Working Group on ILL-DD for REBIUN (Network of Spanish University Libraries) and has been a member of the NILDE Internationalization Working Group since 2019. She participates as a volunteer in RSCVD initiatives and works in the European project "HERMES –Strengthening digital resource sharing during COVID and beyond" ERASMUS PLUS PROGRAMME, led by the Italian CNR (*Consiglio Nazionale delle*

*Ricerche*) 2021-2023. She is a member of the IFLA Document Delivery and Resource Sharing (DDRS) Standing Committee (2021-2025).

## Luca Maggiore, Librarian

Universiteit Antwerpen (University of Antwerp)

https://orcid.org/0000-0001-7827-080X

After high school in Velletri, Italy, Luca Maggiore (1968) graduated in Modern History at KU Leuven (Belgium) and later he obtained a postgraduate degree in Information and library science at the University of Antwerp. He currently works at the University of Antwerp as head of the Acquisition and Metadata Department (Library). He has published in the fields of library science and the history of Italy.

## Stefania Marzocchi, Research Fellow

Consiglio Nazionale delle Ricerche (National Research Council)

https://orcid.org/0000-0001-8777-9720

Stefania Marzocchi (1969) is a librarian and project manager. Her collaboration with the Italian CNR (*Consiglio Nazionale delle Ricerche*) started in 2018 and is related to the implementation of CARONTE, BRAIN@WORK, and HERMES project activities, all funded by the European Commission. She is experienced in the design, management, and evaluation of international technical assistance projects with specific reference to the field of methodological approaches to research activity.

## Debora Mazza, Research Fellow

Biblioteca Dario Nobili - Consiglio Nazionale delle Ricerche (Dario Nobili Library - National Research Council)

https://orcid.org/0000-0001-7842-2922

Debora Mazza (1996) graduated in History and Oriental Studies from the University of Bologna. She has been collaborating with Dario Nobili CNR Library in Bologna since 2019. She is in charge of the library's communication activities (graphic design, social media, website management) and she is involved in front-office, reference services, and document delivery with NILDE. She also develops the TALARIA software as part of the European HERMES Project.

### Isàvena Opisso, Librarian

Universitat Autònoma de Barcelona | Universidad Autónoma de Barcelona (Autonomous University of Barcelona)

https://orcid.org/0000-0003-3337-014X

Isàvena Opisso (1973) became university librarian at the Universitat Autònoma de Barcelona in 2019. Prior to this position, she worked as a university librarian at the Universitat de Barcelona (2009-2018) and public library librarian in the public libraries of Olot and La Garriga (2008-2009). Her current work focuses on interlibrary loan and article processing charges.

### Florian Preiß, Librarian

Deutsches Museum (The Deutsches Museum)

https://orcid.org/0009-0007-0649-192X

Florian Preiß studied medieval and postmedieval archaeology in Bamberg and library sciences in Munich. Since 2016, he has been librarian and head of the reading room department at Deutsches Museum library in Munich. His main responsibilities include the interlibrary loan and document delivery activities of the Museum library.

### Fabienne Prosmans, Fulfilment and ILL Manager

Université de Liège (University of Liège)

https://orcid.org/0000-0002-1408-5207

Fabienne Prosmans (1972) studied mathematics at the University of Liège and obtained a PhD at the University of Paris Nord. Since 2005, she has been working at ULiège Library as a subject librarian in the fields of mathematics and applied sciences. She has been the fulfilment and ILL manager at ULiège Library since 2015 and is now coordinating the "User Services" unit.

### François Renaville, Head of Library Systems

Université de Liège (University of Liège)

https://orcid.org/0000-0003-1453-1040

François Renaville (1976) studied linguistics, literature, and translation. After a two-year experience as a teacher in Finland, he joined the University of Liège Library as a subject librarian. He has been working on library systems since 2005. Since 2022, he has been member of the IGeLU (International Group of Ex Libris Users) Steering Committee and has been coordinating the "Systems & Data" unit at ULiège Library. He is interested in discoverability, integrations, delivery and user services. On a private level, he is a great coffee, chocolate and penguin lover.

### Franciszek Skalski, Junior Librarian

Biblioteka Uniwersytetu Warszawskiego (University of Warsaw Library)

https://orcid.org/0009-0008-5168-0647

Franciszek Skalski (1995) has been employed at the University of Warsaw Library since 2020. Between 2013-2018, he studied archaeology at the University of Warsaw.

### Jennifer Van den Avijle, Head of User Services

Universiteit Antwerpen (University of Antwerp)

https://orcid.org/0000-0001-8545-9168

Jennifer Van den Avijle (1987) is an archaeologist and librarian by training. After three years in the public library sector, Jennifer has been working as head of user services for the Biomedical Library and Library of Exact and Applied Sciences at the University of Antwerp since 2014. She manages the user services team, optimises loan and ILL services, oversees rebuilding and refurbishment projects, looks for new ways to expand our services and functions, all while trying to dissuade the team from adopting a library cat.

### Manon Velasco, Librarian

École Polytechnique Fédérale de Lausanne (Swiss Federal Institute of Technology in Lausanne)

https://orcid.org/0000-0002-6775-7433

Manon Velasco is a multipurpose librarian at EPFL where she is currently in charge of e-book collections. She is also involved in the EPFL's institutional repository and she is a member of the copyright team. Her professional interests include revolutionizing the relationship with publishers, negotiating the best possible agreements with e-book providers, supporting researchers in opening their research and playful learning applied to copyright literacy education.

# Acknowledgements

The editors wish to thank sincerely all librarians and researchers who participated to the 2022 edition of the Erasmus Staff Training Week and shared their experiences and visions in relation to the theme of the Week: Reyes Albo (Universidad Politécnica de Madrid), Joris Baeyens (Ghent University), Léa Bouillet (Université Toulouse Jean Jaurès), Karin Byström (Uppsala University), Camille Carette (École nationale des chartes), Marion Favre (EPFL), Karolina Forma (AGH University of Science and Technology), Veronica Fors (Örebro University), Elisabetta Guerra (Verona University), Martina Kassler (Karlstad University), Susanne Kirchmair (MCI Management Center Innsbruck), Carmen Lomba (Universidad de Cantabria), Debora Mazza (Consiglio Nazionale delle Ricerche), Isàvena Opisso (Universitat Autònoma de Barcelona), Florian Preiß (Deutsches Museum), Franciszek Skalski (University of Warsaw), Frédéric Spagnoli (Université Bourgogne-Franche-Comté), Jennifer Van den Avijle (University of Antwerp), and Ipek Yarar (MEF University).

We express our great appreciation to Paul Thirion, Chief Librarian of the University of Liège, for the unwavering support he has shown to his teams for 20 years.

We also convey our appreciation to all colleagues from ULiège Library who, during the Erasmus Week, kindly accepted to devote their time to the animation and the supervision of workshops and visits of library branches, the heritage collections and the digitization service: Marjorie Bardiau, Dominique Chalono, Cécile Dohogne, Christophe Dony, Marie Goukens, Kevin Jacquet, Marie Jost, Rémy Lhoest, Cécile Nissen, Cécile Oger, Stéphanie Simon, Paul Thirion, Mathieu Uyttebrouck, and Jean-Charles Winkler.

We express our deepest thank to the team of the International Relations Office of the University of Liège, especially to Julie Hollenfeltz and Dominique d'Arripe, for their administrative and financial support. Without their precious help, it would not have been possible to organise this eighth edition of the Erasmus Week.

For their valuable logistic and technical support during the Week, we are very grateful to Aurore Graeven (ULiège Library), Danielle Bartholoméus, Vinciane Godfrind, and Prof. Jean-Pierre Schneiders (ULiège Mathematics Department).

Finally, the editors wish to thank Bernard Pochet for his advice on the use of Pressbooks to get this book into shape and to Sarah Trautes for the design of the book cover.

No library can buy or hold everything its patrons need. At a certain point, librarians need to pool their resources and collaborate to provide access to what they don't have: Collaboration and partnership, centralized and shared collection storage, digitization projects, interlibrary loan and resource sharing, purchase on demand, PDA and EBA are notably key to success.

The 2022 edition of the Erasmus Mobility Staff Training week organized at the University of Liège Library focused on services, projects and policies that libraries can deploy and promote to increase and ease access to materials that do not belong to their print or electronic holdings.

More than 20 librarians, managers, and researchers in library science share their experiences and visions in this book that is available in open access: https://e-publish.uliege.be/beyond-the-library-collections

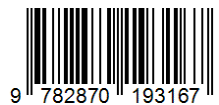